\documentstyle[rmp,harvard,aps,epsf,floats]{revtex}
\newcommand{\be}{\begin{eqnarray}}
\newcommand{\ee}{\end{eqnarray}}

\begin{document}


\title{Instantons in QCD}
\author{T.~Sch\"afer}

\address{Institute for Nuclear Theory, Department of Physics,
         University of Washington,\\ Seattle, WA 98195, USA}

\author{E.V.~Shuryak}

\address{Department of Physics, State University of New York at Stony
         Brook,\\ Stony Brook, New York 11794, USA}

\date{\today}
\maketitle

\begin{abstract}
  We review the theory and phenomenology of instantons in QCD. After a
general overview, we provide a pedagogical introduction to semi-classical
methods in quantum mechanics and field theory. The main part of the 
review summarizes our understanding of the instanton liquid in QCD
and the role of instantons in generating the spectrum of light hadrons.
We also discuss properties of instantons at finite temperature and
how instantons can provide a mechanism for the chiral phase transition.
We give an overview over the role of instantons in some other models,
in particular low dimensional sigma models, electroweak theory and
supersymmetric QCD.

\end{abstract}
\newpage

\tableofcontents

\newpage

\renewcommand{\thefootnote}{\arabic{footnote}}
\setcounter{footnote}{0}


\section{Introduction}
\label{sec_intro} 
\subsection{Motivation}
\label{sec_mot}
 
   QCD, the field theory describing the strong interaction, is now more 
than 20 years old, and naturally it has reached a certain level of maturity. 
Perturbative QCD has been developed in great detail, with most hard processes 
calculated beyond leading order, compared to data and compiled in reviews and 
textbooks. However, the world around us can not be understood on the basis of
perturbative QCD, and the development of non-perturbative QCD has proven to 
be a much more difficult task. 

   This is hardly surprising. While perturbative QCD could build on the 
methods developed in the context of QED, strategies for dealing with
the non-perturbative aspects of field theories first had to be developed.
The gap between hadronic phenomenology on the one side and exactly
solvable model field theories is still huge. While some fascinating
discoveries (instantons among them) have been made, and important 
insights emerged from lattice simulations and hadronic phenomenology,
a lot of work remains to be done in order to unite these approaches
and truly understand the phenomena involved. 

   Among the problems the field is faced with is a difficulty in
communication between researchers working on different aspects 
of non-perturbative field theory, and a shortage of good introductory
material for people interested in entering the field. In this
review, we try to provide an overview over the role of instantons
in field theory, with a particular emphasis on QCD. Such a review
is certainly long overdue. Many readers certainly remember learning
about instantons from some of the excellent papers \cite{CDG_78}
or introductory reviews \cite{Col_77,VZN_82} that appeared in the
late 70's or early 80's, but since then there have been very few
publications addressed at a more general audience\footnote{A reprint 
volume that contains most of the early work and a number of important 
technical papers was recently published by \cite{Shi_94}.}. The 
only exceptions are the book by \cite{Raj_82}, which provides a 
general discussion of topological objects, but without particular 
emphasis on applications and a few chapters in the book by 
\cite{Shu_88b}, which deal with the role of instantons in hadronic 
physics. All of these publications are at least a decade old. 
   
   Writing this review we had several specific goals in mind, which
are represented by different layers of presentation. In the remainder 
of the introduction, we provide a very general introduction into 
instanton physics, aimed at a very broad readership. We will try to 
give qualitative answers to questions like: What are instantons? What 
are their effects? Why are instantons important? What is the main 
progress achieved during the last decade? This discussion is 
continued in Sec. \ref{sec_pheno}, in which we review the current
information concerning the phenomenology of instantons in QCD.

  Section \ref{sec_semicl} is also pedagogical, but the style is 
completely different. The main focus is a systematic development
of the semi-classical approximation. As an illustration, we provide
a detailed discussion of a classic example, the quantum mechanics
of the double-well potential. However, in addition to the well
known derivation of the leading order WKB result, we also deal
with several modern developments, like two-loop corrections, 
instantons and perturbation theory at large orders, as well as
supersymmetric quantum mechanics. In addition to that, we give
an introduction to the semi-classical theory of instantons
in gauge theory. 

   Sections \ref{sec_theory}-\ref{sec_temp} make up the core of the
review. They provide an in-depth discussion of the role of instantons in 
QCD. Specifically, we try to emphasize the connection between the 
structure of the vacuum, hadronic correlation functions, and hadronic 
structure, both at zero and finite temperature. The style is mostly 
that of a review of the modern literature, but we have tried to make 
this part as self-contained as possible. 

   The last two sections, \ref{sec_other},\ref{sec_sum}, deal with 
many fascinating applications of instantons in other theories, and
with possible lessons for QCD. The presentation is rather cursory
and our main motivation is to acquaint the reader with the main
problems and to provide a guide to the available literature.

\subsection{Physics outlook}
\label{sec_outlook}
\subsubsection{Hadronic structure}
\label{sec_intro_had} 

   In this section, we would like to provide a brief outline of 
the theory and phenomenology of hadronic structure and the QCD vacuum. 
We will emphasize the role the vacuum plays in determining the structure 
of the excitations, and explain how instantons come into play in making 
this connection.

  There are two approaches to hadronic structure that predate QCD, the 
quark model and current algebra. The quark model provides a simple (and 
very successful) scheme based on the idea that hadrons can be understood 
as bound states of non-relativistic constituent quarks. Current algebra 
is based on the (approximate) $SU(2)_L\times SU(2)_R$ chiral symmetry of 
the strong interaction. The fact that this symmetry is not apparent in 
the hadronic spectrum led to the important concept that chiral symmetry 
is spontaneously broken in the ground state. Also, since the ``current" 
quark masses appearing as symmetry breaking terms in the effective chiral 
lagrangian are small, it became clear that the constituent quarks of the 
non-relativistic quark model have to be effective, composite, objects. 

   With the advent of QCD, it was realized that current algebra is 
a rigorous consequence of the (approximate) chiral invariance of the
QCD lagrangian. It was also clear that quark confinement and chiral 
symmetry breaking are consistent with QCD, but since these phenomena 
occur in the non-perturbative domain of QCD, there was no 
obvious way to incorporate these features into hadronic models. The 
most popular model in the early days of QCD was the MIT bag model 
\cite{DJJK_75}, which emphasized the confinement property of QCD.
Confinement was realized in terms of a special boundary condition
on quark spinors and a phenomenological bag pressure. However, 
the model explicitly violated the chiral symmetry of QCD. This 
defect was later cured by dressing the bag with a pionic cloud in 
the cloudy \cite{TTM_81} or chiral bag \cite{BR_88} models. The 
role of the pion cloud was found to be quite large, and it was 
suggested that one can do away with the quark core altogether.
This idea leads to (topological) soliton models of the nucleon.
The soliton model was pioneered by Skyrme in a paper that appeared 
long before QCD \cite{Sky_61}. However, the Skyrmion was largely 
ignored at the time and only gained in popularity after Witten 
argued that in the large $N_c$ limit the nucleon becomes a soliton, 
built from non-linear meson fields \cite{Wit_83}.

   While all these models provide a reasonable description of 
static properties of the nucleon, the pictures of hadronic structure 
they suggest are drastically different. Although it is sometimes 
argued that different models provide equivalent, dual, descriptions 
of the same physics, it is clear that not all of these models can be 
right. While in the MIT bag model, for example, everything is
determined by the bag pressure, there is no such thing in the 
Skyrmion and the scale is set by chiral symmetry breaking. Also, 
quark models attribute the nucleon-delta mass splitting to perturbative 
one gluon exchange, while it is due to the collective rotation of the 
pion field in soliton models.

   In order to make progress two shifts of emphasis have to be made. 
First, it is not enough to just reproduce the mass and other static 
properties of the nucleon. A successful model should reproduce the
correlation functions (equivalent to the full spectrum, including 
excited states) in all relevant channels, not just baryons, but also
scalar and vector mesons, etc. Second, the structure of hadrons 
should be understood starting from the structure the QCD vacuum. 
Hadrons are collective excitations, like phonons in solids, so 
one cannot ignore the properties of the ground state when 
studying its excitations.

\subsubsection{Scales of non-perturbative QCD}
\label{sec_intro_scales} 

   In order to develop a meaningful strategy it is important to
establish whether there is a hierarchy of scales that allows the 
problem to be split into several independent parts. In QCD, there
is some evidence that such a hierarchy is provided by the scales
for chiral symmetry breaking and confinement, $\Lambda_{\chi SB}
\gg\Lambda_{conf}$. 
  
 The first hint comes from perturbative QCD. Although the perturbative
coupling constant blows up at momenta given roughly by the scale
parameter $\Lambda_{QCD}\sim 0.2\,{\rm GeV}\simeq (1\,{\rm fm})^{-1}$
(the exact value depends on the renormalization scheme), perturbative
calculations are generally limited to reactions involving a scale
of at least $1\,{\rm GeV}\simeq (0.2\,{\rm fm})^{-1}$.

  A similar estimate of the scale of non-perturbative effects
can be obtained from low-energy effective theories. The first
result of this kind was based on the Nambu and Jona-Lasinio (NJL) model
\cite{NJL_61}. This model was inspired by the analogy between chiral 
symmetry breaking and superconductivity. It postulates a 4-fermion 
interaction which, if it exceeds a certain strength, leads to quark 
condensation, the appearance of pions as Goldstone bosons, etc. The 
scale above which this interactions disappears and QCD becomes
perturbative enters the model as an explicit UV cut-off, $\Lambda_
{\chi SB}\sim 1 GeV$.

  It was further argued that the scales for chiral symmetry breaking 
and confinement are very different \cite{Shu_81}: $\Lambda_{\chi SB}
\gg \Lambda_{conf}\sim \Lambda_{QCD}$. In particular, it was argued 
that constituent quarks (and pions) have sizes smaller than those of 
typical hadrons, explaining the success of the non-relativistic quark 
model. This idea was developed in a more systematic fashion by Georgi 
and Manohar \cite{GM_84}, who argue that $\Lambda_{\chi SB}\simeq 
4\pi f_\pi\simeq 1\,{\rm GeV}$ provides a natural expansion parameter 
in chiral effective Lagrangians. An effective theory using pions and 
``constituent'' quarks is the natural description in the intermediate 
regime $\Lambda_{conf}<Q<\Lambda_{\chi SB}$, where models of hadronic 
structure operate.

  While our understanding of the confinement mechanism in QCD is still
very poor, considerable progress has been made in understanding the 
physics of chiral symmetry breaking. The importance of instantons 
in this context is one of the main points of this review. Instantons
are localized ($\rho\simeq 1/3\,{\rm fm}$) regions of space-time 
with very strong gluonic fields, $G_{\mu\nu}\sim 1/(g\rho^2)$. It
is the large value of this quantity which sets the scale for 
$\Lambda_{\chi SB}$. In the regime $\Lambda_{conf}<Q<\Lambda_
{\chi SB}$, the instanton-induced effective interaction between
quarks is of the form
\be
\label{l_eff_schem}
{\cal L} &=& \bar\psi\left(i\partial\!\!\!\!/\, -m\right)\psi
 +\frac{c_\Gamma}{\Lambda^2}\left(\bar\psi\Gamma\psi\right)^2
 +\frac{d_\Gamma}{\Lambda^5}\left(\bar\psi\Gamma\psi\right)^3
 +\ldots ,
\ee
where $\Gamma$ is some spin-isospin-color matrix, $\Lambda\sim 
\rho^{-1}$ is determined by the chiral symmetry breaking scale 
and higher order terms involve more fermion fields or derivatives.
The mass scale for glueballs is even higher and these states 
decouple in the regime under consideration. The instanton 
induced interaction is non-local and when calculating higher
order corrections with the vertices in (\ref{l_eff_schem}),
loop integrals are effectively cut off at $\Lambda\sim\rho^{-1}$.

  In addition to determining the scale $\Lambda$, instantons also
provide an organizing principle for performing calculations with the
effective lagrangian (\ref{l_eff_schem}). If the instanton liquid
is dilute, $\rho^4(N/V)\ll 1$, where $(N/V)$ is the density of 
instantons, then vertices with more than $2N_f$ legs are suppressed.
In addition to that, the diluteness parameter determines which 
diagrams involving the leading order vertices have to be resummed. 
As a result, one can derive another effective theory valid at 
even smaller scales which describes the interaction of extended,
massive constituent quarks with point-like pions \cite{Dia_96}. 
This theory is of the type considered by \cite{GM_84}.

  Alternatively, one can go beyond leading order in $\rho^4(N/V)$
and study hadronic correlation functions to all orders in the 
instanton-induced interaction. The results are in remarkably 
good agreement with experiment for most light hadrons. Only
when dealing with heavy quarks or high lying resonances do
confinement effects seem to play an important role. We will 
discuss these questions in detail in Sec. \ref{sec_cor}.

\subsubsection{Structure of the QCD vacuum }
\label{sec_intro_vac} 

  The ground state of QCD can be viewed as a very dense state of matter, 
composed of gauge fields and quarks that interact in a complicated 
way. Its properties are not easily accessible in experiments, because
we do not directly observe quark and gluon fields, only the color neutral 
hadrons. Furthermore, we cannot directly determine the interaction between 
quarks, because (unlike for the nuclear force) we cannot measure $qq$ and 
$\bar qq$ scattering amplitudes. Instead, the main tool at our disposal 
are correlation functions of hadronic currents. The phenomenology of these 
functions was recently reviewed in \cite{Shu_93}. Hadronic point-to-point 
correlation functions were first systematically studied in the context of 
``QCD sum rules''. The essential point, originally emphasized by Shifman, 
Vainshtain and Zakharov (SVZ), is that the operator product expansion 
relates the short distance behavior of current correlation function to
the vacuum expectation values of a small set of lowest-dimension quark 
and gluon operators. Using the available phenomenological information 
on hadronic correlation functions, SVZ deduced the quark and gluon 
condensates\footnote{In the operator product expansion one has to
introduce a scale $\mu$ that separates soft and hard contributions. 
The condensates take into account soft fluctuations and the values 
given here correspond to a scale $\mu\simeq 1$ GeV.}
\be
 \langle\bar qq\rangle\, =\, -(230\,{\rm MeV})^3,\hspace{1cm} 
 \langle g^2G^2\rangle\, =\,  (850\,{\rm MeV})^4,
\ee
as well as other, more complicated, expectation values. 

\begin{figure}[t]
\begin{center}
\leavevmode
\epsfxsize=12cm
\epsffile{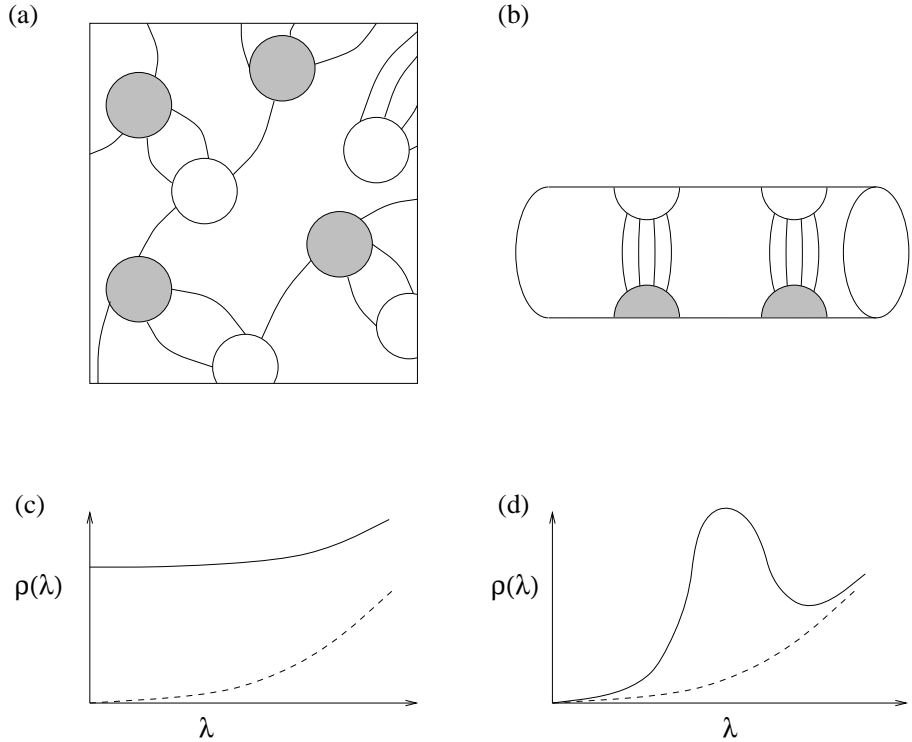}
\caption{\label{fig_liquid}
Schematic picture of the instanton liquid at zero temperature
(a) and above the chiral phase transition (b). Instantons and
anti-instantons are shown as open and shaded circles. The 
lines correspond to fermion exchanges. Figures (c) and (d) 
show the schematic form of the Dirac spectrum in the 
configurations (a) and (b).}
\end{center}
\end{figure}

  The significance of the quark condensate is the fact that it is 
an order parameter for the spontaneous breakdown of chiral symmetry 
in the QCD vacuum. The gluon condensate is important because the 
QCD trace anomaly 
\be
  T_{\mu\mu}&=& \sum_f m_f\langle \bar q_fq_f\rangle 
   -\frac{b}{32\pi^2}\langle g^2G^2\rangle
\ee
relates this quantity to the energy density $\epsilon_0\simeq-500\,{\rm MeV}
/{\rm fm}^3$ of the QCD vacuum. Here, $T_{\mu\nu}$ is the energy momentum 
tensor and $b=11N_c/3-2N_f/3$ is the first coefficient of the beta function. 

   Any model of the QCD vacuum should explain the origin and value of
these condensates, the mechanism for confinement and chiral symmetry
breaking and its relation to the underlying parameters of the theory
(the scale parameter and the matter content of the theory). Most of
the early attempts to understand the QCD ground state were based on
the idea that the vacuum is dominated by classical gauge field 
configurations, for example constant fields \cite{Sav_77} or regions
of constants fields patched together, as in the Spaghetti vacuum 
introduced by the Kopenhagen group \cite{AO_80}. All of these 
attempts were unsuccessful, however, because constant fields were
found to be unstable against quantum perturbations. 

   Instantons are classical solutions to the euclidean equations of
motion. They are characterized by a topological quantum number and
correspond to tunneling events between degenerate classical vacua
in Minkowski space. As in quantum mechanics, tunneling lowers the 
ground state energy. Therefore, instantons provide a simple understanding 
of the negative non-perturbative vacuum energy density. In the presence 
of light fermions, instantons are associated with fermionic zero modes. 
Zero modes not only play a crucial role in understanding the axial anomaly, 
they are also intimately connected with spontaneous chiral symmetry 
breaking. When instantons interact through fermion exchanges, zero modes
can become delocalized, forming a collective quark condensate. 

   A crude picture of quark motion in the vacuum can then be formulated 
as follows (see Fig. \ref{fig_liquid}a). Instantons act as a potential 
well, in which light quarks can form bound states (the zero modes). If 
instantons form an interacting liquid, quarks can travel over large 
distances by hopping from one instanton to another, similar to electrons 
in a conductor. Just like the conductivity is determined by the density
of states near the Fermi surface, the quark condensate is given by the
density of eigenstates of the Dirac operator near zero virtuality. A 
schematic plot of the distribution of eigenvalues of the Dirac operator 
is shown\footnote{We will discuss the eigenvalue distribution of the
Dirac operator in some detail in the main part of the review, see
Figs. \ref{fig_dirac} and \ref{fig_dirac_T}.} in Fig. \ref{fig_liquid}c. 
For comparison, the spectrum of the Dirac operator for non-interacting 
quarks is depicted by the dashed line. If the distribution of instantons
in the QCD vacuum is sufficiently random, there is a non-zero density
of eigenvalues near zero and chiral symmetry is broken. 

  The quantum numbers of the zero modes produce very specific correlations 
between quarks. First, since there is exactly one zero mode per flavor,
quarks with different flavor (say $u$ and $d$) can travel together, but
quarks with the same flavor cannot. Furthermore, since zero modes have
a definite chirality (left handed for instantons, right handed for
anti-instantons), quarks flip their chirality as they pass through an
instanton. This is very important phenomenologically, because it 
distinguishes instanton effects from perturbative interactions, in 
which the chirality of a massless quark does not change. It also 
implies that quarks can only be exchanged between instantons of 
the opposite charge. 

\begin{figure}[t]
\begin{center}
\leavevmode
\epsfxsize=12cm
\epsffile{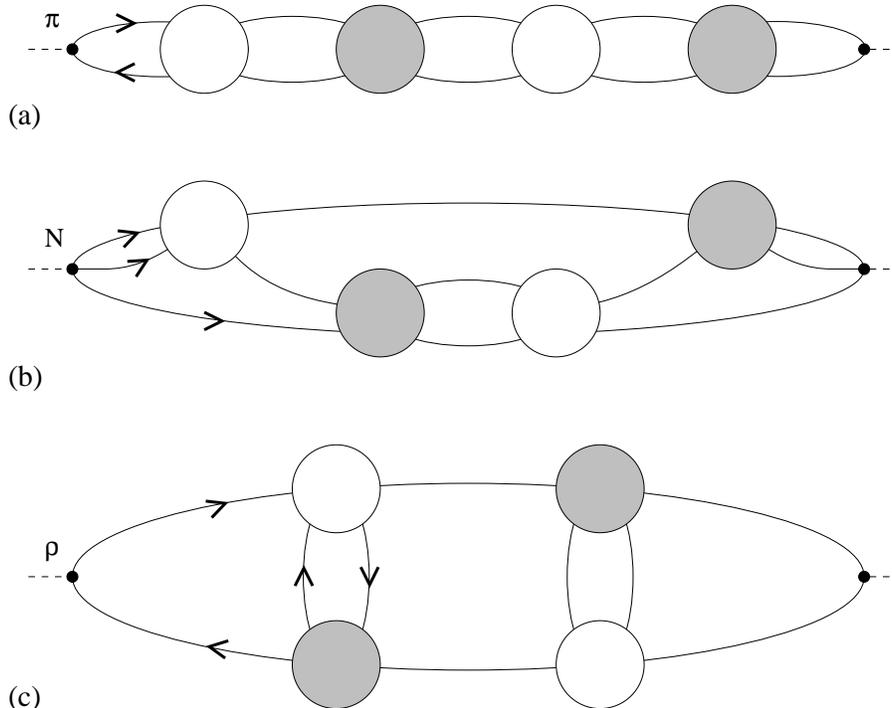}
\caption{\label{fig_cor_schem} 
Instanton contribution to hadronic correlation functions 
Fig. a) shows the pion, b) the nucleon and c) the rho meson
correlator. The solid lines correspond to zero mode contributions 
to the quark propagator.}
\end{center}
\end{figure}

   Based on this picture, we can also understand the formation of
hadronic bound states. Bound states correspond to poles in hadronic
correlation functions. As an example, let us consider the pion, which
has the quantum numbers of the current $j_\pi=\bar u\gamma_5 d$. The
correlation function $\Pi(x)=\langle j_\pi(x)j_\pi(0)\rangle$ is the 
amplitude for an up quark and a down anti-quark with opposite 
chiralities created by a source at point 0 to meet again at the 
point $x$. In a disordered instanton liquid, this amplitude is 
large, because the two quarks can propagate by the process shown 
in Fig. \ref{fig_cor_schem}a. As a result, there is a light pion 
state. For the $\rho$ meson, on the other hand, we need the amplitude 
for the propagation of two quarks with the same chirality. This means 
that the quarks have to be absorbed by different instantons (or 
propagate in non-zero mode states), see Fig. \ref{fig_cor_schem}c.
The amplitude is smaller, and the meson state is much less tightly 
bound. 

  Using this picture, we can also understand the formation of a 
bound nucleon. Part of the proton wave function is a scalar $ud$
diquark coupled to another $u$ quark. This means that the nucleon 
can propagate as shown in Fig. \ref{fig_cor_schem}b. The vertex 
in the scalar diquark channel is identical to the one in the 
pion channel with one of the quark lines reversed\footnote{For
more than three flavors the color structure of the two vertices
is different, so there is no massless diquark in $SU(3)$ color.}
The $\Delta$ resonance has the quantum numbers of a vector diquark 
coupled to a third quark. Just like in the case of the $\rho$ meson, 
there is no first order instanton induced interaction, and we expect 
the $\Delta$ to be less bound than than the nucleon. 

   The paradigm discussed here bears striking similarity to one of 
the oldest approaches to hadronic structure, the Nambu-Jona-Lasinio
(NJL) model \cite{NJL_61}. In analogy with the Bardeen-Cooper-Schrieffer 
theory of superconductivity, it postulates a short-range attractive
force between fermions (nucleons in the original model and light quarks 
in modern versions). If this interaction is sufficiently strong,
it can rearrange the vacuum and the ground state becomes superconducting, 
with a non-zero quark condensate. In the process, nearly massless
current quarks become effectively massive constituent quarks. The
short range interaction can then bind these constituent quarks into
hadrons (without confinement). 

   This brief outline indicates that instantons provide at least a 
qualitative understanding of many features of the QCD ground state
and its hadronic excitations. How can this picture be checked and
made more quantitive? Clearly, two things need to be done. First,
a consistent instanton ensemble has to be constructed in order to
make quantitative predictions for hadronic observables. Second, 
we would like to test the underlying assumption that the topological 
susceptibility, the gluon condensate, chiral symmetry breaking etc.
are dominated by instantons. This can be done most directly on the
lattice. We will discuss both of these issues in the main part of
this review, sections III-VI.

\subsubsection{QCD at finite temperature}
\label{sec_intro_temp}   

  Properties of the QCD vacuum, like the vacuum energy density, the 
quark and gluon condensate are not directly accessible to experiment. 
Measuring non-perturbative properties of the vacuum requires the 
possibility to compare the system with the ordinary, perturbative 
state\footnote{A nice analogy is given by the atmospheric pressure. 
In order to measure this quantity directly one has to evacuate a 
container filled with air. Similarly, one can measure the non-perturbative 
vacuum energy density by filling some volume with another phase, the 
quark-gluon plasma.}. This state of matter has not existed in nature 
since the Big Bang, so experimental attempts at studying the perturbative 
phase of QCD have focused on recreating miniature Big Bangs in  
relativistic heavy ion collisions. 

  The basic idea is that at sufficiently high temperature or
density, QCD will undergo a phase transition to a new state,
referred to as the quark gluon plasma, in which chiral symmetry
is restored and quarks and gluon are deconfined. The temperature
scale for this transition is set by the vacuum energy density 
and pressure of the vacuum. For the perturbative vacuum to have
a pressure comparable to the vacuum pressure $500\,{\rm MeV}/
{\rm fm}^3$, a temperature on the order of $150-200\,{\rm MeV}$ 
is required. According to our current understanding, such 
temperatures are reached in the ongoing or planned experiments
at the AGS (about 2+2 GeV per nucleon in the center of mass system),
CERN SPS (about 10+10 GeV) or RHIC (100+100 GeV).

  In order to interpret these experiments, we need to understand
the properties of hadrons and hadronic matter near and above the phase 
transition. As for cold matter, this requires an understanding
of the ground state and how the rearrangement of the vacuum that 
causes chiral symmetry to be restored takes place. A possible 
mechanism for chiral symmetry restoration in the instanton liquid
is indicated in Fig. \ref{fig_liquid}b,d. At high temperature, 
instantons and anti-instantons have a tendency to bind in pairs 
that are aligned along the (euclidean) time direction. The 
corresponding quark eigenstates are strongly localized and 
chiral symmetry is unbroken. There is some theoretical evidence
for this picture which will be discussed in detail in section VII. 
In particular, there is evidence from lattice simulations that 
instantons do not disappear at the phase transition, but only at 
even higher temperatures. This implies that instantons affect 
properties of the quark gluon plasma at temperatures not too far 
above the phase transition.

\subsection{The history of instantons}
\label{sec_history}

   In books and reviews, physical theories are usually presented as
a systematic development, omitting the often confusing history of the
subject. The history of instantons also did not follow a straight path. 
Early enthusiasm concerning the possibility to understand non-perturbative 
phenomena in QCD, in particular confinement, caused false hopes, which led 
to years of frustration. Only many years later did work on phenomenological 
aspects of instantons lead to breakthroughs. In the following we will try 
to give a brief tour of the two decades that have passed since the discovery 
of instantons. 

\subsubsection{Discovery and early applications}
\label{sec_disc} 

   The instanton solution of the Yang-Mills equations was discovered 
by Polyakov and coworkers \cite{BPS_75}, motivated by the search for 
classical solutions with nontrivial topology in analogy with the 
't Hooft-Polyakov monopole \cite{Pol_75}. Shortly thereafter, a 
number of authors clarified the physical meaning of the instanton 
as a tunneling event between degenerate classical vacua
\cite{JR_76,CDG_76,Pol_77}. These works also introduced the 
concept of $\theta$-vacua in connection with QCD.

   Some of the early enthusiasm was fueled by Polyakov's discovery 
that instantons cause confinement in certain 3-dimensional models
\cite{Pol_77}. However, it was soon realized that this is not the 
case in 4-dimensional gauge theories. An important development 
originated with 't Hooft's classic paper\footnote{In this paper,
't Hooft also coined the term instanton; Polyakov had referred
to the classical solution as a pseudo-particle.} \cite{tHo_76},
in which he calculated the semi-classical tunneling rate. In this
context, he discovered the presence of zero modes in the spectrum 
of the Dirac operator. This result implied that tunneling is 
intimately connected with light fermions, in particular that 
every instanton absorbs one left handed fermion of every species, 
and emits a right handed one (and vice versa for anti-instantons). 
This result also explained how anomalies, for example the violation 
of axial charge in QCD and baryon number in electroweak theory, are 
related to instantons.
   
   In his work, 't Hooft  estimated the tunneling rate in  
electroweak theory, where the large Higgs expectation value 
guarantees the validity of the semi-classical approximation,
and found it to be negligible. Early attempts to study instantons
effects in QCD, where the rate is much larger but also harder
to estimate, were summarized in \cite{CDG_78}. These authors
realized that the instanton ensemble can be described as a 
4-dimensional ``gas" of pseudo-particles that interact via 
forces that are dominantly of dipole type. While they were 
not fully successful in constructing a consistent instanton ensemble,
they nevertheless studied a number of important instanton effects:
the instanton induced potential between heavy quarks, the 
possibility that instantons cause the spontaneous breakdown 
of chiral symmetry, and instanton corrections to the running
coupling constant. 

    One particular instanton-induced effect, the anomalous breaking
of $U(1)_A$ symmetry and the $\eta'$ mass, deserves special
attention\footnote{There is one historical episode that we 
should mention briefly. Crewther and Christos \cite{Chr_84} 
questioned the sign of the axial charge violation caused by 
instantons. A rebuttal of these arguments can be found in \cite{tHo_86}.}. 
Witten and Veneziano wrote down an approximate relation that connects
the $\eta'$ mass with the topological susceptibility \cite{Wit_79,Ven_79}. 
This was a very important step, because it was the first quantitative
result concerning the effect of the anomaly on the $\eta'$ mass. 
However, it also caused some confusion, because the result had been 
derived using the large $N_c$ approximation, which is not easily applied 
to instantons. In fact, both Witten and Veneziano expressed strong doubts 
concerning the relation between the topological susceptibility and 
instantons\footnote{Today, there is substantial evidence from lattice 
simulations that the topological susceptibility is dominated by 
instantons, see Sec. \protect\ref{sec_inst_lat}.}, suggesting 
that instantons are not important dynamically \cite{Wit_79b}. 

\subsubsection{Phenomenology leads to a qualitative picture}

   By the end of the 70's the general outlook was very pessimistic. 
There was no experimental evidence for instanton effects, and no 
theoretical control over semi-classical methods in QCD. If a problem 
cannot be solved by direct theoretical analysis, it often useful to 
turn to a more phenomenological approach. By the early 80's, such an 
approach to the structure of the QCD vacuum became available with the 
QCD sum rule method \cite{SVZ_79}. QCD sum rules relate vacuum parameters, 
in particular the quark and gluon condensates, to the behavior of hadronic 
correlation functions at short distances. Based on this analysis, it was 
realized that ``all hadrons are not alike'' \cite{NSVZ_81}. The Operator 
Product Expansion (OPE) does not give reliable predictions for scalar and 
pseudo-scalar channels ($\pi,\sigma,\eta,\eta'$ as well as scalar and 
pseudo-scalar glueballs). These are precisely the channels that 
receive direct instanton contributions \cite{GI_80,NSVZ_81,Shu_83}. 
 
  In order to understand the available phenomenology, a qualitative 
picture, later termed the instanton liquid model, was proposed in 
\cite{Shu_82}. In this work, the two basic parameters of the instanton 
ensemble were suggested: the mean density of instantons is $n\simeq 
1\,{\rm fm}^{-4}$, while their average size is $\rho\simeq 1/3$ fm. 
This means that the space time volume occupied by instantons $f\sim
n\rho^4$ is small; the instanton ensemble is dilute. This observation
provides a small expansion parameter which we can use in order to
perform systematic calculations. 

  Using the instanton liquid parameters $n\simeq 1\,{\rm fm}^{-4}, 
\,\rho\simeq 1/3$ fm we can reproduce the phenomenological values of 
the quark and gluon condensates. In addition to that, one can calculate 
direct instanton corrections to hadronic correlation functions at short 
distance. The results were found to be in good agreement with experiment 
in both attractive ($\pi,K$) and repulsive ($\eta'$) pseudo-scalar meson 
channels \cite{Shu_83}.

\subsubsection{Technical development during the 80's}

   Despite its phenomenological success, there was no theoretical
justification for the instanton liquid model. The first steps towards
providing some theoretical basis for the instanton model were taken in
\cite{IMP_81,DP_84}. These authors used variational techniques and the 
mean field approximation (MFA) to deal with the statistical mechanics
of the instanton liquid. The ensemble was stabilized using a 
phenomenological core \cite{IMP_81} or a repulsive interaction 
derived from a specific ansatz for the gauge field interaction
\cite{DP_84}. The resulting ensembles were found to be consistent
with the phenomenological estimates.

   The instanton ensemble in the presence of light quarks was studied
in \cite{DP_86}. This work introduced the picture of the quark 
condensate as a collective state built from delocalized zero modes.
The quark condensate was calculated using the mean field approximation
and found to be in agreement with experiment. Hadronic states were
studied in the random phase approximation (RPA). At least in the case 
of pseudo-scalar mesons, the results were also in good agreement with
experiment. 

   In parallel, numerical methods for studying the instanton liquid
were developed  \cite{Shu_88}. Numerical simulations allow one to 
go beyond the MF and RPA approximations and include the 't Hooft 
interaction to all orders. This means that one can also study 
hadronic channels that, like vector mesons, do not have first
order instanton induced interactions, or channels, like the 
nucleon, that are difficult to treat in the random phase 
approximation.

   Nevertheless, many important aspects of the model remain to be 
understood. This applies in particular to the theoretical foundation
of the instanton liquid model. When the instanton-anti-instanton 
interaction was studied in more detail, it became clear that there 
is no classical repulsion in the gauge field interaction. A well
separated instanton-anti-instanton pair is connected to the 
perturbative vacuum by a smooth path \cite{BY_86,Ver_91}. This
means that the instanton ensemble cannot be stabilized by purely
classical interactions. This is related to the fact that in general,
it is not possible to separate non-perturbative (instanton-induced)
and perturbative effects. Only in special cases, like in quantum
mechanics (Sec. \ref{sec_qm_IA}) and supersymmetric field theory
(Sec. \ref{sec_susy_qcd}) has this separation been accomplished.
 
\subsubsection{Recent progress}
   
   In the past few years, a great deal was learned about instantons
in QCD. The instanton liquid model with the parameters mentioned above, 
now referred to as the random instanton liquid model (RILM), was used 
for large-scale, quantitative calculations of hadronic correlation 
functions in essentially all meson and baryon channels \cite{SV_93b,SSV_94}. 
Hadronic masses and coupling constants for most of the low-lying mesons and 
baryon states were shown to be in quantitative agreement with experiment.

   The next surprise came from a comparison of the correlators 
calculated in the random model and first results from lattice 
calculations \cite{CGHN_93a}. The results agree quantitatively 
not only in channels that were already known from phenomenology, 
but also in others (such as the nucleon and delta), were no previous 
information (except for the particle masses, of course) existed.

   These calculations were followed up by direct studies of the 
instanton liquid on the lattice. Using a procedure called cooling, 
one can extract the classical content of strongly fluctuating lattice 
configurations. Using cooled configurations, the MIT group determined 
the main parameters of the instanton liquid \cite{CGHN_94}. Inside the 
accuracy involved ($\sim 10 \% $) the density and average size coincide 
with the values suggested a decade earlier. In the mean time, these 
numbers have been confirmed by other calculations (see Sec. 
\ref{sec_lat}). In addition to that, it was shown that the agreement
between lattice correlation functions and the instanton model
was not a coincidence: the correlators are essentially unaffected
by cooling. This implies that neither perturbative (removed by cooling) 
nor confinement (strongly reduced) forces are crucial for hadronic 
properties.

   Technical advances in numerical simulation of the instanton liquid 
led to the construction of a self-consistent, interacting instanton 
ensemble, which satisfies all the general constraints imposed by the trace 
anomaly and chiral low energy theorems \cite{SV_95,SS_96,SS_96b}. 
The corresponding unquenched (with fermion vacuum bubbles included)
correlation functions significantly improve the description of  
the $\eta'$ and $\delta$ mesons, which are the two channels where  
the random model fails completely. 

   Finally, significant progress was made in understanding the 
instanton liquid at finite temperature and the mechanism for the
chiral phase transition. It was realized earlier that at high 
temperature, instantons should be suppressed by Debye screening 
\cite{Shu_78,PY_80}. Therefore, it was generally assumed that 
chiral symmetry restoration is a consequence of the disappearance 
of instantons at high temperature.
 
   More recently it was argued that up to the critical temperature, 
the density of instantons should not be suppressed \cite{SV_94}. 
This prediction was confirmed by direct lattice measurements of the
topological susceptibility \cite{CS_95}, which indeed found little 
change in the topological susceptibility for $T<T_c$, and the expected 
suppression for $T>T_c$. If instantons are not suppressed around
$T_c$, a different mechanism for the chiral phase transition is
needed. It was suggested that chiral symmetry is restored 
because the instanton liquid is rearranged, going from a random
phase below $T_c$ to a correlated phase of instanton-anti-instanton
molecules above $T_c$ \cite{IS_94,SSV_95}. This idea was confirmed
in direct simulations of the instanton ensemble, and a number of 
consequences of the scenario were explored. 

\subsection{Topics that are not discussed in detail}
\label{sec_not}

   There is a vast literature on instantons (the SLAC database
lists over 3000 references, which probably does not include the
more mathematically oriented works) and limitations of space and time as
well as our expertise have forced us to exclude many interesting
subjects from this review. Our emphasis in writing this review
has been on the theory and phenomenology of instantons in QCD. 
We discuss instantons in other models only to the extent that 
they are pedagogically valuable or provide important lessons 
for QCD. Let us mention a few important omissions and give 
a brief guide to the relevant literature:

\begin{enumerate}
\item{Direct manifestations of small-size instantons in high energy
baryon number violating (BNV) reactions. The hope is that in these 
processes, one may observe rather spectacular instanton effects in 
a regime where reliable semi-classical calculations are possible. 
In the electroweak theory, instantons lead to baryon number 
violation, but the amplitude for this reaction is strongly suppressed 
at low energies. It was hoped that this suppression can be overcome 
at energies on the order of the sphaleron barrier $E\simeq 10$ TeV,
but the emerging consensus is that this dramatic phenomenon will not
be observable. Some of the literature is mentioned in Sec. \ref{sec_ew},
see also the recent review \cite{Aoy_97}.} 

\item{A related problem is the transition from tunneling to thermal
activation and the calculation of the Sphaleron rate at high 
temperature. This question is of interest in connection with 
baryogenesis in the early universe and axial charge fluctuations 
in the quark gluon plasma. A recent discussion can be found in
the review \cite{Smi_96}.}

\item{The decay of unstable vacua in quantum mechanics or field
theory \cite{Col_77}. A more recent review can be found in 
\cite{Aoy_97}.}

\item{Direct instanton contributions to deep inelastic scattering 
and other hard processes in QCD, see \cite{BB_93,BB_95} and the
review \cite{RS_94}.}

\item{Instanton inspired models of hadrons, or phenomenological 
lagrangians supplemented by the 't Hooft interaction. These models
include NJL models \cite{HK_94}, soliton models \cite{DPP_88,CBK_96}, 
potential models \cite{BBH_90}, bag models \cite{DZK_92}, etc.}

\item{Mathematical aspects of instantons \cite{EGH_80}, the ADHM construction
of the most general $n$-instanton solution \cite{AHDM_77}, constrained
instantons \cite{Aff_81}, instantons and four-manifolds \cite{FU_84},
the connection between instantons and solitons \cite{AM_89}. For a
review of known solutions of the classical Yang-Mills field equations
in both Euclidean and Minkowski space we refer the reader to 
\cite{Act_79}.}

\item{Formal aspects of the supersymmetric instanton calculus,
spinor techniques, etc. This material is covered in  
\cite{NSV_83,NSV_83b,Nov_87,AKM_88}.}

\item{The strong CP problem, bounds on the theta parameter,
the axion mechanism \cite{PQ_77}, etc. Some remarks on these
questions can be found in Sec. \ref{sec_theta}.}

\end{enumerate}


\section{Semi-classical theory of tunneling}
\label{sec_semicl}

\subsection{Tunneling in quantum mechanics}
\label{sec_doublewell}

\subsubsection{Quantum mechanics in Euclidean space}
\label{sec_QM}   


  This section serves as a brief introduction into path integral
methods and can easily be skipped by readers familiar with the
subject. We will demonstrate the use of Feynman diagrams
in a simple quantum mechanical problem, which does not suffer
from any of the divergencies that occur in field theory. 
Indeed, we hope that this simple example will find its way
into introductory field theory courses.

  Another point we would like to emphasize in this section is the
similarity between quantum and statistical mechanics. Qualitatively,
both quantum and statistical mechanics deal with variables that
are subject to random fluctuations (quantum or thermal), so that 
only ensemble averaged quantities make sense. Formally the connection
is related to the similarity between the statistical partition function 
${\rm tr}[\exp(-\beta H)]$ and the generating functional (\ref{path_int})
(see below) describing the dynamical evolution of a quantum system. 

  Consider the simplest possible quantum mechanical system, the motion 
of a particle of mass $m$ in a time independent potential $V(x)$. The
standard approach is based on an expansion in terms of stationary states 
$\psi_{n}(x)$, given as solutions of the Schr{\"o}dinger equation
$H\psi_n=E_n\psi_n$. Instead, we will concentrate on another object, 
the Green's function\footnote{We use natural units $\hbar=h/2\pi=1$ and 
$c=1$. Mass, energy and momentum all have dimension of inverse length.}
\be 
G(x,y,t)&=& \langle y|\exp(-i H t)|x\rangle,
\ee
which is the amplitude for a particle to go from point $x$ at time  
$t=0$ to point $y$ at time $t$. The Green's function can be expanded 
in terms of stationary states
\be 
G(x,y,t) &=& \sum_{n=1}^{\infty} 
 \psi_n^*(x) \psi_n(y) \exp(-iE_nt) .
\ee
This representation has many nice features that are described in
standard text books on quantum mechanics. There is, however, another
representation which is more useful in order to introduce semi-classical
methods and to deal with system with many degrees of freedom, the
Feynman path integral \cite{FH_65}
\be 
\label{path_int}
G(x,y,t) &=& \int Dx(t)\, \exp(iS[x(t)]). 
\ee
Here, the Green's functions is given as a sum over all possible paths 
$x(t)$ leading from $x$ at $t=0$ to $y$ at time $t$. The weight for
the paths is provided by the action $S=\int dt [m\dot{x}^2/2-V(x)]$.
One way to provide a more precise definition of the path integral 
is to discretize the path. Dividing the time axis into $N$ intervals, 
$a=t/N$, the path integral becomes an $N$-dimensional integral over 
$x_n=x(t_n=an)\; n=1,\ldots,N$. The discretized action is given by
\be 
\label{S_qm}  
S &=& \sum_{n} \left[\frac{m}{2a}(x_n-x_{n-1})^2-a V(x_n)\right].
\ee
This form is not unique, other discretizations with the same continuum
limit can also be used. The path integral is now reduced to a multiple 
integral over $x_n$, where we have to take the limit $n\to\infty$. In 
general, only Gaussian integrals can be done exactly. In the case of the 
harmonic oscillator $V(x)=m\omega^2 x^2/2$ one finds \cite{FH_65}
\be 
 G_{osc}(x,y,t)&=& \left(\frac{m \omega}{2 \pi i\sin \omega t}\right)^{1/2}
 \exp\left[\left(\frac{i m \omega}{2 \sin\omega t}\right)
 \left( (x^2+y^2)\cos(\omega t)-2xy\right) \right]
\ee
In principal, the discretized action (\ref{S_qm}) should be amenable 
to numerical simulations. In practice, the strongly fluctuating 
phase in (\ref{path_int}) renders this approach completely useless.
There is, however, a simple way to get around the problem. If one 
performs an analytic continuation of $G(x,y,t)$ to imaginary 
(Euclidean) time $\tau=it$, the weight function becomes $\exp(-S_E[x
(\tau)])$ with the Euclidean action $S_E=\int dt [m (dx/d\tau)^2/2 
+V(x)]$. In this case, we have a positive definite weight and 
numerical simulations, even for multidimensional problems, are
feasible. Note that the relative sign of the kinetic and potential
energy terms in the Euclidean action has changed, making it look 
like a Hamiltonian. In Euclidean space, the discretized action 
(\ref{S_qm}) looks like the energy functional of a 1-dimensional 
spin chain with nearest neighbor interactions. This observation 
provides the formal link between an $n$-dimensional statistical 
system and Euclidean quantum (field) theory in $(n-1)$ dimensions. 

  Euclidean Green's functions can be interpreted in terms of ``thermal"
distributions. If we use periodic boundary conditions ($x=y$) and 
integrate over $x$ we obtain the statistical sum 
\be 
\int dx \, G(x,x,\tau) &=& \sum_{n=1}^{\infty} \exp(-E_n \tau),
\ee
where the time interval $\tau$ plays the role of an inverse temperature 
$\tau=T^{-1}$. In particular, $G(x,x,1/T)$ has the physical meaning of 
a probability distribution for $x$ at temperature $T$. For the harmonic
oscillator mentioned above, the Euclidean Green's function is
\be 
\label{G_osc_E}
G_{osc}(x,y,\tau) &=& \left( \frac{m \omega}{2 \pi \sinh\omega\tau}
 \right)^{1/2} \exp\left[ -\left( \frac{m \omega}{2 \sinh \omega 
 \tau}\right) \left( (x^2+y^2)\cosh(\omega \tau)-2xy\right) \right]
\ee
For $x=y$, the spatial distribution is Gaussian at any $T$, with a width
$\langle x^2\rangle =\frac{1}{2m\omega}\coth(\omega/2T)$. If $\tau$ is
very large, the effective temperature is small and the ground state 
dominates. From the exponential decay, we can read off the ground
state energy $E_0=\omega/2$ and from the spatial distribution the
width of the ground state wave function $\langle x^2\rangle =(2m
\omega)^{-1}$. At high $T$ we get the classical result $\langle x^2
\rangle=T/(m\omega^2)$. 

\begin{figure}[t]
\begin{center}
\leavevmode
\epsfxsize=12cm
\epsffile{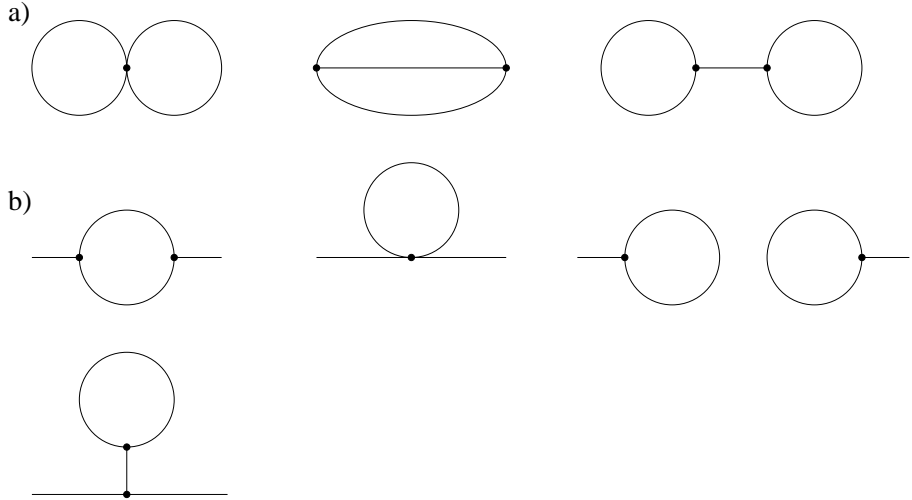}
\end{center}
\caption{\label{fig_qm_graphs} 
Feynman diagrams for the energy (a) and the Green's function
(b) of the anharmonic oscillator.}
\end{figure}

   Non-Gaussian path integrals cannot be done exactly. As long as 
the non-linearities are small, we can use perturbation theory. 
Consider an anharmonic oscillator with (Euclidean) action 
\be
S_E &=& \int d\tau\,
 \left[ {\dot x^2 \over 2}+{\omega^2 x^2 \over 2}+
 \alpha x^3 +\beta x^4 \right]  
\ee 
Expanding the path integral in powers of $\alpha$ and $\beta$ one 
can derive the Feynman rules for the anharmonic oscillator. The
free propagator is given by
\be
\label{qm_prop}
G_0(\tau_1,\tau_2)&=& \langle x(\tau_1)x(\tau_2)\rangle
 \;=\; {1\over 2\omega} \exp(-\omega|\tau_1-\tau_2|).
\ee
In addition to that, there are three and four-point vertices with
coupling constants $\alpha$ and $\beta$. To calculate an $n$-point
Green's function we have to sum over all diagrams with $n$ external legs 
and integrate over the time variables corresponding to internal vertices. 

   The vacuum energy is given by the sum of all closed diagrams. At one 
loop order, there is only one diagram, the free particle loop diagram.
At two loop order, there are two $O(\alpha^2)$ and one $O(\beta)$
diagram, see Fig. \ref{fig_qm_graphs}a. The calculation of the  
diagrams is remarkably simple. Since the propagator is exponentially
suppressed for large times, everything is finite. Summing all the
diagrams, we get
\be
\langle 0|\exp(- H \tau )|0\rangle 
 \;=\; \sqrt{\frac{\omega}{\pi}} \exp\left(-\frac{\omega \tau }{2}\right)
 \left[1-\left(\frac{3\beta}{4\omega^2} - \frac{11\alpha^2}{8\omega^4}
  \right) \tau +\ldots\right]
\ee
For small $\alpha^2,\beta$ and $\tau$ not too large, we can exponentiate 
the result and read off the correction to the ground state energy
\be
\label{delE_osc}
E_0 &=& \frac{\omega}{2}+ \frac{3\beta}{4 \omega^2}-
 \frac{11\alpha^2}{8\omega^4}+\ldots .
\ee
Of course, we could have obtained the result using ordinary 
Rayleigh-Schr{\"o}dinger perturbation theory, but the method
discussed here proves to be much more powerful when we come
to non-perturbative effects and field theory.

   One more simple exercise is worth mentioning: the evaluation of first 
perturbative correction to the Green's function. The diagrams shown in 
Fig. \ref{fig_qm_graphs}b give
\be
\label{delG_osc}
\Delta G_0(0,\tau) &=& {9\alpha^2 \over 4 \omega^6} + 
 {\alpha^2 \over 2\omega^6}e^{-2\omega \tau} + {15\alpha^2 \over 
4\omega^5}\tau e^{-\omega \tau} - {3\beta \over 2 \omega^3}
\tau e^{-\omega \tau} 
\ee
Comparing the result with the decomposition in terms of stationary
states
\be 
 G(0,\tau)&=& \sum_{n=0}^{\infty} e^{-(E_n-E_0)\tau}
    |\langle 0|x|n\rangle |^2 
\ee
we can identify the first (time independent) term with the square of the 
ground state expectation value $\langle 0|x|0\rangle$ (which is non-zero 
due to the tadpole diagram). The second term comes from the excitation of 
2 quanta, and the last two (with extra factors of $\tau$) are the lowest 
order ``mass renormalization",  or corrections to the zero order gap 
between the ground and first excited states, $E_1-E_0=\omega$.

\subsubsection{Tunneling in the double well potential}
\label{sec_qm_tun}

   Tunneling phenomena in quantum mechanics were discovered by George 
Gamow in the late 20's in the context of alpha-decay. He introduced
the exponential suppression factor that explained why a decay 
governed by the Coulomb interaction (with a typical nuclear time
scale of $10^{-22}$ sec) could lead to lifetimes of millions of
years. Tunneling is a quantum mechanical phenomenon, a particle 
penetrating a classically forbidden region. Nevertheless, we will
describe the tunneling process using classical equations of 
motion. Again, the essential idea is to continue the transition
amplitude to imaginary time. 

   Let us give a qualitative argument why tunneling can be treated 
as a classical process in imaginary time. The energy of a particle 
moving in the potential $V(x)$ is given by $E=p^2/(2m)+V(x)$, and in 
classical mechanics only regions of phase space where the kinetic 
energy is positive are accessible. In order to reach the classically 
{\it forbidden} region $E<V$, the kinetic energy would have to be 
negative, corresponding to imaginary momentum $p$. In the semi-classical
(WKB) approximation to quantum mechanics, the wave function is given
by $\psi(x)\sim\exp[i\Phi(x)]$ with $\Phi(x)=\pm\int^x dx'\,p(x')
+O(\hbar)$ where $p(x)=(2m)^{1/2}(E-V(x))^{1/2}$ is the local classical
momentum. In the classically allowed region, the wave function is 
oscillatory, while in the classically forbidden region (corresponding
to imaginary momenta) it is exponentially suppressed. 

  There is another way to introduce imaginary momenta, which is 
more easily generalized to multidimensional problems and field
theory, by considering motion in imaginary time. Continuing $\tau=it$,
the classical equation of motion is given by
\be 
m \frac{d^2x}{d\tau^2}&=&+\frac{d V}{dx} ,
\ee
where the sign of the potential energy term has changed. This means
that classically forbidden regions are now classically allowed. 
The distinguished role of the classical tunneling path becomes clear 
if one considers the Feynman path integral. Although any path is 
allowed in quantum mechanics, the path integral is dominated by
paths that maximize the weight factor $\exp(-S[x_{cl}(\tau)])$, 
or minimize the Euclidean action. The classical path is the 
path with the smallest possible action.

   Let us consider a widely used toy model, the double-well potential
\be
\label{eq_dwpot}
V &=& \lambda (x^{2}-\eta^{2})^{2}
\ee
with minima at $\pm\eta$, the two "classical vacua" of the system. 
Quantizing around the two minima, we would find two degenerate 
states localized at $x=\pm\eta$. Of course, we know that this is not
the correct result. Tunneling will mix the two states, the true ground
state is (approximately) the symmetric combination, while the first 
excited state is the antisymmetric combination of the two states.

  It is easy to solve the equations of motion in imaginary time and obtain
the classical tunneling solution\footnote{This solution is most easily
found using energy conservation $m\dot x^2/2-V(x)={\rm const}$ rather than 
the (second order) equation of motion $m\ddot x=V^\prime$. This is analogous
to the situation in field theory, where it is more convenient to use
self-duality rather than the equations of motion.}
\be 
\label{x_cl}
 x_{cl}(\tau)&=& \eta\tanh\left[ \frac{\omega}{2} (\tau-\tau_0)\right],
\ee
which goes from $x(-\infty)=-\eta$ to $x(\infty)=\eta$. Here, $\tau_0$ 
is a free parameter (the instanton center) and $\omega^{2} =8\lambda 
\eta^{2}$. The action of the solution is $S_0=\omega^3/(12 \lambda)$.  
We will refer to the path (\ref{x_cl}) as the instanton, since 
(unlike a soliton) the solution is localized in time\footnote{
In 1+1 dimensional $\phi^4$ theory, there is a soliton solution with
the same functional form, usually referred to as the kink solution.}.
An anti-instanton solution is given by $x^A_{cl}(\tau)=-x^I_{cl}(\tau)$.
It is convenient to re-scale the time variable such that $\omega=1$ 
and shift $x$ such that one of the minima is at $x=0$. In this case, 
there is only one dimensionless parameter, $\lambda$, and since $S_0
=1/(12\lambda)$, the validity of the semi-classical expansion is 
controlled by $\lambda\ll 1$.

   The semi-classical approximation to the path integral is obtained by
systematically expanding around the classical solution
\be
\label{semicl}
\langle -\eta|e^{-H\tau}|\eta\rangle &=& e^{-S_0}
 \int Dx(\tau) \, \exp\left(-\frac{1}{2}\delta x 
 \left.\frac{\delta^2 S}{\delta x^2}\right|_{x_{cl}}\!\!\!\delta x
 +\ldots \right).
\ee
Note that the linear term is absent, because $x_{cl}$ is a 
solution of the equations of motion. Also note that we implicitly
assume $\tau$ to be large, but smaller than the typical lifetime
for tunneling. If $\tau$ is larger than the lifetime, we have to 
take into account multi-instanton configurations, see below. 
Clearly, the tunneling amplitude is proportional to 
$\exp(-S_{0})$. The pre-exponent requires the calculation of 
fluctuations around the classical instanton solution. We will 
study this problem in the following section. 

\subsubsection{Tunneling amplitude at one loop order}
\label{sec_qm_1loop}

  In order to take into account fluctuations around the classical path, we
have to calculate the path integral
\be
\label{qm_1loop}
 \int [D\delta x] \exp\left( -\frac{1}{2}\int d\tau\;
  \delta x(\tau) O \delta x(\tau) \right)
\ee
where $O$ is the differential operator 
\be 
\label{O_fluc}
O&=&-\frac{1}{2} \frac{d^2}{d\tau^2} +
\left. \frac{d^2 V}{d x^2}\right|_{x=x_{cl}}.
\ee
This calculation is somewhat technical, but it provides a very good
illustration of the steps that are required to solve the more 
difficult field theory problem. We follow here the original work
\cite{Pol_77} and the review \cite{VZN_82}. A simpler method to 
calculate the determinant is described in the appendix of Coleman's 
lecture notes \cite{Col_77}.
      
   The integral (\ref{qm_1loop}) is Gaussian, so it can be done exactly. 
Expanding the differential operator $O$ in some basis $\{x_i(\tau)\}$, we
have
\be 
\int \left(\prod_n dx_n\right)\, 
  \exp\left(-\frac{1}{2}\sum_{ij} x_i O_{ij} x_j\right)
  &=& (2\pi)^{n/2} \left(\det O \right)^{-1/2} 
\ee
The determinant can be calculated by diagonalizing $O$, $O x_n(\tau)=
\epsilon_n x_n(\tau)$. This eigenvalue equation is just a one-dimensional 
Schr\"odinger equation\footnote{This particular Schr\"odinger equation 
is discussed in many text books on quantum mechanics, see e.g. Landau 
and Lifshitz.}
\be    
\label{qm_op}
\left( -{d^2 \over d\tau^2} + \omega^2 
    \left[ 1-\frac{3}{2 \cosh^2(\omega\tau/2)} \right] 
       \right) x_n(\tau) &=&\epsilon_n x_n(\tau) 
\ee
There are two bound states plus a continuum of scattering states.
The lowest eigenvalue is $\epsilon_0=0$, and the other bound state
is at $\epsilon_1={3\over 4}\omega^2$. The eigenfunction of the 
zero energy state is
\be 
\label{qm_zm}
x_0(\tau) &=&\sqrt{\frac{3\omega}{8}}\frac{1}{\cosh^2(\omega\tau/2)},
\ee
where we have normalized the wave function, $\int d\tau x_n^2=1$.
There should be a simple explanation for the presence of a zero 
mode. Indeed, the appearance of a zero mode is related to translational 
invariance, the fact that the action does not depend on the location
$\tau_0$ of the instanton. The zero mode wave function is just the
derivative of the instanton solution over $\tau_0$
\be 
\label{qm_tr_zm}
x_0(\tau)= -S_0^{-1/2} \frac{d}{d\tau_0} x_{cl}(\tau-\tau_0),
\ee
where the normalization follows from the fact that the classical
solution has zero energy. If one of the eigenvalues is zero this  
means that the determinant vanishes and the tunneling amplitude is 
infinite! However, the presence of a zero mode also implies that 
there is one direction in functional space in which fluctuations 
are large, so the integral is not Gaussian. This means that the
integral in that direction should not be performed in Gaussian 
approximation, but has to be done exactly.

   This can be achieved by replacing the integral over the expansion
parameter $c_0$ associated with the zero mode direction (we have 
parameterized the path by $x(\tau)=\sum_{n} c_n x_n(\tau)$) with an
integral over the collective coordinate $\tau_0$. Using
\be 
dx &=& \frac{dx_{cl}}{d\tau_0} d\tau_0 
  \;=\;  - \sqrt{S_0} x_0(\tau) d\tau_0 
\ee
and $dx=x_0dc_0$ we have $dc_0 = \sqrt{S_0}d\tau_0$. The functional 
integral over the quantum fluctuation is now given by
\be 
 \int [D\delta x(\tau)]\, \exp(-S)
 &=& \left[ \prod_{n>0} \left(\frac{2\pi}{\epsilon_n}\right)
  \right]^{1/2} \sqrt{S_0} \int d\tau_0,
\ee
where the first factor, the determinant with the zero mode excluded,
is often referred to as $\det'O$. The result shows that the tunneling 
amplitude grows linearly with time. This is as it should be, there is a 
finite transition probability per unit time.
   
   The next step is the calculation of the non-zero mode determinant.
For this purpose we make the spectrum discrete by considering a finite
time interval $[-\tau_m/2,\tau_m/2]$ and imposing boundary conditions 
at $\pm\tau_m/2$: $x_n(\pm\tau_m/2)=0$. The product of all eigenvalues 
is divergent, but the divergence is related to large eigenvalues, 
independent of the detailed shape of the potential. The determinant 
can be renormalized by taking the ratio over the determinant of the 
free harmonic oscillator. The result is
\be
\label{qm_det_ratio}
\left( \frac{\det\left[ -\frac{d^2}{d\tau^2}+V''(x_{cl})\right]}
            {\det\left[ -\frac{d^2}{d\tau^2}+\omega^2\right]}\right)^{-1/2}
&=& \sqrt{\frac{S_0}{2\pi}}\; \omega\int d\tau_0\;
\left( \frac{\det'\left[ -\frac{d^2}{d\tau^2}+V''(x_{cl})\right]}
 {\omega^{-2}\det\left[ -\frac{d^2}{d\tau^2}+\omega^2\right]}\right)^{-1/2}
\ee
where we have eliminated the zero mode from the determinant and replaced
it by the integration over $\tau_0$. We also have to extract the lowest 
mode from the harmonic oscillator determinant, which is given by $\omega^2$.
The next eigenvalue is $3\omega^2/4$, while the corresponding oscillator
mode is $\omega^2$ (up to corrections of order $1/\tau_m^2$, that are 
not important as $\tau_m\to\infty$). The rest of the spectrum is continuous 
as $\tau_m\to\infty$. The contribution from these states can be calculated 
as follows. 

  The potential $V''(x_{cl})$ is localized, so for $\tau\to\pm\infty$
the eigenfunctions are just plane waves. This means we can take one 
of the two linearly independent solutions to be $x_p(\tau)\sim\exp(
ip\tau)$ as $\tau\to\infty$. The effect of the potential is to give 
a phase shift
\be 
 x_p(\tau) &=& \exp(ip\tau+i\delta_p) \hspace{0.5cm} \tau\to-\infty,
\ee
where, for this particular potential, there is no reflected wave. The 
phase shift is given by \cite{LL}
\be 
\exp(i\delta_p)&=& {1+ip/\omega \over 1-ip/\omega}
                   {1+2ip/\omega \over 1-2ip/\omega} .
\ee
The second independent solution is obtained by $\tau \rightarrow -\tau$.
The spectrum is determined by the quantization condition $x(\pm\tau_m/2)
=0$, which gives 
\be 
\label{quant}
 p_n\tau_m - \delta_{p_n} = \pi n, 
\ee
while the harmonic oscillator modes are determined by $p_n\tau_m=\pi n$.
If we denote the solutions of (\ref{quant}) by $\tilde p_n$, the ratio of 
the determinants is given by
\be 
\prod_n \left[{\omega^2 + \tilde p_n^2 \over \omega^2 +  p_n^2 }\right]
&=& \exp\left( \sum_{n} \log\left[ \frac{\omega^2 + \tilde p_n^2}
 {\omega^2 +  p_n^2}\right]\right)\; =\;
 \exp\left( {1\over \pi}\int_0^\infty 
  {2p dp\, \delta_p \over p^2 +\omega^2}\right) \;= \; \frac{1}{9} .
\ee
where we have expanded the integrand in the small difference 
$\tilde p_n-p_n=\delta_{p_n}/\tau_m$ and changed from summation 
over $n$ to an integral over $p$. In order to perform the integral, 
it is convenient to integrate by part and use the result for 
$(d\delta_p)/(dp)$. Collecting everything, we finally get
\be 
\label{qm_amp}
 \langle -\eta|e^{-H\tau_m}|\eta \rangle &=& 
 \left[\sqrt{\frac{\omega}{\pi}} \exp\left(-\frac{\omega\tau_m}{2}
 \right)\right]
 \left[\sqrt{\frac{6 S_0}{\pi}} \exp\left(-S_0\right)\right]
 \left(\omega \tau_m\right), 
\ee
where the first factor comes from the harmonic oscillator amplitude
and the second is the ratio of the two determinants. 
  
 The result shows that the transition amplitude is proportional to
the time interval $\tau_m$. In terms of stationary states, this is 
due to the fact that the contributions from the two lowest states
almost cancel each other. The ground state wave function is the
symmetric combination $\Psi_0(x)=(\phi_{-\eta}(x)+\phi_\eta(x))/
\sqrt{2}$, while the first excited state $E_1=E_0+\Delta E$ is 
antisymmetric, $\Psi_1(x)=(\phi_{-\eta}(x)-\phi_\eta(x))/\sqrt{2}$. 
Here,  $\phi_{\pm\eta}$ are the harmonic oscillator wave functions
around the two classical minima. For times $\tau\ll 1/\Delta E$, 
the tunneling amplitude is given by 
\be
 \langle -\eta|e^{-H\tau_m}|\eta \rangle &=&
 \Psi_0^*(-\eta)\Psi_0(\eta)e^{-E_0\tau_m}+
 \Psi_1^*(-\eta)\Psi_1(\eta)e^{-E_1\tau_m}+\ldots  \nonumber \\
\label{qm_modes}
 &=& \frac{1}{2}\phi^*_{-\eta}(-\eta)\phi_{\eta}(\eta)
     (\Delta E\tau_m) e^{-\omega\tau_m/2} +\ldots . 
\ee
Note that the validity of the semi-classical approximation requires 
$\tau\gg 1/\omega$.

  We can read off the level-splitting from (\ref{qm_amp}) and 
(\ref{qm_modes}). The result can be obtained in an even more 
elegant way by going to large times $\tau>1/\Delta E$. In this case, 
multi-instanton paths are important. If we ignore the interaction 
between instantons, multi instanton contributions can easily be 
summed up
\be
\langle -\eta|e^{-H\tau_m}|\eta\rangle &=&
  \sqrt{\frac{\omega}{\pi}} e^{-\omega\tau_m/2}  
  \sum_{n\,{\rm odd}}
  \int\limits_{-\tau_m/2<\tau_1<\ldots<\tau_m/2}
 \left[\prod_{i=1}^{n}\omega d\tau_i\right]
 \left(\sqrt{\frac{6S_0}{\pi}}\exp(-S_0)\right)^n \\
 &=&  \sqrt{\frac{\omega}{\pi}} e^{-\omega\tau_m/2}
  \sum_{n\,{\rm odd}}  \frac{(\omega\tau_m d)^n}{n!}
 \;=\;  \sqrt{\frac{\omega}{\pi}} e^{-\frac{\omega\tau_m}{2}}
  \sinh\left(\omega\tau_m d \right), \nonumber
\ee 
where $d=(6S_0/\pi)^{1/2}\exp(-S_0)$. Summing over all instantons
simply leads to the exponentiation of the tunneling rate. Now we 
can directly read off  the level splitting
\be
\label{qm_split}
\Delta E &=& \sqrt{\frac{6 S_0}{\pi}}\omega\exp\left(-S_0\right).
\ee
If the tunneling rate increases, $1/\Delta E\simeq 1/\omega$, interactions
between instantons become important. We will study this problem in Sec.
\ref{sec_qm_IA}.
  
\subsubsection{The tunneling amplitude at two-loop order}
\label{sec_qm_twoloop}

    The WKB method can be used to systematically calculate higher 
orders in $1/S_0$. Beyond leading order, however, the WKB method 
becomes quite tedious, even when applied to quantum mechanics. In 
this subsection we show how the $1/S_0$ correction to the level 
splitting in the double well can be determined using a two loop 
instanton calculation. We follow here \cite{WS_94}, which corrected 
a few mistakes in the earlier literature \cite{AS_87,Ole_89}. 
Numerical simulations were performed in \cite{Shu_88c} while the 
correct result was first obtained using different methods by \cite{ZJ_81}. 

\begin{figure}[t]
\begin{center}
\leavevmode
\epsfxsize=12cm
\epsffile{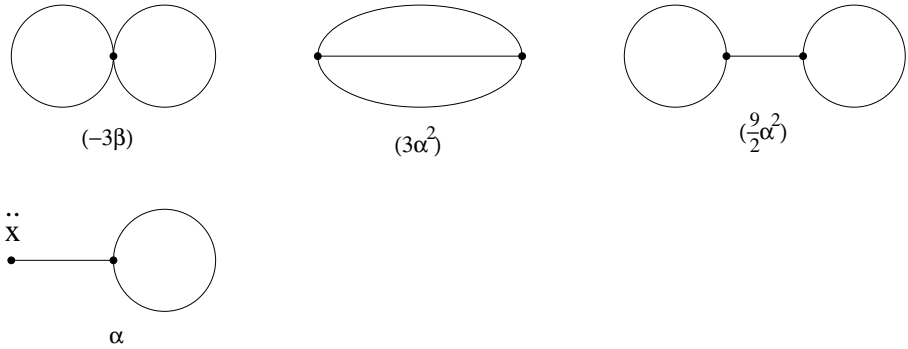}
\end{center}
\caption{\label{fig_twoloop}
Feynman diagrams for the two-loop correction to the tunneling amplitude 
in the quantum mechanical double well potential. The first three correspond 
to the diagrams in Fig. \protect\ref{fig_qm_graphs}a, but with different
propagators and vertices, while the forth diagram contains a new vertex,
generated by the collective coordinate Jacobian.}
\end{figure}

  To next order in $1/S_0$, the tunneling amplitude can be decomposed as
\be
\label{eq_twoloopexp}
\langle -\eta | e^{-H\tau} | \eta \rangle 
 &\simeq& \left|\psi_{0}(\eta)\right|^{2}  
 \left(1 + \frac{2A}{S_{0}} + \ldots\right) 
 \exp\left(-\frac{\omega\tau}{2} \left[1 + \frac{B}{S_{0}} + \ldots
 \right]\right) \Delta E_0 \left(1 + \frac{C}{S_{0}} +\ldots\right) \tau,
\ee
where we are interested in the coefficient $C$, the next order correction
to the level splitting. The other two corrections, $A$ and $B$ are 
unrelated to tunneling and we can get rid of them by dividing the 
amplitude by $\langle \eta|\exp(-H\tau)|\eta\rangle$, see (\ref{delG_osc}).

   In order to calculate the next order correction to the instanton
result, we have to expand the action beyond order $(\delta x)^2$.
The result can be interpreted in terms of a new set of Feynman rules
in the presence of an instanton (see Fig. \ref{fig_twoloop}). The triple 
and quartic coupling constants are $\alpha= 4 \lambda x_{cl}(t)$
and $\beta=\lambda$ (compared to $\alpha_0=4\lambda\eta=\sqrt{2\lambda}$ 
and $\beta_0=\lambda$ for the anharmonic oscillator). The propagator is 
the Green's function of the differential operator (\ref{O_fluc}). There 
is one complication due to the fact that the operator $O$ has a zero mode. 
The Green's function is uniquely defined by requiring it to be 
orthogonal to the translational zero mode. The result is \cite{Ole_89} 
\be
\label{G_qm_inst}
(2\omega) G(x,y) &=& g_{0}(x,y)\left[ 2-xy+ \frac{1}{4} |x-y| (11-3xy) 
  + (x-y)^2\right]
  + \frac{3}{8} \left(1-x^2\right)\left(1-y^2\right)
   \left[\log\left(g_{0}(x,y)\right) - \frac{11}{3}\right] \\
 & & \hspace{1cm} 
 g_{0}(x,y)\;=\; \frac{1-|x-y| - xy}{1 +|x-y|- xy} 
\ee
where $x =\tanh(\omega t/2),\, y=\tanh(\omega t'/2)$ and $g_0(x,y)$
is the Green's function of the harmonic oscillator (\ref{qm_prop}).
There are four diagrams at two-loop order, see Fig. \ref{fig_twoloop}. 
The first three diagrams are of the same form as the anharmonic oscillator 
diagrams. Subtracting these contributions, we get
\be
\label{2loop_diag}
a_{1} &=& -3 \beta_0 \int^{\infty}_{-\infty}dt\, \left( G^{2}(t,t) - 
  G^{2}_{0}(t,t)\right) \;=\; -\frac{97}{1680}S_0^{-1} \\
b_{11} &=& 3 \alpha_0^2\int^{\infty}_{-\infty} \int^{\infty}_{-\infty}
  dt dt'\,\left( \tanh(t/2)\tanh(t'/2)G^{3}(t,t') - G^{3}_{0}(t,t')\right) 
  \;=\;  -\frac{53}{1260}S_0^{-1} \\
b_{12} &=& \frac{9}{2} \alpha_0^2\int^{\infty}_{-\infty}
 \int^{\infty}_{-\infty} dt dt'\,\left( \alpha^{2}\tanh(t/2) \tanh(t'/2)
 G(t,t)G(t,t')G(t',t') - G_{0}(t,t)G_{0}(t,t')G_{0}(t',t')\right) 
 \; =\; -\frac{39}{560}S_0^{-1}.
\ee
The last diagram comes from expanding the Jacobian in $\delta x$. 
This leads to a tadpole graph proportional to $\ddot x_{cl}$,
which has no counterpart in the anharmonic oscillator case. We get
\be
\label{2loop_jac}
c_{1} &=& -9 \beta_0 \int^{\infty}_{-\infty} \int^{\infty}_{-\infty}
 dt dt'\,\frac{\tanh(t/2)}{\cosh^{2}(t/2)}\tanh(t'/2)G(t,t')G(t',t') 
 \;=\; -\frac{49}{60}S_0^{-1}.
\ee
The sum of the four diagrams is $C=(a_{1}+b_{11}+b_{12}+c_{1})S_0=
-71/72$. The two-loop result for the level splitting is 
\be
\label{qm_split_2}
\Delta E &=& \sqrt{\frac{6 S_0}{\pi}}\omega
  \exp\left(-S_0-\frac{71}{72}\frac{1}{S_0}+\ldots\right) ,
\ee
in agreement with the WKB result obtained in \cite{ZJ_81}. The fact that 
the next order correction is of order one and negative is significant.
It implies that the one-loop result becomes inaccurate for moderately 
large barriers ($S\sim 1$), and that it overestimates the tunneling 
probability. We have presented this calculation in detail in order 
to show that the instanton method can be systematically extended
to higher orders in $1/S$. In field theory, however, this calculation
is sufficiently difficult that it has not yet been performed.

\subsubsection{Instanton-anti-instanton interaction and the ground
state energy} 
\label{sec_qm_IA}

   Up to now we focused on the tunneling amplitude for transitions
between the two degenerate vacua of the double well potential. This  
amplitude is directly related to the gap $\Delta E$ between the 
ground state and the first excited state. In this subsection we wish 
to discuss how the semi-classical theory can be used to calculate the
mean $E_{ctr}=(E_0+E_1)/2$ of the two levels. In this context, it is 
customary to define the double well potential by $V=(x^2/2)(1-gx)^2$. 
The coupling constant $g$ is related to the coupling $\lambda$ used
above by $g^2=2\lambda$. Unlike the splitting, the mean energy is 
related to topologically trivial paths, connecting the same vacua. 
The simplest non-perturbative path of this type is an 
instanton-anti-instanton pair. 

  In section \ref{sec_qm_1loop} we calculated the tunneling amplitude
using the assumption that instantons do not interact with each other. 
We found that tunneling makes the coordinates uncorrelated, and leads 
to a level splitting. If we take the interaction among instantons into 
account, the contribution from instanton-anti-instanton pairs is given by
\be 
\label{qm_ia_cont}
\langle\eta|e^{-H\tau_m}|\eta\rangle &=&
 \tau_m \int \frac{d\tau}{\pi g^2} \,\exp\left( S_{IA}(\tau)
 \right),
\ee
where $S_{IA}(\tau)$ is the action of an instanton-anti-instanton pair 
with separation $\tau$ and the prefactor $(\pi g^2)^{-1}$ comes from the 
square of the single instanton density. The action of an instanton
anti-instanton (IA) pair can be calculated given an ansatz for the 
path that goes from one minimum of the potential to the other and
back. An example for such a path is the ``sum ansatz" \cite{ZJ_83}
\be
\label{qm_sum}
 x_{sum}(\tau) &=& \frac{1}{2g} \left( 2 
 -\tanh\left(\frac{\tau-\tau_I}{2}\right)
 +\tanh\left(\frac{\tau-\tau_A}{2}\right)  \right).
\ee
This path has the action $S_{IA}(T)=1/g^2(1/3-2e^{-T}+O(e^{-2T}))$,
where $T=|\tau_I-\tau_A|$. It is qualitatively clear that if the two 
instantons are separated by a large time interval $T \gg 1$, the action 
$S_{IA}(T)$ is close to $2S_0$. In the opposite limit $T\rightarrow 0$, 
the instanton and the anti-instanton annihilate and the action $S_{IA}(T)$ 
should tend to zero. In that limit, however, the IA pair is at best 
an approximate solution of the classical equations of motion and it
is not clear how the path should be chosen. 

\begin{figure}[t]
\begin{center}
\leavevmode
\epsfxsize=12cm
\epsffile{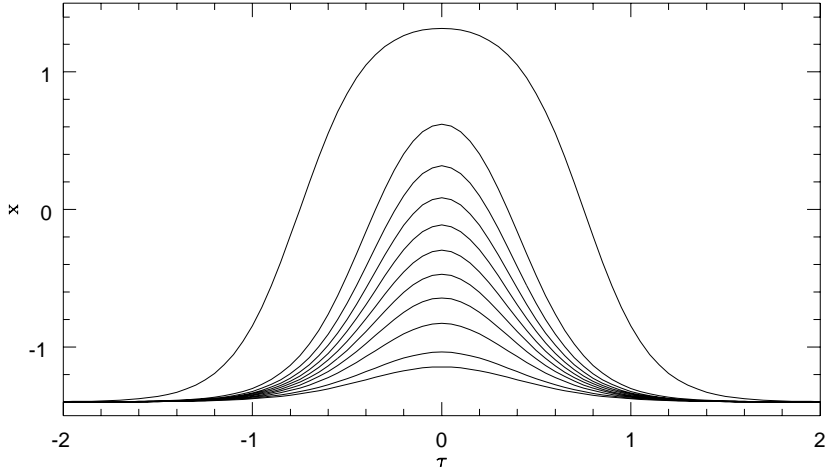}
\end{center}
\vspace*{-2cm}
\caption{\label{fig_dw_stream} 
Streamline configurations in the double well potential for $\eta=1.4$ and 
$\lambda=1$ ($\omega\simeq 4$, $S_0\simeq 5$), adapted from 
\protect\cite{Shu_88c}. 
The horizontal axis shows the time coordinate and the vertical axis the  
amplitude $x(\tau)$. The different paths correspond to different values 
of the streamline parameter $\lambda$ as the configuration evolves from 
a well separated pair to an almost perturbative path. The initial path 
has an action $S=1.99S_0$, the other paths correspond to a fixed 
reductions of the action by $0.2S_0$.}
\end{figure}

\begin{figure}[t]
\begin{center}
\leavevmode
\epsfxsize=12cm
\epsffile{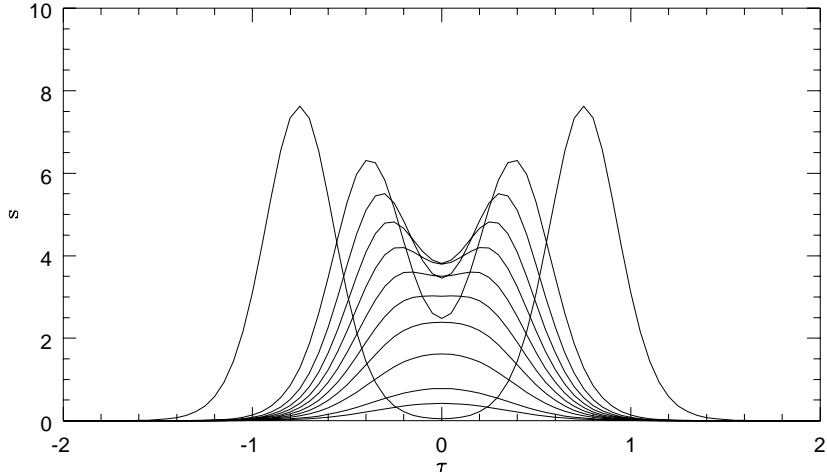}
\end{center}
\vspace*{-2cm}
\caption{\label{fig_dw_act}
Distribution of the action density $s=\dot x^2/2+V(x)$ for the 
streamline configurations shown in Fig. \protect\ref{fig_dw_stream}. }
\end{figure}

  The best way to deal with this problem is the ``streamline" or ``valley"
method \cite{BY_86}. In this approach one starts with a well separated IA 
pair and lets the system evolve using the method of steepest descent. This
means that we have to solve
\be
\label{qm_stream}
f(\lambda)\frac{dx_\lambda(\tau)}{d\lambda} &=& 
  \frac{\delta S}{\delta x_\lambda (\tau)},
\ee
where $\lambda$ labels the path as we proceed from the initial 
configuration $x_{\lambda=0}(\tau)=x_{sum}(\tau)$ down the valley to 
the vacuum configuration and $f(\lambda)$ is an arbitrary function
that reflects the reparametrization invariance of the streamline 
solution. A sequence of paths obtained by solving the
streamline equation (\ref{qm_stream}) numerically is shown in Fig. 
\ref{fig_dw_stream} \cite{Shu_88c}. An analytical solution to first
order in $1/S_0$ can be found in \cite{BY_86}. The action density 
$s=\dot x^2/2+V(x)$ corresponding to the paths in Fig. \ref{fig_dw_stream}
is shown in Fig. \ref{fig_dw_act}. We can see clearly how the two 
localized solutions merge and eventually disappear as the configuration 
progresses down the valley. Using the streamline solution, the 
instanton-anti-instanton action for large $T$ is given by \cite{FS_95} 
\be
\label{eq_dwlargeR}
S(T) &=& \frac{1}{g^2} \Big[\frac{1}{3} -2e^{-T}-12Te^{-2T}+
 O(e^{-2T})\Big] + \Big[24 T e^{-T} +O(e^{-T})\Big]
 + \Big[ \frac{71g^2}{6} +O(g^4) \Big], 
\ee 
where the first term is the classical streamline interaction up to 
next-to-leading order, the second term is the quantum correction 
(one loop) to  the leading order interaction, and the last term is 
the two-loop correction to the single instanton interaction.

  If one tries to use the instanton result (\ref{qm_ia_cont}) in
order to calculate corrections to $E_{ctr}$ one encounters two
problems. First, the integral diverges at large $T$. This is 
simply related to the fact that IA pairs with large separation
should not be counted as pairs, but as independent instantons. 
This problem is easily solved, all one has to do is subtract the 
square of the single instanton contribution. Second, once this
subtraction has been performed, the integral is dominated by the 
region of small $T$, where the action is not reliably calculable. 
This problem is indeed a serious one, related to the fact that 
$E_{ctr}$ is not directly related to tunneling, but is dominated
by perturbative contributions. In general, we expect $E_{ctr}$
to have an expansion
\be
\label{dw_exp}
  E_{ctr} &=& \sum_k g^{2k} E^{(0)}_{ctr,\, k}
       + e^{-2/(6g^2)} \sum_k g^{2k} E^{(2)}_{ctr,\, k} + \ldots ,
\ee
where the first term is the perturbative contribution, the second
corresponds to one IA pair, and so on. However, perturbation theory in $g$ 
is divergent (not even Borel summable), so the calculation of the 
IA contribution requires a suitable definition of perturbation theory. 

  One way to deal with this problem (going back to Dyson's classical 
work on QED) is analytic continuation in $g$. For $g$ imaginary 
($g^2$ negative), the $IA$ contribution is well defined (the integral 
over $T$ is dominated by $T\sim -\log(-g^2)$). The IA contribution to 
$E_{ctr}$ is \cite{Bog_80,ZJ_83}
\be
\label{E_ctr_IA}
E^{(2)}_{ctr} &=& \frac{e^{-1/(3g^2)}}{\pi g^2} \left[
 \log\left(-\frac{2}{g^2}\right) +\gamma + O(g^2\log(g^2))
 \right],
\ee
where $\gamma=0.577\ldots$ is Euler's constant. When we now continue
back to positive $g^2$, we get both real and imaginary contributions
to $E_{ctr}$. Since the sum of all contributions to $E_{ctr}$ is 
certainly real, the imaginary part has to cancel against a small 
$O(e^{-1/(3g^2)})$ imaginary part in the perturbative expansion. This 
allows us to determine the imaginary part ${\rm Im} E_{ctr}^{(0)}$
of the analytically continued perturbative sum\footnote{How can the 
perturbative result develop an imaginary part? After analytic 
continuation, the perturbative sum is Borel summable, because 
the coefficients alternate in sign. If we define $E^{(0)}_{ctr}$
by analytic continuation of the Borel sum, it will have an 
imaginary part for positive $g^2$.}. 

  From the knowledge of the imaginary part of perturbation theory,
one can determine the large order behavior of the perturbation
series $E^{(0)}_{ctr}=\sum_k g^{2k}E^{(0)}_{ctr,\,k}$ \cite{Lip_77,BPZ_77}. 
The coefficients are given by the dispersion integrals
\be 
\label{E_k_disp}
E^{(0)}_{ctr,\,k} &=&  \frac{1}{\pi} \int_0^\infty {\rm Im}\left(
 E^{(0)}_{ctr,\,k}(g^2)\right)\frac{dg^2}{g^{2k+2}}.
\ee
Since the semi-classical result (\ref{E_ctr_IA}) is reliable for 
small $g$, we can calculate the large order coefficients. Including
the corrections calculated in \cite{FS_95}, we have 
\be
E^{(0)}_{ctr,\,k} &=&  {3^{k+1} k\! \over \pi}
  \left( 1-\frac{53}{18k}+\ldots\right).
\ee
The result can be compared with the exact coefficients \cite{BPZ_77}. 
For small $k$ the result is completely wrong, but already for
$k=5,6,7,8$ the ratio of the asymptotic result to the exact 
coefficients is $1.04,\,1.11,\,1.12,\,1.11$. We conclude that
instantons determine the large order behavior of the perturbative 
expansion. This is in fact a generic result: the asymptotic
behavior of perturbation theory is governed by semi-classical 
configurations (although not necessarily involving instantons). 

\begin{table}[t]
\caption{\label{tab_silvestrov}
Exact center of the band energies $E_{ctr}=(E_0+E_1)/2$ for
different values of $g^2$ (expressed in terms of $N=1/(3g^2)$)
compared to the semiclassical estimate discussed in the text.}
\begin{tabular}{c|ccccc}
$N \equiv 1/(3g^2)$ & 4 & 6 & 8 & 10 & 12 \\ \tableline
$E_{ctr}^{ex}$ & 0.4439 & 0.43797 & 0.44832 & 0.459178 & 0.467156 \\
$E_{ctr}^{th}$ & 0.4367 & 0.44367 & 0.44933 & 0.459307 & 0.467173 
\end{tabular}
\end{table}

   In order to check the instanton-anti-instanton result (\ref{E_ctr_IA})
against the numerical value of $E_{ctr}$ for different values of $g$
we have to subtract the perturbative contribution to $E_{ctr}$. This 
can be done using analytic continuation and the Borel transform 
\cite{ZJ_82}, and the result is in very good agreement with the 
instanton calculation. A simpler way to check the instanton result
was proposed by \cite{FS_95}. These authors simply truncate the 
perturbative series at the $N$-th term. In this case, the best 
accuracy occurs when $|N-1/3g^2|\sim 1$ and the estimate for 
$E_{ctr}$ is given by
\be
\label{eq:E_pol}
E_{ctr}&=& \sum_{n=0}^N  g^{2n} E_{ctr,\,n}^{(0)} + \frac{3Ne^{-N}}{\pi}
 \left[ \log(6N)+\gamma+ \frac{1}{3}\sqrt{\frac{2\pi}{N}}\right]
 \left(1-\frac{53}{18N}\right), 
\ee
which is compared to the exact values in the table \ref{tab_silvestrov}. 
We observe that the result (\ref{eq:E_pol}) is indeed very accurate, and 
that the error is on the order of $e^{-N}= e^{-1/3g^2}$.

  In summary: $E_{ctr}$ is related to configurations with no net
topology, and in this case the calculation of instanton effects 
requires a suitable definition of the perturbation series. This
can be accomplished using analytic continuation in the coupling
constant. After analytic continuation, we can perform a reliable 
interacting instanton calculation, but the the result has an 
imaginary part. This shows that the instanton contribution by
itself is not well determined, it depends on the definition of
the perturbation sum. However, the sum of perturbative and 
non-perturbative contributions is well defined (and real)
and agrees very accurately with the numerical value of $E_{ctr}$.

   In gauge theories the situation is indeed very similar: there
are both perturbative and non-perturbative contributions to the 
vacuum energy and the two contributions are not clearly separated.
However, in the case of gauge theories, we do not know how to 
define perturbation theory, so we are not yet able to perform a reliable
calculation of the vacuum energy, similar to eq.(\ref{eq:E_pol}).

\subsection{Fermions coupled to the double well potential}
\label{sec_susy_qm}

   In this section we will consider one fermionic degree of freedom
$\psi_\alpha$ ($\alpha=1,2$) coupled to the double well potential.
This model provides additional insight into the vacuum structure
not only of quantum mechanics, but also of gauge theories: we will
see that fermions are intimately related to tunneling, and that 
the fermion-induced interaction between instantons leads to strong 
instantons-anti-instantons correlations. Another motivation for 
studying fermions coupled to the double well potential is that 
for a particular choice of the coupling constant, the theory is 
supersymmetric. This means that perturbative corrections to the
vacuum energy cancel, and the instanton contribution is more
easily defined.

 The model is defined by the action
\be
\label{susy_qm}
S&=&\frac{1}{2}\int dt\,\left( \dot x^2+ {W^\prime}^2 +\psi\dot\psi 
 + c W^{\prime\prime}\psi\sigma_2\psi \right),
\ee
where $\psi_\alpha\,(\alpha=1,2)$ is a two component spinor,
dots denote time and primes spatial derivatives, and $W^\prime
=x(1-gx)$. We will see that the vacuum structure depends crucially 
on the Yukawa coupling $c$. For $c=0$ fermions decouple and we recover 
the double well potential studied in the previous sections, while 
for $c=1$ the classical action is supersymmetric. The supersymmetry 
transformation is given by
\be 
 \delta x = \zeta \sigma_2 \psi,\hspace{1cm}
 \delta \psi = \sigma_2\zeta\dot x -W^\prime \zeta,
\ee
where $\zeta$ is a Grassmann variable. For this reason, $W$ is usually
referred to as the superpotential. The action (\ref{susy_qm}) can be 
rewritten in terms of two bosonic partner potentials \cite{SH_81,CKS_95}. 
Nevertheless, it is instructive to keep the fermionic degree freedom, 
because the model has many interesting properties that also apply to QCD, 
where the action cannot be bosonized.

   As before, the potential $V=\frac{1}{2}{W^\prime}^2$ has degenerate 
minima connected by the instanton solution. The tunneling amplitude is 
given by
\be
\label{susy_qm_amp}
{\rm Tr}\left(e^{-\beta H}\right) &=& \int d\tau\, J\,
 \frac{\sqrt{\det{\cal O}_F }}{\sqrt{\det{\cal O}^\prime_B}} \, e^{-S_{cl}},
\ee
where $S_{cl}$ is the classical action, ${\cal O}_B$ is the bosonic
operator (\ref{O_fluc}) and ${\cal O}_F$ is the Dirac operator
\be
\label{susy_dirac_op}
{\cal O}_F &=& \frac{d}{dt}+c\sigma_2 W^{\prime\prime}(x_{cl}) .
\ee
As explained in Sec. \ref{sec_qm_1loop}, ${\cal O}_B$ has a zero mode, 
related to translational invariance. This mode has to be treated 
separately, which leads to a Jacobian $J$ and an integral over the 
corresponding collective coordinate $\tau$. The fermion determinant 
also has a zero mode\footnote{In the supersymmetric case, the fermion
zero mode is the super partner of the translational zero mode.}, given by
\be
\label{susy_qm_zm}
\chi^{I,A} &=& N \exp\left(\mp \int_{-\infty}^{t}dt'\, 
  cW^{\prime\prime}(x_{cl}) \right) \frac{1}{\sqrt{2}}
  \left(\begin{array}{c} 1\\ \mp i \end{array}\right).
\ee
Since the fermion determinant appears in the numerator of the tunneling
probability, the presence of a zero mode implies that the tunneling
rate is zero!   
 
  The reason for this is simple: the two vacua have different fermion 
number, so they cannot be connected by a bosonic operator. The tunneling 
amplitude is non-zero only if a fermion is created during the process, 
$\langle 0,+|\psi_+ |0,-\rangle$, where $\psi_\pm=\frac{1}{\sqrt{2}}
(\psi_1\pm i\psi_2)$ and $|0,\pm\rangle$ denote the corresponding 
eigenstates. Formally, we get a finite result because the fermion 
creation operator absorbs the zero mode in the fermion determinant.
As we will see later, this mechanism is completely analogous to the 
axial $U(1)_A$ anomaly in QCD and baryon number violation in electroweak 
theory. For $c=1$, the tunneling rate is given by \cite{SH_81}
\be
\label{susy_qm_rate}
\langle 0,+|\psi_+|0,-\rangle &=& \frac{1}{\sqrt{\pi g^2}}
 e^{-\frac{1}{6g^2}}
\ee
This result can be checked by performing a direct calculation using
the Schr{\"o}dinger equation. 

   Let us now return to the calculation of the ground state energy. 
For $c=0$, the vacuum energy is the sum of perturbative contributions
and a negative non-perturbative shift $O(e^{-1/(6g^2)})$ due to individual
instantons. For $c\neq 0$, the tunneling amplitude (\ref{susy_qm_rate}) 
will only enter squared, so one needs to consider instanton-anti-instanton 
pairs. Between the two tunneling events, the system has an excited
fermionic state, which causes a new interaction between the instantons. 
For $c=1$, supersymmetry implies that all perturbative contributions
(including the zero-point oscillation) to the vacuum energy cancel.
Using supersymmetry, one can calculate the vacuum energy from the 
tunneling rate\footnote{The reason is that for SUSY theories, the 
Hamiltonian is the square of the SUSY generators $Q_\alpha$, $H=
\frac{1}{2}\{Q_+,Q_-\}$. Since the tunneling amplitude $\langle 0,+
|\psi_+|0,-\rangle$ is proportional to the matrix element of $Q_+$ between
the two different vacua, the ground state energy is determined by
the square of the tunneling amplitude.} (\ref{susy_qm_rate}) \cite{SH_81}. 
The result is $O(e^{-1/(3g^2)})$ and positive, which implies that 
supersymmetry is broken\footnote{This was indeed the first known example 
of non-perturbative SUSY breaking \cite{Wit_81}.}. While the dependence on
$g$ is what we would expect for a gas of instanton-anti-instanton
molecules, understanding the sign in the context of an instanton 
calculation is more subtle (see below).

  It is an instructive exercise to calculate the vacuum energy
numerically (which is quite straightforward, we are still dealing 
with a simple quantum mechanical toy model). In general, the vacuum 
energy is non-zero, but for $c\to 1$, the vacuum energy is zero 
up to exponentially small corrections. Varying the coupling constant,
one can verify that the vacuum energy is smaller than any power in
$g^2$, showing that supersymmetry breaking is a non-perturbative 
effect.

  For $c\neq 1$ the instanton-anti-instanton contribution to the 
vacuum energy has to be calculated directly. Also, even for $c=1$,
where the result can be determined indirectly, this is a very 
instructive calculation. For an instanton-anti-instanton path, 
there is no fermionic zero mode. Writing the fermion determinant 
in the basis spanned by the original zero modes of the individual 
pseudo-particles, we have 
\be
\label{qm_det_zmz}
{\det}\left({\cal O}_F\right)_{ZMZ} &=& \left(
 \begin{array}{cc} 0 & T_{IA} \\ T_{AI} & 0 \end{array}\right),
\ee 
where $T_{IA}$ is the overlap matrix element
\be
\label{qm_T_IA}
T_{IA} &=& \int_{-\infty}^{\infty}dt\, \chi_A \left( \partial_t
 + c\sigma_2 W^{\prime\prime}(x_{IA}(t)) \right) \chi_I .
\ee
Clearly, mixing between the two zero modes shifts the eigenvalues
away from zero and the determinant is non-zero. As before, we have 
to choose the correct instanton-anti-instanton path $x_{IA}(t)$ in 
order to evaluate $T_{IA}$. Using the valley method introduced in 
the last section the ground state energy is given by \cite{BY_86}
\be
\label{susy_qm_ia}
 E &=& \frac{1}{2} \left(1-c+O(g^2)\right)
  -\frac{1}{2\pi}e^{-\frac{1}{3g^2}} 
 \left(\frac{g^2}{2}\right)^{c-1}2c^c
 \int_{0}^{\infty} d\tau\,
 \exp\left(-2c(\tau-\tau_0)+\frac{2}{g^2}e^{-2\tau}\right),
\ee
where $\tau$ is the instanton-anti-instanton separation and 
$\exp(-2\tau_0)=cg^2/2$. The two terms in the exponent inside 
the integral corresponds to the fermionic and bosonic interaction 
between instantons. One can see that fermions cut off the integral 
at large $\tau$. There is an attractive interaction which grows 
with distance and forces instantons and anti-instantons to be 
correlated. Therefore, for $c\neq 0$ the vacuum is no longer an 
ensemble of random tunneling events, but consists of correlated 
instanton-anti-instanton molecules.

   The fact that both the bosonic and fermionic interaction is 
attractive means that the integral (\ref{susy_qm_ia}), just like
(\ref{qm_ia_cont}), is dominated by small $\tau$ where the integrand 
is not reliable. This problem can be solved as outlined in the last
section, by analytic continuation in the coupling constant. As an
alternative, Balitsky and Yung suggested to shift the integration
contour in the complex $\tau$-plane, $\tau\rightarrow \tau+i\pi/2$. 
On this path, the sign of the bosonic interaction is reversed
and the fermionic interaction picks up a phase factor $\exp(ic\pi)$.
This means that there is a stable saddle point, but the instanton
contribution to the ground state energy is in general complex. 
The imaginary part cancels against the imaginary part of the 
perturbation series, and only the sum of the two contributions
is well defined.

  A special case is the supersymmetric point $c=1$. In this 
case, perturbation theory vanishes and the contribution from 
instanton-anti-instanton molecules is real,
\be
E = \frac{1}{2\pi} e^{-\frac{1}{3g^2}}\left(1+O(g^2)\right).
\ee  
This implies that at the SUSY point $c=1$, there is a well 
defined instanton-anti-instanton contribution. The result 
agrees with what one finds from the $H=\frac{1}{2}\{Q_+,Q_-\}$
relation or directly from the Schr\"odinger equation. 

  In summary: In the presence of light fermions, tunneling is 
possible only if the fermion number changes during the transition.
Fermions create a long-range attractive interaction between instantons 
and anti-instantons and the vacuum is dominated by instanton-anti-instanton 
``molecules''. It is non-trivial to calculate the contribution of these
configurations to the ground state energy, because topologically trivial
paths can mix with perturbative corrections. The contribution of 
molecules is most easily defined if one allows the collective coordinate 
(time separation) to be complex. In this case, there exists a saddle point 
where the repulsive bosonic interaction balances the attractive fermionic 
interaction and molecules are stable. These objects give a non-perturbative 
contribution to the ground state energy, which is in general complex, except 
in the supersymmetric case where it is real and positive. 

\subsection{Tunneling in Yang-Mills theory}
\label{sec_tunelingYM}
\subsubsection{Topology and classical vacua}
\label{sec_topology}

   Before we study tunneling phenomena in Yang-Mills theory, we have 
to become more familiar with the classical vacuum of the theory. In
the Hamiltonian formulation, it is convenient to use the temporal
gauge $A_0=0$ (here we use matrix notation $A_i=A_i^a\lambda^a/2$, where 
the $SU(N)$ generators satisfy $[\lambda^a,\lambda^b]=2if^{abc}\lambda^c$
and are normalized according to ${\rm Tr}(\lambda^a\lambda^b)=2\delta^{ab}$). 
In this case, the conjugate momentum to the field variables $A_i(x)$ is 
just the electric field $E_i= \partial_0 A_i$. The Hamiltonian is given by  
\be
\label{H_QCD}
 H&=&\frac{1}{2g^2}\int d^3x\,(E_i^2+B_i^2), 
\ee
where $E_i^2$ is the kinetic and $B_i^2$ the potential energy term. 
The classical vacuum corresponds to configurations with zero field 
strength. For non-abelian gauge fields this does not imply that the 
potential has to be constant, but limits the gauge fields to be 
``pure gauge"
\be 
\label{pure_gauge}
A_i &=& i U(\vec x) \partial_i U(\vec x)^\dagger .
\ee
In order to enumerate the classical vacua we have to classify all
possible gauge transformations $U(\vec x)$. This means that we have 
to study equivalence classes of maps from 3-space $R^3$ into the
gauge group $SU(N)$. In practice, we can restrict ourselves to
matrices satisfying $U(\vec x)\to 1$ as $x\to\infty$ 
\cite{CDG_78}. Such mappings can be classified using an integer 
called the winding (or Pontryagin) number, which counts how many times 
the group manifold is covered
\be 
\label{win_num}
n_{W} &=& \frac{1}{24\pi^2}\int d^3x\, \epsilon^{ijk}{\rm Tr}
\left[(U^\dagger \partial_i U)(U^\dagger \partial_j U)
      (U^\dagger \partial_k U)\right] .
\ee
In terms of the corresponding gauge fields, this number is the 
Chern-Simons characteristic
\be
\label{cs_num}
n_{CS} &=& \frac{1}{16\pi^2} \int d^3x\, \epsilon^{ijk}\left(
 A_i^a\partial_j A_k^a +\frac{1}{3}f^{abc}A_i^a A_j^b
 A_k^c \right).
\ee
Because of its topological meaning, continuous deformations of the 
gauge fields do not change $n_{CS}$. In the case of $SU(2)$, an 
example of a mapping with winding number $n$ can be found from
the ``hedgehog" ansatz
\be 
\label{hdg_hg}
U(\vec x) &=& \exp(if(r) \tau^a \hat x^a).
\ee
where $r=|\vec x|$ and $\hat x^a=x^a/r$. For this mapping, we find
\be 
n_{W}&=& \frac{2}{\pi}\int dr\, \sin^2(f) \frac{df}{dr} \,=\, 
  \frac{1}{\pi}\left[f(r)-\frac{\sin(2f(r))}{2}\right]_0^\infty .
\ee
In order for $U(\vec x)$ to be uniquely defined, $f(r)$ has to be 
a multiple of $\pi$ at both zero and infinity, so that $n_W$ is 
indeed an integer. Any smooth function with $f(r\to\infty)=0$ 
and $f(0)=n\pi$ provides an example for a function with winding 
number $n$. 

   We conclude that there is an infinite set of classical vacua 
enumerated by an integer $n$. Since they are topologically different,  
one cannot go from one vacuum to another by means of a continuous gauge 
transformation. Therefore, there is no path from one vacuum to another,
such that the energy remains zero all the way.

\subsubsection{Tunneling and the BPST instanton}
\label{sec_BPST}

  Two important questions concerning the classical vacua immediately
come to mind. First, is there some physical observable which distinguishes 
between them? Second, is there any way to go from one vacuum to another? 
The answer to the first one is positive, but most easily demonstrated in 
the presence of light fermions, so we will come to it later. Let us now 
concentrate on the second one.

  We are going to look for a tunneling path in gauge theory, which connects
topologically different classical vacua. From the quantum mechanical example
we know that we have to look for classical solutions of the euclidean
equations of motion. The best tunneling path is the solution with minimal
euclidean action connecting vacua with different Chern-Simons number.
To find these solutions, it is convenient to exploit the following identity
\be 
\label{Bog_ineq}
S &=& \frac{1}{4g^2} \int d^4x\, G^a_{\mu\nu} G^a_{\mu\nu} \;=\;
 \frac{1}{4g^2}\int d^4x\, \left[\pm G^a_{\mu\nu} \tilde G^a_{\mu\nu}
 + \frac{1}{2} \left( G^a_{\mu\nu}\mp \tilde G^a_{\mu\nu}\right)^2
  \right],
\ee
where $\tilde G_{\mu\nu}=1/2\epsilon_{\mu\nu\rho\sigma}G_{\rho\sigma}$
is the dual field strength tensor (the field tensor in which the roles
of electric and magnetic fields are interchanged). Since the first term
is a topological invariant (see below) and the last term is always 
positive, it is clear that the action is minimal if the field 
is (anti) self-dual
\be 
\label{self_dual}
G^a_{\mu \nu}&=&\pm\tilde G^a_{\mu \nu},
\ee
One can also show directly that the self-duality condition implies the
equations of motion\footnote{The reverse is not true, but one can show
that non self-dual solutions of the equations of motion are saddle 
points, not local minima of the action.}, $D_\mu G_{\mu\nu}=0$. This
is a useful observation, because in contrast to the equation of 
motion, the self-duality equation (\ref{self_dual}) is a first order
differential equation. In addition to that, one can show that the energy 
momentum tensor vanishes for self-dual fields. In particular, self-dual 
fields have zero (Minkowski) energy density. 

  The action of a self-dual field configuration is determined by the 
topological charge (or 4 dimensional Pontryagin index)
\be 
\label{top_charge}
Q&=& {1\over 32\pi^2}\int d^4x\, G^a_{\mu\nu} \tilde G^a_{\mu\nu} 
\ee
 From (\ref{Bog_ineq}), we have $S=(8\pi^2|Q|)/g^2$ for self-dual fields. For 
finite action field configurations, $Q$ has to be an integer. This can be 
seen from the fact that the integrand is a total derivative
\be 
\label{Q_surf}
Q &=& {1\over 32\pi^2}\int d^4x\, G^a_{\mu\nu} \tilde G^a_{\mu\nu}
 \;=\; \int d^4x \,\partial_\mu K_\mu  
 \;=\; \int d\sigma_\mu K_\mu ,\\
 & & \hspace{1cm} K_\mu \;=\; {1\over 16\pi^2}
 \epsilon_{\mu\alpha\beta\gamma}
 \left(A^a_\alpha \partial_\beta A^a_\gamma+ 
 {1\over 3} f^{abc} A^a_\alpha A^b_\beta A^c_\gamma\right).
\ee
For finite action configurations, the gauge potential has to be 
pure gauge at infinity $A_\mu \to iU\partial_\mu U^\dagger$. Similar
to the arguments given in the last section, all maps from the 
three sphere $S_3$ (corresponding to $|x|\to\infty$) into the 
gauge group can be classified by a winding number $n$. Inserting
$A_\mu = iU\partial_\mu U^\dagger$ into (\ref{Q_surf}) one finds  
that $Q=n$.

  Furthermore, if the gauge potential falls off sufficiently rapid 
at spatial infinity, 
\be 
Q&=& \int dt {d \over dt} \int d^3x K_0
 \;=\; n_{CS}(t=\infty)-n_{CS}(t=-\infty)
\ee
which shows that field configurations with $Q\neq 0$ connect 
different topological vacua. In order to find an explicit solution
with $Q=1$, it is useful to start from the simplest winding number 
$n=1$ configuration. Similar to (\ref{hdg_hg}), we can take 
$A_\mu=iU\partial_\mu U^\dagger$ with $U=i\hat x_\mu\tau_\mu^+$, 
where $\tau_\mu^\pm=(\vec\tau,\mp i)$. Then $A_\mu^a=2 \eta_{a\mu\nu}
x_\nu/x^2$, where we have introduced the 't Hooft symbol $\eta_{a\mu\nu}$. 
It is defined by 
\be 
\label{eta_def}
\eta_{a\mu\nu}&=&\left\{ \begin{array}{rcl}
 \epsilon_{a\mu\nu} &\hspace{0.5cm}& \mu,\,\nu=1,\,2,\,3, \\
 \delta_{a\mu}      &              & \nu=4,  \\
-\delta_{a\nu}      &              & \mu=4.
\end{array}\right.
\ee
We also define $\overline\eta_{a\mu\nu}$ by changing the sign of
the last two equations. Further properties of $\eta_{a\mu\nu}$ are 
summarized in appendix \ref{app_eta}. We can now look for a solution of 
the self-duality equation (\ref{self_dual}) using the ansatz $A_\mu^a
=2\eta_{a\mu\nu} x_\nu f(x^2)/x^2$, where $f$ has to satisfy the
boundary condition $f\to 1$ as $x^2\to\infty$. Inserting the ansatz 
in (\ref{self_dual}), we get
\be
 f(1-f)-x^2f^\prime &=& 0.
\ee
This equation is solved by $f=x^2/(x^2+\rho^2)$, which gives the BPST 
instanton solution \cite{BPS_75}
\be  
\label{BPST_inst} 
A^a_\mu(x)= \frac{2\eta_{a\mu\nu}x_\nu}
  {x^2+\rho^2}.
\ee 
Here $\rho$ is an arbitrary parameter characterizing the size of
the instanton. A solution with topological charge $Q=-1$ can be
obtained by replacing $\eta_{a\mu\nu}\to\overline\eta_{a\mu\nu}$.
The corresponding field strength is
\be 
\label{g2_inst}
(G^a_{\mu\nu})^2 &=& \frac{192\rho^4}{(x^2+\rho^2)^4} .
\ee 
In our conventions, the coupling constant only appears as a factor 
in front of the action. This convention is very convenient in
dealing with classical solutions. For perturbative calculations,
it is more common to rescale the fields as $A_\mu\to gA_\mu$. In
this case, there is a factor $1/g$ in the instanton gauge potential,
which shows that the field of the instanton is much stronger than
ordinary, perturbative, fields. 

   Also note that $G^a_{\mu\nu}$ is well localized (it falls off as 
$1/x^4$) despite the fact that the gauge potential is long-range, 
$A_\mu\sim 1/x$. The invariance of the Yang-Mills equations under 
coordinate inversion \cite{JR_76b} implies that the singularity of 
the potential can be shifted from infinity to the origin by means 
of a (singular) gauge transformation $U=i\hat x_\mu\tau^+$. The 
gauge potential in singular gauge is given by
\be 
\label{inst_sing} 
A^a_\mu(x)= 2 
 \frac{x_\nu}{x^2}\frac{\overline\eta_{a\mu\nu}\rho^2}{x^2+\rho^2}. 
\ee 
This singularity at the origin is not physical, the field strength 
and topological charge density are smooth. However, in order to  
calculate the topological charge from a surface integral over 
$K_\mu$, the origin has to be surrounded by a small sphere. The 
topology of this configuration is therefore located at the 
origin, not at infinity. In order to study instanton-anti-instanton
configurations, we will mainly work with such singular configurations.

  The classical instanton solution has a number of degrees of freedom, 
known as collective coordinates. In the case of $SU(2)$, the solution 
is characterized by the instanton size $\rho$, the instanton position 
$z_\mu$, and three parameters which determine the color orientation of 
the instanton. The group orientation can be specified in terms of the 
$SU(2)$ matrix $U,\;A_\mu\to UA_\mu U^\dagger$, or the corresponding 
rotation matrix $R^{ab}=\frac{1}{2}{\rm tr}(U\tau^aU^\dagger\tau^b)$, 
such that $A_\mu^a\to R^{ab}A_\mu^b$. Due to the symmetries of the
instanton configuration, ordinary rotations do not generate new
solutions. 

  $SU(3)$ instantons can be constructed by embedding the $SU(2)$ solution.
For $|Q|=1$, there are no genuine $SU(3)$ solutions. The number of 
parameters characterizing the color orientation is seven, not eight, 
because one of the $SU(3)$ generators leaves the instanton invariant.
For $SU(N)$, the number of collective coordinates (including position
and size) is $4N$. There exist exact $n$-instanton solutions with
$4nN$ parameters, but they are difficult to construct in general 
\cite{AHDM_77}. A simple solution where the relative color orientations
are fixed was given by 't Hooft (unpublished), see \cite{Wit_77,JNR_77} 
and appendix \ref{app_inst_pot}. 

 To summarize: We have explicitly constructed the tunneling path that
connects different topological vacua. The instanton action is given 
by $S=(8\pi^2|Q|)/g^2$, implying that the tunneling probability is 
\be 
P_{tunneling} \sim \exp(-8 \pi^2/g^2) .
\ee
As in the quantum mechanical example, the coefficient in front of the  
exponent is determined by a one-loop calculation.

\subsubsection{The theta vacua}
\label{sec_theta}

   We have seen that non-abelian gauge theory has a periodic potential,
and that instantons connect the different vacua. This means that the
ground state of QCD cannot be described by any of the topological 
vacuum states, but has to be a superposition of all vacua. This 
problem is similar to the motion of an electron in the periodic
potential of a crystal. It is well known that the solutions form
a band $\psi_\theta$, characterized by a phase $\theta\in [0,2\pi]$ 
(sometimes referred to as quasi-momentum). The wave functions are 
Bloch waves satisfying the periodicity condition $\psi_\theta(x+n)
=e^{i\theta n}\psi_\theta(x)$.

   Let us see how this band arises from tunneling events. If instantons
are sufficiently dilute, then the amplitude to go from one topological
vacuum $|i\rangle$ to another $|j\rangle$ is given by
\be 
\label{sum_SIA}
 \langle j | \exp(-H \tau) | i\rangle  &=& \sum_{N_+} \sum_{N_-} 
 \frac{\delta_{N_+-N_- -j+i}}{N_+! N_-!} 
 \left(K \tau e^{-S}\right)^{N_+ +N_-},
\ee 
where $K$ is the pre-exponential factor in the tunneling amplitude
and $N_\pm$ are the numbers of instantons and ant-instantons. Using 
the identity
\be  
\delta_{ab} &=& \frac{1}{2\pi }\int_0^{2\pi} d\theta\, e^{i \theta (a-b)} 
\ee 
the sum over instantons and anti-instantons can rewritten as 
\be 
 \langle j | \exp(-H \tau) | i\rangle  
 &=& \frac{1}{2\pi} \int_0^{2\pi} d\theta\, 
 e^{i \theta (i-j)} \exp\left[2K \tau \cos(\theta) \exp(-S)\right] .
\ee
This result shows that the true eigenstates are the theta vacua 
$|\theta\rangle =\sum_n e^{in\theta}|n\rangle$. Their energy is 
\be 
\label{E_theta}
E(\theta) &=&  - 2 K \cos(\theta) \exp(-S) 
\ee
The width of the zone is on the order of tunneling rate. The lowest 
state corresponds to $\theta=0$ and has negative energy. This is as
it should be, tunneling lowers the ground state energy. 

   Does this result imply that in QCD there is a continuum of states,
without a mass gap? Not at all: Although one can construct 
stationary states for any value of $\theta$, they are not excitations 
of the $\theta=0$ vacuum, because in QCD the value of $\theta$ cannot 
be changed. As far as the strong interaction is concerned, different
values of $\theta$ correspond to different worlds. Indeed, we can 
fix the value of $\theta$ by adding an additional term 
\be
\label{L_theta}
{\cal L}&=& \frac{i\theta}{32\pi^2}G^a_{\mu\nu}\tilde G^a_{\mu\nu}
\ee
to the QCD Lagrangian. 

   Does physics depend on the value of $\theta$? Naively, the 
interaction (\ref{L_theta}) violates both T and CP invariance.
On the other hand, (\ref{L_theta}) is a surface term and one 
might suspect that confinement somehow screens the effects of 
the $\theta$-term. A similar phenomenon is known to occur in 
3-dimensional compact electrodynamics \cite{Pol_77}. In QCD,
however, one can show that if the $U(1)_A$ problem is solved 
(there is no massless $\eta'$ state in the chiral limit) and
none of the quarks is massless, a non-zero value of $\theta$ 
implies that $CP$ is broken \cite{SVZ_80c}.

    Consider the expectation value of the CP violating observable
$\langle G\tilde G\rangle$. Expanding the partition function in 
powers of $\theta$, we have $\langle G\tilde G\rangle =\theta
(32\pi^2)\chi_{top}$. Furthermore, in Sec. \ref{sec_screen}
we will prove an important low energy theorem that determines
the topological susceptibility for small quark masses. Using
these results, we have
\be
\label{Q_let}
\langle G\tilde G\rangle &=& -\theta (32\pi^2)f_\pi^2 m_\pi^2
\frac{m_um_d}{(m_u+m_d)^2}
\ee
for two light flavors to leading order in $\theta$ and the quark
masses. Similar estimates can be obtained for $CP$ violating
observables that are directly accessible to experiment. The 
most severe limits on CP violation in the strong interaction 
come from the electric dipole of the neutron. Current experiments  
imply that \cite{Bal_79,CVV_79}
\be 
\theta &<& 10^{-9}  .
\ee 
The question why $\theta$ is so small is known as the strong CP  
problem. The status of this problem is unclear. As long as we do 
not understand the source of CP violation in nature, it is not 
clear whether the strong CP problem is expected to have a solution 
within the standard model, or whether there is some mechanism 
outside the standard model that adjusts $\theta$ to be small. 

  One possibility is provided by the fact that the state with 
$\theta=0$  has the lowest energy. This means that if $\theta$ 
becomes a dynamical variable, the vacuum can relax to the $\theta=0$
state (just like electrons can drop to the bottom of the conduction 
band by emitting phonons). This is the basis of the axion mechanism 
\cite{PQ_77}. The axion is a hypothetical pseudo-scalar particle, 
which couples to $G\tilde G$. The equations of motion for the 
axion field automatically remove the effective $\theta$ term, 
which is now a combination of $\theta_{QCD}$ and the axion 
expectation value. Experimental limits on the axion coupling
are very severe, but an ``invisible axion" might still exist 
\cite{Kim_79,SVZ_80,Zhi_80,DF_83,PWW_83}.

  The simplest way to resolve the strong CP problem is to assume 
that the mass of the $u$-quark vanishes (presumably because of
a symmetry not manifest in the standard model). Unfortunately, 
this possibility appears to be ruled out phenomenologically,
but there is no way to know for sure before this scenario is
explored in more detail on the lattice. More recently, it was 
suggested that QCD might undergo a phase transition near $\theta
=0,\pi$. In the former case  some support for this idea from lattice 
simulations \cite{Sch_94}, but the instanton model and lattice measurements
of the topological susceptibility etc do not suggest any singularity around
 $\theta=0$. 
The latter limit $\theta=\pi$ also conserves CP and has 
a number of interesting properties \cite{SG_81}:  in this 
world the instanton-induced interaction between quarks would 
change sign. Clearly, it is important to understand the properties 
of QCD with a non-zero $\theta$-angle in more detail.

\subsubsection{The tunneling amplitude}
\label{sec_thooft}

   The next natural step is the one-loop calculation of the pre-exponent 
in the tunneling amplitude. In gauge theory, this is a rather tedious 
calculation which was done in the classic paper by \cite{tHo_76b}.
Basically, the procedure is completely analogous to what we did in the 
context of quantum mechanics. The field is expanded around the 
classical solution, $A_\mu=A_\mu^{cl}+\delta A_\mu$. In QCD, we have
to make a gauge choice. In this case, it is most convenient to work
in a background field gauge $D_\mu (A_\nu^{cl})\delta A_\mu=0$. 
 
  We have to calculate the one-loop determinants for gauge fields,
ghosts and possible matter fields (which we will deal with later). The
determinants are divergent both in the ultraviolet, like any other 
one-loop graph, and in the infrared, due to the presence of zero modes. 
As we will see below, the two are actually related. In fact, the QCD 
beta function is partly determined by zero modes (while in certain 
supersymmetric theories, the beta function is completely determined by 
zero modes, see Sec. \ref{sec_NSVZ_beta}).

  We already know how to deal with the $4N_c$ zero modes of the system.
The integral over the zero mode is traded for an integral over 
the corresponding collective variable. For each zero mode, we get one
factor of the Jacobian $\sqrt{S_0}$. The integration over the color 
orientations is compact,
so it just gives a factor, but the integral over size and position we 
have to keep. As a result, we get a differential tunneling rate 
\be
 dn_I \sim  \left(\frac{8\pi^2}{g^2}\right)^{2N_c}
 \exp\left(-\frac{8\pi^2}{g^2}\right) \rho^{-5}d\rho dz,
\ee
where the power of $\rho$ can be determined from dimensional considerations. 

   The ultraviolet divergence is regulated using the Pauli-Vilars
scheme, which is the most convenient method when dealing with 
fluctuations around non-trivial classical field configurations (the
final result can be converted into any other scheme). This means 
that the determinant $\det O$ of the differential 
operator $O$ is divided by $\det(O+M^2)$, where $M$ is the regulator mass. 
Since we have to extract $4N_c$ zero modes from $\det O$, this gives a 
factor $M^{4N_c}$ in the numerator of the tunneling probability. 

   In addition to that, there will be a logarithmic dependence on $M$
coming from the ultraviolet divergence. To one loop order, it is just
the logarithmic part of the polarization operator. For any classical
field $A^{cl}_{\mu}$ the result can be written as a contribution to 
the effective action \cite{BC_78,VZN_82}
\be 
\delta S_{NZM}= {2\over 3} {g^2 \over 8\pi^2} \log(M \rho)S(A^{cl})
\ee
In the background field of an instanton the classical action cancels 
the prefactor ${g^2 \over 8\pi^2}$, and $\exp(-\delta S_{NZM})\sim 
(M\rho)^{-2/3}$. Now, we can collect all terms in the exponent of
the tunneling rate
\be
 dn_I &\sim&  \exp\left(-\frac{8\pi^2}{g^2}+4N_c\log(M\rho)
  - \frac{N_c}{3}\log(M\rho) \right) \rho^{-5}d\rho dz_\mu
 \;\equiv\; \exp\left(-\frac{8\pi^2}{g^2(\rho)}\right) 
\rho^{-5}d\rho dz_\mu,
\ee
where we have recovered the running coupling constant $(8\pi^2)/
g^2(\rho) = (8\pi^2)/g^2 - (11N_c/3)\log(M\rho)$. Thus, the infrared 
and ultraviolet divergent terms combine to give the coefficient 
of the one-loop beta function, $b=11N_c/3$, and the bare charge 
and the regulator mass $M$ can be combined into to a running coupling 
constant. At two loop order, the renormalization group requires the
miracle to happen once again, and the non-zero mode determinant 
can be combined with the bare charge to give the two-loop beta function
in the exponent, and the one-loop running coupling in the pre-exponent. 

  The remaining constant was calculated in \cite{tHo_76b,BSG_77}. The 
result is
\be 
\label{eq_d(rho)} 
dn_I &=& \frac{0.466\exp(-1.679 N_c)}{(N_c-1)!(N_c-2)!}
 \left(\frac{8 \pi^2}{g^2}\right)^{2 N_c} 
 \exp\left(-\frac{8\pi^2}{g^2(\rho)}\right) 
 \frac{d^4zd\rho}{\rho^5}. 
\ee
The tunneling rate $dn_A$ for anti-instantons is of course identical. 
Using the one-loop beta function the result can also be written as
\be 
\frac{dn_I}{d^4z} &\sim&  \frac{d\rho}{\rho^5} (\rho \Lambda)^b 
\ee
and because of the large coefficient $b=(11 N_c/3)=11$, the 
exponent overcomes the Jacobian and small size instantons are
strongly suppressed. On the other hand, there appears to be 
a divergence at large $\rho$. This is related to the fact that 
the perturbative beta function is not applicable in this regime.
We will come back to this question in Sec. III.

\subsection{Instantons and light quarks}
\label{sec_fermions}
\subsubsection{Tunneling and the $U(1)_A$ anomaly}
\label{sec_anomaly}

  When we considered the topology of gauge fields and the appearance
of topological vacua, we posed the question whether the different vacua 
can be physically distinguished. In the presence of light fermions, there
is a simple physical observable that distinguishes between the topological 
vacua, the axial charge. This observation helped to clarify the mechanism 
of chiral anomalies and showed how perturbative and non-perturbative
effects in the breaking of classical symmetries are related. 

   Anomalies first appeared in the context of perturbation theory
\cite{Adl_69,BJ_69}, when it became apparent that loop diagrams 
involving external vector and axial-vector currents cannot be regulated 
in such a way that all the currents remain conserved. From the 
triangle diagram involving two gauge fields and the flavor singlet 
axial current one finds 
\be 
\label{u1a_anom}
\partial_\mu j_\mu^5 = \frac{N_F}{16\pi^2} 
G^a_{\mu\nu}\tilde G^a_{\mu\nu},
\ee 
where $j_\mu^5=\bar q\gamma_\mu\gamma_5 q$ with $q=(u,d,s,\dots)$.
This result is not modified at higher orders in the perturbative
expansion. At this point, the gauge field on the rhs is some arbitrary 
background field. The fact that the flavor singlet current has an
anomalous divergence was quite welcome in QCD, because it seemed to 
explain the absence of a ninth Goldstone boson, the so-called $U(1)_A$ 
puzzle. 

  Nevertheless, there are two apparent problems with (\ref{u1a_anom})
if we want to understand the $U(1)_A$ puzzle. The first one is that 
the right hand side is proportional to the divergence of the topological 
current, $\partial_\mu K_\mu$, so it appears that we can still define a 
conserved axial current. The other is that, since the rhs of the anomaly 
equation is just a surface term, it seems that the anomaly does not 
have any physical effects. 

  The answer to the first problem is that while the topological charge
is gauge invariant, the topological current is not. The appearance
of massless poles in correlation functions of $K_\mu$ does not 
necessarily correspond to massless particles. The answer to the second
question is of course related to instantons. Because QCD has 
topologically distinct vacua, surface terms are relevant. 

  In order to see how instantons can lead to the non-conservation
of axial charge, let us calculate the change in axial charge 
\be
 \Delta Q_5 &=& Q_5(t=+\infty)-Q_5(t=-\infty)
 \;=\; \int d^4x\, \partial_\mu j_\mu^5.
\ee
In terms of the fermion propagator, $\Delta Q_5$ is given by
\be
 \Delta Q_5 &=& \int d^4x\, N_f\partial_\mu{\rm tr}
 \left(S(x,x)\gamma_\mu\gamma_5 \right).
\ee
The fermion propagator is the inverse of the Dirac operator,
$S(x,y)=\langle x|(iD\!\!\!\!/\,)^{-1}|y\rangle$. For any given 
gauge field, we can determine the propagator in terms of the 
eigenfunctions $iD\!\!\!\!/\,\psi_\lambda=\lambda\psi_\lambda$
of the Dirac operator 
\be
 S(x,y)&=&\sum_\lambda \frac{\psi_\lambda(x)\psi^\dagger_\lambda(y)}
 {\lambda}.
\ee
Using the eigenvalue equation, we can now evaluate $\Delta Q_5$
\be
\label{q5_sum}
 \Delta Q_5 &=& N_f \int d^4x \, {\rm tr}\left(\sum_\lambda
 \frac{\psi_\lambda(x)\psi^\dagger_\lambda(x)}{\lambda}
 2\lambda\gamma_5 \right).
\ee
For every non-zero $\lambda$, $\gamma_5\psi_\lambda$ is an eigenvector
with eigenvalue $-\lambda$. But this means, that $\psi_\lambda$ and 
$\gamma_5\psi_\lambda$ are orthogonal, so only {\em zero modes} can
contribute to (\ref{q5_sum})
\be
\label{q5_zm}
 \Delta Q_5 &=& 2N_f (n_L-n_R),
\ee
where $N_{L,R}$ is the number of left (right) handed zero modes and we 
have used the fact that the eigenstates are normalized.

  The crucial property of instantons, originally discovered by 't Hooft, 
is that the Dirac operator has a zero mode $iD\!\!\!\!/\,\psi_0(x)=0$
in the instanton field. For an instanton in the singular gauge, the
zero mode wave function is
\be  
\label{eq_zm}
\psi_0(x)={\rho \over \pi} \frac{1}{(x^2+\rho^2)^{3/2}} 
 \frac{\gamma\cdot x}{\sqrt{x^2}}\frac{1+\gamma_5}{2} \phi 
\ee
where $\phi^{\alpha m}=\epsilon^{\alpha m}/\sqrt{2}$ is a constant 
spinor in which the $SU(2)$ color index $\alpha$ is coupled to the 
spin index $m=1,2$. Let us briefly digress in order to show that
(\ref{eq_zm}) is indeed a solution of the Dirac equation. First
observe that\footnote{We use Euclidean Dirac matrices that 
satisfy $\{\gamma_\mu,\gamma_\nu\}=2\delta_{\mu\nu}$. We also
have $\sigma_{\mu\nu}=i/2[\gamma_\mu,\gamma_\nu]$ and 
$\gamma_5=\gamma_1\gamma_2\gamma_3\gamma_4$.}
\be
\label{dslash2}
 (iD\!\!\!\!/\,)^2 &=& 
      \left( -D^2+\frac{1}{2}\sigma_{\mu\nu}G_{\mu\nu} \right) .
\ee
We can now use the fact that $\sigma_{\mu\nu}G_{\mu\nu}^{(\pm)}=\mp
\gamma_5 \sigma_{\mu\nu}G_{\mu\nu}^{(\pm)}$ for (anti) self-dual fields 
$G_{\mu\nu}^{(\pm)}$. In the case of a self-dual gauge potential the 
Dirac equation $iD\!\!\!\!/\,\psi=0$ then implies ($\psi=\chi_L+\chi_R$)
\be
\label{dslash_LR}
 \left( -D^2 +\frac{1}{2}\sigma_{\mu\nu}G_{\mu\nu}^{(+)}\right)\chi_L\;=\;0,
 \hspace{1cm}  -D^2\chi_R \;=\; 0,
\ee
and vice versa ($+\leftrightarrow -,\,L\leftrightarrow R$) for 
anti-self-dual fields. Since $-D^2$ is a positive operator, $\chi_R$
has to vanish and the zero mode in the background field of an instanton 
has to be left-handed, while it is right handed in the case of an 
anti-instanton\footnote{This result is not an accident. Indeed, 
there is a mathematical theorem (the Atiyah-Singer index theorem), 
that requires that $Q=n_L-n_R$ for every species of chiral fermions. 
In the case of instantons, this relation was proven by 
\protect\cite{Sch_77}, see also the discussion in \protect\cite{Col_77}.}.
A general analysis of the solutions of (\ref{dslash_LR}) was given in
\cite{tHo_76b,JR_77}. In practice, the zero mode is most easily 
found by expanding the spinor $\chi$ as $\chi^m_\alpha= M_\mu
(\tau^{(+)}_\mu)^{\alpha m}$. For (multi) instanton gauge 
potentials of the form $A_\mu^a=\bar\eta_{\mu\nu}^a\partial_\nu
\log\Pi(x)$ (see App. \ref{app_inst_pot}) it is convenient to
make the ansatz \cite{Gro_77}
\be
\label{zm_ansatz}
 \chi_\alpha^m &=& \sqrt{\Pi(x)}\partial_\mu 
   \left(\frac{\Phi(x)}{\Pi(x)}\right) (\tau_\mu^{(+)})^{\alpha m}.
\ee
Substituting this ansatz in Equ. (103) shows that $\Phi(x)$ has to
satisfy the Laplace equation $\Box\Phi(x)=0$. A solution
that leads to a normalizable zero mode is given by $\Phi(x)=\rho^2
/x^2$, from which we finally obtain Equ. (\ref{eq_zm}). Again, we 
can obtain an $SU(3)$ solution by embedding the $SU(2)$ result. 

    We can now see how tunneling between topologically different 
configurations (described semi-classically by instantons) explains 
the axial anomaly. Integrating the anomaly equation (\ref{u1a_anom}), 
we find that $\Delta Q_5$ is related to the topological charge $Q$. On 
the other hand, from equation (\ref{q5_zm}) we know that $\Delta Q_5$ 
counts the number of left handed zero modes minus the number of right 
handed zero modes. But this is exactly what instantons do: every 
instanton contributes one unit to the topological charge and has a 
left handed zero mode, while anti-instantons have $Q=-1$ and a right 
handed zero mode. 

  There is another way to look at this process, known as the ``infinite 
hotel story" \cite{Gri_81}, see also \cite{Shi_89}. Let us consider the 
gauge potential of an instanton as a fixed background field and diagonalize 
the time dependent Dirac Hamiltonian $i\vec\alpha\cdot\vec D$ (again, it
is most convenient to work in the temporal gauge). The presence of a 
4-dimensional normalizable zero mode implies that there is one left 
handed state that crosses from positive to negative energy during the 
tunneling event, while one right handed state crosses the other way.
This can be seen as follows: In the adiabatic approximation,
solutions of the Dirac equation are given by
\be
 \psi_i(\vec x,t)&=& \psi_i(\vec x,t=-\infty)
  \exp\left(-\int_{-\infty}^{t}dt'\, \epsilon_i(t')\right).
\ee 
The only way we can have a 4-dimensional normalizable wave function
is if $\epsilon_i$ is positive for $t\to\infty$ and negative for $t\to
-\infty$. This explains how axial charge can be violated during 
tunneling. No fermion ever changes its chirality, all states simply
move one level up or down. The axial charge comes, so to say, from 
the ``bottom of the Dirac sea". 

  In QCD, the most important consequence of the anomaly is the fact 
that the would-be ninth Goldstone boson, the $\eta'$, is massive even in 
the chiral limit. The way the $\eta'$ acquires its mass is also intimately
related with instantons and we will come back to this topic a number
of times during this review. Historically, the first attempt to 
understand the origin of the $\eta'$ mass from the anomaly was based 
on anomalous Ward identities \cite{Ven_79}, see Sec. \ref{sec_screen}. 
Saturating these Ward identities with hadronic resonances and using 
certain additional assumptions, one can derive the Witten-Veneziano 
relation \cite{Wit_79b,Ven_79}
\be 
\label{Wit-Ven}
\chi_{top} &=& \int d^4x\,\langle Q(x)Q(0)\rangle
 \;=\; \frac{f^2_\pi}{2N_F} \left(m^2_\eta +m^2_{\eta'}-2m^2_K\right) . 
\ee
In this relation, we have introduced an important new quantity, 
the topological susceptibility $\chi_{top}$, which measures 
fluctuations of the topological charge in the QCD vacuum. The
combination of meson masses on the rhs corresponds to the part
of the $\eta'$ mass which is not due to the strange quark mass.

  There are several subtleties in connection with the Witten-Veneziano
relation. In Sec. (\ref{sec_screen}), we will show that in QCD
with massless flavors the topological charge is screened and $\chi_{top}
=0$. This means that the quantity on the lhs of the Witten-Veneziano
relation is the topological susceptibility in pure gauge theory\footnote{
It is usually argued that the Witten-Veneziano relation is derived 
in the large $N_c$ approximation to QCD and that $\chi_{top}=O(1)$
in this limit. That does not really solve the problem, however. In 
order to obtain a finite topological susceptibility, one has to set 
$N_f=0$, even if $N_c\to\infty$.} (without quarks). But in pure  
gauge theory, the $\eta'$ correlation function is pathological
(see Sec. \ref{sec_cor_iilm}), so the $\eta'$ mass on the rhs
of the Witten-Veneziano relation has to be determined in full QCD. 
This means that the lhs and the rhs of the Witten-Veneziano relation
are actually defined in different theories\footnote{Which means that
a priori it is not even defined how the two numbers should be compared}.
Nevertheless, the Witten-Veneziano relation provides a reasonable 
estimate of the quenched topological susceptibility (Sec. 
\ref{sec_top_sus}), and effective lagrangians that incorporate
the Witten-Veneziano relation provide a good description of the 
pseudo-scalar meson spectrum. Also, we will show that the Witten-Veneziano 
relation can be derived within the instanton liquid model using the
mean field approximation (Sec. \ref{sec_bos}). An analog of the
Witten-Veneziano relation in full QCD is described in Sec. 
\ref{sec_screen}. A more detailed study of the $\eta'$ correlation 
function in different instanton ensembles will be given in Sec. 
\ref{sec_cor_iilm}.

\subsubsection{Tunneling amplitude in the presence of light fermions}
\label{sec_tun_qu}

   The previous subsection was a small detour from the calculation of 
the tunneling amplitude. Now we return to our original problem and study
the effect of light quarks on the tunneling rate. Quarks modify the 
weight in the euclidean partition function by the fermionic determinant
\be
 \prod_f \det\left[ D\!\!\!\!/\,(A_\mu)+m_f\right],
\ee
which depends on the gauge fields through the covariant derivative.
Using the fact that non-zero eigenvalues come in pairs, the determinant
can be written as
\be
 \det\left[ D\!\!\!\!/\,+m\right] &=& m^\nu
 \prod_{\lambda > 0}(\lambda^2+m^2),
\ee
where $\nu$ is the number of zero modes. Note that the integration over 
fermions gives a determinant in the numerator. This means that (as $m
\to 0$) the fermion zero mode makes the tunneling amplitude vanish and
individual instantons cannot exist! 

   We have already seen what the reason for this phenomenon is: During
the tunneling event, the axial charge of the vacuum changes by two units,
so instantons have to be accompanied by fermions. Indeed, it was pointed
out by 't Hooft that the tunneling amplitude is non-zero in the presence
of {\em external quark sources}, because zero modes in the denominator
of the quark propagator may cancel against zero modes in the determinant.
This is completely analogous to the situation in the quantum mechanical
toy model of Sec. \ref{sec_susy_qm}. Nevertheless, there are important
differences. In particular, we will see that in QCD, chiral symmetry
breaking implies that isolated instantons can have a non-zero amplitude
(Sec. \ref{sec_qbarq}).

  Consider the fermion propagator in the instanton field
\be
\label{S_inst}
S(x,y)&=&\frac{\psi_0(x)\psi^+_0(y)}{im}
 +\sum_{\lambda\neq 0}\frac{\psi_\lambda(x)\psi^+_\lambda(y)}{\lambda+im} 
\ee
where we have written the zero mode contribution separately. Suppose 
there are $N_f$ light quark flavors, so that the instanton amplitude 
is proportional to $m^{N_f}$ (or, more generally, to $\prod_f m_f$). 
Instead of the tunneling amplitude, let us calculate a $2N_f$-quark 
Green's function $\langle \prod_f \bar\psi_f(x_f)\Gamma \psi_f(y_f)
\rangle$, containing one quark and anti-quark of each flavor. Contracting 
all the quark fields, the Green's function is given by the tunneling 
amplitude multiplied by $N_f$ fermion propagators. Every propagator 
has a zero mode contribution with one power of the fermion mass in 
the denominator. As a result, the zero mode contribution to the 
Green's function is finite in the chiral limit\footnote{Note that 
Green's functions involving more than $2N_f$ legs are not singular
as $m\to 0$. The Pauli principle always ensures that no more than 
$2N_f$ quarks can propagate in zero mode states.}. 

    The result can be written in terms of an effective Lagrangian 
\cite{tHo_76b} (see Sec. \ref{sec_hfa}, where we give a more detailed 
derivation). It is a non-local $2N_f$-fermion interaction, where  
the quarks are emitted or absorbed in zero mode wave functions. In
general, it has a fairly complicated structure, but under certain
assumptions, it can be significantly simplified. First, if we limit 
ourselves to low momenta, the interaction is effectively local. 
Second, if instantons are uncorrelated we can average over their 
orientation in color space. For $SU(3)$ color and $N_f=1$ the result 
is \cite{SVZ_80b}
\be 
\label{Leff_nf1}
{\cal L}_{N_f=1}= \int d\rho\, n_0(\rho) \left( m\rho- \frac{4}{3}
 \pi^2\rho^3 \bar q_R q_L \right), 
\ee
where $n_0(\rho)$ is the tunneling rate without fermions. Note that 
the zero mode contribution acts like a mass term. This is quite natural, 
because for $N_f=1$, there is only one chiral $U(1)_A$ symmetry, which is 
anomalous. Unlike the case $N_f>0$, the anomaly can therefore generate 
a fermion mass term.

For $N_f=2$, the result is 
\be
\label{Leff_nf2}
{\cal L}_{N_f=2}= \int d\rho\, n_0(\rho) \left[  \prod_f
 \left( m\rho- \frac{4}{3}\pi^2\rho^3 \bar q_{f,R} q_{f,L} \right)  
 +  \frac{3}{32}\left(\frac{4}{3}\pi^2\rho^3\right)^2
 \left( \bar u_R\lambda^a u_L \bar d_R\lambda^a d_L
  - \frac{3}{4}\bar u_R\sigma_{\mu\nu}\lambda^a u_L 
    \bar d_R\sigma_{\mu\nu}\lambda^a d_L \right) \right],
\ee
where the $\lambda^a$ are color Gell-Mann matrices. One can easily 
check that the interaction is $SU(2)\times SU(2)$ invariant, but 
$U(1)_A$ is broken. This means that the 't Hooft Lagrangian provides 
another derivation of the $U(1)_A$ anomaly. Furthermore, in Secs. III 
and IV we will argue that the importance of this interaction goes much 
beyond the anomaly, and that it explains the physics of chiral symmetry 
breaking and the spectrum of light hadrons. 

  Finally, we need to include the effects of non-zero modes on the 
tunneling probability. One effect is that the coefficient in the 
beta function is changed to $b=11N_c/3 -2N_f/3$. In addition to 
that there is an overall constant that was calculated in 
\cite{tHo_76b,CC_79b}
\be 
 n(\rho) &\sim& 1.34 (m\rho)^{N_f} \left( 1 + N_f (m\rho)^2 \log
 (m\rho) + \ldots \right),
\ee 
where we have also specified the next order correction in the quark 
mass. Note that at two loop order, we not only get the two-loop 
beta function in the running coupling, but also one-loop anomalous
dimensions in the quark masses.

   We should emphasize that in this section, we have only studied 
the effect of fermions on the tunneling rate for widely separated,
individual instantons. But light fermions induce strong correlations
between instantons and the problem becomes very complicated. Many
statements that are correct in the case of pure gauge theory, e.g. 
the fact that tunneling lowers the ground state energy, are no longer 
obvious in the theory with quarks. But before we try to tackle these
problems, we would like to review what is known phenomenologically about 
instantons in QCD.


\section{Phenomenology of instantons}
\label{sec_pheno}

\subsection{How often does the tunneling occur in the QCD vacuum?}
\label{sec_density}

   In order to assess the importance of instantons in the QCD vacuum, 
we have to determine the total tunneling rate in QCD. Unfortunately,
the standard semi-classical theory discussed in the last section is 
not able to answer this question. The problem is related to the
the presence of large-size instantons for which the action is of 
order one. The naive use of the one-loop running coupling leads 
to an infrared divergence in the semi-classical rate, which is 
a simple consequence of the Landau pole. Before we discuss any
attempts to improve on the theoretical estimate, we would like
to study phenomenological estimates of the instanton density 
in QCD.

   The first attempt along this line \cite{SVZ_78} was based on
information on properties of the QCD vacuum obtained from QCD sum 
rules. We cannot go into details of the sum rule method, which is 
based on using dispersion theory to match experimental information 
with the OPE (operator product expansion) prediction for hadronic 
correlation functions, see the reviews \cite{RRY_85,Nar_89,Shi_92}.
The essential point is that the method provides an estimate for the 
gluon condensate\footnote{Again, we use conventions appropriate for 
dealing with classical fields. In standard perturbative notations, 
the fields are rescaled by a factor $g$ and the condensate is given 
by $\langle 0|(gG^{a}_{\mu\nu})^2|0\rangle$.} 
$\langle 0|(G^a_{\mu\nu})^2|0\rangle$. From an analysis of the 
charmonium spectrum, \cite{SVZ_79} obtained
\be
\label{svz_value}
\langle 0|(G^a_{\mu\nu})^2|0\rangle  &\simeq&  0.5\,{\rm GeV}^4.
\ee
It is difficult to assess the accuracy of this number. The analysis
has been repeated many times, including many more channels. Reinders,
Rubinstein and Yazaki agree with the value (\ref{svz_value}) and
quote an error of $25\%$ \cite{RRY_85}. On the other hand, a recent 
analysis gives $\langle 0|(G^a_{\mu\nu})^2|0\rangle = (1.02\pm 0.1) 
{\rm GeV}^4$, about twice the original SVZ value \cite{Nar_95}.     

\begin{figure}[t]
\begin{center}
\leavevmode
\epsfxsize=10cm
\epsffile{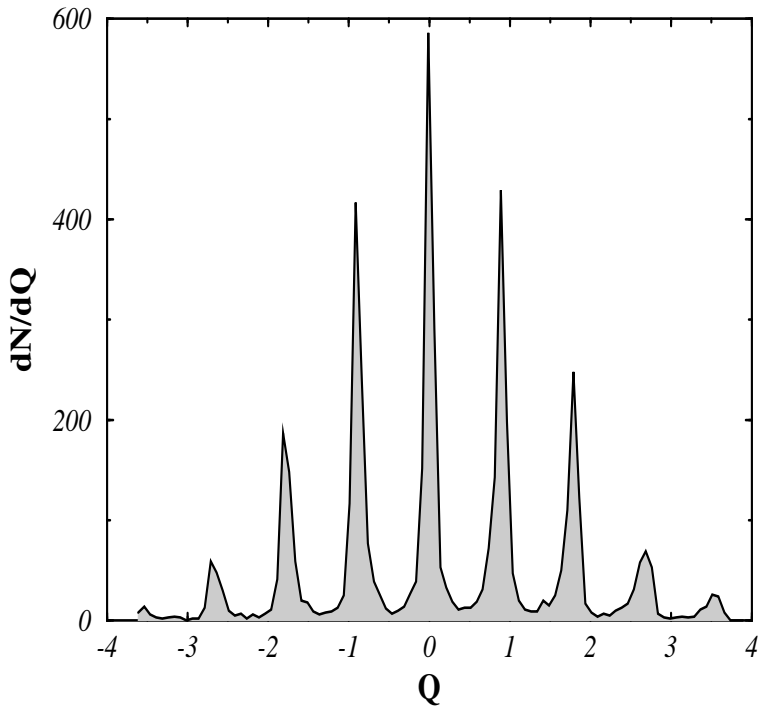}
\end{center}
\caption{\label{alles_fig4}
Distribution of topological charges for an ensemble of 5000
thermalized configurations in pure gauge $SU(3)$ at $\beta=6.1$, 
from \protect\cite{ADD_96} .}
\end{figure}
  
  The tunneling rate can now be estimated using the following simple
idea. {\em If} the non-perturbative fields contributing to the gluon 
condensate are dominated by (weakly interacting) instantons, the 
condensate is simply proportional to their density, because every 
single instanton contributes a finite amount $\int d^4x (G^a_{\mu\nu})^2
=32\pi^2$. Therefore, the value of the gluon condensate provides an 
{\em upper limit} for the instanton density
\be
\label{n_g2}
 n&=&{dN_{I+A}\over d^4x} \;\leq\;
 \frac{1}{32\pi^2}\langle(G^a_{\mu\nu})^2\rangle 
 \;\simeq\; 1\,{\rm fm}^{-4} . 
\ee
Another estimate of the instanton density can be obtained from the 
topological susceptibility. This quantity measures fluctuations
of the topological charge in a 4-volume $V$,
\be 
\label{eq_chi}
 \chi_{top} &=& \lim_{V\rightarrow \infty}\frac{\langle Q^2\rangle }{V} .
\ee
On average, the topological charge vanishes, $\langle Q\rangle =0$,
but in a given configuration $Q\neq 0$ in general (see Fig. 
\ref{alles_fig4}). Topological fluctuations provide an important 
characteristic of the vacuum in pure gauge QCD. However, in the 
presence of massless quarks, the topological charge is screened 
and $\chi_{top}=0$ (see Sec. \ref{sec_screen}). The value of 
$\chi_{top}$ in quenched QCD can be estimated using the Witten-Veneziano 
relation (\ref{Wit-Ven}) 
\be 
\chi_{top} &=& \frac{f^2_\pi}{2N_F} 
  \left(m^2_\eta +m^2_{\eta'}-2m^2_K\right)  
  \;=\; (180\,{\rm MeV})^4 .
\ee
If one assumes that instantons and anti-instantons are uncorrelated, 
the topological susceptibility can be estimated as follows. The 
topological charge in some volume $V$ is $Q=N_I-N_A$. For a system 
with Poissonian statistics, the fluctuations in the particle numbers 
are $\Delta N_{I,A} \simeq \sqrt{N_{I,A}}$. This means that for a 
random system of instantons, we expect $\chi_{top}=N/V$. Using the 
phenomenological estimate from above, we again get $(N/V)\simeq 
1\,{\rm fm}^{-4}$.

   How reliable are these estimates? Both methods suffer from 
uncertainties that are hard to assess. For the gluon condensate,
there is no systematic method to separate perturbative and 
non-perturbative contributions. The sum rule method effectively
determines contributions to the gluon condensate with momenta below 
a certain separation scale. This corresponds to instanton fluctuations 
above a given size. This in itself is not a problem, since the 
rate of small instantons can be determined perturbatively, but
the value of the separation scale ($\mu\sim 1\,{rm GeV}$) is 
not very well determined. In any case, the connection of the 
gluon condensate with instantons is indirect, other fluctuations
might very well play a role. The estimate of the instanton 
density from the (quenched) topological susceptibility relies
on the assumption that instantons are distributed randomly. In 
the presence of light quarks, this assumption is certainly 
incorrect (that is why an extrapolation from quenched to real 
QCD is necessary).

\subsection{The typical instanton size and the instanton liquid model}
\label{sec_sizes}

   Next to the tunneling rate, the typical instanton size is the
most important parameter characterizing the instanton ensemble. If 
instantons are too large, it does not make any sense to speak of 
individual tunneling events and semi-classical theory is inapplicable.
If instantons are too small, then semi-classical theory is good, but 
the tunneling rate is strongly suppressed. The first estimate of the 
typical instanton size was also made in \cite{SVZ_78}, based on the 
estimate of the tunneling rate given above. 

   {\em If} the total tunneling rate can be calculated from the 
semi-classical 't Hooft formula, we can ask up to what critical 
size we have to integrate the rate in order to get the phenomenological
instanton density\footnote{This procedure cannot be entirely consistent, 
since simply cutting off the size integration violates many exact
relations such as the trace anomaly (see Sec. \protect\ref{sec_MFA}). 
Nevertheless, given all the other uncertainties, this method provides 
a reasonable first estimate.}   
\be 
\label{svz_crit}
\int_0^{\rho_{max}} d\rho\, n_{0}(\rho) &=& n_{phen} 
\ee
Using $n_{phen}=1\,{\rm fm}^{-4}$, Shifman et al. concluded 
that $\rho_{max} \simeq 1\,{\rm fm}$. This is a very pessimistic 
result, because it implies that the instanton action is not large, 
so the semi-classical approximation is useless. Also, if instantons
are that large, they overlap strongly and it makes no sense to 
speak of individual instantons. 

  There are two possible ways in which the semi-classical 
approximation can break down as the typical size becomes 
large. One possibility is that (higher loop) perturbative 
fluctuations start to grow. We have discussed these effects
in the double well potential (see Sec. \ref{sec_qm_twoloop}),
but in gauge theory, the two-loop ($O(1/S_0)$) corrections to 
the semi-classical result are not yet known. Another possibility
is that non-perturbative (multi instanton etc.) effects 
become important. These effects can be estimated from the 
gluon condensate. The interaction of an
instanton with an arbitrary weak external field $G^{a\,ext}_{\mu\nu}$
is given by (see Sec. \ref{sec_int_bos})
\be 
\label{dip_int}
S_{int}&=& {2\pi^2\rho^2 \over g^2}\bar\eta^a_{\mu\nu} U^{ab}   
G^{b\,ext}_{\mu\nu} 
\ee
where $U$ is the matrix that describes the instanton orientation
in color space. This is a dipole interaction, so to first order 
it does not contribute to the average action. To second order
in $G^{a\,ext}_{\mu\nu}$, one has \cite{SVZ_80}
\be 
\label{n_scl_int}
n(\rho) &=& n_{0}(\rho) \left[ 1+ \frac{\pi^4 \rho^4}{2 g^4} 
  \langle(G^a_{\mu\nu})^2\rangle  +\ldots \right].
\ee
From the knowledge of the gluon condensate we can now estimate for
what size non-perturbative effects become important. Using the 
SVZ value, we see that for $\rho > 0.2$ fm the interaction with 
vacuum fields (of whatever origin) is not negligible. Unfortunately, 
for $\rho < 0.2$ fm the total density of instantons is too small 
as compared to the phenomenological estimate.

  However, it is important to note the sign of the correction. The 
non-perturbative contribution leads to a tunneling rate that grows 
even faster than the semi-classical rate.  Accounting for higher 
order effects by exponentiating the second order contribution, 
\cite{Shu_82} suggested to estimate the critical size from the 
modified condition
\be 
\label{mod_crit}
\int_0^{\rho_{max}} d\rho\, n_{0}(\rho) \exp\left[
 \frac{\pi^4 \rho^4}{2 g^4} \langle(G_{a\mu\nu})^2\rangle \right] 
 &=& n_{phen} .
\ee
Since the rate grows faster, the critical size is shifted 
to a smaller value
\be 
\label{eq_rho}
\rho_{max}&\sim& 1/3\,{\rm fm} .
\ee
If the typical instanton is indeed small, we obtain a completely
different perspective on the QCD vacuum:
\begin{enumerate}

\item Since the instanton size is significantly smaller than
the typical separation $R$ between instantons, $\rho/R \sim 
1/3$, the vacuum is fairly dilute. The fraction of spacetime
occupied by strong fields is only a few per cent.

\item The fields inside the instanton are very strong, $G_{\mu\nu}  
\gg \Lambda^2$. This means that the semi-classical approximation
is valid, and the typical action is large
\be
\label{S_typ}
 S_0 = 8\pi^2/g^2(\rho) \sim 10-15 \gg 1 .
\ee
Higher order corrections are proportional to $1/S_0$ and presumably
small.

\item Instantons retain their individuality and are not destroyed 
by interactions. From the dipole formula, one can estimate
\be
\label{S_int_typ}
 |\delta S_{int}| \sim (2-3) \ll S_0  .
\ee

\item Nevertheless, interactions are important for the structure 
of the instanton ensemble, since
\be
\label{exp_S_int}
  \exp|\delta S_{int}| \sim 20 \gg 1 .
\ee
This implies that interaction have a significant effect on correlations
among instantons, the instanton ensemble in QCD is not a dilute gas, 
but an interacting liquid. 
\end{enumerate}

   Improved estimates of the instanton size can be obtained from  
phenomenological applications of instantons. The average instanton 
size determines the structure of chiral symmetry breaking, in 
particular the values of the quark condensate, the pion mass,
its decay constant and its form factor. We will discuss these 
observables in more detail in the next sections. 

In particular, the consequences of the vacuum structure advocated here 
were studied in the context of the ``Random Instanton Liquid Model'' 
(RILM). The idea is to fix $N/V=1\,{\rm fm}^{-4}$ and $\rho=1/3$ fm and
add the assumption that the distribution of instanton positions 
as well as color orientations is completely random. This is not
necessarily in contradiction with the observation (\ref{exp_S_int})
that interactions are important, as long as they do not induce 
strong correlations among instantons. The random model is sufficiently
simple that one can study a large number of hadronic observables.
The agreement with experimental results is quite impressive, thus
providing support for the underlying parameters.


\subsection{Instantons on the lattice}
\label{sec_lat}
\subsubsection{The topological charge and susceptibility}
\label{sec_top_sus}

   The most direct way to determine the parameters of the instanton
liquid is provided by numerical simulations on the lattice. Before 
we come to direct instantons searches, we would like to discuss 
the determination of the topological susceptibility, which requires
measurements of the total topological charge inside a given volume.
This has become a very technical subject, and we will not be able to 
go into much detail (see the nice, albeit somewhat dated review 
\cite{Kro_88}). The standard techniques used to evaluate the 
topological charge are the (i) (naive) field-theoretical, (ii) 
geometrical and (iii) fermionic methods. Today, these methods 
are usually used in conjunction with various improvements, like
cooling, blocking or improved actions.  

  The field theoretic method is based on the naive lattice discretization
of the topological charge density $G\tilde G\sim\epsilon_{\mu\nu\rho
\sigma}{\rm tr}[U_{\mu\nu}U_{\rho\sigma}]$, where $U_{\mu\nu}$ is 
the elementary plaquette in the $\mu\nu$ plane. The method is simple 
to implement, but it has no topological meaning and the naive topological 
charge $Q$ is not even an integer. In addition to that, the topological
susceptibility suffers from large renormalization effects and mixes with 
other operators, in particular the unit operator and the gluon condensate 
\cite{CDP_88}.

    There are a number of ``geometric" algorithms, that ensure that 
$Q$ has topological significance \cite{Lue_82,Woi_83,PS_86}. This means 
that $Q$ is always an integer and that the topological charge can be 
expressed as a surface integral. All these methods are based on fixing 
the gauge and using some interpolation procedure to reconstruct a smooth 
gauge potential from the discrete lattice data. For a finite lattice 
with the topology of a 4-dimensional torus, the topology of the gauge 
fields resides in the transition functions that connect the gauge 
potential on the boundaries. The geometric method provides a well 
defined topological charge for almost all gauge configurations. In the 
continuum, different topological sectors are separated by configurations 
with infinite action. On the lattice however, different sectors
are separated by exceptional finite action configurations called
dislocations. Although expected to be unimportant for sufficiently
smooth fields, dislocations may spoil the continuum limit on the
lattice \cite{PT_89}. 

   Fermionic methods for calculating the topological charge rely on
the connection between instantons and fermionic zero modes. Exceptionally
small eigenvalues of the Dirac operator on the lattice have been 
identified \cite{SV_87b,SV_88,LSV_90}. Furthermore, it was 
demonstrated that the corresponding Dirac eigenvectors have the correct 
chirality and are spatially correlated with instantons. 
Fermionic methods are not sensitive to dislocations, but they
suffer from problems connected with the difficulty of defining
chiral fermions on the lattice. In particular, the (almost)
zero-modes connected with instantons are not exactly chiral
and the topological charge defined through a fermionic expectation
value does suffer from renormalization (for both Wilson and 
staggered fermions). For this reason, fermionic methods have 
never been pursued very vigorously (see \cite{Vin_88} for a rare
exception). Recently, some progress in constructing chiral 
fermions on the lattice has been made, see for example 
\cite{Kap_92,NN_95}. These methods may provide improved
measurements of the topological susceptibility \cite{NV_97}.

   Since most of the difficulties with the field theoretical and 
geometrical algorithms are related to fluctuations on very short 
scales, it is natural to supplement these algorithms with some
sort of smoothing procedure. The simplest method of this type
is {\em cooling} \cite{Hoe_86,HTW_87}. In the cooling method, one 
locally minimizes the action. This operation quickly eliminates short 
range quantum fluctuations and eventually leads to a smooth 
configuration, corresponding to the classical content of the 
original field. It has been verified that these configurations
are indeed dominated by instantons\footnote{This ``experimental" 
result also shows that all stable classical solutions of the Yang-Mills
equations are of multi-instanton type.} \cite{HTW_87,CGHN_94}. 
It has also been checked that in the cooled configurations,
the field-theoretic definition of the topological charge agrees
with the more sophisticated, geometrical methods \cite{Wie_90,AGG_90}.  
Unfortunately, the cooling method also suffers from systematic 
uncertainties. If the simplest Wilson action is used, instantons 
gradually shrink and finally fall through the lattice. Improved
lattice actions can make instantons stable \cite{FGS_95}, but 
instanton-anti-instanton pairs  still annihilate during the
cooling process. 
 
   The study of topological objects on the lattice is part of a larger
effort to find improved or even ``perfect" lattice actions and 
operators\footnote{Classically improved (perfect) actions have no 
discretization errors of order $a^n$ for some (all) $n$, where $a$ is the 
lattice spacing.}. An example for such a method is the ``inverse blocking''
procedure considered in \cite{HDZ_96}. From the field configuration
on the original coarse lattice, one constructs a smoother configuration
on a finer lattice by an approximate inverse renormalization group 
transformation. The method has the advantage that it gives a larger 
action for dislocations than the standard Wilson action does, thus
suppressing undesirable contributions to the topological charge.
First tests of improved topological charges are very promising, correctly
recovering instantons with sizes as small as $0.8 a$. Significantly
improved values for the topological susceptibility will hopefully
be forthcoming. 
 
  Whatever method is used to define $Q$ and measure the topological 
susceptibility on the lattice, one has to test the behavior of  
$\chi_{top}$ as the lattice spacing is taken to zero, $a\rightarrow 0$. 
If the topological susceptibility is a physical quantity, it has to
exhibit scaling towards the continuum limit. In the geometrical method,
$Q$ itself is not renormalized, and $\chi_{top}$ is expected to show 
logarithmic scaling (if it were not for the contribution of dislocations).
If the field theoretic method is used, $\chi_{top}$ mixes with the unit 
operator and the gluon condensate. Early studies of the scaling 
behavior of $\chi_{top}$ can be found in \cite{Woi_83,ISS_83,FGLS_85}.
A more detailed study of scaling and the role of mixing effects (in 
pure gauge $SU(2)$) was recently performed in \cite{ACG_93}. These
authors use ``heating" to investigate the effect of fluctuations on
classical instanton configurations. As quantum fluctuations are turned
on, one can study the renormalization of the topological charge. For 
comparison, if a configuration with no topology is heated, one is 
mainly sensitive to mixing. Alles et al. conclude that in their window 
$\beta=(2/g^2)=2.45-2.525$ everything scales correctly and that 
$\chi_{top}=3.5(4)\cdot 10^5 \Lambda_L^4\simeq (195\,{\rm MeV})^4$
(where we have used $\Lambda_{LAT}=8$ MeV). For comparison, the result 
from cooling is about 30\% larger. This discrepancy gives a rough estimate 
of the uncertainty in the calculation. 

\begin{figure}[t]
\begin{center}
\leavevmode
\epsfxsize=10cm
\epsffile{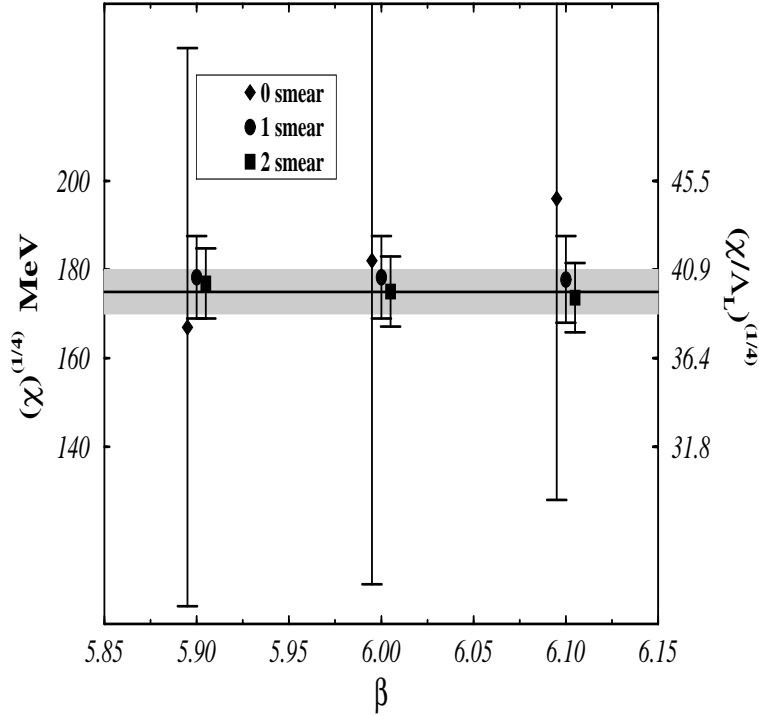}
\end{center}
\caption{\label{alles_fig3}
Scaling test of the topological susceptibility in pure 
gauge $SU(3)$, from \protect\cite{ADD_96}. The improvement 
from $Q^{(0)}_L$ to $Q^{(2)}_L$ is clearly visible. Inside 
errors, the results are independent on the bare charge 
$\beta=6/g^2$.}
\end{figure}

   An example for the use of an improved topological charge operator
is shown in figs. \ref{alles_fig4} and \ref{alles_fig3}. Figure 
\ref{alles_fig4} shows the distribution of topological charges in
pure gauge $SU(3)$. As expected, the distribution is a Gaussian
with zero mean. A scaling test of the topological susceptibility
is shown in Fig. \ref{alles_fig3}. The result clearly shows 
the reduction in the statistical error achieved using the improved
operator and the quality of the scaling behavior. The topological
susceptibility is $\chi_{top}=(175(5)\,{\rm MeV})^4$. On the other 
hand, the geometric method (and preliminary results from inverse 
blocking) gives larger values, for example $\chi_{top}=(260\,{\rm MeV})^4$ 
in pure gauge $SU(3)$ simulations \cite{GG_94}. We conclude that 
lattice determinations are consistent with the phenomenological 
value of $\chi_{top}$, but that the uncertainty is still rather 
large. Also, the scaling behavior of the topological susceptibility 
extracted from improved (perfect) operators or fermionic definitions
still needs to be established in more detail. 

   A byproduct of the measurements of Alles et al. is a new determination 
of the gluon condensate in pure gauge $SU(2)$
\be 
\label{g2_latg2}
 \langle G^2\rangle  &=& (4\pi^2)\cdot 0.38(6)\cdot 10^8 \Lambda_L^4 
 \;\simeq\; (1.5\,{\rm GeV})^4 .
\ee
This number is significantly larger than the SVZ value given in the 
last section, but it is consistent with other lattice data, for example 
\cite{CGG_89}. However, when comparing the lattice data to the SVZ
estimate one should keep in mind that the two results are based on
very different physical observables. While the SVZ value is based 
on the OPE of a hadronic correlator and the value of the separation
scale is fairly well determined, lattice results are typically 
based on the value of the average plaquette and the separation scale 
is not very well defined.

\subsubsection{The instanton liquid on the lattice} 
\label{sec_inst_lat}

   The topological susceptibility only provides a global characterization
of the instanton ensemble. In addition to that, we would also like to 
identify individual instantons and study their properties. This is true 
in particular for theories with light quarks, where the total charge is 
suppressed because of screening effect (see Sec. \ref{sec_screen}).

   Most studies of instantons on the lattice are based on the cooling 
method. As already mentioned, cooling is not an ideal method for this
purpose, because  instantons and anti-instantons annihilate during 
cooling. While this process does not affect the total charge, it does 
affect the density of instantons and anti-instantons. Ultimately,
improved operators are certainly the method of choice. Nevertheless, 
cooling has the advantage of providing smooth gauge configurations 
that are easily interpreted in terms of continuum fields.

\begin{figure}[t]
\begin{center}
\leavevmode
\epsfxsize=12cm
\epsffile{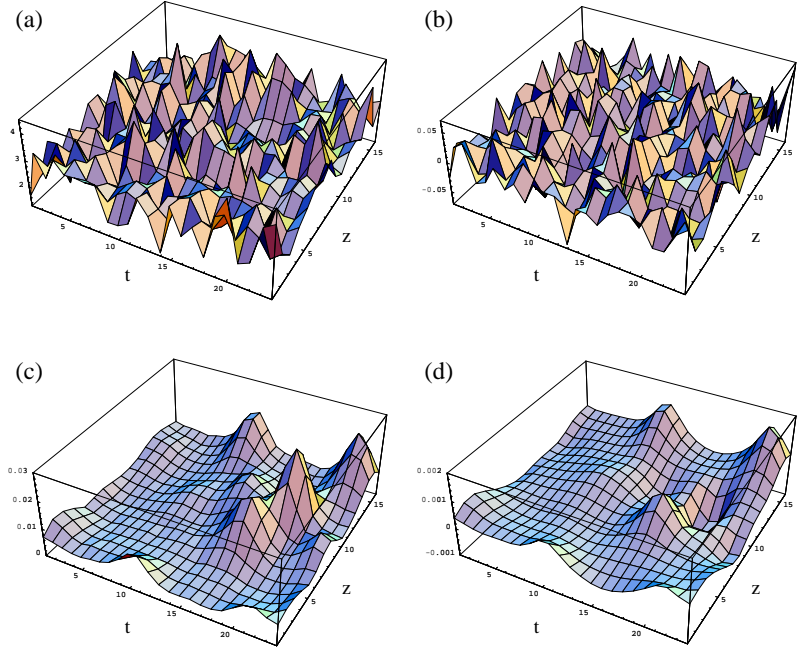}
\end{center}
\caption{\label{fig_cooled}
A typical slice through a lattice configuration before (a,b)
and after 25 cooling sweeps (c,d), from \protect\cite{CGHN_94}. 
Figs. (a,c) show the fields strength $G^2$ and Figs. (b,d)
the topological charge density $G\tilde G$. The lattice units 
are $a=0.16$ fm before and $a=0.14$ fm after 25 cooling sweeps.}
\end{figure}

   A typical field configuration after the quantum noise has 
disappeared is shown in Fig. \ref{fig_cooled} \cite{CGHN_94}. The
left panel shows the field strength on a slice through the 
lattice. One can clearly identify individual classical objects.
The right panel shows the distribution of the topological charge,
demonstrating that the classical configurations are indeed
instantons and anti-instantons. Fixing physical units from the 
(quenched) rho meson mass, Chu et al. conclude that the instanton 
density in pure gauge QCD is $(1.3-1.6)\,{\rm fm}^{-4}$. This 
number is indeed close to the estimate presented above. 

    The next question concerns the average size of an instanton. A 
qualitative confirmation of the instanton liquid value $\rho\sim 1/3
\,{\rm fm}$ was first obtained by \cite{PV_88}. More quantitative 
measurements \cite{CGHN_94} are based on fitting of the topological 
charge correlation function after cooling. The fit is very good and 
gives an average size $\rho=0.35 fm$. 

\begin{figure}[t]
\begin{center}
\leavevmode
\epsfxsize=8cm
\epsffile{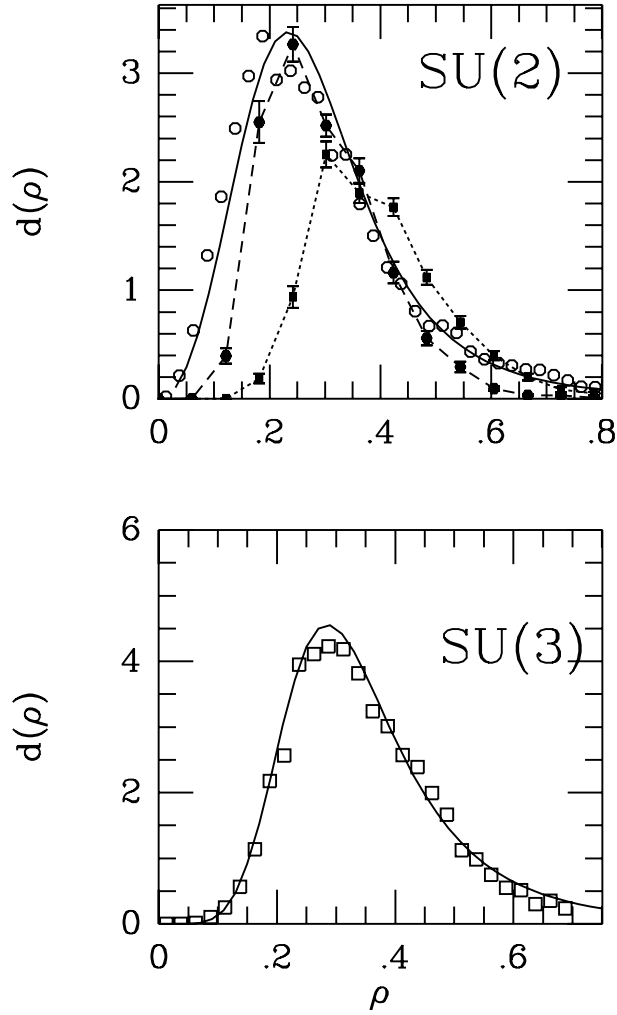}
\end{center}
\caption{\label{fig_sizes} 
Instanton size distribution in purge gauge $SU(2)$, from 
\protect\cite{MS_95,Shu_95}. The size $\rho$ is 
given in $fm$, where lattice units have been fixed from the 
glueball mass $m_{0^{++}}=1.7$ GeV. Solid squares and dots 
correspond to $16^4,\,4/g^2=2.4$ and $24^4,\,4/g^2=2.5$, respectively 
(the dotted and dashed lines simply serve to guide the eye). 
The open points come from an interacting instanton calculation, 
while the solid curve corresponds to the parameterization 
discussed in the text.} 
\end{figure}

    More detail is provided by measurements of the instanton size 
distribution. Lattice studies of this type were recently performed 
for purge gauge $SU(2)$ by \cite{MS_95} and \cite{FPS_97}. The 
results obtained by Michael and Spencer on two different size 
lattices are shown in Fig. \ref{fig_sizes}(a). The agreement 
between the two measurements is best for large instantons, while 
it is not so good for small $\rho$. That is of course exactly as 
expected; instantons of size $\rho\sim a$ fall through the lattice 
during cooling. The most important result is the existence of a 
relatively sharp maximum in the size distribution together with
a strong suppression of large-size instantons. The physical mechanism
for this effect is still not adequately understood. In Fig. \ref{fig_sizes} 
we compare the lattice data with an instanton liquid calculation, where 
large instantons are suppressed by a repulsive core in the instanton 
interaction \cite{Shu_95}.
   
\begin{figure}[t]
\begin{center}
\leavevmode
\epsfxsize=8cm
\epsffile{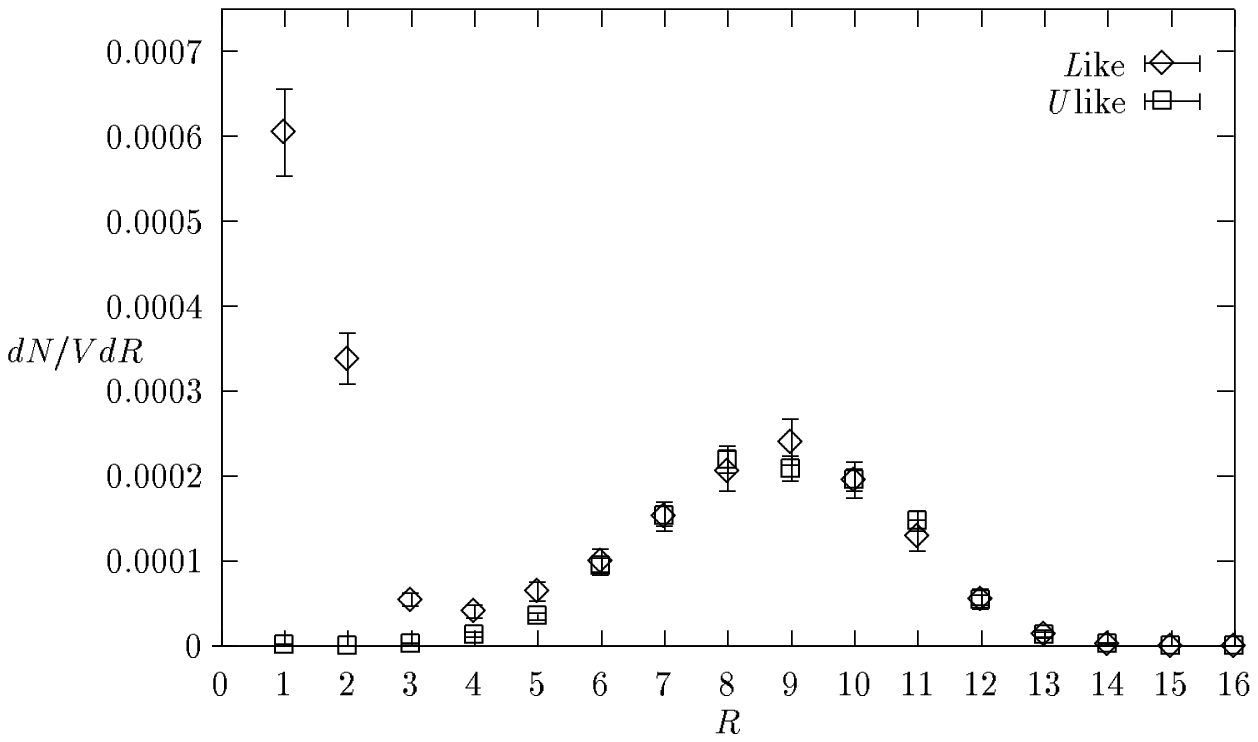}
\end{center}
\caption{\label{fig_coolcorrelations}
The distribution of the separation of like and unlike instanton pairs
in pure gauge $SU(2)$, from \protect\cite{MS_95}. The distance to the
closest neighbor (after cooling) is given in lattice units $a\simeq
0.08$ fm.}
\end{figure}

   In addition to the size distribution, \cite{MS_95} also studied 
correlations among instantons. They find that, on average, unlike 
pairs are closer than like pairs, $\langle R(IA)\rangle \simeq 0.7
\langle R(II)\rangle$. This clearly suggests an attractive interaction
between instantons and anti-instantons. The distribution of pseudo-particle
separations is shown in Fig. \ref{fig_coolcorrelations}. There are very 
few IA-pairs with very small separation, because these pairs easily 
annihilate during cooling. Like pairs show an enhancement at small $R$, 
a feature which is not understood at present.

  After identifying instantons on the lattice, the next step is to 
study the importance of instantons for physical observables. We will
discuss an example for this line of research in Sec. \ref{sec_cool_cor}, 
where we present hadronic correlation functions in cooled configurations.
Another example can be found in \cite{FMT_97}, which shows that there
is a strong correlation between the quark condensate and the location
of instantons after cooling.

\subsubsection{The fate of large-size instantons and the beta function} 
\label{sec_beta}

   We have seen that both phenomenology and the available lattice data
suggest that instantons larger than $\rho\simeq 1/3$ fm are strongly
suppressed in QCD. In Sec. \ref{sec_thooft} we saw that this 
result can not be understood from the leading order semi-classical 
formula. This leaves essentially three possibilities: The instanton
distribution is regulated by higher order quantum effects, by 
classical instanton interactions, or by the interaction of instantons
with other classical objects (e.g. monopoles or strings).

  The possible role of repulsive interactions between instantons will
be discussed in Sec. \ref{sec_theory} (this is also what the open
dots in Fig. \ref{fig_sizes} are based on). It is hard to speculate 
on the role of other classical fields, although we will try to summarize
some work in this direction in the next section. Here, we would like 
to discuss the possibility that the size distribution is regulated 
by quantum fluctuations \cite{Shu_95}. If this is the case, gross 
features of the size distribution can be studies by considering a 
single instanton rather than an interacting ensemble.

  The Gell-Mann-Low beta function is defined as a derivative of the
coupling constant $g$ over the logarithm of the normalization scale
$\mu$ at which $g$ is determined
\be
\beta(g) = \frac{\partial g}{\partial\log\mu} =
 -b \frac{g^3}{16\pi^2} - b'\frac{g^5}{(16\pi^2)^2} + \ldots
\ee
In QCD with $N_c$ colors and $N_f$ light flavors we have $b=11N_c/3-
2N_f/3$ and $b'=34N_c^2/3-13N_cN_f/3+N_f/N_c$. Remember that the 
tunneling amplitude is $n(\rho)\sim\rho^{-5} \exp[-(8\pi^2)/g^2(\rho)]$. 
In the weak coupling domain one can use the one-loop running coupling 
and $n(\rho)\sim \rho^{b-5}\Lambda^b$. This means that the strong 
growth of the size distribution in QCD is related to the large value 
of $b\simeq 9$. 

For pedagogical reasons, we would like to start 
our discussion in the domain of the phase diagram where $b$ is small 
and this dangerous phenomenon is absent. For $N_c=3$ colors, $b$ is
zero if $N_f=33/2\simeq 16$. When $b$ is small and positive, it turns 
out that the next coefficient $b'$ is negative \cite{BM_74,BZ_82} and 
therefore beta function has a zero at 
\be 
\label{g_IR}
\frac{g^2_*}{16\pi^2}=-\frac{b}{b'}.
\ee
Since $b$ is small, so is $g_*$, and the perturbative calculation  
is reliable. If the beta function has a zero, the coupling constant 
will first grow as one goes to larger distances but then stop 
running (``freeze") as the critical value $g_*$ is approached. 
In this case, the large distance behavior of all correlation 
functions is governed by the infrared fixed point and the 
correlators decay as fractional powers of the distance.
In this domain the instanton contribution to the partition function 
is of the order $\exp(-16\pi^2/g^2_*) \sim \exp(-|b'/b|)$, so it is 
exponentially small if $b$ is small\footnote{Since $N_f$ is large,
the vacuum consists of a dilute gas of instanton-anti-instanton 
molecules, so the action is twice that for a single instanton.}.

  What happens if $N_f$ is reduced, so that we move away from the $b=0$ 
line? The infrared fixed point survives in some domain (the conformal 
region), but the value of the critical coupling grows. Eventually,
non-perturbative effects (instantons in particular) grow exponentially 
and the perturbative result (\ref{g_IR}) is unreliable. As we will 
discuss in Sec. \ref{sec_big_pic}, existing lattice data suggests 
that the boundary of the conformal domain is around $N_f=7$ for $N_c=3$
\cite{IKK_96}. 

   There is no unique way to define the beta function in the 
non-perturbative regime. In our context, a preferred definition is 
provided by the value of instanton action $S_{inst}=8\pi^2/g^2(\rho)$
as a function of the instanton size. This quantity can be studied 
directly on the lattice by heating (i.e. adding quantum fluctuations) 
a smooth instanton of given size. This program is not yet implemented, 
but one can get some input from the measured instanton size distribution. 
If $S_{inst}\gg 1$, the semiclassical expression $n(\rho)\sim \rho^{-5}
\exp(-S_{inst})$ should be valid and we can extract the effective charge
from the measured size distribution.

   We have already discussed lattice measurements of $n(\rho)$ in Sec. 
\ref{sec_sizes}. These results are well reproduced using the semi-classical 
size distribution with a modified running coupling constant \cite{Shu_95}
\be
\label{g_reg}
\frac{8\pi^2}{g^2(\rho)} &=& b L+ \frac{b'}{b} \log L,
\ee
where (for $N_f=0$) the coefficients are $b=11N_c/3,\, b'=17N_c^2/3$ as 
usual, but the logarithm is regularized according to
\be
\label{l_reg}
L &=& \frac{1}{p}\log\left[ \left(\frac{1}{\rho\Lambda}\right)^p
  + C^p\right].
\ee
For small $\rho$, Eq.(\ref{g_reg}) reduces to the perturbative 
running coupling, but for large $\rho$ the coupling stops to run
in a manner controlled by the two parameters $C$ and $p$. A good
description of the measured size distribution can be obtained 
with\footnote{The agreement is even more spectacular in the 
$O(3)$ model. In this case the instanton size distribution 
is measured over a wider range in $\rho$, and shows a very 
nice $n(\rho)\sim 1/\rho^3$ behavior, which is what one would 
expect if the coupling stops to run.} $\Lambda=0.66\,{\rm fm}^{-1},
\, p=3.5,\, C=4.8$, shown by the solid line in Fig. \ref{fig_sizes}.
In real QCD, the coupling cannot freeze, because the theory 
has certainly no infrared fixed point. Nevertheless, in order to
make the instanton density convergent one does not need the 
beta function to vanish. It is sufficient that the coupling 
constant runs more slowly, with an effective $b<5$.

  There are some indications from lattice simulations that this is indeed
the case, and that there is a consistent trend from $N_f=0$ to $N_f=16$. 
These results are based on a  definition of the non-perturbative 
beta function
 based on pre-asymptotic scaling of hadronic observables. The idea
is the following: Ideally, the lattice scale is determined by performing 
simulations at different couplings $g$, fixing the scale $a$ from the 
asymptotic (perturbative) relation $g(a)$. Scaling behavior is established
by studying many different hadronic observables. Although asymptotic 
scaling is often violated, a weaker form of scaling might still be
present. In this case, the lattice scale $a$ at a given coupling $g$ 
is determined by measuring a hadronic observable in units of $a$ and 
then fixing $a$ to give the correct experimental value. This procedure 
makes sense as long as the result is universal (independent of the 
observable). Performing simulations at different $g$, one can determine 
the function $g(a)$ and the beta function.

\begin{figure}[t]
\begin{center}
\leavevmode
\epsfxsize=8cm
\epsffile{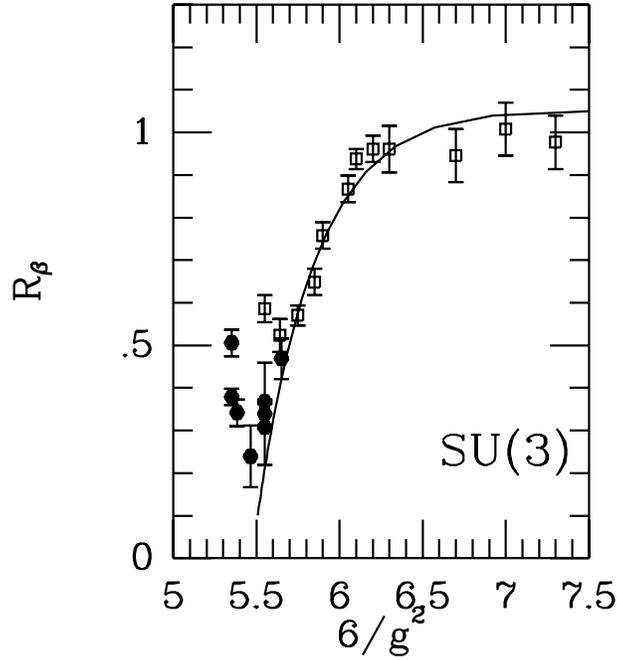}
\end{center}
\caption{\label{fig_beta} 
Non-perturbative beta function from $SU(3)$ lattice gauge theory 
with Wilson action. Open points are from \protect\cite{Gup_92b}, 
solid points from \protect\cite{BKT_95}. The solid line shows the fit 
discussed in the text.}
\end{figure}

  Lattice results for both pure gauge \cite{Gup_92b} (open points) and 
$N_f=2$ \cite{BKT_95} $SU(3)$ are shown in Fig. \ref{fig_beta}. In 
order to compare different theories, it is convenient to normalize the 
beta function to its asymptotic ($g\rightarrow 0$) value. The ratio
\be
R_{\beta}(g) &=& \frac{d(1/g^2)}{d\log(a)} \left(\left.
 \frac{d(1/g^2)}{d\log(a)}\right|_{g\to 0}\right)^{-1}
\ee
tends to 1 as $g\to 0$ ($6/g^2\to\infty$). The two-loop correction is 
positive, so $R_\beta\to 1$ from above. However, in the non-perturbative 
region the results show the opposite trend, with $R_{\beta}$ dropping
significantly below 1 around $(6/g^2)\simeq 6$. This means that the 
coupling constant runs slower than the one-loop formula suggests. 
At somewhat smaller $6/g^2$, $R_\beta$ displays a rapid upward turn 
which is known as the transition to the strong coupling regime. This 
part of the data is clearly dominated by lattice artefacts. The significant 
reduction of the beta function (by about 50\%) observed for intermediate 
values of $g$ is precisely what is is needed to explain the suppression 
of large size instantons. Furthermore, the reduction of $R_\beta$ 
is even larger for $N_f=2$. This might very well be a precursor of 
the infrared fixed point. At some critical $N_f$ we expect $R_\beta$
to touch zero. As $N_f$ is further increased, the zero should move 
to weaker coupling (to the right) and reach infinity at $N_f=33/2$.

\subsection{Instantons and confinement}
\label{sec_conf}

   After the discovery of instantons, it was hoped that instantons
might help to understand confinement in QCD. This hope was mainly
inspired by Polyakov's proof that instantons lead to confinement 
in 3 dimensional compact QED \cite{Pol_87}). However, there are 
important difference between 3 and 4 dimensional theories. In 3 
dimensions, the field of an instanton looks like a magnetic monopole 
(with $B\sim 1/r^2$), while in 4 dimensions it is a dipole field which 
falls off as $1/r^4$. 

  For a random ensemble of instantons one can calculate the instanton 
contribution to the heavy quark potential. In the dilute gas approximation
the result is determined by the Wilson loop in the field of an individual 
instanton \cite{CDG_78b}. The corresponding potential is $V\sim x^2$ for 
small $x$, but tends to a constant at large distances. This result was 
confirmed by numerical simulations in the random ensemble \cite{Shu_89}, 
as well as the mean field approximation \cite{DPP_89}. The main instanton 
effect is a renormalization of the heavy quark mass $\delta M_Q = 50-70 
MeV$. The force $|dV/dx|$, peaks at $x\simeq 0.5$ fm, but even at this 
point it is almost an order of magnitude smaller than the string tension 
\cite{Shu_89}. 

  Later attempts to explain confinement in terms of instantons (or 
similar objects) fall roughly into three different categories: objects 
with fractional topological charge, strongly correlated instantons
and the effects of very large instantons.
  
  Classical objects with fractional topological charge were first 
seriously considered in \cite{CDG_78}, which proposed a liquid 
consisting of instantons and merons. Merons are singular 
configurations with topological charge $Q=1/2$ \cite{AFF_76}. 
Basically, one can interpret merons as the result of splitting 
the dipole field of an instanton into two halves. This means 
that merons have long-range fields.
 
  Another way to introduce fractional charge is by considering
twisted boundary conditions \cite{tHo_81b}. In this case, one
finds fracton solutions (also known as 't Hooft fluxes) with
topological charges quantized in units $1/N_c$. Gonzales-Arroyo 
and Martinez suggested that confinement is produced by an ensemble 
of these objects glued together \cite{GAM_96}. In order to study
this hypothesis, they measured the string tension in cooled and 
uncooled configurations with twisted boundary conditions. They
find that fractionally charged objects can indeed be identified
and that their number roughly scales with the string tension. 
Clearly, there are many problems that need to be understood, in
particular what the role of the boundary condition is and how 
regions with twisted boundary conditions can be glued together. 

   An important attempt at understanding confinement is based
on the role of magnetic monopoles (see below). One can make 
monopole-like configuration (more precisely, dyon-like, since
the fields are self-dual) from instantons by lining up identically 
oriented instantons\footnote{An example for this is the finite 
temperature caloron solution, see Sec. \ref{sec_caloron}.}. 
Another possibility is a chain of strongly correlated 
instanton-anti-instanton pairs which might create an infinitely 
long monopole loop. Under normal circumstances, however, these 
objects have very small entropy and they are not found in the
simulations discussed in Sec. \ref{sec_sim}. 

   The possible role of very large instantons was discussed in 
\cite{DP_96}. These authors propose that instantons can cause 
confinement if the size distribution behaves as $n(\rho)\sim 
1/\rho^3$. This can be understood as follows. In Sec. 
\ref{sec_V_ss}, we will show that the mass renormalization of 
a heavy quark due to instantons is $\Delta M_Q\sim (N/V)\rho^3$. 
For typical instanton radii this contribution is not very important, 
but if the size distribution has a $1/\rho^3$ tail, then $\Delta M_Q$ 
is linearly divergent, a possible signature of confinement. This is 
very intriguing, but again, there are a number of open questions.
For one, the $1/\rho^3$ distribution implies that the total 
volume occupied by instantons is infinite. Also, very large 
instantons would introduce long range correlations which
are inconsistent with the expected exponential decay of 
gluonic correlation functions.

\begin{figure}[t]
\begin{center}
\leavevmode
\epsfxsize=12cm
\epsffile{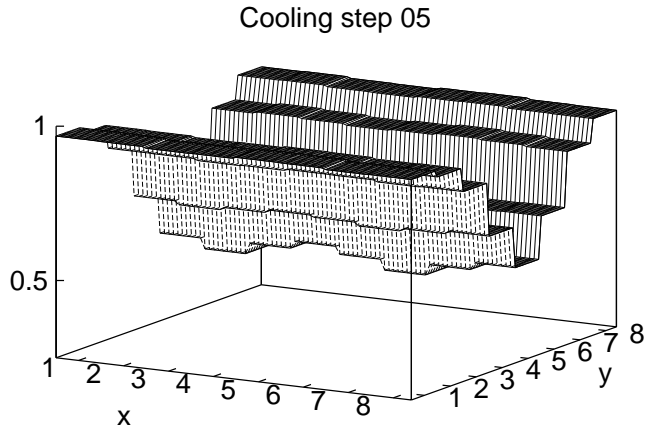}
\end{center}
\caption{\label{fig_staticquark}
Squared topological charge density around a static heavy quark-anti-quark 
pair in $SU(3)$ with $N_f=3$, from \protect\cite{FMO_95}. The results
were obtained on an $8^3\times 4$ at $6/g^2=5.6$. The topological charge 
was measured after 5 cooling steps.} 
\end{figure}

   Whatever the mechanism of confinement may turn out to be, it is
clear that instantons should be affected by confinement is some way.
One example for this line of reasoning is the idea that confinement
might be the reason for the suppression of large size instantons. 
A related, more phenomenological suggestion is that instantons 
provide a dynamical mechanism for bag formation, see 
\cite{Shu_78,CDG_79}. A lattice measurement of the suppression of 
instantons in the field of a static quark was recently performed 
by the Vienna group (see \cite{FMO_95} and references therein). 
The measured distribution of the topological charge is shown in 
Fig. \ref{fig_staticquark}. The instanton density around the 
static charge is indeed significantly suppressed, by about a 
factor of two. Since instantons give a vacuum energy density on 
the order of $-1\,{\rm GeV/fm}^3$, this effect alone generates a 
significant difference in the non-perturbative energy density (bag 
constant) inside and outside of a hadron. 

   A very interesting picture of confinement in QCD is based on 
the idea that the QCD vacuum behaves like a dual superconductor, formed 
by the condensation of magnetic monopoles \cite{Man_76}. Although 
the details of this mechanism lead to many serious questions \cite{DFGO},
there is some evidence in favor of this scenario from lattice simulations.
There are no semi-classical monopole solutions in QCD, but monopoles 
can arise as singularities after imposing a gauge condition \cite{tHo_81}.
The number and the location of monopole trajectories will then depend
on the gauge choice. In practice, the so called maximal abelian gauge
(MAG) has turned out to be particularly useful \cite{KSW_87}. The MAG 
is specified by the condition that the norm of the off-diagonal components 
of the gauge fields are minimal, e.g. ${\rm tr}((A\mu^1)^2+(A_\mu^2)^2)=
{\rm min}$ in $SU(2)$. This leaves one $U(1)$ degree of freedom unfixed, 
so in this preferred subgroup one can identify magnetic charges, study their 
trajectories and evaluate their contribution to the (abelian) string 
tension. The key observation is that the abelian string tension (in MAG) 
is numerically close to the full non-abelian string tension, and that it 
is dominated by the longest monopole loops \cite{SS_89,Suz_93}. 

   We can certainly not go into details of these calculations, which
would require a review in itself, but only mention some ideas how instantons
and monopoles might be correlated. If monopoles are responsible for color
confinement, then their paths should wind around the color flux tubes, 
just like electric current lines wind around magnetic flux tubes in 
ordinary superconductors. We have already mentioned that color flux
tubes expel topological charges, see Fig. \ref{fig_staticquark}. This 
suggests that instantons and monopoles should be (anti?) correlated. 
Lattice simulations that combine abelian projection with the cooling
technique or improved topological charges have indeed observed a
strong positive correlation between monopoles and instantons
\cite{FMT_97b}. In order to understand this phenomenon in more
detail, a number of authors have studied monopole trajectories
associated with individual instantons or instanton pairs
\cite{HT_95,SIT_95,CG_95}. It was observed that each instanton 
is associated with a monopole loop. For instanton-anti-instanton pairs, 
these loops may either form two separate or one common loop, depending 
on the relative color orientation of the two instantons. In the mean
time it was shown that the small loops associated with individual 
instantons do not correspond to global minima of the gauge fixing 
functional and should be discarded \cite{BOT_96}. On the other hand,
monopole trajectories connecting two or more instantons appear to be 
physical. The main physical question is then whether these monopole 
loops can percolate to form the long monopole loops responsible for 
confinement.


\section{Towards a theory of the instanton ensemble}
\label{sec_theory}

\subsection{The instanton interaction}
\label{sec_int}

\subsubsection{The gauge interaction}
\label{sec_int_bos}

   In the last chapter we argued that the density of instantons in the
QCD vacuum is quite significant, implying that interactions among them 
are essential in understanding the instanton ensemble. In general, it 
is clear that field configurations containing both instanton and 
anti-instantons are not exact solutions of the equations of motion 
and that the action of an instanton anti-instanton pair is not equal
to twice the single instanton action. The interaction is defined by 
\be
\label{s_int}
S_{int} &=& S(A_\mu^{IA}) - 2S_0,
\ee
where $A_\mu^{IA}$ is the gauge potential of the instanton-anti-instanton
pair. Since $A_\mu^{IA}$ is not an exact solution, there is some freedom
in choosing an ansatz for the gauge field. This freedom corresponds to
finding a convenient parametrization of the field configurations in the
vicinity of an approximate saddle point. In general, we have to integrate 
over all field configurations anyway, but the ansatz determines the way
we split coordinates into approximate zero modes and non-zero modes.

  For well separated IA pairs, the fields are not strongly distorted 
and the interaction is well defined. For very close pairs, on the other 
hand, the fields are strongly modified and the interaction is not well 
determined. In addition to that, if the instanton and anti-instanton 
begin to annihilate, the gauge fields become perturbative and should 
not be included in semi-classical approximations. We will comment on 
the present understanding of these questions below. 

  The interaction between instantons at large distances was derived by 
Callan, Dashen and Gross \cite{CDG_78} (see also \cite{For_77}). They began
by studying the interaction of an instanton with a weak, slowly varying 
external field $(G^a_{\mu\nu})_{ext}$. In order ensure that the gauge 
field is localized, the instanton is put in the singular gauge. One 
then finds
\be
\label{int_ext}
S_{int} &=& -\frac{2\pi^2}{g^2} \rho^2\bar\eta^a_{\mu\nu} 
(G^a_{\mu\nu})_{ext}.
\ee
This result can be interpreted as the external field interacting with
the color magnetic ``dipole moment" $\frac{2\pi^2}{g^2}\rho^2\bar\eta^a_
{\mu\nu}$ of the instanton. Note that the interaction vanishes if the 
external field is self-dual. If the external field is taken to be an 
anti-instanton at distance $R$, the interaction is
\be
\label{int_dip}
S_{int} &=& \frac{32\pi^2}{g^2} \rho_I^2\rho_A^2
 \bar\eta^a_{\mu\rho}\eta^b_{\nu\rho} R^{ab}
  \frac{\hat R_\mu\hat R_\nu}{R^4},
\ee
where $R^{ab}$ is the relative color orientation matrix 
$R^{ab} =\frac{1}{2}{\rm tr}(U\tau^a U^\dagger\tau^b)$ and $\hat R$
is the unit vector connecting the centers of the instanton and 
anti-instanton. The dipole-dipole form of the interaction is 
quite general, the interaction of topological objects in other
theories, such as skyrmions or $O(3)$ instantons, is of similar
type. 

   Instead of considering the dipole interaction as a 
classical phenomenon, we can also look at the interaction as 
a quantum effect, due to gluon exchanges between the instantons
\cite{Zak_92}. The linearized interaction (\ref{int_ext}) 
corresponds to an instanton vertex that describes the emission 
of a gluon from the classical instanton field. The amplitude 
for the exchange of one gluon between an instanton and an 
anti-instanton is given by
\be
\label{oge_int}
 \frac{4\pi^4}{g^2} \rho_1^2\rho_2^2 R_I^{ab}R_A^{cd}
 \eta^b_{\mu\nu}\eta^d_{\alpha\beta} 
 \langle G^a_{\mu\nu}(x)G^c_{\alpha\beta}(0)\rangle ,
\ee
where $\langle G^a_{\mu\nu}(x)G^c_{\alpha\beta}(0)\rangle$ is the 
free gauge field propagator
\be
\label{g_prop}
 \langle G^a_{\mu\nu}(x)G^c_{\alpha\beta}(0)\rangle  &=&
 \frac{2\delta^{ac}}{\pi^2 x^6}(g_{\nu\alpha}x_\mu x_\beta
 +g_{\mu\beta}x_\nu x_\alpha - g_{\nu\beta}x_\mu x_\alpha
 -g_{\mu\alpha}x_\nu x_\beta).
\ee
Inserting (\ref{g_prop}) into (\ref{oge_int}) gives the dipole
interaction to first order. Summing all $n$ gluon exchanges allows
one to exponentiate the result and reproduce the full dipole 
interaction (\ref{int_dip}).

  In order to study this interaction in more detail, it is 
useful to introduce some additional notation. We can characterize
the relative color orientation in terms of the four vector $u_\mu
=\frac{1}{2i}{\rm tr}(U\tau^+_\mu)$, where $\tau^+_\mu=(\vec\tau,-i)$. 
Note that for the gauge group $SU(2)$, $u_\mu$ is a real vector 
with $u^2=1$, whereas for $SU(N>2)$, $u_\mu$ is complex and $|u|^2\leq 
1$. Also note that for $SU(N>2)$, the interaction is completely 
determined by the upper $2\times 2$ block of the $SU(N)$ matrix 
$U$. We can define a relative color orientation angle $\theta$
by
\be
\label{def_theta}
\cos^2\theta &=& \frac{|u\cdot\hat R|^2}{|u|^2}.
\ee
In terms of this angle, the dipole interaction is given by
\be
\label{int_dip2}
S_{int} &=& -\frac{32\pi^2}{g^2} \frac{\rho_1^2\rho_2^2}{R^4}
 |u|^2 \left( 1- 4\cos^2\theta \right).
\ee
The orientational factor $d=1-4\cos^2\theta$ varies between 1 and 
$-3$. We also observe that the dipole interaction vanishes upon
averaging over all angles $\theta$. 

\begin{figure}[t]
\begin{center}
\leavevmode
\epsfxsize=7cm
\epsffile{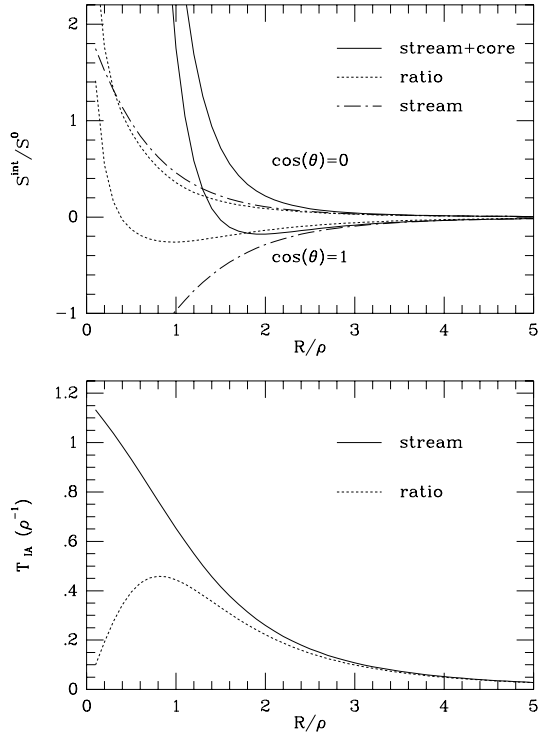}
\end{center}
\caption{\label{fig_SIA}
a) Classical instanton-anti-instanton interaction in the 
streamline (dash-dotted) and ratio ansatz (short dashed 
line). The interaction is given in units of the single 
instanton action $S_0$ for the most attractive ($\cos
\theta=1$ and most repulsive ($\cos\theta=0$) orientations. 
The solid line shows the original stream line interaction 
supplemented by the core discussed in Sec. \ref{sec_sim_res}.
b) Fermionic overlap matrix element in the stream line 
(solid) and ratio ansatz (dotted line). The matrix 
elements are given in units of the geometric mean of the 
instanton radii.} 
\end{figure}

  The dipole interaction is the leading part of the interaction 
at large distances. In order to determine the interaction at 
intermediate separation, we have to evaluate the total action 
for an IA pair. This is most easily done for the so called sum ansatz
\be
\label{sum_ans}
  A_\mu^{IA} &=& A_\mu^I + A_\mu^A,
\ee
where again both the instanton and the anti-instanton are in the 
singular gauge. The action $S=\frac{1}{4}\int d^4x\, G^2$ can easily 
be split into the free part $2S_0$ and the interaction. Systematically 
expanding in $1/R^2$, one finds \cite{DP_84}
\be 
\label{int_sum}
S^{IA}_{int} &=& -\frac{8\pi^2}{g^2}\left\{ 
 \left(|u|^2-4|u\cdot\hat R|^2\right) \left[ \frac{4\rho_1^2\rho_2^2}{R^4}
 -\frac{15\rho_1^2\rho_2^2(\rho_1^2+\rho_2^2)}{2R^6} \right]\right. \\
 & &  \hspace{2cm}+ \left.|u|^2\frac{9\rho_1^2\rho_2^2
  (\rho_1^2+\rho_2^2)}{2R^6}+ O(R^{-8}) \right\}\nonumber
\ee
for the $IA$ interaction and
\be
\label{int_sum_II}
S^{II}_{int} &=& \frac{8\pi^2}{g^2} \left( 9 + 3|u|^2 - 4|\vec u|^2
 \right) \frac{\rho_1^2\rho_2^2(\rho_1^2+\rho_2^2)}{2R^6}
 + O(R^{-8}) 
\ee
for the $II$ interaction. Here, $\vec u=$ denotes the spatial 
components of the four-vector $u_\mu$. To first order, we find 
the dipole interaction in the $IA$ channel and no interaction 
between two instantons. To next order, there is a repulsive 
interaction for both $IA$ and $II$. 

  Clearly, it is important to understand to what extent this interaction is 
unique, or if it depends on details of the underlying ansatz for the gauge
potential. It was also realized that the simple sum ansatz leads to certain 
artefacts, for example that the field strength at the centers of the two 
instantons becomes infinite and that there is an interaction between 
pseudo-particles of the same charge. One of us \cite{Shu_88} therefore 
proposed the ``ratio" ansatz
\be 
\label{ratio}
A_\mu^a &=&  \frac{
2R_I^{ab}\bar\eta^b_{\mu\nu}\frac{\rho_I^2(x-z_I)_\nu}{(x-z_I)^4}
+2R_A^{ab}\eta^b_{\mu\nu}\frac{\rho_A^2(x-z_A)_\nu}{(x-z_A)^4} }
{1+\frac{\rho_I^2}{(x-z_I)^2}+\frac{\rho_A^2}{(x-z_A)^2} },
\ee
whose form was inspired by 't Hooft's exact multi-instanton 
solution. This ansatz ensures that the field strength is regular 
everywhere and that (at least if they have the same orientation)
there is no interaction between pseudo-particles of the same charge.
Deriving an analytic expression for the interaction in the ratio 
ansatz is somewhat tedious. Instead, we have calculated the interaction
numerically and parameterized the result, see \cite{SV_91} and
Fig. \ref{fig_SIA}. We observe that both the sum and the ratio ansatz 
lead to the dipole interaction at large distances, but the amount of 
repulsion at short distance is significantly reduced in the ratio ansatz. 

   Given these ambiguities, the question is whether one can find    
the best possible ansatz for the gauge field of an $IA$ pair. This
question was answered with the introduction of the streamline or
valley method \cite{BY_86}. The basic idea is to start from a
well separated instanton-anti-instanton pair, which is an 
approximate solution of the equations of motion, and then let 
the system evolve under the force given by the variation of the
action \cite{Shu_88c}. For a scalar field theory, the streamline 
equation is given by
\be
\label{def_stream}
 f(\lambda)\frac{d\phi}{d\lambda}&=& \frac{\delta S}{\delta\phi},
\ee
where $\lambda$ is the collective variable corresponding to the  
evolution of the pair and $f(\lambda)$ is a function that depends
on the parametrization of the streamline. The initial conditions
are $\phi(\infty)=\phi_c$ and $\phi'(\infty)=\phi_0$, where $\phi_c$
is the classical solution corresponding to a well separated
instanton anti-instanton pair and $\phi_0$ is the translational
zero mode. 

    In QCD, the streamline equations were solved by \cite{Ver_91} 
using the conformal invariance of the classical equations of motion. 
Conformal invariance implies that an instanton anti-instanton pair 
in the singular gauge at some distance $R$ is equivalent to a regular 
gauge instanton and a singular gauge anti-instanton with the same center, 
but different sizes. We do not want to go into details of the construction, 
but refer the reader to the original work. It turns out that the resulting 
gauge configurations are very close to the following ansatz originally 
proposed by Yung \cite{Yun_88}
\be
\label{Yung_ans}
A_\mu^a &=& 2\eta^a_{\mu\nu} x_\nu \frac{1}{x^2+\rho^2\lambda}
+ 2R^{ab}\eta^b_{\mu\nu}x_\nu\frac{\rho^2}{\lambda}
\frac{1}{x^2(x^2+\rho^2/\lambda)},
\ee
where $\rho=\sqrt{\rho_1\rho_2}$ is the geometric mean of the two
instanton radii and $\lambda$ is the conformal parameter
\be
\label{conf_par}
\lambda &=& \frac{R^2+\rho_1^2+\rho_2^2}{2\rho_1\rho_2}
+ \left( \frac{(R^2+\rho_1^2+\rho_2^2)^2}{4\rho_1^2\rho_2^2}-1
\right)^{1/2}.
\ee
Note that the conformal parameter is large not only if the 
instanton-antiinstanton separation $R$ is much bigger than the 
mean (geometric) size, but also if $R$ is small and one of the 
instantons is much bigger than the other. This situation is 
important when we consider the suppression of large-size
instantons.

The interaction for this ansatz is given by \cite{Ver_91}
\be
\label{S_IA,Yung}
 S_{IA} &=& \frac{8\pi^2}{g^2}
   \frac{4}{(\lambda^2-1)^3}  \bigg\{
    -4\left( 1-\lambda^4 + 4\lambda^2\log(\lambda) \right)
         \left[ |u|^2-4|u\cdot\hat R|^2 \right] \\
   & & \hspace{1cm}\mbox{} +
     2\left( 1-\lambda^2 + (1+\lambda^2)\log(\lambda) \right)
         \left[ (|u|^2-4|u\cdot\hat R|^2)^2
                + |u|^4 + 2(u)^2(u^*)^2 \right]
       \bigg\}\nonumber ,
\ee
which is also shown in Fig. \ref{fig_SIA}. For the repulsive orientation,
the interaction is similar to the ratio ansatz,
and the average interaction is repulsive. For the most attractive
orientation, however, the interaction approaches $-2S_0$ at short distance,
showing that the instanton and anti-instanton annihilate each other and the
total action of the pair approaches zero. 

   It is interesting to note that the Yung ansatz, at least to order 
$1/R^6$, leads to the same interaction as the perturbative method 
(\ref{oge_int}) if carried out to higher order \cite{AM_91,BS_93,DP_94}.  
This problem is related to the calculation of instanton induced processes 
at high energy (like multi gluon production in QCD, or the baryon number 
violating cross section in electroweak theory). To leading order in 
$E/M_{sph}$, where $E$ is the bombarding energy and $M_{sph}$ the       
sphaleron barrier (see Sec. \ref{sec_ew}), the cross section is given by
\be 
  \sigma_{tot} &=& {\rm disc} \int d^4R\, e^{ip\cdot R}
  \int\int d\Omega_1d\Omega_2\, n(\rho_1)n(\rho_2)
  e^{-S_{int}(R,\Omega_{12})},
\ee
where $\Omega_i=(z_i,\rho_i,U_i)$ are the collective coordinates 
associated with the instanton and anti-instanton and ${\rm disc}\,f(s)
=\frac{1}{2i}(f(s+i\epsilon)-f(s-i\epsilon))$ denotes the discontinuity
of the amplitude after continuation to Minkowski space, $s=p^2>0$. Given 
the agreement of the streamline method and the perturbative calculation 
to leading and next to leading order, it has been suggested to use
the behavior of $\sigma_{tot}$ at high energy in order to define the
interaction $S_{int}$ \cite{DP_93}. The behavior of this quantity 
is still debated, but it is generally believed that the baryon
number violating cross section does not reach the unitarity bound.
In that case, the interaction would have to have some repulsion 
at short distance, unlike the streamline solution. 

   Another possibility to make sense of the short distance part 
of the $IA$ interaction in the streamline method is to use analytic
continuation in the coupling constant, as we have discussed
in Sec. \ref{sec_doublewell}. Allowing $g^2\to -g^2$ gives
a new saddle point in the complex $g$ plane at which the $IA$ 
interaction is repulsive and the semi-classical method is under 
control.  

\subsubsection{The fermionic interaction}          
\label{sec_tia}

   In the presence of light fermions, instantons interact with each 
other not only through their gauge fields, but also through 
fermion exchanges. The basic idea in dealing with the fermionic
interaction is that the Dirac spectrum can be split into 
quasi-zero modes, linear combinations of the zero modes of 
the individual instantons, and non-zero modes. Here we will
focus on the interaction due to approximate zero modes. The 
interaction due to non-zero modes and the corrections due to
interference between zero and non-zero modes were studied in
\cite{BC_78,LB_79}.

   In the basis spanned by the zero modes, we can write the 
Dirac operator as
\be 
\label{D_zmz}
i D\!\!\!\!/ &=& \left (
\begin{array}{cc}  0 & T_{IA}\\
                   T_{AI} & 0 
\end{array} \right ),
\ee
where we have introduced the overlap matrix elements $T_{IA}$
\newcommand{\dslash}{D\!\!\!\!/\,}
\be
\label{def_TIA}
T_{IA} &=& \int d^4 x\, \psi_{0,I}^\dagger (x-z_I)i\dslash
\psi_{0,A}(x-z_A),
\ee
Here, $\psi_{0,I}$ is the fermionic zero mode (\ref{eq_zm}). The matrix 
elements have the meaning of a hopping amplitude for a quark from 
one pseudo-particle to another. Indeed, the amplitude for an instanton
to emit a quark is given by the amputated zero mode wave function
$i\dslash \psi_{0,I}$. This shows that the matrix element (\ref{def_TIA}) 
can be written as two quark-instanton vertices connected by a 
propagator, $\psi_{0,I}^\dagger i\dslash(i\dslash)^{-1}i\dslash\psi_{0,A}$.  
At large distance, the overlap matrix element decreases as $T_{IA}\sim 
1/R^3$, which corresponds to the exchange of a massless quark. 
The determinant of the matrix (\ref{D_zmz}) can be interpreted 
as the sum of all closed diagrams consisting of (zero mode) quark 
exchanges between pseudo-particles. In other words, the logarithm
of the Dirac operator (\ref{D_zmz}) is the contributions of the 
't Hooft effective interaction to the vacuum energy.

   Due to the chirality of the zero modes, the matrix elements $T_{II}$ 
and $T_{AA}$ between instantons of the same charge vanish. In the 
sum ansatz, we can use the equations of motion to replace the 
covariant derivative in (\ref{def_TIA}) by an ordinary one. 
The dependence on the relative orientation is then given by
$T_{IA}=(u\cdot\hat R)f(R)$. This means that, like the gluonic
dipole interaction, the fermion overlap is maximal if $\cos\theta
=1$. The matrix element can be parameterized by
\be
\label{TIA_sum_par}
  T_{IA} &=& i(u\cdot R)\frac{1}{\rho_I\rho_A}
  \frac{4.0}{\left(2.0+R^2/(\rho_I\rho_A)\right)^2}.
\ee
A parametrization of the overlap matrix element for the streamline
gauge configuration can be found in \cite{SV_92}. The result is 
compared with the sum ansatz in Fig. \ref{fig_SIA} (for $\cos\theta=1$). 
We observe that the matrix elements are very similar at large distance, 
but differ at short distance.  

   Using these results, the contribution of an instanton-anti-instanton 
pair to the partition function may be written as 
\be
\label{Z_IA}
Z_{IA} = V_4\int d^4zdU\, \exp\left( N_f 
 \log |T_{IA}(U,z)|^2-S_{int}(U,z) \right).
\ee
Here, $z$ is the distance between the centers and $U$ is the relative
orientation of the pair. The fermionic part is attractive, while the 
bosonic part is either attractive or repulsive, depending on the 
orientation. If the interaction is repulsive, there is a real saddle
point for the $z$-integral, whereas for the attractive orientation
there is only a saddle point in the complex $z$ plane (as in Sec.
\ref{sec_susy_qm}). 

  The calculation of the partition function (\ref{Z_IA}) in the 
saddle point approximation was recently attempted by \cite{SV_97}. 
They find that for a large number of flavors, $N_f>5$, the ground
state energy oscillates as a function of $N_f$. The period of the
oscillation is 4, and the real part of the energy shift vanishes
for even $N_f=6,8,\ldots$. The reason for these oscillations is
exactly the same as in the case of SUSY quantum mechanics: the 
saddle point gives a complex contribution with a phase that is
proportional to the number of flavors. 

\subsection{Instanton ensembles}
\label{sec_ens}

   In Sec. \ref{sec_thooft} we studied the semi-classical theory of 
instantons, treating them as very rare (and therefore independent)
tunneling events. However, as emphasized in Sec. \ref{sec_density}, in
QCD instantons are not rare, so one cannot just exponentiate the 
results obtained for a single instanton. Before we study the instanton 
ensemble in QCD, we would like to discuss a simple physical analogy.
In this picture, we think of instantons as atoms and light quarks 
as valence electrons.

   A system like this can exist in many different phases,
e.g. as a gas, liquid or solid. In theories with massless quarks 
instantons  have ``unsaturated bonds" and cannot appear individually.
Isolated instantons in the form of an atomic gas can only exist
if there is a non zero quark mass. The simplest ``neutral" object 
is an instanton-anti-instanton ($IA$) molecule. Therefore, if the 
instanton density is low (for whatever reason: high temperature, 
a large Higgs expectation value, etc.) the system should be in a phase 
consisting of $IA$ molecules. Below, we will also argue that this 
is the case if the density is not particularly small, but the 
interactions are strong and favor the formation of molecules, for 
example if the number of light quarks $N_f$ exceeds some critical 
value. If the instanton ensemble is in the molecular phase, then 
quarks are bound to molecules and cannot propagate over large 
distances. This means that there are no eigenmodes with almost
zero virtuality, the ``conductivity" is zero, or chiral symmetry 
remains unbroken.

     The liquid phase differs from the gas phase in many important 
respects. The density is determined by the interactions and cannot 
be arbitrary small. A liquid has a surface tension, etc. As we will
see below, the instanton ensemble in QCD has all these properties. 
In QCD we also expect that chiral symmetry is spontaneously broken. 
This means that in the ground state, there is a preferred direction in 
flavor space, characterized by the quark condensate $\langle\bar qq
\rangle$. This preferred orientation can only be established if quarks 
form an infinite cluster. In terms of our analogy, this means that 
electrons are delocalized and the conductivity is non-zero. In the 
instanton liquid phase, quarks are delocalized because the instanton 
zero modes become collective.

    At very high density, interactions are strong and the ensemble is
expected to form a 4-dimensional crystal. In this case, both Lorentz 
and gauge invariance would be broken spontaneously: clearly, this 
phase is very different from the QCD vacuum. Fortunately however, explicit 
calculations \cite{DP_84,SV_90} show that the instanton liquid crystalizes
only if the density is pushed to about two orders of magnitude larger 
than the phenomenological value. At physical densities, the crystaline
phase is not favored: although the interaction is attractive, the 
entropy is too small as compared to the random liquid\footnote{In 
\protect\cite{DP_84} it was suggested that the instanton liquid 
crystalizes in the limit of a large number of colors, $N_c\to\infty$. 
The reason is that the interaction is proportional to the charge 
$\frac{8\pi^2}{g^2}$, which is of order $N_c$. As long as instantons 
do not disappear altogether (the action is of order 1), the interaction 
becomes increasingly important. However, little is known about the 
structure of large $N_c$ QCD. 
}. The electronic structure 
of a crystal consists of several bands. In our case, the Fermi surface 
lies between different bands, so the crystal is an insulator. In QCD, 
this means that chiral symmetry is not broken. 

   Another analogy with liquids concerns the question of density 
stabilization. In order for the system to saturate, we need a 
short range repulsive force in addition to the attractive long 
range dipole interaction. We have already mentioned that the 
nature of the short range interaction and the (possibly related)
question of the fate of large instantons in QCD are not well
understood. Lattice simulations indicate that large instantons 
are strongly suppressed, but for the moment we have to include
this result in a phenomenological manner.

   Let us summarize this qualitative discussion. Depending on 
the density, the instanton ensemble is expected to be in a gas, 
liquid or solid phase. The phase boundaries will in general 
depend on the number of colors and flavors. Large $N_c$ favors 
a crystalline phase, while large $N_f$ favors a molecular gas. 
Both cases are not phenomenologically  acceptable as a description 
of real QCD, with two light and one intermediate mass flavor. We 
therefore expect (and will show below) that the instanton ensemble 
in QCD is in the liquid phase.   
 
\subsection{The mean field approximation}
\label{sec_MFA}

  In order to study the structure of the instanton ensemble in a 
more quantitative fashion, we consider the partition function 
for a system of instantons in pure gauge theory
\be
\label{Z_glue}
 Z &=& \frac{1}{N_+!N_-!}\prod_i^{N_++N_-}\int [d\Omega_i\, n(\rho_i)]
 \, \exp(-S_{int}).
\ee
Here, $N_{\pm}$ are the numbers of instantons and anti-instantons, 
$\Omega_i=(z_i,\rho_i,U_i)$ are the collective coordinates of the 
instanton $i$, $n(\rho)$ is the semi-classical instanton distribution
function (\ref{eq_d(rho)}) and $S_{int}$ is the bosonic instanton 
interaction. In general, the dynamics of a system of pseudo-particles
governed by (\ref{Z_glue}) is still quite complicated, so we have 
to rely on approximation schemes. There are a number of techniques
well known from statistical mechanics that can be applied to the 
problem, for example the mean field approximation or the variational
method. These methods are expected to be reliable as long correlations 
between individual instantons are weak. 
  
   The first such attempt was made by Callan, Dashen and Gross 
\cite{CDG_78}. These authors only included the dipole interaction, 
which, as we noted above, vanishes on average and produces an 
attractive interaction to next order in the density. In this case, 
there is nothing to balance the growth $n(\rho)\sim\rho^{b-5}$ of 
the semi-classical instanton distribution. In order to deal with 
this problem, Ilgenfritz and M\"uller-Preu{\ss}ker \cite{IMP_81} 
introduced an ad hoc ``hard core" in the instanton interaction, 
see also \cite{Mun_82}. The hard core automatically excludes large 
instantons and leads to a well behaved partition function. It is 
important to note that one cannot simply cut the size integration 
at some $\rho_c$, but has to introduce the cutoff in a way that does 
not violate the conformal invariance of the classical equations 
of motion. This guarantees that the results do not spoil the 
renormalization properties of QCD, most notably the trace 
anomaly (see below). In practice, they chose 
\be
\label{core}
S_{int} &=& \left\{ \begin{array}{cc}
 \infty & |z_I-z_A|< (a\rho_I^2\rho_A^2)^{1/4} \\
   0    & |z_I-z_A|> (a\rho_I^2\rho_A^2)^{1/4} 
\end{array}\right.
\ee
which leads to an excluded volume in the partition function,
controlled by the dimensionless parameter $a$.

   We do not go into details of their results, but present the next
development \cite{DP_84}. In this work, the arbitrary core is replaced 
by the interaction in the sum ansatz, see (\ref{int_sum},\ref{int_sum_II}). 
The partition function is evaluated using a trial distribution function. 
If we assume that correlations between instantons are not very important, 
then a good trial function is given by the product of single instanton 
distributions $\mu(\rho)$   
\be
\label{var_ans}
 Z_1 &=& \frac{1}{N_+!N_-!}\prod_i^{N_++N_-}\int d\Omega_i\, 
 \mu(\rho_i)= \frac{1}{N_+!N_-!}(V\mu_0)^{N_++N_-}
\ee
where $ \mu_0 = \int d\rho\,\mu(\rho)$. The distribution $\mu(\rho)$ 
is determined from the variational principle $\delta\log Z_1/\delta\mu
=0$. In quantum mechanics a variational wave functions always provides
an upper bound on the true ground state energy. The analogous 
statement in statistical mechanics is known Feynman's variational
principle. Using convexity 
\be 
\label{convex}
   Z &=& Z_1 \langle\exp(-(S-S_1))\rangle  
   \;\geq\; Z_1 \exp(-\langle S-S_1\rangle ),
\ee
where $S_1$ is the variational action, one can see that the
variational vacuum energy is always higher than the true one.
 
  In our case, the single instanton action is given by $S_1=\log
(\mu(\rho))$ while $\langle S\rangle $ is the average action calculated 
from the variational distribution (\ref{var_ans}). Since the 
variational ansatz does not include any correlations, only 
the average interaction enters
\be
\label{av_int}
 \langle S_{int}\rangle  &=& \gamma^2\rho_1^2\rho_2^2, \hspace{1cm}
  \gamma^2 = \frac{27}{4}\frac{N_c}{N_c^2-1}\pi^2
\ee
for both $IA$ and $II$ pairs. Clearly, (\ref{av_int}) is of the 
same form as the hard core (\ref{core}) discussed above, only 
that the dimensionless parameter $\gamma^2$ is determined from
the interaction in the sum ansatz. Applying the variational 
principle, one finds \cite{DP_84}
\be
\label{reg_dis}
\mu(\rho) &=& n(\rho)\exp\left[ -\frac{\beta\gamma^2}{V}N
\overline{\rho^2}\rho^2\right],
\ee
where $\beta=\beta(\overline{\rho})$ is the average instanton
action and $\overline{\rho^2}$ is the average size. We observe
that the single instanton distribution is cut off at large sizes
by the average instanton repulsion. The average size $\overline{
\rho^2}$ is determined by the self consistency condition $\overline{
\rho^2}=\mu_0^{-1}\int d\rho\mu(\rho)\rho^2$. The result is
\be
\label{rho2_av}
\overline{\rho^2} &=& \left(\frac{\nu V}{\beta\gamma^2 N}\right)^{1/2},
\hspace{1cm}\nu = \frac{b-4}{2},
\ee
which determines the dimensionless diluteness of the ensemble,
$\rho^4(N/V)=\nu/(\beta\gamma^2)$. Using the pure gauge 
beta function $b=11$, $\gamma^2\simeq 25$ from above and $\beta
\simeq 15$, we get $\rho^4(N/V)=0.01$, even more dilute than
phenomenology requires. The instanton density can be fixed from
the second self-consistency requirement, $(N/V)=2\mu_0$ (the factor
2 comes from instantons and anti-instantons). We get 
\be
\label{dens_av}
\frac{N}{V} &=& \Lambda_{PV}^4 \left[ C_{N_c}\beta^{2N_c} \Gamma(\nu)
(\beta\nu\gamma^2)^{-\nu/2}\right]^{\frac{2}{2+\nu}},
\ee
where $C_{N_c}$ is the prefactor in Equ. (\ref{eq_d(rho)}). The
formula shows that $\Lambda_{PV}$ is the only dimensionful 
parameter. The final results are
\be
\label{MFA_res}
\chi_{top}\simeq\frac{N}{V} = (0.65\Lambda_{PV})^4,\hspace{0.5cm}
(\overline{\rho^2})^{1/2}= 0.47 \Lambda_{PV}^{-1} \simeq{1\over 3} R,
\hspace{0.5cm} \beta=S_0\simeq 15,
\ee
consistent with the phenomenological values for $\Lambda_{PV}
\simeq 300$ MeV. It is instructive to calculate the free energy
as a function of the instanton density. Using $F=-1/V\cdot\log Z$, 
we have
\be
\label{F_MFA}
F &=& \frac{N}{V} \left\{ \log\left(\frac{N}{2V\mu_0}\right)
 -\left(1+\frac{\nu}{2}\right) \right\}.
\ee
The instanton density is determined by the minimizing the free energy,
$\partial F/(\partial (N/V))=0$. The vacuum energy density is given by 
the value of the free energy at the minimum, $\epsilon=F_0$. We find 
$N/V=2\mu_0$ as above and
\be
\label{eps_MFA}
 \epsilon &=& -\frac{b}{4} \left(\frac{N}{V}\right)
\ee
Estimating the value of the gluon condensate in a dilute instanton gas,
$\langle G^2\rangle =32\pi^2(N/V)$, we see that (\ref{eps_MFA}) is 
consistent with the trace anomaly. Note that for non-interacting
instantons (with the size integration regularized in some fashion), 
one would expect $\epsilon\simeq-(N/V)$, which is inconsistent with 
the trace anomaly and shows the importance of respecting classical
scale invariance.

  The second derivative of the free energy with respect to the instanton 
density, the compressibility of the instanton liquid, is given by
\be
\label{comp_MFA}
 \left.\frac{\partial^2 F}{\partial (N/V)^2}\right|_{n_0} 
 &=& \frac{4}{b}\left(\frac{N}{V}\right) ,
\ee
where $n_0$ is the equilibrium density. This observable is also determined 
by a low energy theorem based on broken scale invariance \cite{NSVZ_81}
\be
\label{scal_let}
\int d^4x\; \langle G^2(0)G^2(x)\rangle  
 &=& (32\pi^2)\frac{4}{b}\langle G^2\rangle .
\ee
Here, the left hand side is given by an integral over the field strength
correlator, suitably regularized and with the constant disconnected term
$\langle G^2\rangle^2$ subtracted. For a dilute system of instantons, 
the low energy theorem gives 
\be
\label{n_fluc}
 \langle N^2\rangle -\langle N\rangle ^2
 &=&\frac{4}{b}\langle N\rangle .
\ee
Here, $\langle N\rangle$ is the average number of instantons in 
a volume $V$. The result (\ref{n_fluc}) shows that 
density fluctuations in the instanton liquid are not Poissonian. 
Using the general relation between fluctuations and the 
compressibility gives the result (\ref{comp_MFA}). This 
means that the form of the free energy near the minimum is 
determined by the renormalization properties of the theory.
Therefore, the functional form (\ref{F_MFA}) is more general 
than the mean field approximation used to derive it. 

   How reliable are the numerical results derived from the mean 
field approximation? The accuracy of the MFA can be checked by
doing statistical simulations of the full partition function. We 
will come to this approach in Sec. \ref{sec_sim}). The other question 
concerns the accuracy of the sum ansatz. This can be checked explicitly 
by calculating the induced current $j_\mu=D_\mu G^{cl}_{\mu\nu}$ in the 
classical gauge configurations \cite{Shu_85}. This current measures
the failure of the gauge potential to be a true saddle point. In the 
sum ansatz, the induced current gives a sizeable contribution to 
the action which means that this ansatz is not a good starting point 
for a self-consistent solution. 

  In principle, this problem is solved in the streamline method, because 
by construction $j_\mu$ is orthogonal to quantum fluctuations\footnote{
Except for the soft mode leading to the trivial gauge configuration, but 
the integration over this mode can be done explicitly.}. However, applying
the variational method to the streamline configurations \cite{Ver_91} is
also not satisfactory, because the ensemble contains too many close 
pairs and too many large instantons. 

  In summary: Phenomenology and the lattice seem to favor a fairly 
dilute instanton ensemble. This is well reproduced by the mean field 
approximation based on the sum ansatz, but the results are not really
self-consistent. How to generate a fully consistent ensemble in which
large instantons are automatically suppressed remains an open problem. 
Nevertheless, as long as large instantons are excluded in a way that
does not violate the symmetries of QCD, the results are likely to 
be independent of the precise mechanism that leads to the suppression
of large instantons.

\subsection{The quark condensate in the mean field approximation}
\label{sec_qbarq}

   Proceeding from pure glue theory to QCD with light quarks, one has
to deal with the much more complicated problem of quark-induced 
interactions. Not only does the fermion determinant induce a very
non-local interaction, but the very presence of instantons cannot
be understood in the single instanton approximation. Indeed, 
as discussed in Sec. \ref{sec_fermions}, the semi-classical instanton 
density is proportional to the product of fermion masses, and 
therefore vanishes in the chiral limit $m\rightarrow 0$. In the 
QCD vacuum, however, chiral symmetry is spontaneously broken and the 
quark condensate $\langle\bar qq\rangle$ is non-zero. The quark
condensate is the amplitude for a quark to flip its chirality,
so we expect that the instanton density is not controlled by the 
current masses, but by the quark condensate, which does not vanish 
as $m\to 0$.

  Given the importance of chiral symmetry breaking, we will discuss
this phenomenon on a number of different levels. In this section, we 
would like to give a simple qualitative derivation following 
\cite{Shu_82a}. Earlier works on chiral symmetry breaking by 
instantons are \cite{Cal_77,CC_79,CC_79b}, see also the review
\cite{Dia_95}. 

  The simplest case is QCD with just one light flavor, $N_f=1$. In 
this theory, the only chiral symmetry is the axial $U(1)_A$ symmetry 
which is broken by the anomaly. This means that there is no spontaneous 
symmetry breaking, and the quark condensate appears at the level of a 
single instanton. The condensate is given by
\be
\label{def_qq}
 \langle \bar qq\rangle  &=& i\int d^4x \,{\rm tr}\left( S(x,x)\right) .
\ee
In the chiral limit, non-zero modes do not contribute to the quark
condensate. Using the zero mode propagator $S(x,y)=-\psi_0(x)\psi^
\dagger_0(y)/(im)$ the contribution of a single instanton to the  
quark condensate is given by $-1/m$. Multiplying this result by 
the density of instantons, we have $\langle\bar qq\rangle=-(N/V)/m$. 
Since the instanton density is proportional to the quark mass $m$, 
the quark condensate is finite in the chiral limit.

  The situation is different for $N_f>1$. The theory still
has an anomalous $U(1)_A$ symmetry which is broken by instantons.
The corresponding order parameter $\det_f(\bar q_{f,L}q_{f,R})$ 
(where $f$ is the flavor index) appears already at the one-instanton 
level. But in addition to that, there is a chiral $SU(N_f)_L \times 
SU(N_f)_R$ symmetry which is spontaneously broken to $SU(N_f)_V$. 
This effect cannot be understood on the level of a single instanton: 
the contribution to $\langle\bar qq\rangle$ is still $(N/V)/m$, but 
the density of instantons is proportional to $(N/V)\sim m^{N_f}$. 

   Spontaneous symmetry breaking has to be a collective effect involving 
infinitely many instantons. This effect is most easily understood 
in the context of the mean field method. For simplicity, we consider
small-size instantons. Then the tunneling rate is controlled by the 
vacuum expectation value (VEV) of the $2N_f$-fermion operator 
(\ref{Leff_nf2}) in the 't Hooft effective lagrangian. This VEV 
can be estimated using the ``vacuum dominance'' (or factorization)
approximation
\be
\label{vac_dom}
\langle \bar\psi\Gamma_1\psi\bar\psi\Gamma_2\psi\rangle  
 &=& \frac{1}{N^2}\left(
{\rm Tr}\left[\Gamma_1\right] {\rm Tr}\left[\Gamma_2\right] 
- {\rm Tr}\left[\Gamma_1\Gamma_2\right] \right)
\langle \bar qq\rangle ^2,
\ee
where $\Gamma_{1,2}$ is a spin, isospin, color matrix and $N=4N_fN_c$
is the corresponding degeneracy factor. Using this approximation, we find
that the factor $\prod_f m_f$ in the instanton density should be 
replaced by $\prod_f m^*_f$, where the effective quark mass is given by
\be  
\label{eff_mass}
m_f^* &=& m_f- {2\over 3}\pi^2\rho^2 \langle \bar q_f q_f\rangle  .
\ee  
Thus, if chiral symmetry is broken, the instanton density is $O((m^*)^
{N_f})$, finite in the chiral limit. 
 
  The next question is whether it is possible to make this estimate 
self-consistent and calculate the quark condensate from instantons. 
If we replace the current mass by the effective mass also in the 
quark propagator\footnote{We will give a more detailed explanation 
for this approximation in Sec. \ref{sec_cor_sia}.}, the contribution 
of a single instanton to the quark condensate is given by $1/m^*$ and, 
for a finite density of instantons, we expect 
\be 
\label{qbarq_MFA}
\langle \bar qq\rangle  &=& -\frac{(N/V)}{m^*} 
\ee 
This equation, taken together with (\ref{eff_mass}), gives a self-consistent
value of the quark condensate
\be  
\label{qbarq_selfc}
\langle \bar qq\rangle  &=& -\frac{1}{\pi\rho}\sqrt{\frac{3N}{2V}}
\ee  
Using the phenomenological values $(N/V)=1\,{\rm fm}^{-4}$ and $\rho=
0.33\,{\rm fm}$, we get $\langle\bar qq\rangle\simeq-(215 {\rm MeV})^3$,
quite consistent with the experimental value $\langle\bar qq\rangle\simeq
-(230 {\rm MeV})^3$. The effective quark mass is given by $m^* =\pi
\rho(2/3)^{1/2}(N/V)^{1/2}\simeq 170$ MeV. The self-consistent pair of 
equations (\ref{eff_mass}, \ref{qbarq_MFA}) has the general form of 
a gap equation. We will provide a more formal derivation of the gap 
equation in Sec. \ref{sec_hfa}.

\subsection{Dirac eigenvalues and chiral symmetry breaking}
\label{sec_dirac_spec}

  In this section we will get a different and more microscopic
perspective on the formation of the quark condensate. The main idea 
is to study the motion of quarks in a fixed gauge field, and then 
average over all gauge field configurations. This approach is 
quite natural from the point of view of the path integral (and lattice 
gauge theory). Since the integral over the quark fields can always 
be performed exactly, quark observables are determined by the exact 
quark propagator in a given gauge configuration, averaged over all 
gauge fields.

  In a given gauge field configuration, we can determine the spectrum
of the Dirac operator $i\dslash = (i\partial_\mu+A_\mu(x))\gamma_\mu$
\be
\label{D_eigen} 
i\dslash \psi_\lambda &=& \lambda\psi_\lambda,
\ee
where $\psi_\lambda$ is an eigenstate with ``virtuality" $\lambda$.
In terms of the spectrum, the quark propagator $S(x,y)=-\langle x|
i{\dslash}^{-1}|y\rangle$ is given by
\be
\label{S_rep} 
S(x,y) &=& -\sum_\lambda \frac{\psi_\lambda(x)\psi^\dagger_\lambda(y)}
{\lambda+im} 
\ee
Using the fact that the eigenfunctions are normalized, the quark 
condensate is 
\be 
\label{qbarq_rep}
i\int d^4x \,{\rm tr}\left(S(x,x)\right) &=& 
  -i \sum_\lambda \frac{1}{\lambda+im} .
\ee
We can immediately make a few important observations concerning
the spectrum of the Dirac operator:
\begin{enumerate}

\item Since the Dirac operator is hermitean, the eigenvalues $\lambda$ 
are real. The inverse propagator $(i\dslash+im)^{-1}$, on the other hand, 
consists of a hermitean and an antihermitean piece.

\item For every non-zero eigenvalue $\lambda$ with eigenvector $\psi_
\lambda$, there is another eigenvalue $-\lambda$ with eigenvector 
$\gamma_5\psi_\lambda$.

\item This implies that the fermion determinant is positive: combining
the non-zero eigenvalues in pairs, we get 
\be
\label{det_rep}
\prod_{\lambda\neq 0} (i\lambda - m) &=&
\prod_{\lambda\geq 0} (i\lambda - m)(-i\lambda-m)\,=\,
\prod_{\lambda\geq 0} (\lambda^2+m^2)
\ee

\item Only zero modes can be unpaired. Since $\gamma_5\psi_0=\pm\psi_0$,
zero mode wave functions have to have a definite chirality. We have 
already seen this in the case of an instantons, where the Dirac
operator has a left-handed zero mode.
\end{enumerate}

  Using the fact that non-zero eigenvalues are paired, the trace of the 
quark propagator can be written as
\be
\label{qbarq_rep2}
i\int d^4x\,{\rm tr}\left(S(x,x)\right) &=& -\sum_{\lambda\geq 0} 
 \frac{2m}{\lambda^2+m^2}.
\ee
We have excluded zero modes since they do not contribute to the 
quark condensate in the limit $m\to 0$ (for $N_f>1$). In order 
to determine the average quark condensate, we introduce the
spectral density $\rho(\nu)=\langle\sum_\lambda \delta (\nu-
\lambda)\rangle$. We then have
\be
\label{qbarq_rep3}
\langle \bar qq\rangle &=& -\int_0^\infty d\lambda\,\rho(\lambda) 
\frac{2m}{\lambda^2+m^2}.
\ee
This result shows that the order in which we take the chiral and 
thermodynamic limits is crucial. In a finite system, the integral 
is well behaved in the infrared and the quark condensate vanishes 
in the chiral limit. This is consistent with the observation that 
there is no spontaneous symmetry breaking in a finite system.
A finite spin system, for example, cannot be magnetized if there 
is no external field. If the energy barrier between states
with different magnetization is finite and there is no external 
field that selects a preferred magnetization, the system will tunnel
between these states and the average magnetization is zero. Only in 
the thermodynamic limit can the system develop a spontaneous 
magnetization. 

   However, if we take the thermodynamic limit first, we can have a
finite density of eigenvalues arbitrarily close to zero. In this
case, the $\lambda$ integration is infrared divergent as $m\to 0$
and we get a finite quark condensate
\be 
\langle\bar qq\rangle  &=& -\pi \rho(\lambda=0),
\ee
a result which is known as the Banks-Casher relation \cite{BC_80}.
This expression shows that quark condensation is connected with
quark states of arbitrarily small virtuality.

   Studying chiral symmetry breaking requires an understanding
of quasi-zero modes, the spectrum of the Dirac operator near 
$\lambda=0$. If there is only one instanton the spectrum consists 
of a single zero mode, plus a continuous spectrum of non-zero modes. 
If there is a finite density of instantons, the spectrum is complicated, 
even if the ensemble is very dilute. In the chiral limit, fluctuations 
of the topological charge are suppressed, so one can think of the system 
as containing as many instantons as anti-instantons. The zero modes are
expected to mix, so that the eigenvalues spread over some range of 
virtualities $\Delta\lambda$. If chiral symmetry is broken, the natural 
value of the quark condensate is of order $(N/V)/\Delta \lambda$. 

  There is a useful analogy from solid state physics. In condensed
matter, atomic bound states may become delocalized and form a band. 
The material is a conductor if the fermi surface lies inside a band. 
Such zones also exist in disordered systems like liquids, but in this 
case they do not have a sharp boundary. 

    In the basis spanned by the zero modes of the individual instantons 
the Dirac operator reduces to the matrix
\be 
\label{D_zmz_2}
iD\!\!\!\!/ &=& \left (
\begin{array}{cc}  0 & T_{IA}\\
                   T_{AI} & 0 
\end{array} \right ) ,
\ee
already introduced in Sec. \ref{sec_tia}. The width of the zero mode 
zone in the instanton liquid is governed by the off-diagonal matrix 
elements $T_{IA}$ of the Dirac operator. The matrix elements depend 
on the relative color orientation of the pseudo-particles. If the 
interaction between instantons is weak, the matrix elements are 
distributed randomly with zero average, but their variance is non-zero. 
Averaging $T_{IA}T^*_{IA}$ over the positions and orientations of a pair 
of pseudo-particles, one gets 
\be
\label{TIA_var}
\langle |T_{IA}|^2\rangle &=& \frac{2\pi^2}{3N_c}\frac{N \rho^2}{V}.
\ee 
If the matrix elements are distributed according to a Gaussian 
unitary  ensemble\footnote{In the original work \cite{DP_85a} from
which these arguments are taken, it was assumed that the spectrum 
has a Gaussian shape, with the width given above. However, for a 
random matrix the correct result is a semicircle. In reality, the 
spectrum of the Dirac operator for a system of randomly distributed
instantons is neither a semicircle nor a Gaussian, it has a non-analytic
peak at $\lambda=0$ \cite{SV_90}. This does not qualitatively change 
the estimate quark condensate.}, the spectral density is a semi-circle
\be
\label{semi_circ}
\rho(\lambda) &=& \frac{N}{\pi\sigma V} \left( 1 - 
\frac{\lambda^2}{4\sigma^2} \right)^{1/2}.
\ee
 From the Casher-Banks formula, we then get the following result for the 
quark condensate
\be  
\langle \bar qq\rangle  &=& -\frac{1}{\pi\rho} \left( \frac{3N_c}{2}
\frac{N}{V}\right)^{1/2} \simeq -(240\,{\rm MeV})^3, 
\ee 
which has the same parametric dependence on $(N/V)$ and $\rho$ as 
the result in the previous section, only the coefficient is slightly 
different. In addition to that, we can identify the effective mass 
$m^*$ introduced in the previous section with a weighted average
of the eigenvalues, ${(m^*)}^{-1} = N^{-1} \sum\lambda^{-1}$.

  It is very interesting to study the spectral density of the 
Dirac operator at small virtualities. Similar to the density of
states near the Fermi surface in condensed matter problems, this
quantity controls the low energy excitations of the system. If
chiral symmetry is broken, the spectral density at $\lambda=0$
is finite. In addition to that, chiral perturbation theory predicts 
the slope of the spectrum \cite{SS_93}
\be 
\label{eq_smilga_stern}
\rho(\lambda)= -\frac{1}{\pi} \langle\bar qq\rangle +
   \frac{\langle\bar qq\rangle^2}{32\pi^2f_\pi^4}
   \frac{N_f^2-4}{N_f} |\lambda|+ \ldots \, ,
\ee
which is valid for $N_f\geq 2$. The second term is connected with 
the fact that for $N_f>2$ there is a Goldstone boson cut in the 
scalar-isovector ($\delta$ meson) correlator, while the decay 
$\delta \to \pi\pi$ is not allowed for two flavors. The result 
implies that the spectrum is flat for $N_f=2$, but has a cusp 
for $N_f>2$.

\subsection{Effective interaction between quarks and the mean field 
approximation}
\label{sec_hfa}

   In this section we would like to discuss chiral symmetry breaking
in terms of an effective, purely fermionic theory that describes the
effective interaction between quarks generated by instantons \cite{DP_86}. 
For this purpose, we will have to reverse the strategy used in the last
section and integrate over the gauge field first. This will leave us
with an effective theory of quarks that can be treated with standard 
many-body techniques. Using these methods allows us not only to study
chiral symmetry breaking, but also the formation of quark-anti-quark 
bound states in the instanton liquid. 

  For this purpose we rewrite the partition function of the instanton  
liquid 
\be
\label{Z_ferm}
 Z &=& \frac{1}{N_+!N_-!}\prod_i^{N_++N_-}\int [d\Omega_i\, n(\rho_i)]
 \, \exp(-S_{int})\det(D\!\!\!\!/\,+m)^{N_f}
\ee
in terms of a fermionic effective action
\be
\label{eff_act}
Z &=& \int d\psi d\psi^\dagger \exp\left(\int d^4x\,
\psi^\dagger (i\partial\!\!\!/+im)\psi \right)
\left\langle \prod_{I,f} \left(\Theta_I-im_f\right) 
\prod_{A,f} \left(\Theta_A -im_f\right)
\right\rangle, \\
\label{tHooft_vert}
& &\Theta_{I,A} = \int d^4x\,\left( \psi^\dagger(x)i\partial
\!\!\!/\phi_{I,A}(x-z_{I,A})\right)
 \int d^4y\,\left( \phi^\dagger_{I,A}(y-z_{I,A})i\partial
\!\!\!/\psi(y)\right),
\ee
which describes quarks interacting via the 't Hooft vertices $\Theta_
{I,A}$. The expectation value $\langle.\rangle$ corresponds to an 
average over the distribution of instanton collective coordinates.
Formally, (\ref{eff_act}) can be derived by ``fermionizing" the original 
action, see e.g. \cite{Now_91}. In practice, it is easier check the result 
by performing the integration over the quark fields and verifying that one 
recovers the fermion determinant in the zero mode basis.

   Here, however, we want to use a different strategy and exponentiate 
the 't Hooft vertices $\Theta_{I,A}$ in order to derive the effective 
quark interaction. For this purpose we calculate the average in Eq. 
(\ref{eff_act}) with respect to the variational single instanton 
distribution (\ref{reg_dis}). There are no correlations, so only the 
average interaction induced by a single instanton enters. For simplicity, 
we only average over the position and color orientation and keep the 
average size $\rho=\overline\rho$ fixed
\be
\label{eff_ver}
Y_{\pm} &=& \int d^4z \int dU \,\prod_f \Theta_{I,A} .
\ee
In order to exponentiate $Y_\pm$ we insert factors of unity  $\int 
d\Gamma_\pm \int d\beta_\pm/(2\pi)\exp(i\beta_\pm(Y_\pm-\Gamma_\pm))$ 
and integrate over $\Gamma_\pm$ using the saddle point method
\be
\label{eff_spi}
Z &=& \int d\psi d\psi^\dagger \exp\left(\int d^4x\,
\psi^\dagger i\partial\!\!\!/\psi \right)
\int \frac{d\beta_\pm}{2\pi} \exp\left( i\beta_\pm Y_\pm\right)
\exp\left[
 N_+ \left( \log\left( \frac{N_+}{i\beta_+ V}\right) - 1\right)
 \; + \; (+\leftrightarrow -) \right], \nonumber
\ee
where we have neglected the current quark mass. In this partition 
function, the saddle point parameters $\beta_\pm$ play the role of 
an activity for instantons and anti-instantons. 

\subsubsection{The gap equation for $N_f$=1}

   The form of the saddle point equations for $\beta_\pm$ depends 
on the number of flavors. The simplest case is $N_f=1$, where the 
Grassmann integration is quadratic. The average over the 't Hooft 
vertex is most easily performed in momentum space
\be
\label{ver_mom}
  Y_\pm &=& \int \frac{d^4k}{(2\pi)^4}\int dU
  \psi^\dagger (k) \left[ k\!\!\!/ \phi_{I,A}(k)\phi^\dagger_{I,A}
 (k)k\!\!\!/ \right] \psi(k),
\ee
where $\phi(k)$ is the Fourier transform of the zero mode profile
(see appendix). Performing the average over the color orientation, 
we get
\be
\label{ver_nf1}
Y_\pm &=& \int \frac{d^4k}{(2\pi)^4} \frac{1}{N_c}
k^2\varphi^{\prime 2}(k)\psi^\dagger(k)\gamma_\pm\psi(k),
\ee
where $\gamma_\pm=(1\pm\gamma_5)/2$ and $\varphi^\prime(k)$ is defined
in the appendix. Clearly, the saddle point equations are symmetric 
in $\beta_\pm$, so that the average interaction is given by $Y_++Y_-$, 
which acts like a mass term. This can be seen explicitly by first 
performing the Grassmann integration
\be
\label{seff_nf1}
Z &=&\int \frac{d\beta_\pm}{2\pi} \exp\left[
 N_\pm\left(\log\frac{N_\pm}{i\beta_\pm V}-1 \right)
 +N_c V\int\frac{d^4k}{(2\pi)^4} {\rm tr}\log
 \left(k\!\!\!/ +\gamma_\pm\beta_\pm
 \frac{k^2\varphi^{\prime 2}(k)}{N_c} \right)\right]
\ee
and then doing the saddle point integral. Varying with respect to 
$\beta=\beta_\pm$ gives the gap equation \cite{DP_86}
\be 
\label{gap}
\int \frac{d^4k}{(2\pi)^4} \frac{M^2(k)}{k^2+M^2(k)} &=&
\frac{N}{4N_cV},
\ee
where $M(k)=\beta k^2\varphi^{\prime 2}(k)/N_c$ is the momentum 
dependent effective quark mass. The gap equation determines 
the effective constituent mass $M(0)$ in terms of the instanton 
density $N/V$. For the parameters (\ref{MFA_res}), the effective 
mass is $M\simeq 350$ MeV. We can expand the gap equation in the instanton 
density \cite{Pob_89}. For small $N/V$, one finds $M(0)\sim \rho
(\frac{N}{2VN_c})^{1/2}$, which parametrically behaves like the 
effective mass $m^*$ introduced above. Note that a dynamical mass 
is generated for arbitrarily small values of the instanton density. 
This is expected for $N_f=1$, since there is no spontaneous symmetry 
breaking and the effective mass is generated by the anomaly at 
the level of an individual instanton.

\subsubsection{The effective interaction for two or more flavors}

  In the context of QCD, the more interesting case is the one of
two or more flavors. For $N_f=2$, the effective 't Hooft vertex 
is a four-fermion interaction
\be
\label{vert_nf2}
 Y_\pm &=& \left[ \prod_{i=1,4} \int \frac{d^4k_i}{(2\pi)^4}k_i
 \varphi^{\prime}(k_i) \right] \,
 (2\pi)^4 \delta^4\left( {\textstyle \sum_i k_i} \right)\,
 \frac{1}{4(N_c^2-1)}
 \left( \frac{2N_c-1}{2N_c}(\psi^\dagger\gamma_\pm\tau_a^{-}\psi)^2
 + \frac{1}{8N_c}(\psi^\dagger\gamma_\pm\sigma_{\mu\nu}\tau_a^{-}\psi)^2 
   \right),
\ee
where $\tau_a^{-}=(\vec\tau,i)$ is an isospin matrix and we have 
suppressed the momentum labels on the quark fields. In the long 
wavelength limit $k\to 0$, the 't Hooft vertex (\ref{vert_nf2}) 
corresponds to a local four-quark interaction
\be
\label{leff_nf2}
{\cal L} &=& \beta(2\pi\rho)^4 \frac{1}{4(N_c^2-1)}
 \left( \frac{2N_c-1}{2N_c}\left[
  (\psi^\dagger\tau_a^{-}\psi)^2 +
  (\psi^\dagger\gamma_5\tau_a^{-}\psi)^2\right]
 + \frac{1}{4N_c}(\psi^\dagger\sigma_{\mu\nu}\tau_a^{-}\psi)^2 \right).
\ee
The structure of this interaction is identical to the one given in 
Equ. (\protect\ref{Leff_nf2}), as one can check using Fierz identities.
The form given here is a little more convenient in order to read off 
the instanton interaction in color singlet meson states. The only new 
ingredient is that the overall constant $\beta$ is determined 
self-consistently from a gap equation. In fact, as we will see below, 
$\beta$ is not simply proportional to the instanton density, but goes
to a constant as $(N/V)\to 0$.

  The Lagrangian (\ref{leff_nf2}) is of the type first studied by Nambu 
and Jona-Lasinio \cite{NJL_61} and widely used as a model for chiral 
symmetry breaking and as an effective description for low energy chiral
dynamics, see the reviews \cite{VW_91,Kle_92,HK_94}. Unlike the NJL model, 
however, the interaction has a natural cut-off parameter $\Lambda\sim
\rho^{-1}$, and the coupling constants in (\ref{leff_nf2}) are determined 
in terms of a physical parameter, the instanton density $(N/V)$. 
The interaction is attractive 
for quark-anti-quark pairs with the quantum numbers of the $\pi$ and 
$\sigma$ meson, and repulsive in the pseudoscalar-isoscalar (the $SU(2)$ 
singlet $\eta'$) and scalar-isovector $\delta$ channels, showing the 
effect of the $U(1)_A$ anomaly. Note that to first order in the instanton 
density, there is no interaction in the vector $\rho,\omega, a_1,f_1$ 
channels. 


   In the case of two (or more) flavors the Grassmann integration 
cannot be done exactly, since the effective action is more
than quadratic in the fermion fields. Instead, we perform the
integration over the quark fields in mean field approximation.
This procedure is consistent with the approximations used to
derive the effective interaction (\ref{eff_ver}). The MFA is 
most easily derived by decomposing fermion bilinears into a 
constant and a fluctuating part. The integral over the fluctuations 
is quadratic and can be done exactly. Technically, this can be 
achieved by introducing auxiliary scalar fields $L_a,R_a$ into 
the path integral\footnote{In the MFA, we do not need to introduce 
auxiliary fields $T_{\mu\nu}$ in order to linearize the tensor part 
of the interaction, since $T_{\mu\nu}$ cannot have a vacuum 
expectation value.} and then shifting the integration variables
in order to linearize the interaction. Using this method, the
four-fermion interaction becomes
\be
 (\psi^\dagger\tau^{-}_a\gamma_-\psi)^2 &\to &
2(\psi^\dagger\tau^{-}_a\gamma_-\psi)L_a -L_aL_a, \\
 (\psi^\dagger\tau^{-}_a\gamma_+\psi)^2 &\to &
2(\psi^\dagger\tau^{-}_a\gamma_+\psi)R_a -R_aR_a.
\ee
In the mean field approximation, the $L_a,R_a$ integration can be  
done using the saddle point method. Since isospin and parity are 
not broken, only $\sigma=L_4=R_4$ can have a non-zero value. At the 
saddle point, the free energy $F=-1/V\cdot\log Z$ is given by 
\be
\label{F_nf2}
 F &=& 4N_c \int\frac{d^4k}{(2\pi)^4} \log \left( 
  k^2 +\beta\sigma k^2\varphi^{\prime 2}(k)\right)
  -2\frac{2N_c(N_c^2-1)}{2N_c-1} \beta\sigma^2
  -\frac{N}{V}\log\left( \frac{\beta V}{N} \right).
\ee
Varying with respect to $\beta\sigma$ gives the same gap equation
as in the $N_f=1$ case, now with $M(k)=\beta\sigma k^2\varphi^{\prime 2}
(k)$. We also find $(N/V)=2f\sigma^2\beta$ where $f=2N_c(N_c^2-1)/
(2N_c-1)$. Expanding everything in $(N/V)$ one can show that 
$M(0)\sim (N/V)^{1/2}$, $\sigma\sim (N/V)^{1/2}$ and $\beta\sim
const$. 

  The fact that the gap equation is independent of $N_f$ is a 
consequence of the mean field  approximation. It implies that 
even for $N_f=2$, chiral symmetry is spontaneously broken for 
arbitrarily small values of the instanton density. As we will 
see in the next chapter, this is not correct if the full fermion 
determinant is included. If the instanton density is too small, 
the instanton ensemble forms a molecular gas and chiral symmetry 
is unbroken. However, as we will show in Sec. \ref{sec_sim}, the 
mean field approximation is quite useful for physically interesting 
values of the instanton density. 

  The quark condensate is given by
\be
\label{qbarq_hfa}
\langle \bar qq\rangle  &=& 
 -4N_c \int\frac{d^4k}{(2\pi)^4} \frac{M(k)}{M^2(k)+k^2}.
\ee
Solving the gap equation numerically, we get $\langle\bar qq\rangle
\simeq -(255\,{\rm MeV})^3$. It is easy to check that $\langle\bar 
qq\rangle \sim (N/V)^{1/2}\rho^{-1}$, in agreement with the results 
obtained in Secs. \ref{sec_qbarq} and \ref{sec_dirac_spec}. The 
relation (\ref{qbarq_hfa}) was first derived by Diakonov and Petrov 
using somewhat different techniques, see Sec. \ref{sec_prop_mfa}.

   The procedure for three flavors is very similar, so we do not
need to go into detail here. Let us simply quote the effective
lagrangian for $N_f=3$ \cite{NVZ_89c}
\be
\label{leff_nf3}
{\cal L} &=& \beta(2\pi\rho)^6 \frac{1}{6N_c(N_c^2-1)}
 \epsilon_{f_1f_2f_3}\epsilon_{g_1g_2g_3}
 \left( \frac{2N_c+1}{2N_c+4}
  (\psi_{f_1}^\dagger \gamma_+ \psi_{g_1})
  (\psi_{f_2}^\dagger \gamma_+ \psi_{g_2})
  (\psi_{f_3}^\dagger \gamma_+ \psi_{g_3}) \right. \\
& & \left. + \frac{3}{8(N_c+2)} 
  (\psi_{f_1}^\dagger \gamma_+ \psi_{g_1})
  (\psi_{f_2}^\dagger \gamma_+ \sigma_{\mu\nu} \psi_{g_2})
  (\psi_{f_3}^\dagger \gamma_+ \sigma_{\mu\nu}\psi_{g_3})
  + ( +\leftrightarrow - ) \right) \nonumber ,
\ee  
which was first derived in slightly different form in \cite{SVZ_80b}. 
So far, we have neglected the current quark mass dependence and 
considered the $SU(N_f)$ symmetric limit. Real QCD is intermediate 
between the $N_f=2$ and $N_f=3$ case. Flavor mixing in the instanton 
liquid with realistic values of quark masses was studied in \cite{NVZ_89c} 
to which we refer the reader for more details. 

  Before we discuss the spectrum of hadronic excitations let us
briefly summarize the last three subsections. A random system of
instantons leads to spontaneous chiral symmetry breaking. If the 
system is not only random, but also sufficiently dilute, this 
phenomenon ist most easily studied using the mean field approximation. 
We have presented the MFA in three slightly different versions:
using a schematic model in Sec. \ref{sec_qbarq}, using random 
matrix arguments in Sec. \ref{sec_dirac_spec}, and using an
effective quark model in this section. The results are consistent
and illustrate the phenomenon of chiral symmetry breaking in 
complementary ways. Of course, the underlying assumptions of
randomness and diluteness have to be checked. We will come 
back to this problem in Sec. \ref{sec_sim}.

\subsection{Bosonization and the spectrum of pseudo-scalar mesons}
\label{sec_bos}

   We have seen that the 't Hooft interaction leads to the spontaneous
breakdown of chiral symmetry and generates a strong interaction in
the pseudo-scalar meson channels. We will discuss mesonic correlation 
functions in great detail in chapter \ref{sec_cor}, and now consider 
only the consequences for the spectrum of pseudo-scalar mesons. The
pseudo-scalar spectrum is most easily obtained by bosonizing the 
effective action. In order to correctly describe the $U(1)_A$ 
anomaly and the scalar $\sigma$ channel, one has to allow for 
fluctuations of the number of instantons . The fluctuation 
properties of the instanton ensemble can be described by the 
following ``coarse grained" partition function \cite{NVZ_89b}
\be
\label{seff_cg}
 S_{eff} &=& \frac{b}{4} \int d^4z  \left( n^+(z)+n^-(z) \right)
 \left( \log\left(\frac{n^+(z)+n^-(z)}{n_0}\right) - 1 \right)
 + \frac{1}{2n_0} \int d^4z  \left( n^+(z)-n^-(z) \right)^2 \\
 & & + \int d^4z \left( \psi^\dagger ( i\partial\!\!\!/ +im)\psi
  - n^+(z)\overline\Theta_I(z) - n^-(z)\overline\Theta_A(z) \right), 
\nonumber 
\ee
where $n^\pm (z)$ is the local density of instantons/anti-instantons
and $\overline\Theta_{I,A}(z)$ is the 't Hooft interaction 
(\ref{tHooft_vert}) averaged over the instanton orientation. This 
partition function reproduces the low energy theorem (\ref{n_fluc})
as well as the relation $\chi_{top}=(N/V)$ expected for a dilute 
system in the quenched approximation. In addition to that, the 
divergence of the flavor-singlet axial current is given by $\partial_\mu 
j_\mu^5= 2N_f(n^+(z)-n^-(z))$, consistent with the axial $U(1)_A$
anomaly. 

   Again, the partition function can be bosonized by introducing 
auxiliary fields, linearizing the fermion interaction and performing
the integration over the quarks. In addition to that, we expand the
fermion determinant in derivatives of the meson fields in order 
to recover kinetic terms for the meson fields. This procedure
gives the standard nonlinear $\sigma$-model lagrangian. To leading 
order in the quark masses, the pion and kaon masses satisfy the 
Gell-Mann, Oakes, Renner relations
\be
\label{GMOR}
 f_\pi^2 m_\pi^2 &=& -2m \langle \bar qq\rangle , \\
 f_K^2 m_K^2   &=& -(m+m_s) \langle \bar qq\rangle ,
\ee
with the pion decay constant
\be
\label{fpi}
 f_\pi^2 &=& 4N_c \int \frac{d^4k}{(2\pi)^4} \frac{M^2(k)}
 {\left( k^2+M^2(k)\right)^2} \simeq (100 \,{\rm MeV})^2 .
\ee
To this order in the quark masses, $f_K^2=f_\pi^2$. The mass matrix
in the flavor singlet and octet sector is more complicated. One
finds
\be 
\label{V_eta_etap}
V &=& \frac{1}{2} \left( \frac{4}{3}m_K^2-\frac{1}{3}m_\pi^2\right)
 \eta_8^2 +\frac{1}{2}\left(\frac{2}{3}m_K^2+\frac{1}{3}m_\pi^2\right)
 \eta_0^2 + \frac{1}{2}\frac{4\sqrt{2}}{3}\left(m_\pi^2-m_K^2
 \right)\eta_0\eta_8 + \frac{N_f}{f_\pi^2} 
   \left(\frac{N}{V}\right) \eta_0^2.
\ee
The last term gives the anomalous contribution to the $\eta'$ mass. It  
agrees with the effective lagrangian originally proposed in \cite{Ven_79} 
and leads to the Witten-Veneziano relation (with $\chi_{top}=(N/V)$)
\be
f_\pi^2\left(m_{\eta'}^2+m_\eta^2-2m_K^2\right)= 2N_f (N/V) .
\ee
Diagonalizing the mass matrix for $m=5$ MeV and $m_s=120$ MeV, we 
find $m_\eta=527$ MeV, $m_{\eta'}=1172$ MeV and a mixing angle 
$\theta = -11.5^\circ$. The $\eta'$ mass is somewhat too heavy, 
but given the crude assumptions, the result is certainly not bad.
One should also note that the result corresponds to an ensemble 
of uncorrelated instantons. In full QCD, however, the topological 
susceptibility is zero and correlations between instantons have 
to be present, see Sec. \ref{sec_screen}.

\subsection{Spin dependent interactions induced by instantons}
\label{sec_V_ss}

   The instanton-induced effective interaction between light quarks 
produces large spin-dependent effects. In this section, we wish to
compare these effects with other spin-dependent interactions in 
QCD and study the effect of instantons on spin-dependent forces 
in heavy quark systems. In QCD, the simplest source of spin-dependent
effects is the hyperfine interaction generated by the one-gluon 
exchange (OGE) potential
\be 
\label{V_OGE}
 V^{OGE}_{ij} &=& -\frac{\alpha_s}{m_i m_j} \frac{\pi}{6}
  (\lambda^a_i\lambda^a_j)(\vec\sigma_i\vec\sigma_j)
  \delta^3(\vec r).
\ee
This interaction has at least two phenomenologically important
features: (a) The $(\vec\sigma\vec\sigma)(\lambda^a\lambda^a)$ term is 
twice bigger in mesons than in baryons and (b) the ratio of spin-dependent
forces in strange and non-strange system is controlled by the inverse
(constituent) quark mass.

  For comparison, the non-relativistic limit of the 't Hooft effective 
interaction is
\be
\label{V_inst}
  V^{inst}_{ij} &=& -\frac{\pi^2\rho^2}{6} \frac{(m^*)^2}{m_i^*m_j^*}
  \left( 1 + \frac{3}{32} \left(1+3\vec\sigma_i\vec\sigma_j\right)
  \lambda^a_i\lambda^a_j \right) \left( \frac{1-\tau^a_i\tau^a_j}{2}
  \right) \delta^3(\vec{r}).
\ee
The spin-dependent part of $V^{inst}$ clearly shares the attractive 
features mentioned above. The dependence on the effective mass
comes from having to close two of the external legs in the 
three flavor interaction (\ref{leff_nf3}). However, there are 
important differences in the flavor dependence of the OGE and
instanton interactions. In particularly, there is no 't Hooft 
interaction between quarks of the same flavor ($uu,dd$ or $ss$). 
Nevertheless, as shown in \cite{SR_89}, the potential provides 
a description of spin splittings in the octet and decuplet
which is as good as the OGE. The instanton induced potential has
two additional advantages over the OGE potential \cite{DZK_92}.
First, we do not have to use an uncomfortably large value of the
strong coupling constant $\alpha_s$, and the instanton potential 
does not have a (phenomenologically unwanted) large spin-orbit
part. 

  In addition to that, instantons provide genuine three-body forces. 
These forces only act in $uds$ singlet states, like the flavor singlet 
$\Lambda$ (usually identified with the $\Lambda(1405)$) or the 
hypothetical dilambda (H-dibaryon)\footnote{In \protect\cite{TO_90}
it was found that instanton-induced forces make the H unbound, a 
conclusion which is quite welcome since so far it has eluded all 
searches.}.

  Another interesting question concerns instanton induced 
forces between heavy quarks \cite{CDG_78b}. For heavy quarks,
the dominant part of the interaction is due to non-zero modes,
which we have completely neglected in the discussion above. 
These effects can be studied using the propagator of an
infinitely heavy quark
\be
\label{S_Q}
  S(x) &=& \frac{1+\gamma_4}{2} \delta^3 (\vec r) \Theta(\tau)
 P\exp\left(i\int A_\mu dx_\mu\right)
\ee
in the field of an instanton. Here, $P$ denotes a path ordered 
integral and we have eliminated the mass of the heavy quark using
a Foldy-Wouthuysen transformation. The phase accumulated by a heavy 
quark in the field of a single instanton (in singular gauge) is
\be 
 U(\vec{r}) &=& P\exp\left(i\int_{-\infty}^{\infty} A_4 dx_4\right)
 \;=\;  \cos(F(r)) + i\vec r\cdot\vec\tau \sin(F(r)) , \\
 & & \hspace{1cm} 
  F(r) \;=\; \pi\left(1-\frac{r}{\sqrt{r^2+\rho^2}}\right),
  \nonumber  
\ee
where $\vec r$ is the spatial distance between the instanton
and the heavy quark and $r=|\vec r|$. From this result, we can
determine the mass renormalization of the heavy quark due to 
the interaction with instantons in the dilute gas approximation
\cite{CDG_78b,DPP_89,CNZ_95}
\be
\label{mass_renorm}
\Delta M_Q &=& \frac{16\pi}{N_c} \left(\frac{N}{V}\right)\rho^3
  \cdot 0.552 \simeq 70 {\rm MeV}. 
\ee
In a similar fashion, one can determine the spin-independent part of
the heavy quark potential (for color singlet $\bar qq$ systems)
\be   
\label{V_QQ}
  V_{QQ}(\vec r_i,\vec r_j) &=& \int d\rho\, n(\rho) \int d^3r\; 
  \frac{1}{3} {\rm Tr} \left[U(\vec r_i-\vec r)
       U^\dagger(\vec r_j-\vec r)-1 \right].
\ee 
The potential (\ref{V_QQ}) rises quadratically at short distance but 
levels of at a value of $2\Delta M_Q$ for $r>5\rho$. This is a reflection 
of the fact that dilute instantons do not confine. Also, the magnitude of 
the potential is on the order of 100 MeV, too small to be of much importance 
in charmonium or bottonium systems. The spin dependent part of the heavy 
quark potential is
\be
\label{V_QQ_ss}
 V_{QQ}^{ss}(\vec r_i,\vec r_j) &=& -\frac{1}{4M_iM_j} 
 (\vec\sigma_i\vec\nabla_i)(\vec\sigma_j\vec\nabla_j) V(\vec r_i-\vec r_j).
\ee
and it is also too small to be important phenomenologically. More important 
is the instanton induced interaction in heavy-light systems. This problem 
was studied in some detail in \cite{CNZ_95}. The effective potential between 
the heavy and the light quark is given by
\be
\label{V_qQ}
 V_{qQ}(\vec r) &=& \frac{\Delta M_Q m^*_q}{2(N/V)N_c} \left[
  \left(1+\frac{\lambda^a_Q\lambda^a_q}{4} \right)
  - \frac{\Delta M_Q^{spin}}{\Delta M_Q} \vec\sigma_q\vec\sigma_Q
  \frac{\lambda^a_Q\lambda^a_q}{4} \right] \delta^3 (\vec r),
\ee
where $\Delta M_Q$ is the mass renormalization of the heavy quark and 
\be 
\Delta M_Q^{spin} &=& \frac{16\pi}{N_c} \left(\frac{N}{V}\right)
 \frac{\rho^2}{M_Q} \cdot 0.193 
\ee
controls the hyperfine interaction. This interaction gives very 
reasonable results for spin splittings in heavy-light mesons. In 
addition to that, instantons generate many-body forces which might 
be important in heavy-light baryons. We conclude that instantons
can account for spin-dependent forces required in light quark 
spectroscopy without the need for large hyperfine interactions. 
Instanton induced interactions are not very important in heavy 
quark systems, but may play a role in heavy-light systems. 

\section{The interacting instanton liquid }
\label{sec_sim}


\subsection{Numerical simulations}
\label{sec_method}

    In the last section we discussed an analytic approach 
to the statistical mechanics of the instanton ensemble based 
on the mean field approximation. This approach provides important
insights into the structure of the instanton ensemble and the 
qualitative dependence on the interaction. However, the method 
ignores correlations among instantons which are important 
for a number of phenomena, such as topological charge screening 
(Sec. \ref{sec_screen}), chiral symmetry restoration (Sec. 
\ref{sec_chi_res}) and hadronic correlation functions (Sec. 
\ref{sec_cor}).

   In order to go beyond the mean field approximation and study 
the instanton liquid with the 't Hooft interaction included to 
all orders, we have performed numerical simulations of the interacting 
instanton liquid \cite{Shu_88,SV_90,NVZ_89c,SS_96}. In these 
simulations, we make use of the fact that the quantum field 
theory problem is analogous to the statistical mechanics of 
a system of pseudo-particles in 4 dimensions. The distribution of 
the $4N_c N$ collective coordinates associated with a system of 
$N$ pseudo-particles can be studied using standard Monte Carlo 
techniques (e.g. the Metropolis algorithm), originally developed 
for simulations of statistical systems containing atoms or molecules. 

   These simulations have a number of similarities to lattice 
simulations of QCD, see the textbooks \cite{Cre_83,Rot_92,Mon_95}.
Like lattice gauge theory, we consider systems in a finite 
4-dimensional volume, subject to periodic boundary conditions. 
This means that both approaches share finite size problems,
especially the difficulty to work with realistic quark masses.
Also, both methods are formulated in euclidean space which means
that it is difficult to extract real time (in particular 
non-equilibrium) information. However, in contrast to the lattice,
space-time is continuous, so we have no problems with chiral 
fermions. Furthermore, the number of degrees is drastically 
reduced and meaningful (unquenched!) simulations of the
instanton ensemble can be performed in a few minutes on an
average workstation. Finally, using the analogy with an 
interacting liquid, it is easier to develop an intuitive
understanding of the numerical simulations.
   
  The instanton ensemble is defined by the partition function
\be
\label{part_fct}
Z =   \sum_{N_+,\, N_-} {1 \over N_+ ! N_- !}\int
    \prod_i^{N_+ + N_-} [d\Omega_i\; n(\rho_i) ]
    \exp(-S_{int})\prod_f^{N_f} \det(D\!\!\!\!/\,+m_f) \, ,
\ee
describing a system of pseudo-particles interacting via the bosonic
action and the fermionic determinant. Here, $d\Omega_i=dU_i\, d^4z_i\,
d\rho_i$ is the measure in the space of collective coordinates (color 
orientation, position and size) associated with a single instantons and 
$n(\rho)$ is the single instanton density (\ref{eq_d(rho)}). 

  The gauge interaction between instantons is approximated by a sum
of pure two-body interaction $S_{int}=\frac{1}{2}\sum_{I\neq J}S_{int}
(\Omega_{IJ})$. Genuine three-body effects in the instanton interaction
are not important as long as the ensemble is reasonably dilute. This 
part of the interaction is fairly easy to deal with. The computational
effort is similar to a statistical system with long range forces.

  The fermion determinant, however, introduces non-local interactions 
among many instantons. Changing the coordinates of a single instanton 
requires the calculation of the full $N$-instanton determinant, not 
just $N$ two-body interactions. Evaluating the determinant exactly 
is a quite formidable problem. In practice we proceed as in Sec. 
\ref{sec_hfa} and factorize the determinant into a low and a high 
momentum part 
\be
\label{det}
  \det(D\!\!\!\!/\,+m_f) = \left( \prod_{i}^{N_++N_-}\hspace{-0.3cm}
  1.34 \rho_i\right) \;\det(T_{IA}+ m_f),
\ee
where the first factor, the high momentum part, is the product of
contributions from individual instantons calculated in Gaussian
approximation, whereas the low momentum part associated with the
fermionic zero modes of individual instantons is calculated exactly.
$T_{IA}$ is the $N_+\times N_-$ matrix of overlap matrix elements
introduced in \ref{sec_tia}. As emphasized before, this determinant
contains the 't Hooft interaction to all orders. 

  In practice, it is simpler to study the instanton ensemble for a 
fixed particle number $N=N_++N_-$. This means that instead of the 
grand canonical partition function (\ref{part_fct}), we consider the 
canonical ensemble 
\be
\label{can_part_fct}
Z_N =  {1 \over N_+ ! N_- !} \prod_i^{N_+ + N_-} [d\Omega_i\; n(\rho_i) ]
    \exp(-S_{int})\prod_f^{N_f} \det(D\!\!\!\!/\,+m_f) \, .
\ee
for different densities and determine the ground state by minimizing 
the free energy $F=-1/V\log Z_N$ of the system. Furthermore, we will 
only consider ensembles with no net topology. The two constraints 
$N/V={\rm const}$ and $Q=N_+-N_-=0$ do not affect the results in the 
thermodynamic limit. The only exceptions are of course fluctuations
of $Q$ and $N$ (the topological susceptibility and the compressibility
of the instanton liquid). In order to study these quantities one has
to consider appropriately chosen subsystems, see Sec. \ref{sec_screen}.

  In order to simulate the partition function (\ref{can_part_fct}) 
we generate a sequence $\{\Omega_i\}_j$ ($i=1,\ldots,N;\, j=1,\ldots,
N_{conf}$) of configurations according to the weight function 
$p(\{\Omega_i\}) \sim \exp(-S)$, where  
\be
\label{s_tot}
   S = -\sum_{i=1}^{N_++N_-}\log (n(\rho_i)) + S_{int}
               +{\rm tr}\log (D\!\!\!\!/\,+m_f)
\ee
is the total action of the configuration. This is most easily accomplished
using the Metropolis algorithm: A new configuration is generated using
some micro-reversible procedure $\{\Omega_i\}_j\to\{\Omega_i\}_{j+1}$.
The configuration is always accepted if the new action is smaller than 
the old one, and it is accepted with the probability $\exp(-\Delta S)$ 
if the new action is larger. Alternatively, one can generate the ensemble 
using other techniques, e.g. the Langevin \cite{NVZ_89c}, heat bath
or microcanonical methods. 

\subsection{The free energy of the instanton ensemble}
\label{sec_free}

  Using the sequence of configurations generated by the Metropolis
algorithm, it is straightforward to determine expectation values 
by averaging measurements in many configurations
\be
\label{exp_val}
  \langle {\cal O}\rangle  &=& \lim_{N\to\infty} \frac{1}{N} 
          \sum_{j=1}^N {\cal O}(\{\Omega_i\}) .
\ee  
This is how the quark and gluon condensates, as well as the hadronic 
correlation functions discussed in this and the following section
have been determined. However, more work is required to determine  
the partition function, which gives the overall normalization of
the instanton distribution. The knowledge of the partition function
is necessary in order to calculate the free energy and the 
thermodynamics of the system. In practice, the partition function
is most easily evaluated using the thermodynamic integration method 
\cite{Kir_31}. For this purpose we write the total action as
\be
S(\alpha) = S_{0}+\alpha \Delta S,
\ee
which interpolates between a solvable action $S_0$ and the full
action $S(\alpha\!=\! 1)=S_0+\Delta S$. If the partition function for 
the system governed by the action $S_0$ is known, the full partition 
function can be determined from
\be
\label{int_coup}
  \log Z(\alpha\! =\! 1) &=& \log Z(\alpha\! =\! 0)
  - \int_0^1 d\alpha'\, \langle 0| \Delta S |0\rangle_{\alpha'},
\ee
where the expectation value $\langle 0|.|0\rangle_{\alpha}$ depends on
the coupling constant $\alpha$. The obvious choice for decomposing the 
action of the instanton liquid would be to use the single-instanton 
action, $S_0=\sum_i\log(n(\rho_i))$, but this does not work since the 
$\rho$ integration in the free partition function is not convergent.
We therefore consider the following decomposition
\be
\label{S_var}
   S(\alpha) = \sum_{i=1}^{N_++N_-}\left(- \log (n(\rho_i)) +
           (1-\alpha)\nu\frac{\rho_i^2}{\,\overline{\rho^2}\,}\right)
           + \alpha \left( S_{int}
               +{\rm tr}\log (D\!\!\!\!/\,+m_f) \right),
\ee
where $\nu=(b-4)/2$ and $\overline{\rho^2}$ is the average size squared
of the instantons with the full interaction included. The $\rho_i^2$
term serves to regularize the size integration for $\alpha=0$. It does
not affect the final result for $\alpha=1$. The specific form 
of this term is irrelevant, our choice here is motivated by the fact
that $S(\alpha\! =\! 0)$ gives a single instanton distribution with 
the correct average size $\overline{\rho^2}$. The $\alpha=0$ partition 
function corresponds to the variational single instanton distribution 
\be
\label{Z_free}
   Z_0 = \frac{1}{N_+!\, N_-!} (V\mu_0)^{N_++N_-}, \hspace{1cm}
         \mu_0 = \int_0^\infty  d\rho \, n(\rho)
           \exp(-\nu\frac{\rho^2}{\,\overline{\rho^2}\,} ) ,
\ee
where $\mu_0$ is the normalization of the one-body distribution.
The full partition function obtained from integrating over the coupling  
$\alpha$ is
\be
\label{int_coup2}
  \log Z &=& \log (Z_0)
  + N \int_0^1 d\alpha'\,  \langle 0|
      \nu\frac{\rho^2}{\,\overline{\rho^2}\,} - \frac{1}{N}
      \left( S_{int}+{\rm tr}\log(D\!\!\!\!/\,+m_f) \right)
      |0\rangle_{\alpha'},
\ee
where $N=N_++N_-$. The free energy density is finally given by $F=-1/V\cdot
\log Z$ where $V$ is the four-volume of the system. The pressure and the
energy density are related to $F$ by $p=-F$ and $\epsilon = T \frac{dp}{dT}
-p$.

\subsection{The instanton ensemble}
\label{sec_sim_res}

    If correlations among instantons are important, the variational
(or mean field) partition function $Z_0$ is not expected to provide 
an accurate estimate for the partition function. This is certainly
the case in the presence of light fermions, in particular at finite 
temperature. In this section we want to present numerical results 
obtained from simulations of the instanton liquid at zero temperature.

\begin{figure}[t]
\begin{center}
\leavevmode
\epsfxsize=8cm
\epsffile{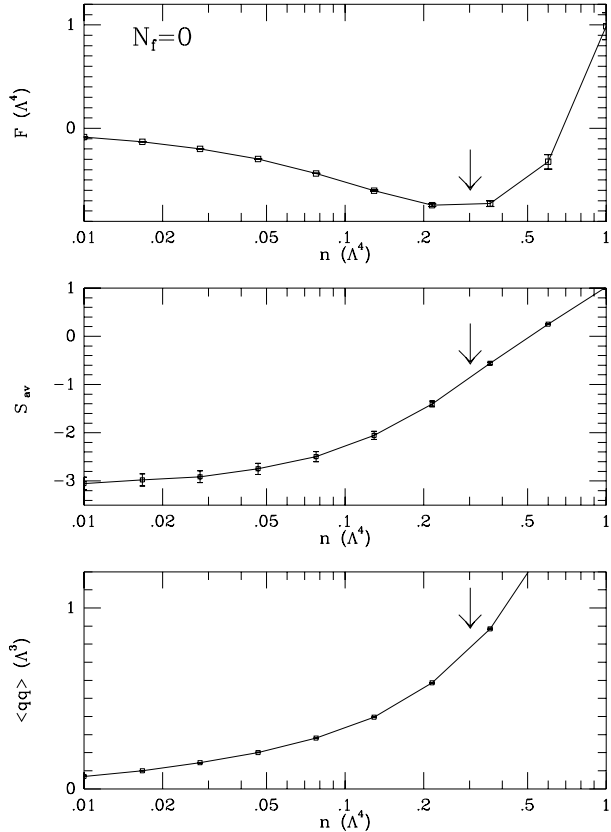}
\end{center}
\caption{\label{fig_F_nf0}   
Free energy, average instanton action and quark condensate as a 
function of the instanton density in the pure gauge theory,
from \protect\cite{SS_96}. All dimensionful quantities are 
given in units of the scale parameter $\Lambda_{QCD}$.}
\end{figure}

  As discussed in section \ref{sec_int_bos}, a general problem in 
the interacting instanton model is the treatment of very close 
instanton-anti-instanton pairs. In practice we have decided to deal 
with this difficulty by introducing a phenomenological short range 
repulsive core
\be
   S_{\rm core} = \frac{8\pi^2}{g^2} \frac{A}{\lambda^4}
   |u|^2, \hspace{1cm}
 \lambda = \frac{R^2+\rho_I^2+\rho_A^2}{2\rho_I\rho_A}
  + \left( \frac{(R^2+\rho_I^2+\rho_A^2)^2}{4\rho_I^2\rho_A^2}
   - 1\right)^{1/2}
\ee
into the streamline interaction. Here, $\lambda$ is the conformal
parameter (\ref{conf_par}) and $A$ controls the strength of the 
core. This parameter essentially governs the dimensionless diluteness 
$f=\rho^4(N/V)$ of the ensemble. The second parameter of the 
instanton liquid is the scale $\Lambda_{QCD}$ in the instanton
size distribution, which fixes the absolute units. 

  We have defined the scale parameter by fixing the instanton density 
to be $N/V=1\,{\rm fm}^{-4}$. This means that in our units, the 
average distance between instantons is 1 fm by definition. 
Alternatively, one can proceed as in lattice gauge simulations, 
and use an observable such as the $\rho$ meson mass to set the scale.
Using $N/V$ is very convenient and, as we will see in the next 
section, using the $\rho$ or nucleon mass would not make much 
of a difference. We use the same scale setting procedure for
all QCD-like theories, independent of $N_c$ and $N_f$. This
provides a simple prescription to compare dimensionful quantities
in theories with different matter content.

  The remaining free parameter is the (dimensionless) strength of
the core $A$, which determines the (dimensionless) diluteness of
the ensemble. In \cite{SS_96}, we chose $A=128$ which gives 
$f=\bar\rho^4(N/V)=0.12$ and $\bar\rho=0.43$ fm. As a result, the 
ensemble is not quite as dilute as phenomenology seems to demand
($(N/V)=1\,{\rm fm}^{-4}$ and $\bar\rho=0.33$ fm), but comparable 
to the lattice result $(N/V)=(1.4$-$1.6)\,{\rm fm}^{-4}$ and $\bar
\rho=0.35$ fm \cite{CGHN_94}. The average instanton action is 
$S\simeq 6.4$ while the average interaction is $S_{int}/N\simeq1.0$, 
showing that the system is still semi-classical and that interactions 
among instantons are important, but not dominant.

   Detailed simulations of the instanton ensemble in QCD are 
discussed in \cite{SS_96}. As an example, we show the free energy 
versus the instanton density in pure gauge theory (without 
fermions) in Fig. \ref{fig_F_nf0}. At small density the free energy 
is roughly proportional to the density, but at larger densities 
repulsive interactions become important, leading to a well-defined 
minimum. We also show the average action per instanton as a function 
of density. The average action controls the probability $\exp(-S)$ 
to find an instanton, but has no minimum in the range of densities 
studied. This shows that the minimum in the free energy is a compromise 
between maximum entropy and minimum action.

\begin{table}[t]
\caption{\label{tab_liquid_par}
Bulk parameters density $n=N/V$, average size $\bar\rho$, diluteness
$\bar\rho^4(N/V)$ and quark condensate $\langle\bar qq\rangle$ in the
different instanton ensembles. We also give the value of the 
Pauli-Vilars scale parameter $\Lambda^4$ that corresponds to 
our choice of units, $n\equiv 1\,{\rm fm}^{-4}$.}
\begin{tabular}{crrrr}
          & unquenched         & quenched           & RILM    
          & ratio ansatz (unqu.) \\  \tableline
$N/V$     & 0.174$\Lambda^4$   & 0.303$\Lambda^4$   & 1.0 ${\rm fm}^4$ 
          &  0.659$\Lambda^4$    \\
$\bar\rho$& 0.64$\Lambda^{-1}$ & 0.58$\Lambda^{-1}$ & 0.33 ${\rm fm}$  
          & 0.66$\Lambda^{-1}$   \\
          & (0.42 fm)          & (0.43 fm)          &     
          & (0.59 fm)            \\
$\bar\rho^4 (N/V)$ 
          & 0.029              & 0.034              & 0.012   
          & 0.125                \\
$\langle\bar qq\rangle$ 
          & 0.359$\Lambda^3$   & 0.825$\Lambda^3$   & $(264\,{\rm MeV})^3$ 
          & 0.882$\Lambda^3$     \\
          &$(219\,{\rm MeV})^3$&$(253\,{\rm MeV})^3$&          
          &  $(213\,{\rm MeV})^3$ \\
$\Lambda$ & 306 MeV            &    270 MeV         &  -    
          &   222 MeV           
\end{tabular}
\end{table}

  Fixing the units such that $N/V=1\,{\rm fm}^{-4}$, we have $\Lambda=270$
MeV and the vacuum energy density generated by instantons is $\epsilon=-526\,
{\rm MeV}/{\rm fm}^3$. We have already stressed that the vacuum energy 
is related to the gluon condensate by the trace anomaly. Estimating the
gluon condensate from the instanton density, we have $\epsilon=-b/4(N/V)
=-565\,{\rm MeV}/{\rm fm}^3$, which is in good agreement with the direct
determination of the energy density. Not only the depth of the free energy, 
but also its curvature (the instanton compressibility) is fixed from the
low energy theorem (\ref{comp_MFA}). The compressibility determined from 
Fig. \ref{fig_F_nf0} is $3.2(N/V)^{-1}$, to be compared with $2.75
(N/V)^{-1}$ from the low energy theorem. At the minimum of the free energy 
we can also determine the quark condensate (see Fig. \ref{fig_F_nf0}c). In 
quenched QCD, we have $\langle\bar qq\rangle=-(251 {\rm MeV})^3$, while a 
similar simulation in full QCD gives $\langle\bar qq\rangle=-(216 {\rm 
MeV})^3$, in good agreement with the phenomenological value.

\subsection{Dirac Spectra}
\label{sec_spectra}

\begin{figure}[t]
\begin{center}
\leavevmode
\epsfxsize=12cm
\epsffile{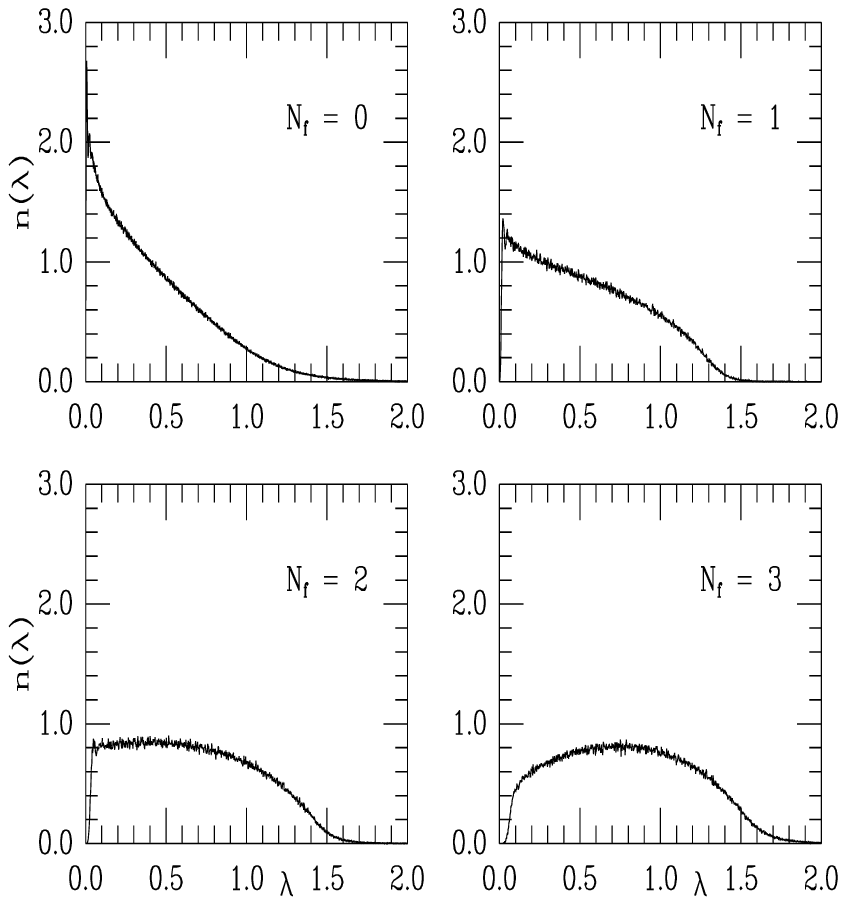}
\end{center}
\caption{\label{fig_dirac}   
Spectrum of the Dirac operator for different values of the 
number of flavors $N_f$, from \protect\cite{Ver_94b}. The 
eigenvalue is given in units of the scale parameter $\Lambda_{QCD}$
and the distribution function is normalized to one.}
\end{figure}

   We have already emphasized that the distribution of eigenvalues of the 
Dirac operator $iD\!\!\!\!/\,\psi_\lambda=\lambda\psi_\lambda$, is of great 
interest for many phenomena in QCD. In this section, we wish to study the 
spectral density of $iD\!\!\!\!/\,$ in the instanton liquid for different 
numbers of flavors. Before we present the results, let us review a few 
general arguments. First, since the weight function contains the fermion
determinant $\det_f(iD\!\!\!\!/\,)=(\prod_i\lambda_i)^{N_f}$, it is clear 
that small eigenvalues will be suppressed if the number of flavors is 
increased. This can also be seen from the Smilga-Stern result 
(\ref{eq_smilga_stern}). For $N_f=2$ the spectral density at $\lambda=0$ 
is flat, while for $N_f>2$ the slope of spectrum is positive. 

  In general, one might therefore expect that there is a critical number 
of flavors (smaller than $N_f=17$, where asymptotic freedom is lost) 
chiral symmetry is restored. There are a number of arguments that this 
indeed happens in non-abelian gauge theories, see \ref{sec_big_pic}. 
Let us only mention the simplest one here. The number of quark 
degrees of freedom is $N_q=4N_cN_f$, while, if chiral symmetry 
is broken, the number of low energy degrees of freedom (``pions") 
is $N_\pi=N_f^2-1$. If chiral symmetry is still broken for $N_f>12$, 
this leads to the unusual situation that the effective number of 
degrees of freedom at low energy is larger than the number of 
elementary degrees of freedom at high energy. In this case it 
is hard to see how one could ever have a transition to a phase
of weakly interacting quarks and gluons, since the pressure of
the low temperature phase is always bigger. 

   In Fig. \ref{fig_dirac} we show the spectrum of the Dirac operator
in the instanton liquid for $N_f=0,1,2,3$ light flavors \cite{Ver_94}.   
Clearly, the results are qualitatively consistent with the Smilga-Stern 
theorem\footnote{Numerically, the slope in the $N_f=3$ spectrum appears
to be too large, but it is not clear how small $\lambda$ has to be 
for the theorem to be applicable.} for $N_f\geq 2$. In addition to
that, the trend continues for $N_f<2$, where the result is not applicable. 
We also note that for $N_f=3$ (massless!) flavors a gap starts to open 
up in the spectrum. In order check whether this gap indicates chiral 
symmetry restoration in the infinite volume limit, one has to investigate 
finite size scaling. The problem was studied in more detail in \cite{SS_96}, 
where it was concluded that chiral symmetry is restored in the instanton 
liquid between $N_f=4$ and $N_f=5$. Another interesting problem is the
dependence on the dynamical quark mass in the chirally restored phase 
$N_f>N_f^{crit}$. If the quark mass is increased, the influence of the 
fermion determinant is reduced, and eventually ``spontaneous" symmetry 
breaking is recovered. As a consequence, QCD has an interesting phase 
structure as a function of the number of flavors and their masses, even 
at zero temperature.

\subsection{Screening of the topological charge}
\label{sec_screen}

   Another interesting phenomenon associated with dynamical quarks
is topological charge screening. This effect is connected with 
properties of the $\eta'$ meson, strong CP violation, and the 
structure of QCD at finite $\theta$ angle. 

   Topological charge screening can be studied in a number of 
complementary ways. Historically, it was first discussed on the 
basis of Ward identities associated with the anomalous $U(1)_A$
symmetry \cite{Ven_79}. Let us consider the flavor singlet 
correlation function $\Pi_{\mu\nu}=\langle \bar q\gamma_\mu
\gamma_5q(x)\bar q\gamma_\nu \gamma_5 q(0)\rangle$. Taking two
derivatives and using the the anomaly relation (\ref{u1a_anom}),  
we can derive the low energy theorem
\be
\label{ua1_wi}
\chi_{top} &=& \int d^4x\, \langle Q(x)Q(0)\rangle  \;=\;
 -\frac{m\langle\bar qq\rangle}{2N_f} 
 +\frac{m^2}{4N_f^2} \int d^4x\,
 \langle \bar q\gamma_5q(x)\bar q\gamma_5 q(0)\rangle .
\ee
Since the correlation function on the rhs does not have any massless
poles in the chiral limit, the topological susceptibility $\chi_{top}
\sim m$ as $m\to 0$. More generally, $\chi_{top}$ vanishes if there is
at least one massless quark flavor. 

   Alternatively, we can use the fact that the topological susceptibility
is the second derivative of the vacuum energy with respect to the $\theta$
angle. Writing the QCD partition function as a sum over all topological
sectors and extracting the zero modes from the fermion determinant, we
have
\be
\label{Z_nu}
 Z &=& \sum_\nu e^{i\theta\nu} \int_\nu dA_\mu\, e^{-S}
       {\det}_f(D\!\!\!\!/\, +m_f ) 
  \;=\; \sum_\nu \left(e^{i\theta}\det{\cal M}\right)^\nu \int_\nu dA_\mu\, 
    e^{-S} {\prod_{n,f}}^\prime \left(\lambda_n^2+m_f^2 \right) .
\ee
Here, $\nu$ is the winding number of the configuration, ${\cal M}$ is
the mass matrix and $\Pi^\prime$ denotes the product of all eigenvalues
with the zero modes excluded. The result shows that the partition 
function depends on $\theta$ only through the combination $e^{i\theta}
\det{\cal M}$, so the vacuum energy is independent of $\theta$ if one 
of the quark masses vanishes. 

\begin{figure}[t]
\begin{center}
\leavevmode
\epsfxsize=8cm
\epsffile{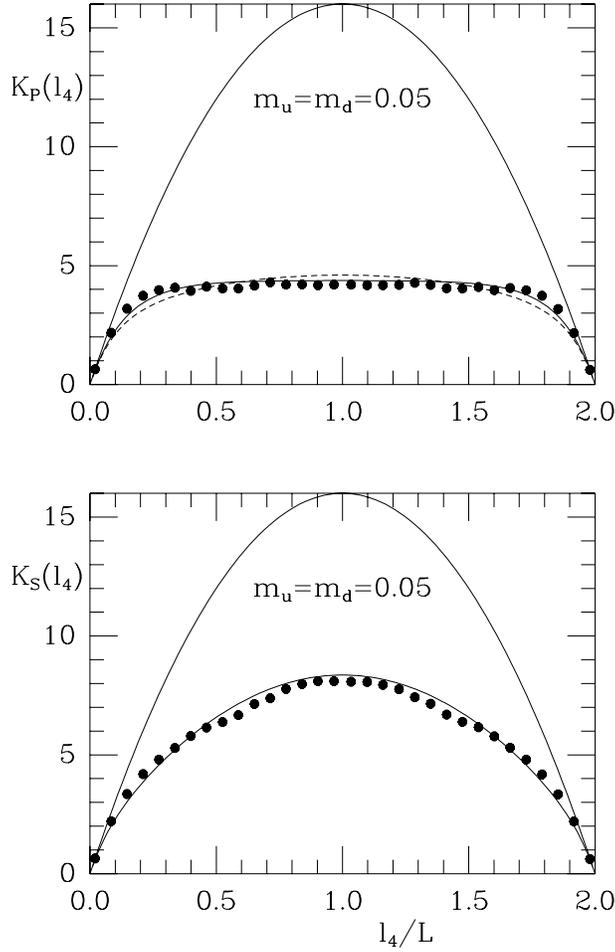}
\end{center}
\caption{\label{fig_screen}   
Pseudoscalar correlator $K_P(l_4)$ (upper panel) and scalar 
gluonic correlator $K_S(l_4)$ (lower panel) as a function of
the length $l_4$ of the subvolume $l_4\times L^3$, from 
\protect\cite{SV_95}. Screening implies that the correlator 
depends only on the surface, not on the volume of the torus. 
This means that in the presence of screening, $K(l_4)$ goes
to a constant. The results were obtained for $N_c=3$
and $m_u=m_d=10$ MeV and $m_s=150$ MeV. The upper solid
lines correspond to a random system of instantons, while 
the other solid line shows the parametrization discussed 
in text (the dashed line in the upper panel shows a slightly
more sophisticated parametrization). Note the qualitative 
difference between the data for topological and number 
fluctuations.}
\end{figure}

   The fact that $\chi_{top}$ vanishes implies that fluctuations in 
the topological charge are suppressed, so instantons and anti-instantons 
have to be correlated. Every instanton is surrounded by a cloud of 
anti-instantons which completely screen its topological charge, 
analogous to Debye screening in ordinary plasmas. This fact can be 
seen most easily by using the bosonized effective Lagrangian 
(\ref{V_eta_etap}). For simplicity we consider instantons to be 
point-like, but in contrast to the procedure in Sec. \ref{sec_bos}
we do allow the positions of the pseudo-particles to be correlated. 
The partition function is given by \cite{NVZ_89b,KW_92,DM_93,SV_95}
\be
\label{Z_Q_mes}
 Z &=& \sum_{N_+,N_-} \frac{\mu_0^{N_++N_-}}{N_+!N_-!}
       \prod_i^{N_++N_-}d^4z_i\,\exp\left(-S_{eff}\right), 
\ee
where $\mu_0$ is the single instanton normalization (\ref{var_ans}) 
and the effective action is given by 
\be
 S_{eff} &=& i\int d^4x\, \frac{\sqrt{2N_f}}{f_\pi} \eta_0 Q
            + \int d^4x\,{\cal L}(\eta_0,\eta_8).
\ee
Here, the topological charge density is $Q(x)=\sum Q_i\delta(x-z_i)$ and
${\cal L}(\eta_0,\eta_8)$ is the non-anomalous part of the pseudo-scalar 
meson lagrangian with the mass terms given in (\ref{V_eta_etap}). We can 
perform the sum in (\ref{Z_Q_mes}) and, keeping only the quadratic terms, 
integrate out the meson fields and determine the topological charge 
correlator in this model. The result is
\be 
\label{top_cor}
 \langle Q(x)Q(0)\rangle  &=& \left(\frac{N}{V}\right) \left\{
  \delta^4(x) -\frac{2N_f}{f_\pi^2}\frac{N}{V} 
  \left[ \cos^2(\phi) D(m_{\eta'},x) + \sin^2(\phi) D(m_\eta,x)
  \right] \right\},
\ee 
where $D(m,x)=m/(4\pi^2x)K_1(mx)$ is the (euclidean) propagator of
a scalar particle and $\phi$ is the $\eta-\eta'$ mixing angle. The
correlator (\ref{top_cor}) has an obvious physical interpretation. 
The local terms is the contribution from a single instanton located
at the center, while the second term is the contribution from the
screening cloud. One can easily check that the integral of the 
correlator is of order $m_\pi^2$, so $\chi_{top}\sim m$ in the 
chiral limit. We also observe that the screening length is given 
by the mass of the $\eta'$.

   Detailed numerical studies of topological charge screening in
the interacting instanton model were performed in \cite{SV_95}. 
The authors verified that complete screening takes place if one 
of the quark masses goes to zero and that the screening length 
is consistent with the $\eta'$ mass. They also addressed the 
question how the $\eta'$ mass can be extracted from topological
charge fluctuations. The main idea is not to study the limiting
value of $\langle Q^2\rangle/V$ for large volumes, but determine
its dependence on $V$ for small volumes $V<1\,{\rm fm}^4$. In 
this case, one has to worry about possible surface effects. It 
is therefore best to consider the topological charge in a segment 
$H(l_4)=l_4\times L^3$ of the torus $L^4$ (a hypercube with periodic 
boundary conditions). This construction ensures that the surface 
area of $H(l_4)$ is independent of its volume. Using the effective 
meson action introduced above, we expect (in the chiral limit)
\be
\label{box_cor}
 K_P(l_4) &\equiv & \langle Q(l_4)^2\rangle \;=\;
 L^3\left( \frac{N}{V} \right)
 \frac{1}{m_{\eta'}} \left( 1 - e^{-m_{\eta'}l_4} \right).
\ee
Numerical results for $K_P(l_4)$ are shown in Fig. \ref{fig_screen}.
The full line shows the result for a random system of instantons 
with a finite topological susceptibility $\chi_{top}\simeq (N/V)$
and the dashed curve is a fit using the parametrization (\ref{box_cor}).
Again, we clearly observe topological charge screening. Furthermore,
the $\eta'$ mass extracted from the fit is $m_{\eta'}=756$ MeV
(for $N_f=2$), quite consistent with what one would expect. The
figure also shows the behavior of scalar correlation function,
related to the compressibility of the instanton liquid. The instanton
number $N=N_++N_-$ is of course not screened, but the fluctuations
in $N$ are reduced by a factor $4/b$ due to the interactions, see
Eq. (\ref{comp_MFA}). For a more detailed analysis of the correlation
function, see \cite{SV_95}.

   We conclude that the topological charge in the instanton liquid
is completely screened in the chiral limit. The $\eta'$ mass is not
determined by the topological susceptibility, but by fluctuations
of the charge in small subvolumes.


\section{Hadronic correlation functions}
\label{sec_cor}

\subsection{Definitions and Generalities}
\label{sec_cor_def}

  In a relativistic field theory, current correlation functions 
determine the spectrum of hadronic resonances. In addition to that, 
hadronic correlation functions provide a bridge between hadronic 
phenomenology, the underlying quark-gluon structure and the 
structure of the QCD vacuum. The available theoretical and 
phenomenological information was recently reviewed in \cite{Shu_93}, 
so we will only give a brief overview here. 
 
  In the following, we consider hadronic point-to-point correlation 
functions 
\be
\label{cor_gen}
 \Pi_h(x) &=& \langle 0|j_h(x)j_h(0)|0\rangle  .
\ee
Here, $j_h(x)$ is a local operator with the quantum numbers of a hadronic 
state $h$. We will concentrate on mesonic and baryonic currents of the type
\be 
\label{cur}
 j_{mes}(x) &=& \delta^{ab}\bar\psi^a(x)\Gamma \psi^b(x) ,\\
 j_{bar}(x) &=& \epsilon^{abc}({\psi^a}^T(x)C\Gamma\psi^b(x))
                \Gamma'\psi^c(x).
\ee
Here, $a,b,c$ are color indices and $\Gamma,\Gamma'$ are isospin
and Dirac matrices. At zero temperature, we will focus exclusively
on correlators for spacelike (or euclidean) separation $\tau=\sqrt{-x^2}$.
The reason is that spacelike correlators are exponentially suppressed 
rather than oscillatory at large distance. In addition to that, most  
theoretical approaches, like the Operator Product Expansion
(OPE), lattice calculations or the instanton model deal with
euclidean correlators. 

  Hadronic correlation functions are completely determined by the 
spectrum (and the coupling constants) of the physical excitations 
with the quantum numbers of the current $j_h$. For a scalar correlation 
function, we have the standard dispersion relation
\be 
\label{disp_rel}
 \Pi(Q^2) &=& \frac{(-Q^2)^n}{\pi} \int ds\,
  \frac{{\rm Im}\Pi(s)}{s^n(s+Q^2)} + a_0 + a_1 Q^2 + \ldots ,
\ee
where $Q^2=-q^2$ is the euclidean momentum transfer and we have
indicated possible subtraction constants $a_i$. The spectral 
function $\rho(s)=\frac{1}{\pi}{\rm Im}\Pi(s)$ can be expressed 
in terms of physical states
\be
\label{spec_rep}  
\rho(s=-q^2) &=& (2\pi)^3 \sum_n \delta^4(q-q_n) 
  \langle 0|j_h(0)|n\rangle \langle n|j_h^\dagger(0)|0\rangle  ,
\ee
where $|n\rangle$ is a complete set of hadronic states. Correlation
functions with non-zero spin can be decomposed into Lorentz
covariant tensors and scalar functions. Fourier transforming
the relation (\ref{disp_rel}) gives a spectral representation
of the coordinate space correlation function
\be
\label{coord_rep}
 \Pi(\tau) &=& \int ds\,\rho(s) D(\sqrt{s},\tau).
\ee
Here, $D(m,\tau)$ is the euclidean propagator of a scalar particle
with mass $m$,
\be
  D(m,\tau) &=& \frac{m}{4\pi^2\tau} K_1(m\tau) .
\ee
Note that, except for possible contact terms, subtraction constants 
do not affect the coordinate space correlator. For large arguments,
the correlation function decays exponentially, $\Pi(\tau)\sim \exp
(-m\tau)$, where the decay is governed by the lowest pole in the 
spectral function. This is the basis of hadronic spectroscopy on the 
lattice. In practice, however, lattice simulations often rely on
more complicated sources in order to improve the suppression of 
excited states. While useful in spectroscopy, these correlation
functions mix short and long distance effects and are not as 
interesting theoretically.

  Correlation functions of currents built from quarks fields only 
(like the meson and baryon currents introduced above) can be expressed 
in terms of the full quark propagator. For an isovector meson current 
$j_{I=1} =\bar u\Gamma d$ (where $\Gamma$ is only a Dirac matrix), the 
correlator is given by the ``one-loop'' term 
\be
\label{mes_cor}
 \Pi_{I=1}(x) &=& \langle {\rm Tr}\left[ S^{ab}(0,x)\Gamma
 S^{ba}(x,0)\Gamma \right] \rangle .
\ee
The averaging is performed over all gauge configurations, with the
weight function $\det(D\!\!\!\!/\,+m)\exp(-S)$. Note that the quark 
propagator is not translation invariant before the vacuum average is 
performed, so the propagator depends on both arguments. Also note that 
despite the fact that (\ref{mes_cor}) has the appearance of a one-loop 
(perturbative) graph, it includes arbitrarily complicated, multi-loop, 
gluon exchanges as well as non-perturbative effects. All of these effects 
are hidden in the vacuum average. Correlators of isosinglet meson 
currents $j_{I=0}=\frac{1}{\sqrt{2}}(\bar u\Gamma u+\bar d\Gamma d)$ 
receive an additional two-loop, or disconnected, contribution
\be
\label{sing_cor}
 \Pi_{I=0}(x) &=& \langle {\rm Tr}\left[ S^{ab}(0,x)\Gamma
 S^{ba}(x,0)\Gamma \right] \rangle  - 
 2 \langle {\rm Tr}\left[ S^{aa}(0,0)\Gamma
 \right]\, {\rm Tr}\left[ S^{bb}(x,x)\Gamma \right]\rangle .
\ee
In an analogous fashion, baryon correlators can be expressed as 
vacuum averages of three quark propagators. 

  At short distance, asymptotic freedom implies that the
correlation functions are determined by free quark propagation.
The free quark propagator is given by
\be
\label{s_free}
  S_0(x) &=& \frac{i}{2\pi^2}\frac{\gamma\cdot x}{x^4}.
\ee
This means that mesonic and baryonic correlation functions
at short distance behave as $\Pi_{mes}\sim 1/x^6$ and
$\Pi_{bar}\sim 1/x^9$, respectively. Deviations from 
asymptotic freedom at intermediate distances can be studied 
using the operator product expansion (OPE). The basic idea 
\cite{Wil_69} is to expand the product of currents in (\ref{cor_gen}) 
into a series of coefficient functions $c_n(x)$ multiplied by local 
operators ${\cal O}_n(0)$
\be
\label{ope}
  \Pi(x) &=& \sum_n c_n(x) \langle {\cal O}_n(0)\rangle  .
\ee
From dimensional considerations it is clear that the most 
singular contributions correspond to operators of the 
lowest possible dimension. Ordinary perturbative contributions
are contained in the coefficient of the unit operator. The
leading non-perturbative corrections are controlled by the
quark and gluon condensates of dimension three and four. 

  In practice, the OPE for a given correlation function
is most easily determined using the short distance expansion 
of the propagator in external, slowly varying quark and gluon 
fields \cite{NSVZ_83}\footnote{This expression was derived in 
the Fock-Schwinger gauge $x_\mu A_\mu^a=0$. The resulting hadronic 
correlation functions are of course gauge invariant.}
\be 
\label{s_ope}
  S_f^{ab}(x) = \frac{i\delta^{ab}}{2\pi^2}\frac{\gamma\cdot x}{x^4}
   -\frac{\delta^{ab}}{4\pi^2}\frac{m_f}{x^2}
   + q^a(0)\bar q^b(0)
   - \frac{i}{32\pi^2}(\lambda^k)^{ab} G^k_{\alpha\beta}
   \frac{\gamma\cdot x \sigma_{\alpha\beta} + \sigma_{\alpha\beta}
   \gamma\cdot x}{x^2} + \ldots \; .
\ee
The corrections to the free propagator have an obvious interpretation
in terms of the interaction of the quark with the external quark and 
gluon fields. There is an enormous literature about QCD sum rules based 
on the OPE, see the reviews \cite{Shi_92,Nar_89,RRY_85}. The general
idea is easily explained. If there is a window in which both  
the OPE (\ref{ope}) has reasonable accuracy and the spectral
representation (\ref{coord_rep}) is dominated by the ground state,
one can match the two expressions in order to extract ground state 
properties. In general, the two requirements are in conflict  
with each other, so the existence of a sum rule window has to be 
established in each individual case.

    The main  sources of phenomenological information about
the correlation functions are \cite{Shu_93}:
\begin{enumerate}
\item Ideally, the spectral function is determined from an 
      experimentally measured cross section using the optical
      theorem. This is the case, for example, in the vector-isovector 
      (rho meson) channel, where the necessary input is provided
      by the ratio
      \be
       R(s) = \frac{\sigma (e^+e^-\to (I=1\,{\rm hadrons}))}
                   {\sigma (e^+e^-\to \mu^+\mu^-)},
      \ee
      where $s$ is the invariant mass of the lepton pair. Similarly, 
      in the axial-vector ($a_1$ meson channel) the spectral function 
      below the $\tau$ mass can be determined from the hadronic decay 
      width of the $\tau$ lepton $\Gamma(\tau\to\nu_\tau +{\rm hadrons})$.

\item In some cases, the coupling constants of a few resonances
      can be extracted indirectly, for example using low energy theorems.
      In this way, the approximate shape of the pseudo-scalar $\pi,K,\eta,
      \eta'$ and some glueball correlators can be determined.

\item Ultimately, the best source of information about hadronic 
      correlation functions is the lattice. At present most lattice 
      calculations use complicated non-local sources. Exceptions 
      can be found in \cite{CGHN_93a,CGHN_93b,Lei_95a,Lei_95b}. So 
      far, all results have been obtained in the quenched approximation.
\end{enumerate}

In general, given the fundamental nature of hadronic correlators,
all models of hadronic structure or the QCD vacuum should be tested
against the available information on the correlators. We will discuss
some of these models as we go along. 

\subsection{The quark propagator in the instanton liquid}
\label{sec_prop}

  As we have emphasized above, the complete information about 
mesonic and baryonic correlation functions is encoded in the 
quark propagator in a given gauge field configurations. Interactions
among quarks are represented by the failure of expectation values
to factorize, e.g. $\langle S(\tau)^2\rangle \neq \langle S(\tau)
\rangle ^2$. In the following, we will construct the quark propagator 
in the instanton ensemble, starting from the propagator in the 
background field of a single instanton.  

  From the quark propagator, we calculate the ensemble averaged 
meson and baryon correlation functions. However, it is also 
interesting to study the vacuum expectation value of the 
propagator\footnote{The quark propagator is of course not a 
gauge invariant object. Here, we imply that a gauge has been 
chosen or the propagator is multiplied by a gauge string. Also 
note that before averaging, the quark propagator has a more general
Dirac structure $S(x)=E+P\gamma_5+V_\mu\gamma_\mu+A_\mu\gamma_\mu
\gamma_5+T_{\mu\nu}\sigma_{\mu\nu}$. This decomposition, together
with positivity, is the basis of a number of exact results about 
correlation functions \protect\cite{Wei_83,VW_84}.}  
\be
 \langle S(x)\rangle &=& S_S (x) + \gamma\cdot x S_V(x),
\ee
From the definition of the quark condensate, we have $\langle
\bar qq\rangle = S_S(0)$, which means that the scalar component of 
the quark propagator provides an order parameter for chiral symmetry 
breaking. To obtain more information, we can define a gauge 
invariant propagator by adding a Wilson line 
\be
\label{s_inv}
 S_{inv}(x) &=& \langle  \psi(x)P\exp(\int_0^x A_\mu(x')\, dx'_\mu)
                \bar\psi(0) \rangle .
\ee
This object has a direct physical interpretation, because
it describes the propagation of a light quark coupled to an 
infinitely heavy, static, source \cite{Shu_82,SV_93a,CNZ_95}. 
It therefore determines the spectrum of heavy-light mesons 
(with the mass of the heavy quark subtracted) in the limit where 
the mass of the heavy quark goes to infinity. 
 
\subsubsection{The propagator in the field of a single instanton}
\label{sec_prop_sia}

  The propagators of massless scalar bosons, gauge fields and 
fermions in the background field of a single instanton can be
determined analytically\footnote{The result is easily generalized 
to 't Hoofts exact multi-instanton solution, but much more effort
is required to construct the quark propagator in the most general
(ADHM) instanton background \cite{CGT_79}.} \cite{BCC_78,LY_79}. 
We do not go into details of the construction, which is quite 
technical, but only provide the main results. 

  We have already seen that the quark propagator in the field
of an instanton is ill behaved because of the presence of a 
zero mode. Of course, the zero mode does not cause any harm, 
since it is compensated by a zero in the tunneling probability. 
The remaining non-zero mode part of the propagator satisfies 
the equation
\be
\label{s_nz_def}
 iD\!\!\!\!/\, S^{nz}(x,y) &=& \delta (x-y) 
       -\psi_0(x)\psi_0^\dagger (y),
\ee
which ensures that all modes in $S^{nz}$ are orthogonal to the 
zero mode. Formally, this equation is solved by
\be 
\label{s_nz_con}
 S^{nz}(x,y) &=& \stackrel{\rightarrow}{D\!\!\!\!/\,}_x \Delta (x,y) 
 \left( \frac{1+\gamma_5}{2}\right) +
 \Delta (x,y)\stackrel{\leftarrow}{D\!\!\!\!/\,}_y \left(
  \frac{1-\gamma_5}{2}\right) ,
\ee
where $\Delta(x,y)$ is the propagator of a scalar quark in the 
fundamental representation, $-D^2 \Delta (x,y) = \delta (x,y)$.
Equation (\ref{s_nz_con}) is easily checked using the methods
we employed in order to construct the zero mode solution, see
Eq. (\ref{dslash2}). The scalar propagator does not have any zero 
modes, so it can be constructed using standard techniques. The 
result (in singular gauge) is \cite{BCC_78}
\be
\label{s_scal}
 \Delta (x,y) &=& \frac{1}{4\pi^2 (x-y)^2}
 \frac{1}{\sqrt{1+\rho^2/x^2}}\frac{1}{\sqrt{1+\rho^2/y^2}}
 \left( 1 + \frac{\rho^2\tau^-\!\cdot x\, \tau^+\!\cdot y}{x^2y^2}
 \right) .
\ee
For an instanton located at $z$, one has to make the obvious
replacements $x\to (x-z)$ and $y\to(y-z)$. The propagator in 
the field of an anti-instanton is obtained by interchanging 
$\tau^+$ and $\tau^-$. If the instanton is rotated by the 
color matrix $R^{ab}$, then $\tau^\pm$ have to be replaced
by $(R^{ab}\tau^b,\mp i)$. 

  Using the result for the scalar quark propagator and the 
representation (\ref{s_nz_con}) of the spinor propagator
introduced above, the non-zero mode propagator is given by
\be
\label{s_nz}
S^{nz}(x,y) &=& \frac{1}{\sqrt{1+\rho^2/x^2}}\frac{1}{\sqrt{1+\rho^2/y^2}}
\left( S_0(x,y)\left( 1 + 
   \frac{\rho^2\tau^-\!\cdot x\, \tau^+\!\cdot y}
     {x^2y^2} \right) \right.\nonumber \\
& & \hspace{1cm}-D_0(x,y)\frac{\rho^2}{x^2y^2} \left.\left( 
\frac{\tau^-\!\cdot x\, 
   \tau^+\!\cdot\gamma\, 
      \tau^-\!\cdot(x-y)\,
         \tau^+\!\cdot y}
            {\rho^2+x^2}\gamma_+ 
\;+\; \frac{\tau^-\!\cdot x\, 
         \tau^+\!\cdot(x-y)\,
            \tau^-\!\cdot\gamma\,
              \tau^+\!\cdot y}
                 {\rho^2+x^2}\gamma_- \right)\right),
\ee
where $\gamma_\pm=(1\pm\gamma_5)/2$. The propagator can be generalized 
to arbitrary instanton positions and color orientations in the same 
way as the scalar quark propagator discussed above. 

  At short distance, as well as far away from the instanton,  
the propagator reduces to the free one. At intermediate distance, 
the propagator is modified due to gluon exchanges with the 
instanton field
\be
\label{s_nz_exp}
 S^{nz}(x,y) &=& -\frac{\gamma\cdot (x-y)}{2\pi^2(x-y)^4}
 - \frac{1}{16\pi^2(x-y)^2}(x-y)_\mu\gamma_\nu\gamma_5
   \tilde G_{\mu\nu} + \ldots  \; .
\ee
This result is consistent with the OPE of the quark propagator
in a general background field, see eq.(\ref{s_ope}). It is
interesting to note that all the remaining terms are regular 
as $(x-y)^2\to 0$. This has important consequences for the OPE 
of hadronic correlators in a general self-dual background field 
\cite{DS_81}.

  Finally, we need the quark propagator in the instanton field for 
small but non-vanishing quark mass. Expanding the quark propagator 
for small $m$, we get
\be
\label{s_m}
S(x,y) &=& \frac{1}{iD\!\!\!\!/ +im} \; =\;
  \frac{\psi_0(x)\psi^\dagger_0(y)}{m}
+ S^{nz}(x,y)  + m\Delta(x,y) + \ldots .
\ee

\subsubsection{The propagator in the instanton ensemble}
\label{sec_prop_ens}

  In this section we generalize the results of the last section to 
the more general case of an ensemble consisting of many pseudo-particles.
The quark propagator in an arbitrary gauge field can always be expanded as 
\be
\label{prop_exp} 
 S &=& S_0 + S_0 A\!\!\!/ \, S_0 + S_0 A\!\!\!/\,
  S_0 A\!\!\!/ S_0 + \ldots ,
\ee
where the individual terms have an obvious interpretation
as arising from multiple gluon exchanges with the background
field. If the gauge field is a sum of instanton contributions,
$A_\mu = \sum_I A_{I\,\mu}$, then (\ref{prop_exp}) becomes
\be
\label{prop_sum} 
 S &=& S_0 + \sum_I S_0 {A\!\!\!/}_I S_0
 + \sum_{I,J} S_0 {A\!\!\!/}_I
  S_0 {A\!\!\!/}_J S_0 + \ldots \\
\label{prop_hop}
 &=& S_0 + \sum_I (S_I - S_0) + \sum_{I\neq J} 
 (S_I - S_0) S_0^{-1} (S_J - S_0) \\
 & & + \sum_{I\neq J,\,J\neq K}  (S_I - S_0) S_0^{-1} 
 (S_J - S_0)S_0^{-1}(S_K-S_0) + \ldots \; .\nonumber
\ee
Here, $I,J,K,\ldots$ refers to both instantons and anti-instantons. 
In the second line, we have resummed the contributions corresponding 
to an individual instanton. $S_I$ refers to the sum of zero and 
non-zero mode components. At large distance from the center of the 
instanton, $S_I$ approaches the free propagator $S_0$. Thus Eq.
(\ref{prop_hop}) has a nice physical interpretation: Quarks propagate 
by jumping from one instanton to the other. If $|x-z_I|\ll \rho_I,\,
|y-z_I| \ll \rho_I$ for all $I$, the free propagator dominates. At 
large distance, terms involving more and more instantons become 
important.

  In the QCD ground state, chiral symmetry is broken. The 
presence of a condensate implies that quarks can propagate 
over large distances. Therefore, we cannot expect that 
truncating the series (\ref{prop_hop}) will provide a useful 
approximation to the propagator at low momenta. Furthermore, 
we know that spontaneous symmetry breaking is related to small 
eigenvalues of the Dirac operator. A good approximation to the 
propagator is obtained by assuming that $(S_I-S_0)$ is dominated 
by fermion zero modes
\be
\label{zm_dom}
 \left(S_I-S_0\right)(x,y) &\simeq& 
   \frac{\psi_I(x)\psi_I^\dagger(y)}{im}.
\ee
In this case, the expansion (\ref{prop_hop}) becomes
\be
 S(x,y) &\simeq& S_0(x,y) + \sum_I\frac{\psi_I(x)
  \psi_I^\dagger(y)}{im}
   + \sum_{I\neq J} \frac{\psi_I(x)}{im}
      \left( \int d^4r\,\psi_I^\dagger(r)
      (-i\partial\!\!\!/ -im)\psi_J(r) \right)
     \frac{\psi_J^\dagger(y)}{im} + \ldots ,
\ee
which contains the overlap integrals $T_{IJ}$ defined in Eq. 
(\ref{def_TIA}). This expansion can easily be summed to give
\be
\label{prop_zmz}
 S(x,y) &\simeq& S_0(x,y) + \sum_{I,J}\psi_I(x)
    \frac{1}{T_{IJ}+im D_{IJ} -im \delta_{IJ}}
     \psi_J^\dagger(y).
\ee
Here, $D_{IJ}= \int d^4r\,\psi_I^\dagger(r)\psi_J(r)-\delta_{IJ}$ 
arises from the restriction $I\neq J$ in the expansion (\ref{prop_hop}). 
The quantity $mD_{IJ}$ is small in both the chiral expansion and in 
the packing fraction of the instanton liquid and will be neglected in 
what follows. Comparing the resummed propagator (\ref{prop_zmz}) with 
the single instanton propagator (\ref{zm_dom}) shows the importance of 
chiral symmetry breaking. While (\ref{zm_dom}) is proportional to $1/m$, 
the diagonal part of the full propagator is proportional to $(T^{-1})_{II}
=1/m^*$.

    The result (\ref{prop_zmz}) can also be derived by 
inverting the Dirac operator in the basis spanned by the 
zero modes of the individual instantons
\be
\label{prop_zmz2}
 S(x,y) &\simeq& S_0(x,y) + \sum_{I,J}|I\rangle \langle I|
    \frac{1}{iD\!\!\!\!/ + im}|J\rangle \langle J| .
\ee
The equivalence of (\ref{prop_zmz}) and (\ref{prop_zmz2})
is easily seen using the fact that in the sum ansatz, the 
derivative in the overlap matrix element $T_{IJ}$ can be
replaced by a covariant derivative. 

  The propagator (\ref{prop_zmz}) can be calculated either
numerically or using the mean field approximation introduced
in Sec. \ref{sec_hfa}. We will discuss the mean field propagator 
in the following section. For our numerical calculations, we have
improved the zero mode propagator by adding the contributions
from non-zero modes to first order in the expansion (\ref{prop_hop}). 
The result is 
\be
\label{s_num}
 S(x,y) &=& S_0(x,y) + S^{ZMZ}(x,y) + 
  \sum_I (S^{NZM}_I(x,y)-S_0(x,y) ) .
\ee
How accurate is this propagator? We have seen that the propagator 
agrees with the general OPE result at short distance. We also
know that it accounts for chiral symmetry breaking and spontaneous 
mass generation at large distances. In addition to that, we have
performed a number of checks on the correlation functions that
are sensitive to the degree to which (\ref{s_num}) satisfies the
equations of motion, for example by testing whether the vector 
correlator is transverse (the vector current is conserved).

\subsubsection{The propagator in the mean field approximation}
\label{sec_prop_mfa}

\begin{figure}[t]
\begin{center}
\leavevmode
\epsfxsize=12cm
\epsffile{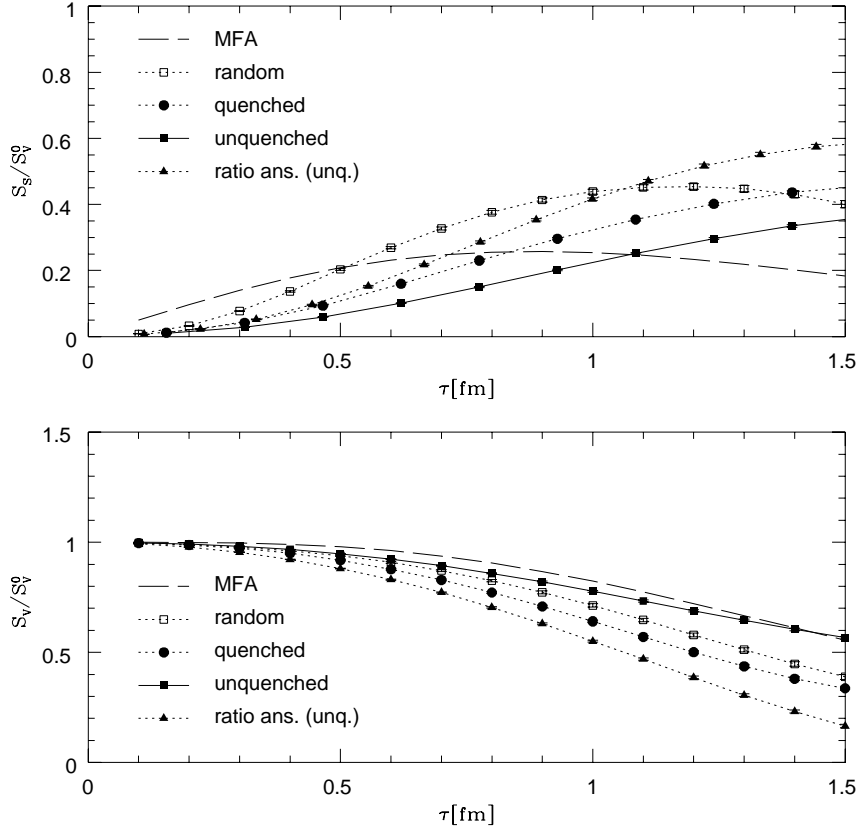}
\end{center}
\caption{\label{fig_quark_prop}
Scalar and vectors components of the quark propagator, normalized
to the free vector propagator. The dashed lines show the result of 
the mean field approximation while the data points were obtained 
in different instanton ensembles.}
\end{figure}

   In order to understand the propagation of quarks in the 
``zero mode zone" it is very instructive to construct the 
propagator in the mean field approximation. The mean field
propagator can be obtained in several ways. Most easily, we 
can read off the propagator directly from the effective 
partition function (\ref{seff_nf1}). We find
\be
\label{s_mfa}
 S(p) &=& \frac{p\!\!\!/\, + iM(p)}{p^2+M^2(p)}
\ee
with the momentum dependent effective quark mass
\be 
\label{m_mfa}
  M(p) &=& \frac{\beta}{2N_c}\frac{N}{V} p^4{\varphi'}^2(p).
\ee
Here, $\beta$ is the solution of the gap equation (\ref{gap}).
Originally, the result (\ref{s_mfa}) was obtained by Diakonov
and Petrov from the Dyson-Schwinger equation for the quark 
propagator in the large $N_c$ limit \cite{DP_86}. At small
momenta, chiral symmetry breaking generates an effective 
mass $M(0)=\frac{\beta}{2N_c} \frac{N}{V}(2\pi\rho)^2$. The 
quark condensate was already given in Eq. (\ref{qbarq_hfa}). 
At large momenta, we have $M(p)\sim 1/p^6$ and constituent 
quarks become free current quarks. 

   For comparison with our numerical results, it is useful 
to determine the mean field propagator in coordinate space.
Fourier transforming the result (\ref{s_mfa}) gives
\be
  S_V(x) &=& \frac{1}{4\pi^2x}\int dp\,
  \frac{p^4}{p^2+M^2(p)}J_2(px) , \\
  S_S(x) &=& \frac{1}{4\pi^2x}\int dp\,
  \frac{p^3 M(p)}{p^2+M^2(p)}J_1(px) .
\ee
The result is shown in Fig. \ref{fig_quark_prop}. The scalar and
vector components of the propagator are normalized to the free
propagator. The vector component of the propagator is exponentially
suppressed at large distance, showing the formation of a constituent
mass. The scalar component again shows the breaking of chiral 
symmetry. At short distance, the propagator is consistent with 
the coordinate dependent quark mass $m=\pi^2/3\cdot x^2\langle
\bar\psi\psi\rangle$ inferred from the OPE expression (\ref{s_ope}).
At large distance the exponential decay is governed by the
constituent mass $M(0)$.

   For comparison, we also show the quark propagator in different 
instanton ensembles. The result is very similar to the mean field 
approximation, the differences are mainly due to different values 
of the quark condensate. The quark propagator is not very sensitive
to correlations in the instanton liquid. In \cite{SV_93a,CNZ_95},
the quark propagator was also used to study heavy-light mesons in
the $1/M_Q$ expansion. The results are very encouraging, and we refer 
to the reader to the original literature for details.

\subsection{Mesonic correlators}
\label{sec_mes_cor}

\subsubsection{General results and the OPE}

\begin{figure}[t]
\begin{center}
\leavevmode
\epsfxsize=12cm
\epsffile{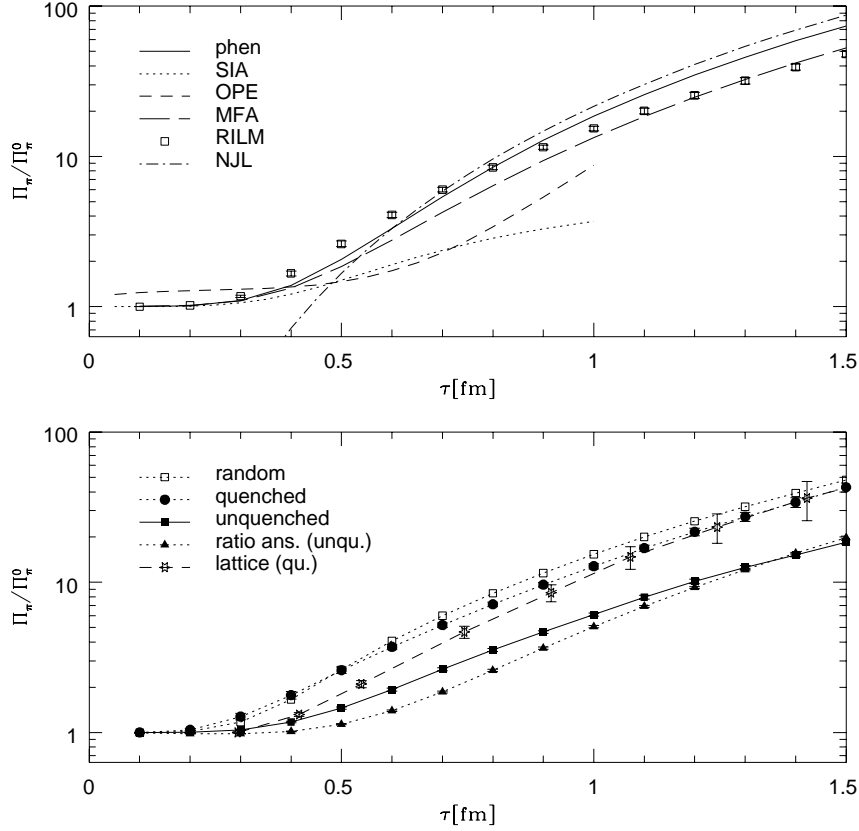}
\end{center}
\caption{\label{fig_pi_cor}
Pion correlation function in various approximations and instanton
ensembles. In Fig. a) we show the phenomenological expectation
(solid), the OPE (dashed), the single instanton (dash-dotted) and mean 
field approximations (dashed) as well as data in the random instanton 
ensemble. In Fig. b) we compare different instanton ensembles,
random (open squares), quenched (circles) and interacting
(streamline: solid squares, ratio ansatz solid triangles).}
\end{figure}

  A large number of mesonic correlation functions have been studied
in the instanton model, and clearly this is not the place to list 
all of them. Instead, we have decided to discuss three examples 
that are illustrative of the techniques and the most important
effects. We will consider the $\pi$, $\rho$ and $\eta'$ channels, 
related to the currents
\be
  j_\pi  = \bar q\tau^a\gamma_5 q, \hspace{1cm}
  j_{\rho} = \bar q\frac{\tau^a}{2}\gamma_\mu q ,\hspace{1cm}
  j_{\eta_{ns}}  = \bar q\gamma_5 q,
\ee 
where $q=(u,d)$. Here, we only consider the non-strange $\eta'$ 
and refer to \cite{Sch_96,SV_93b} for a discussion of $SU(3)$ flavor 
symmetry breaking and $\eta-\eta'$ mixing. The three channels 
discussed here are instructive because the instanton-induced interaction 
is attractive for the pion channel, repulsive for the $\eta'$ and 
(to first order in the instanton density) does not affect the $\rho$ 
meson. The $\pi,\rho$ and $\eta'$ are therefore representative for 
a much larger set of correlation functions. In addition to that, 
these three mesons have special physical significance. The pion is 
the lightest hadron, connected with the spontaneous breakdown of 
$SU(N_f)_L\times SU(N_f)_R$ chiral symmetry. The $\eta'$ is 
surprisingly heavy, a fact related to the anomalous $U(1)_A$ 
symmetry. The $\rho$ meson, finally, is the lightest non-Goldstone 
particle and the first $\pi\pi$ resonance.

\begin{figure}[t]
\begin{center}
\leavevmode
\epsfxsize=12cm
\epsffile{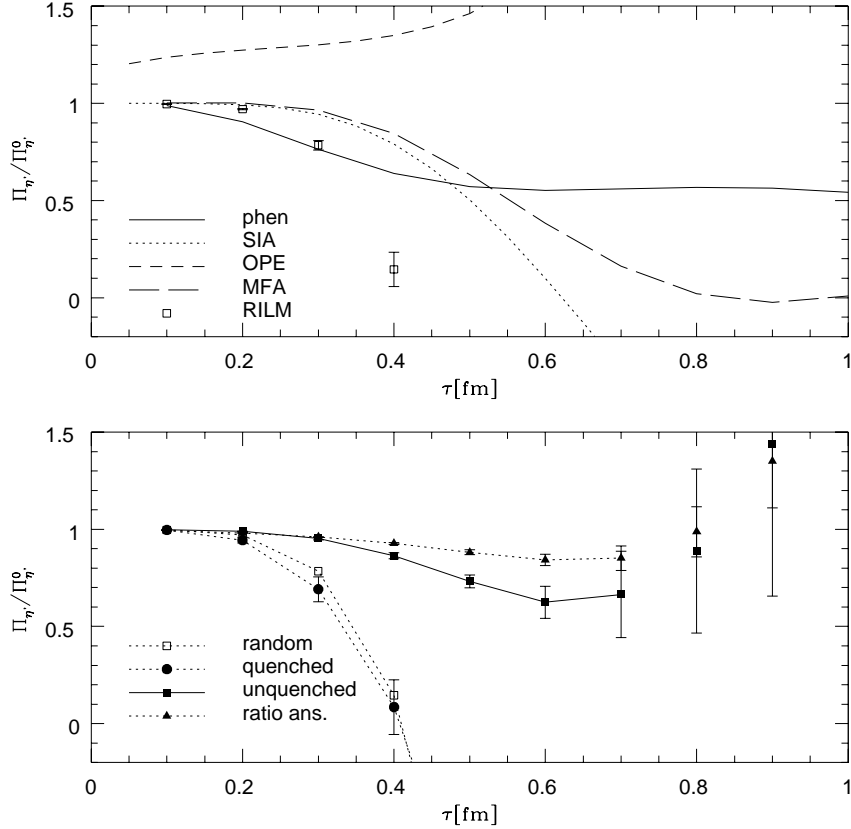}
\end{center}
\caption{\label{fig_eta_cor}
Eta prime meson correlation functions. The various curves and data sets
are labeled as in in Fig. \ref{fig_pi_cor}.} 
\end{figure}

   Phenomenological predictions for the correlation functions are 
shown in Figs. \ref{fig_pi_cor}-\ref{fig_rho_cor} \cite{Shu_93}. All 
correlators are normalized to the free ones, $R(\tau)=\Pi_\Gamma(\tau)
/\Pi_\Gamma^0(\tau)$, where $\Pi_\Gamma^0(\tau) = {\rm Tr} [\Gamma 
S_0(\tau)\Gamma S_0(-\tau)]$. At short distance, asymptotic freedom 
implies that this ratio approaches one. At large distance, the 
correlators are exponential and $R$ is small. At intermediate distance, 
$R$ depends on the quark-quark interaction in that channel. If $R>1$, 
we will refer to the correlator as attractive, while $R<1$ implies 
repulsive interactions. The normalized pion correlation function 
$R_\pi$ is significantly larger than one, showing a strongly
attractive interaction, and a light bound state. The rho meson
correlator is close to one out to fairly large distances $x\simeq 
1.5$ fm, a phenomenon referred to as ``superduality" in \cite{Shu_93}. 
The $\eta'$ channel is strongly repulsive, only showing an enhancement 
at intermediate distances due to mixing with the $\eta$.

   Theoretical information about the short distance behavior 
of the correlation functions comes from the operator product
expansion \cite{SVZ_79}. For the $\pi$ and $\rho$ channels, we 
have
\be
\label{ps_ope}
\Pi^{OPE}_\pi(x) &=& \Pi^0_\pi(x) \left( 
  1 + \frac{11}{3}\frac{\alpha_s(x)}{\pi} - \frac{\pi^2}{3}
  m\langle\bar qq\rangle x^4 + \frac{1}{384}
  \langle G^2\rangle\, x^4
    + \frac{11\pi^3}{81}\alpha_s(x)
      \langle\bar\psi\psi\rangle^2 \log (x^2) x^6\,
    +\ldots \right) , \\
\label{vec_ope}
\Pi^{OPE}_\rho(x) &=& \Pi^0_\rho(x) \left( 
  1 + \frac{\alpha_s(x)}{\pi} - \frac{\pi^2}{4}
  m\langle\bar qq\rangle x^4 - \frac{1}{384}
  \langle G^2\rangle\, x^4
    + \frac{7\pi^3}{81}\alpha_s(x)
      \langle\bar\psi\psi\rangle^2 \log(x^2) x^6\,
    +\ldots \right)   ,
\ee
where we have restricted ourselves to operators of dimension up
to six and the leading order perturbative corrections. We have also
used the factorization hypothesis for the expectation value of
four-quark operators. Note that to this order, the non-strange
eta prime and pion correlation functions are identical. This result 
shows that QCD sum rules cannot account for the $U(1)_A$ anomaly. 

   The OPE predictions (\ref{ps_ope},\ref{vec_ope}) are also 
shown in Figs. \ref{fig_pi_cor}-\ref{fig_rho_cor}. The leading 
quark and gluon power corrections in the $\pi$ channel are both 
attractive, so the OPE prediction has the correct tendency, 
but underpredicts the rise for $x>0.25$ fm. In the $\rho$ meson 
channel, the leading corrections have a tendency to cancel each 
other, in agreement the superduality phenomenon mentioned above

\subsubsection{The single-instanton approximation}
\label{sec_cor_sia}

    After this prelude we come to the instanton model. We start by 
considering instanton contributions to the short distance behavior 
of correlation functions in the single-instanton approximation\footnote{
We would like to distinguish this method from the dilute gas approximation
(DIGA). In the DIGA, we systematically expand the correlation functions
in terms of the one (two, three, etc.) instanton contribution. In
the presence of light fermions (for $N_f>1$), however, this method is 
useless, because there is no zero mode contribution to chirality 
violating operators from any finite number of instantons.} (SIA) 
\cite{Shu_83}. The main idea is that if $x-y$ is small compared to 
the typical instanton separation $R$, we expect that the contribution 
from the instanton $I=I_*$ closest to the points $x$ and $y$ will dominate
over all others. For the propagator in the zero mode zone, this 
implies 
\be
\label{S_SIA}
S(x,y)\; = \;\sum_{IJ}\psi_I(x)\left(\frac{1}{T+im}\right)_{IJ}\!\!
        \psi^\dagger_J(y)
      \;\simeq\;\psi_{I_*}(x)\left(\frac{1}{T+im}\right)_{I_*I_*}
      \!\!\psi^\dagger_{I_*}(y)
      \;\simeq\;\frac{\psi_{I_*}(x)\psi^\dagger_{I_*}(y)}{m^*}
\ee
where we have approximated the diagonal matrix element by its average, 
$(T+im)^{-1}_{I_*I_*}\simeq N^{-1}\sum_I (T+im)^{-1}_{II}$, and introduced 
the effective mass $m^*$ defined in Sec. \ref{sec_dirac_spec}, $(m^*)^{-1}
=N^{-1}\sum \lambda^{-1}$. In the following we will use the mean field 
estimate $m^*=\pi\rho(2n/3)^{1/2}$. As a result, the propagator in the 
SIA looks like the zero mode propagator of a single instanton, but for 
a particle with an effective mass $m^*$. 

  The $\pi$ and $\eta'$ correlators receive zero mode contributions.
In the single instanton approximation, we find \cite{Shu_83}
\be
\label{ps_SIA}
\Pi^{SIA}_{\pi,\eta'}(x) &=& \pm\int d\rho\, n(\rho)
 \frac{6\rho^4}{\pi^2}\frac{1}{(m^*)^2}
 \frac{\partial^2}{\partial (x^2)^2}
 \left\{ \frac{4\xi^2}{x^4} \left(
 \frac{\xi^2}{1-\xi^2} +\frac{\xi}{2}
 \log\frac{1+\xi}{1-\xi}\right)\right\},
\ee
where $\xi^2=x^2/(x^2+4\rho^2)$. There is also a non-zero mode contribution 
to these correlation functions. It was calculated in \cite{Shu_89}, but 
numerically it is not very important.

  We show the result (\ref{ps_SIA}) in Fig. \ref{fig_pi_cor}. For 
simplicity, we have chosen $n(\rho)=n_0\delta(\rho-\rho_0)$ with the 
standard parameters $n_0=1\,{\rm fm}^{-4}$ and $\rho_0=0.33$ fm. The pion 
correlator is similar to the OPE prediction at short distance $x\lesssim 
0.25$ fm, but follows the phenomenological result out to larger distances 
$x\simeq 0.4$ fm. In particular, we find that even at very short distances
$x\simeq 0.3$ fm, the regular contributions to the correlator coming 
from instanton zero modes are larger than the singular contributions 
included in the OPE. The fact that non-perturbative corrections 
not accounted for in the OPE are particularly large in spin zero
channels ($\pi,\eta',\sigma$ and scalar glueballs) was first 
emphasized by \cite{NSVZ_81}. For the $\eta'$, the instanton 
contribution is strongly repulsive. This means that in contrast 
to the OPE, the single instanton approximation at least qualitatively 
accounts for the $U(1)_A$ anomaly. The SIA was extended to the full 
pseudo-scalar nonet in \cite{Shu_83}. It was shown that replacing 
$m^*\to m^*+m_s$ gives a good description of $SU(3)$ flavor symmetry 
breaking and the $\pi,K,\eta$ correlation functions. 

  In the $\rho$ meson correlator zero modes cannot contribute since 
the chiralities do not match. Non-vanishing contributions come from 
the non-zero mode propagator (\ref{s_nz}) and from interference between 
the zero mode part and the leading mass correction in (\ref{s_m})
\be
\Pi^{SIA}_\rho(x,y) &=& {\rm Tr}\left[ \gamma_\mu S^{nz}(x,y)
\gamma_\mu S^{nz}(y,x) \right] + 2{\rm Tr}\left[ \gamma_\mu
\psi_0(x)\psi_0^\dagger(y)\gamma_\mu \Delta(y,x)\right]
\ee
The latter term survives even in the chiral limit, because the factor 
$m$ in the mass correction is cancelled by the $1/m$ from the zero mode.
Also note that the result corresponds to the standard DIGA, so true
multi-instanton effects are not included. After averaging over the 
instanton coordinates, we find\footnote{There is a mistake by an
overall factor 3/2 in the original work.} \cite{AG_78}
\be
\label{vec_SIA}
\Pi^{SIA}_\rho(x) &=&  \Pi^0_\rho + \int d\rho\, n(\rho)
 \frac{12}{\pi^2}\frac{\rho^4}{x^2}\frac{\partial}{\partial (x^2)}
 \left\{ \frac{\xi}{x^2} \log\frac{1+\xi}{1-\xi}\right\}
\ee
The result is also shown in Fig. \ref{fig_rho_cor}. Similar to the OPE, 
the correlator is attractive at intermediate distances. The correlation
function does not go down at larger distances, since the DIGA does 
not account for a dynamically generated mass. It is very instructive 
to compare the result to the OPE in more detail. Expanding (\ref{vec_SIA}), 
we get
\be
\label{vec_SIA_exp}
\Pi^{SIA}_\rho(x) &=& \Pi^0_\rho(x) \left( 1 + \frac{\pi^2x^4}{6}
 \int d\rho n(\rho) \right). 
\ee
This agrees exactly with the OPE expression, provided we use the 
average values of the operators in the dilute gas approximation
\be
 \langle G^2\rangle  \;=\; 32\pi^2 \int d\rho\, n(\rho)\, ,\hspace{1cm} 
 m\langle\bar qq\rangle \;=\;  - \int d\rho\, n(\rho) \, .
\ee 
Note, that the value of $m\langle\bar qq\rangle$ is ``anomalously" 
large in the dilute gas limit. This means that the contribution from
dimension 4 operators is attractive, in contradiction with the OPE
prediction based on the canonical values of the condensates.

   An interesting observation is the fact that (\ref{vec_SIA_exp})
is the only singular term in the DIGA correlation function. In fact,
the OPE of {\em any} mesonic correlator in {\em any} self-dual field 
contains only dimension 4 operators \cite{DS_81}. This means that for 
all higher order operators either the Wilson coefficient vanishes 
(as it does, for example, for the triple gluon condensate $\langle 
f^{abc}G^a_{\mu\nu} G^b_{\nu\rho}G^c_{\rho\mu}\rangle$) or the 
matrix elements of various operators of the same dimension cancel 
each other\footnote{Here, we do not consider radiative corrections 
like $\alpha_s\langle\bar\psi\psi\rangle^2$. Technically, this is 
because we evaluate the OPE in a fixed (classical) background 
field without taking into account radiative corrections to the
background field.}. This is a very remarkable result, because it 
helps to explain the success of QCD sum rules based on the OPE
in many channels. In the instanton model, the gluon fields 
are very inhomogeneous, so one would expect that the OPE fails 
for $x>\rho$. The Dubovikov-Smilga result shows that quarks 
can propagate through very strong gauge fields (as long as 
they are self-dual) without suffering strong interactions.

\subsubsection{The random phase approximation}
\label{sec_cor_rpa}

\begin{figure}[t]
\begin{center}
\leavevmode
\epsfxsize=12cm
\epsffile{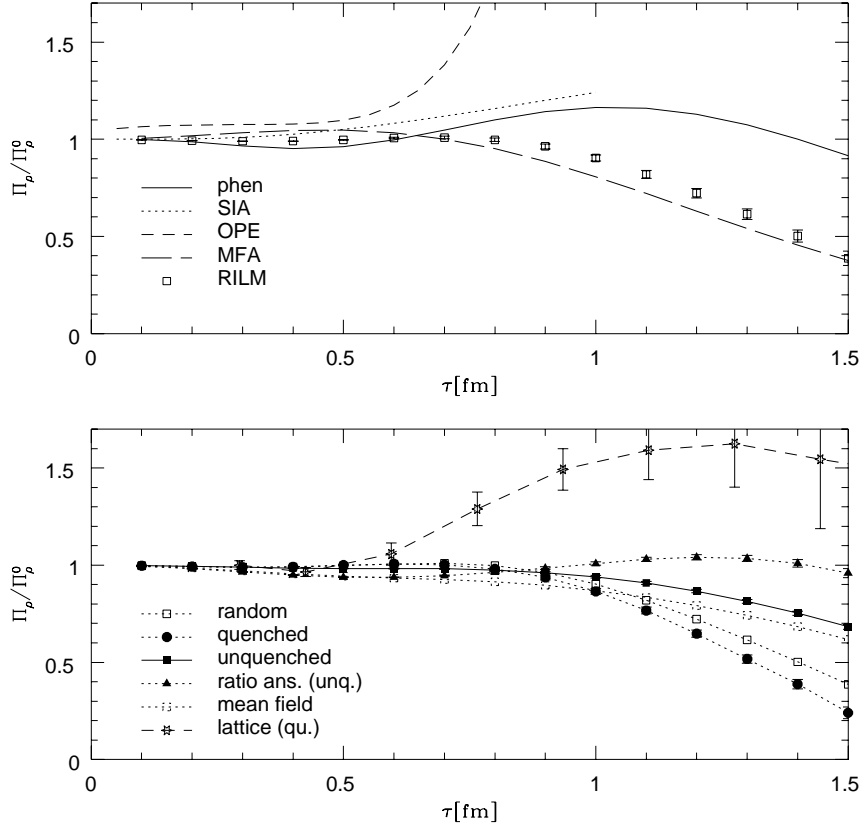}
\end{center}
\caption{\label{fig_rho_cor}
Rho meson correlation functions. The various curves and data sets
are labeled as in in Fig. \ref{fig_pi_cor}. The dashed squares show
the non-interacting part of the rho meson correlator in the interacting
ensemble.} 
\end{figure}

   The SIA clearly improves on the short distance behavior 
of the $\pi,\eta'$ correlation functions as compared to the
OPE. However, in order to describe a true pion bound state
one has to resum the attractive interaction generated by 
the 't Hooft vertex. This is most easily accomplished using
the random phase (RPA) approximation, which corresponds 
to iterating the average 't Hooft vertex in the $s$-channel 
(see Fig. \ref{fig_cor_schem}). The solution of the Bethe-Salpeter 
can be written as \cite{DP_86,Hut_95,KPZ_96}
\be
\label{cor_rpa}
\Pi^{RPA}_\pi(x)  &=& \Pi_\pi^{MFA}(x) + \Pi_\pi^{int} \\ 
\Pi^{RPA}_\rho(x) &=& \Pi_\rho^{MFA}(x)\nonumber .
\ee
Here $\Pi^{MFA}_\Gamma$, denotes the mean field (non-interacting)
part of the correlation functions
\be
\label{cor_fact}
\Pi^{MFA}_\Gamma(x) &=& {\rm Tr}\left[ \overline S(x)\Gamma
  \overline S(-x)\Gamma\right] ,
\ee
where $\overline S(x)$ is the mean field propagator discussed 
in Sec. \ref{sec_prop_mfa}. In the $\rho$ meson channel, the 
't Hooft vertex vanishes and the correlator is given by the 
mean field contribution only. The interacting part of the $\pi,
\eta'$ correlation functions is given by
\be
\label{pscor_int}
\Pi^{int}_{\pi,\eta'}(x) &=& \int d^4q\, e^{iq\cdot x}\;\;
           \Gamma_5(q)\frac{\pm 1}{1\mp C_5(q)}\Gamma_5(q) 
\ee
where the elementary loop function $C_5$ and the vertex 
function $\Gamma_5$ are given by
\be
\label{ps_bubble_mfa}
 C_5 (q) &=& 4N_c\left(\frac{V}{N}\right)
  \int \frac{d^4p}{(2\pi)^4} 
  \frac{M_1M_2(M_1M_2-p_1\cdot p_2)}
       {(M_1^2+p_1^2)(M_2^2+p_2^2)} ,\\
\label{ps_vertex_mfa}
\Gamma_5 (q) &=& \;\; 4\; 
  \int \frac{d^4p}{(2\pi)^4} 
  \frac{\sqrt{M_1M_2}(M_1M_2-p_1\cdot p_2)}
       {(m_1^2+p_1^2)(M_2^2+p_2^2)}  ,
\ee
where $p_1=p+q/2$, $p_2=p-q/2$ and $M_{1,2}=M(p_{1,2})$ are the momentum 
dependent effective quark masses. We have already stressed that in the 
long wavelength limit, the effective interaction between quarks is of 
NJL type. Indeed, the pion correlator in the RPA approximation to the 
NJL model is given by
\be
\label{pscor_njl}
\Pi^{NJL}_\pi(x) &=& \int d^4q\, e^{iq\cdot x}\;\;
           \frac{J_5(q)}{1-GJ_5(q)},
\ee
where $J_5(q)$ is the pseudo-scalar loop function
\be 
\label{ps_bubble_njl}
J_5 (q) &=& 4N_c
  \int^\Lambda \frac{d^4p}{(2\pi)^4} 
  \frac{(M^2-p_1\cdot p_2)}
       {(M^2+p_1^2)(M^2+p_2^2)},  
\ee
$G$ is the four fermion coupling constant and the (momentum
independent) constituent mass $M$ is the solution of the 
NJL gap equation. The loop function (\ref{ps_bubble_njl}) is
divergent and is regularized using a cutoff $\Lambda$.

  Clearly, the RPA correlation functions (\ref{pscor_int}) and 
(\ref{pscor_njl}) are very similar, the only difference is 
that in the instanton liquid both the quark mass and the vertex 
are momentum dependent. The momentum dependence of the vertex
ensures that no regulator is required, the loop integral is 
automatically cut off at $\Lambda\sim\rho^{-1}$. We also 
note that the momentum dependence of the effective interaction
at the mean field level is very simple, the vertex function 
is completely separable. 

   The mean field correlation functions are shown in Figs. 
\ref{fig_pi_cor}-\ref{fig_rho_cor}. The coordinate space 
correlators are easily determined by Fourier transforming 
the result (\ref{pscor_int}). We also show the NJL result 
for the pion correlation function. In this case, the correlation 
function at short distance is not very meaningful, since the 
cutoff $\Lambda$ eliminates all contributions to the spectral 
function above that scale. 

   The pion correlator nicely reproduces the phenomenological
result. The $\eta'$ correlation function has the correct tendency
at short distance but becomes unphysical for $x>0.5$ fm. The
result in the $\rho$ meson channel is also close to phenomenology,
despite the fact that it corresponds to two unbound constituent 
quarks.

\subsubsection{The interacting instanton liquid}
\label{sec_cor_iilm}

   What is the quality and the range of validity of the RPA? The 
RPA is usually motivated by the large $N_c$ approximation, and the 
dimensionless parameter that controls the expansion is $\rho^4(N/V)
/(4N_c)$. This parameter is indeed very small, but in practice
there are additional parameters that determine the size of 
corrections to the RPA.

\begin{table}[t]
\caption{\label{tab_cur_def}
Definition of various currents and hadronic matrix elements  
referred to in this work.}
\begin{tabular}{llll}
channel         &    current         
                & matrix element     
                & experimental value    \\  \tableline
$\pi$           &  $j^a_\pi=\bar q\gamma_5\tau^a q$           
                &  $\langle 0|j_\pi^a|\pi^b\rangle=\delta^{ab}\lambda_\pi$             
                &  $\lambda_\pi\simeq (480\,{\rm MeV})^3$     \\
                &  $j^a_{\mu\,5}=\bar q\gamma_\mu\gamma_5\frac{\tau^a}{2}q$
                &  $\langle 0|j^a_{\mu\,5}|\pi^b\rangle
                                    =\delta^{ab}q_\mu f_\pi$
                &  $f_\pi=93$ MeV \\
$\delta$        &  $j^a_\delta=\bar q\tau^a q$           
                &  $\langle 0|j_\delta^a|\delta^b\rangle
                                    =\delta^{ab}\lambda_\delta$         
                &       \\
$\sigma$        &  $j_\sigma=\bar q q$           
                &  $\langle 0|j_\sigma|\sigma\rangle=\lambda_\sigma$                   
                &       \\
$\eta_{ns}$     &  $j_{\eta_{ns}}=\bar q\gamma_5  q$           
                &  $\langle 0|j_{\eta_{ns}}|\eta_{ns}\rangle
                                    =\lambda_{\eta_{ns}}$           
                &       \\
$\rho$          &  $j^a_{\mu}=\bar q\gamma_\mu\frac{\tau^a}{2}q$
                &  $\langle 0|j^a_{\mu}|\rho^b\rangle
                      =\delta^{ab}\epsilon_\mu\frac{m_\rho^2}{g_\rho}$ 
                &  $g_\rho=5.3$  \\
$a_1$           &  $j^a_{\mu\,5}=\bar q\gamma_\mu\gamma_5\frac{\tau^a}{2}q$
                &  $\langle 0|j^a_{\mu\,5}|a_1^b\rangle
                      =\delta^{ab}\epsilon_\mu\frac{m_{a_1}^2}{g_{a_1}}$
                &  $g_{a_1}=9.1$  \\
$N$             &  $\eta_1 = \epsilon^{abc}(u^aC\gamma_\mu u^b)\gamma_5
                   \gamma_\mu d^c$
                &  $\langle 0|\eta_1 |N(p,s)\rangle =\lambda_1^N u(p,s)$ 
                &             \\
$N$             &  $\eta_2 = \epsilon^{abc}(u^aC\sigma_{\mu\nu} u^b)
                   \gamma_5\sigma_{\mu\nu} d^c$
                &  $\langle 0|\eta_2 |N(p,s)\rangle =\lambda_2^N u(p,s)$ 
                &             \\
$\Delta$        &  $\eta_\mu = \epsilon^{abc}(u^aC\gamma_\mu u^b) u^c$
                &  $\langle 0|\eta_\mu |N(p,s)\rangle 
                           =\lambda^\Delta u_\mu(p,s)$ 
                &             \\
\end{tabular}
\end{table}

   First of all, $\rho^4(N/V)$ is a useful expansion parameter only
if the instanton liquid is random and the role of correlations is 
small. As discussed in \ref{sec_ens},
if the density is very small the instanton liquid is in a 
molecular phase, while it is in a crystaline 
phase if the density is large. Clearly, the RPA is expected 
to fail in both of these limits. In general, the RPA corresponds to 
using linearized equations for the fluctuations around a mean field 
solution. In our case, low lying meson states are collective 
fluctuations of the chiral order parameter. For isovector scalar 
mesons, the RPA is expected to be good, because the scalar mean 
field (the condensate) is large and the masses are light. For 
isosinglet mesons, the fluctuations are much larger, and the 
RPA is likely to be less useful.

   In the following we will therefore discuss results from 
numerical calculations of hadronic correlators in the instanton
liquid. These calculations go beyond the RPA in two ways: (i) 
the propagator includes genuine many instanton effects and 
non-zero mode contributions; (ii)  the ensemble is determined 
using the full (fermionic and bosonic) weight function, so it 
includes correlations among instantons. In addition to that,
we will also consider baryonic correlators and three point
functions that are difficult to handle in the RPA.

   We will discuss correlation function in three different
ensembles, the random ensemble (RILM), the quenched (QILM)
and fully interacting (IILM) instanton ensembles. In the random
model, the underlying ensemble is the same as in the mean field
approximation, only the propagator is more sophisticated. In 
the quenched approximation, the ensemble includes correlations
due to the bosonic action, while the fully interacting ensemble 
also includes correlations induced by the fermion determinant.
In order to check the dependence of the results on the instanton
interaction, we study correlation functions in two different
unquenched ensembles, one based on the streamline interaction
(with a short-range core) and one based on the ratio ansatz  
interaction. The bulk parameters of these ensembles are 
compared in Tab. \ref{tab_liquid_par}. We note that the ratio 
ansatz ensemble is more dense than the streamline ensemble.

   In the following we are not only interested in the behavior
of the correlation functions, but also in numerical results for 
the ground state masses and coupling constants. For this purpose
we have fitted the correlators using a simple ``pole plus continuum" 
model for the spectral functions. In the case of the pion, this
leads to the following parametrization of the correlation function
\be 
\label{pi_rep}
  \Pi_\pi(x)&=& \lambda^2_\pi D(m_\pi,x) + \frac{3}{8\pi^2}
     \int_{s_0}^\infty ds\, s D(\sqrt{s},x),
\ee
where $\lambda_\pi$ is the pion coupling constant defined
in Tab. \ref{tab_cur_def} and $s_0$ is the continuum threshold. 
Physically, $s_0$ roughly represents the position of the 
first excited state. Resolving higher resonances requires 
high quality data and more sophisticated techniques. The model 
spectral function used here is quite popular in connection with 
QCD sum rules. It provides a surprisingly good description of 
all measured correlation functions, not only in the instanton 
model, but also on the lattice \cite{CGHN_93b,Lei_95a}.

\begin{table}[t]
\caption{\label{tab_mes_res}
Meson parameters in the different instanton ensembles. All
quantities are given in units of GeV. The current quark mass is $m_u
=m_d=0.1\Lambda$. Except for the pion mass, no attempt has been made to
extrapolate the parameters to physical values of the quark mass.}
\begin{tabular}{crrrr}
               & unquenched         & quenched           & RILM       
               & ratio ansatz (unqu.)  \\  \tableline
$m_\pi$        &  0.265             &   0.268            &  0.284            
               &  0.128      \\
$m_\pi$ (extr.)&  0.117             &   0.126            &  0.155
               &  0.067      \\
$\lambda_\pi$  &  0.214             &   0.268            &  0.369             
               &  0.156      \\
$f_\pi$        &  0.071             &   0.091            &  0.091            
               &  0.183 \\
$m_\rho$       &  0.795             &   0.951            &  1.000             
               &  0.654      \\
$g_\rho$       &  6.491             &   6.006            &  6.130            
               &  5.827      \\
$m_{a_1}$      &  1.265             &   1.479            &  1.353             
               &  1.624      \\
$g_{a_1}$      &  7.582             &   6.908            &  7.816             
               &  6.668      \\
$m_{\sigma}$   &  0.579             &   0.631            &  0.865             
               &  0.450      \\
$m_{\delta}$   &  2.049             &   3.353            &  4.032            
               &  1.110      \\
$m_{\eta_{ns}}$&  1.570             &   3.195            &  3.683             
               &  0.520 \\
\end{tabular}
\end{table}

   Correlation functions in the different instanton ensembles 
were calculated in \cite{SV_93b,SSV_94,SS_96} to which we refer 
the reader for more details. The results are shown in 
Fig. \ref{fig_pi_cor}-\ref{fig_rho_cor} and summarized 
in Tab. \ref{tab_mes_res}. The pion correlation functions 
in the different ensembles are qualitatively very similar. The
differences are mostly due to different values of the quark
condensate (and the physical quark mass) in the different
ensembles. Using the Gell-Mann, Oaks, Renner relation, one can 
extrapolate the pion mass to the physical value of the quark 
masses, see Tab. \ref{tab_mes_res}. The results are consistent with 
the experimental value in the streamline ensemble (both quenched 
and unquenched), but clearly too small in the ratio ansatz ensemble. 
This is a reflection of the fact that the ratio ansatz ensemble is 
not sufficiently dilute.

   In Fig. \ref{fig_rho_cor} we also show the results in the $\rho$ 
channel. The $\rho$ meson correlator is not affected by instanton zero 
modes to first order in the instanton density. The results in the 
different ensembles are fairly similar to each other and all fall 
somewhat short of the phenomenological result at intermediate distances 
$x\simeq 1$ fm. We have determined the $\rho$ meson mass and coupling 
constant from a fit similar to (\ref{pi_rep}). The results are given 
in Tab. \ref{tab_mes_res}. The $\rho$ meson mass is somewhat too heavy 
in the random and quenched ensembles, but in good agreement with 
the experimental value $m_\rho=770$ MeV in the unquenched ensemble.

    Since there are no interactions in the $\rho$ meson channel 
to first order in the instanton density, it is important to study 
whether the instanton liquid provides any significant binding. In 
the instanton model, there is no confinement, and $m_\rho$ is close 
to the two (constituent) quark threshold. In QCD, the $\rho$ meson 
is also not a true bound state, but a resonance in the 2$\pi$ continuum. 
In order to determine whether the continuum contribution in the 
instanton liquid is predominantly from 2-$\pi$ or 2-quark states 
would require the determination of the corresponding three point 
functions, which has not been done yet. Instead, we have compared 
the full correlation function with the non-interacting (mean field)
correlator (\ref{cor_fact}), where we use the average (constituent 
quark) propagator determined in the same ensemble, see Fig. 
\ref{fig_rho_cor}). This comparison provides a measure of the 
strength of interaction. We observe that there is an attractive 
interaction generated in the interacting liquid. The interaction
is due to correlated instanton-anti-instanton pairs, see the 
discussion in Sec. \ref{sec_mol}, in particular Eq. \ref{lmol}. 
This is consistent with the fact that the interaction is considerably 
smaller in the random ensemble. In the random model, the strength of 
the interaction grows as the ensemble becomes more dense. However, 
the interaction in the full ensemble is significantly larger than 
in the random model at the same diluteness. Therefore, most of the 
interaction is due to dynamically generated pairs.

   The situation is drastically different in the $\eta'$ channel. Among
the $\sim 40$ correlation functions calculated in the random ensemble, 
only the $\eta'$ (and the isovector-scalar $\delta$ discussed in the 
next section) are completely unacceptable: The correlation function 
decreases very rapidly and becomes negative at $x\sim 0.4$ fm. This 
behaviour is incompatible with the positivity of the spectral function. 
The interaction in the random ensemble is too repulsive, and the 
model ``overexplains" the $U(1)_A$ anomaly. 

    The results in the unquenched ensembles (closed and open points) 
significantly improve the situation. This is related to dynamical
correlations between instantons and anti-instantons (topological charge
screening). The single instanton contribution is repulsive, but 
the contribution from pairs is attractive \cite{SSV_95}. Only if 
correlations among instantons and anti-instantons are sufficiently strong, 
the correlators are prevented from becoming negative. Quantitatively,
the $\delta$ and $\eta_{ns}$ masses in the streamline ensemble are still 
too heavy as compared to their experimental values. In the ratio ansatz,
on the other hand, the correlation functions even shows an enhancement 
at distances on the order of 1 fm, and the fitted masses is too light. 
This shows that the $\eta'$ channel is very sensitive to the strength 
of correlations among instantons.

    In summary, pion properties are mostly sensitive to global
properties of the instanton ensemble, in particular its diluteness.
Good phenomenology demands $\bar\rho^4 n\simeq 0.03$, as originally
suggested in \cite{Shu_82b}. The properties of the $\rho$ meson are 
essentially independent of the diluteness, but show  sensitivity 
to $IA$ correlations. These correlations become crucial
in the $\eta'$ channel.  

    Finally, we compare the correlation functions in the instanton 
liquid to lattice measurements reported in \cite{CGHN_93a,CGHN_93b},
see Sec. \ref{sec_cool_cor}. These correlation functions were 
measured in quenched QCD, so they should be compared to the random
or quenched instanton ensembles. The results agree very well 
in the pion channel, while the lattice correlation function 
in the rho meson channel is somewhat more attractive than the  
correlator in the instanton liquid. 

\subsubsection{Other mesonic correlation functions}
\label{sec_mes_rem}

   After discussing the $\pi,\rho,\eta'$ in some detail we only 
briefly comment on other correlation functions. The remaining 
scalar states are the isoscalar $\sigma$ and the isovector
$\delta$ (the $f_0$ and $a_0$ according to the notation of the 
particle data group). The sigma correlator has a disconnected 
contribution, which is proportional to $\langle\bar qq\rangle^2$
at large distance. In order to determine the lowest resonance 
in this channel, the constant contribution has to
be subtracted, which makes it difficult to obtain reliable 
results. Nevertheless, we find that the instanton liquid favors
a (presumably broad) resonance around 500-600 MeV. The isovector
channel is in many ways similar to the $\eta'$. In the random
ensemble, the interaction is too repulsive and the correlator 
becomes unphysical. This problem is solved in the interacting 
ensemble, but the $\delta$ is still very heavy, $m_\delta>1$
GeV. 

   The remaining non-strange vectors are the $a_1,\omega$ and $f_1$. 
The $a_1$ mixes with the pion, which allows a determination of
the pion decay constant $f_\pi$ (as does a direct measurement of
the $\pi-a_1$ mixing correlator). In the instanton liquid, disconnected
contributions in the vector channels are small. This is consistent
with the fact that the $\rho$ and the $\omega$, as well as the $a_1$
and the $f_1$ are almost degenerate.   
 
   Finally, we can also include strange quarks. $SU(3)$ flavor
breaking in the 't Hooft interaction nicely accounts for the 
masses of the $K$ and the $\eta$. More difficult is a correct 
description of $\eta-\eta'$ mixing, which can only be achieved
in the full ensemble. The random ensemble also has a problem
with the mass splittings among the vectors $\rho,K^*$ and $\phi$
\cite{SV_93b}. This is related to the fact that flavor symmetry 
breaking in the random ensemble is so strong that the strange 
and non-strange constituent quark masses are almost degenerate. 
This problem is improved (but not fully solved) in the interacting 
ensemble.

\subsection{Baryonic correlation functions}
\label{sec_bar_cor}

   After discussing quark-anti-quark systems in the last section, we
now proceed to three quark (baryon) channels. As emphasized in
\cite{SR_89}, the existence of a strongly attractive interaction
in the pseudo-scalar quark-anti-quark (pion) channel also implies an
attractive interaction in the scalar quark-quark (diquark) channel. 
This interaction is phenomenologically very desirable, because it
not only explains why the spin 1/2 nucleon is lighter than the 
spin 3/2 Delta, but also why Lambda is lighter
than  Sigma.

\subsubsection{Nucleon correlation functions}
\label{sec_nuc_cor}

\begin{table}[t]
\caption{\label{tab_bar_def}
Definition of nucleon and delta correlation functions.}
\begin{tabular}{ll|ll}
correlator          &  definition       & 
correlator          &  definition       \\ \tableline
$\Pi_1^N(x)$        & $\langle {\rm tr}(\eta_1(x)\bar\eta_1(0))\rangle$  &
$\Pi_1^\Delta(x)$   & $\langle {\rm tr}
                        (\eta_\mu(x)\bar\eta_\mu(0))\rangle$  \\
$\Pi_2^N(x)$        & $\langle {\rm tr}
                        (\gamma\cdot\hat x\eta_1(x)\bar\eta_1(0))\rangle$& 
$\Pi_2^\Delta(x)$   & $\langle {\rm tr}(\gamma\cdot\hat x\eta_\mu(x)
                        \bar\eta_\mu(0))\rangle$ \\                 
$\Pi_3^N(x)$        & $\langle {\rm tr}(\eta_2(x)\bar\eta_2(0))\rangle$  &
$\Pi_3^\Delta(x)$   & $\langle {\rm tr}(\hat x_\mu \hat x_\nu \eta_\mu(x)
                        \bar\eta_\nu(0))\rangle$ \\                 
$\Pi_4^N(x)$        & $\langle {\rm tr}
                        (\gamma\cdot\hat x\eta_2(x)\bar\eta_2(0))\rangle$ &
$\Pi_4^\Delta(x)$   & $\langle {\rm tr}
                        (\gamma\cdot\hat x\hat x_\mu \hat x_\nu 
                               \eta_\mu(x)\bar\eta_\nu(0))\rangle$ \\
$\Pi_5^N(x)$        & $\langle {\rm tr}
                        (\eta_1(x)\bar\eta_2(0))\rangle$ &  & \\
$\Pi_6^N(x)$        & $\langle {\rm tr}(\gamma\cdot\hat x
                         \eta_1(x)\bar\eta_2(0))\rangle$ & & \\
\end{tabular}
\end{table}

  The proton current can be constructed by coupling a $d$-quark to a  
$uu$-diquark. The diquark has the structure $\epsilon_{abc} u_b C\Gamma 
u_c$ which requires that the matrix $C\Gamma$ is symmetric. This condition 
is satisfied for the $V$ and $T$ gamma matrix structures. The two possible 
currents (with no derivatives and the minimum number of quark fields) 
with positive parity and spin $1/2$ are given by \cite{Iof_81}
\be
\label{ioffe}
\eta_1 \;=\; \epsilon_{abc} (u^a C\gamma_\mu u^b) 
\gamma_5 \gamma_\mu d^c, \hspace{1cm}
\eta_2 \;=\; \epsilon_{abc} (u^a C\sigma_{\mu\nu} u^b) 
\gamma_5 \sigma_{\mu\nu} d^c .
\ee
It is sometimes useful to rewrite these currents in terms of scalar
and pseudo-scalar diquarks
\be
\label{ioffe_ps}
\eta_{1,2} &=& (2,4) \left\{ \epsilon_{abc} (u^a C d^b)\gamma_5 u^c 
 \mp \epsilon_{abc} (u^a C\gamma_5 d^b) u^c \right\}.
\ee
Nucleon correlation functions are defined by $\Pi^N_{\alpha\beta}(x) 
= \langle \eta_\alpha(0)\bar\eta_\beta(x) \rangle$, where $\alpha,\beta$ 
are the Dirac indices of the nucleon currents. In total, there are 
six different nucleon correlators: the diagonal $\eta_1\bar\eta_1,\,
\eta_2\bar\eta_2$ and off-diagonal $\eta_1\bar\eta_2$ correlators, 
each contracted with either the identity or $\gamma\cdot x$, see Tab. 
\ref{tab_bar_def}. Let us focus on the first two of these correlation 
functions. For more detail, we refer the reader to \cite{SSV_94} and 
references therein. The OPE predicts \cite{BI_82})
\be                                 
\label{pi1n_vac_dom}
\frac{\Pi_1^N}{\Pi_2^{N0}} &=& \;\;\,\frac{\pi^2}{12}
            |\langle\bar qq\rangle| \tau^3, \\
\frac{\Pi_2^N}{\Pi_2^{N0}} &=& 1 + \frac{1}{768}
   \left\langle G^2\right\rangle\tau^4
  +\frac{\pi^4}{72} |\langle\bar qq\rangle|^2 \tau^6 .
\ee
The vector components of the diagonal correlators receive perturbative 
quark-loop contributions, which are dominant at short distance. The 
scalar components of the diagonal correlators, as well as the 
off-diagonal correlation functions, are sensitive to chiral symmetry 
breaking, and the OPE starts at order $\langle\bar qq\rangle$ or higher.
Single instanton corrections to the correlation functions were 
calculated in \cite{DK_90,FB_93}\footnote{The latter paper corrects 
a few mistakes in the original work by Dorokhov and Kochelev.}. Instantons
introduce additional, regular, contributions in the scalar channel
and violate the factorization assumption for the 4-quark condensates. 
Similar to the pion case, both of these effects increase the amount of 
attraction already seen in the OPE. 

\begin{figure}[t]
\begin{center}
\leavevmode
\epsfxsize=12cm
\epsffile{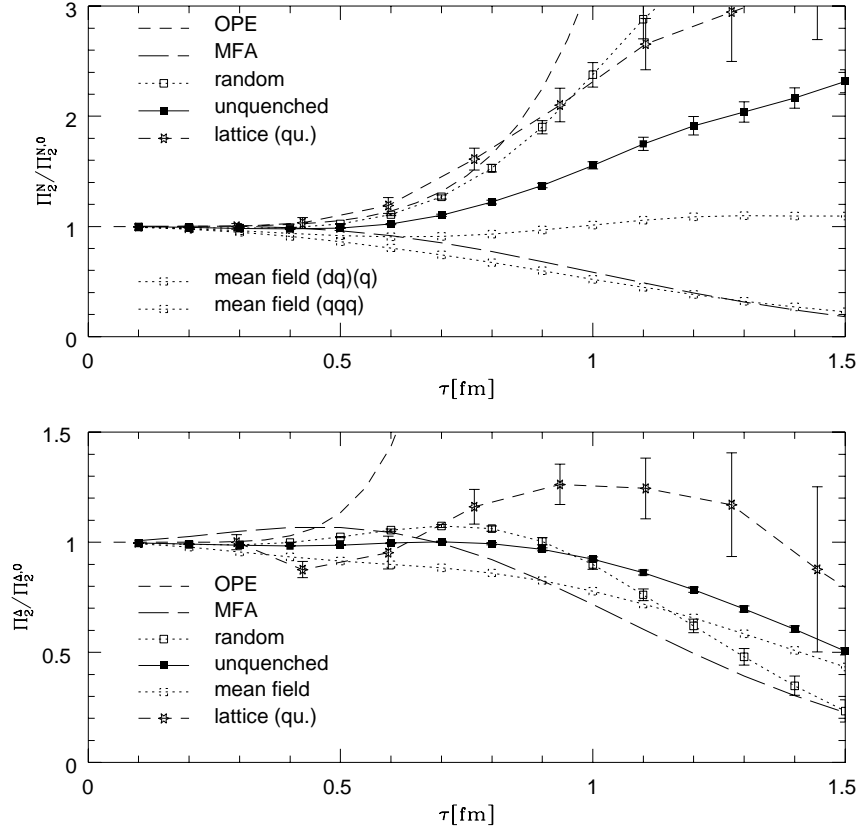}
\end{center}
\caption{\label{fig_bar_cor}
Nucleon and Delta correlation functions $\Pi_2^N$ and $\Pi_2^\Delta$.
Curves labeled as in Fig. \ref{fig_pi_cor}.}
\end{figure}

  The correlation function $\Pi_2^N$ in the interacting ensemble
is shown in Fig. \ref{fig_bar_cor}. There is a significant enhancement
over the perturbative contribution which corresponds to a tightly 
bound nucleon state with a large coupling constant. Numerically, 
we find\footnote{Note that this value corresponds to a relatively 
large current quark mass $m=30$ MeV.} $m_N=1.019$ GeV (see Tab. 
\ref{tab_bar_res}). In the random ensemble, we have measured the 
nucleon mass at smaller quark masses and found $m_N=0.960\pm 0.30$ 
GeV. The nucleon mass is fairly insensitive to the instanton ensemble. 
However, the strength of the correlation function depends on the 
instanton ensemble. This is reflected by the value of the nucleon 
coupling constant, which is smaller in the interacting model.

   Fig. \ref{fig_bar_cor} also shows the nucleon correlation 
function measured in a quenched lattice simulation \cite{CGHN_93b}.
The agreement with the instanton liquid results is quite impressive,
especially given the fact that before the lattice calculations 
were performed, there was no phenomenological information on the 
value of the nucleon coupling constant and the behavior of the 
correlation function at intermediate and large distances. 

   The fitted position of the threshold is $E_0\simeq 1.8$ GeV,
larger than the mass of the first nucleon resonance, the Roper 
$N^*(1440)$, and above the $\pi\Delta$ threshold $E_0=1.37$ GeV.
This might indicate that the coupling of the nucleon current to 
the Roper resonance is small. In the case of the $\pi\Delta$
continuum, this can be checked directly using the phenomenologically
known coupling constants. The large value of the threshold energy
also implies that there is little strength in the (unphysical)
three-quark continuum. The fact that the nucleon is deeply bound can 
also be demonstrated by comparing the full nucleon correlation function 
with that of three non-interacting quarks, see Fig. \ref{fig_bar_cor}). 
The full correlator is significantly larger than the non-interacting 
(mean field) result, indicating the presence of a strong, attractive 
interaction. 

    Some of this attraction is due to the scalar diquark content
of the nucleon current. This raises the question whether the nucleon 
(in our model) is a strongly bound diquark very loosely coupled to a 
third quark. In order to check this, we have decomposed the nucleon
correlation function into quark and diquark components. Using the
mean field approximation, that means treating the nucleon as a 
non-interacting quark-diquark system, we get the correlation 
function labeled (q)(dq) in Fig. \ref{fig_bar_cor}. We observe that 
the quark-diquark model explains some of the attraction seen in 
$\Pi_2^N$, but falls short of the numerical results. This means 
that while diquarks may play some role in making the nucleon bound, 
there are substantial interactions in the quark-diquark system. 
Another hint for the qualitative role of diquarks is provided 
by the values of the nucleon coupling constants $\lambda^{1,2}_N$.
Using (\ref{ioffe_ps}), we can translate these results into the 
coupling constants $\lambda^{s,p}_N$ of nucleon currents built 
from scalar or pseudo-scalar diquarks. We find that the coupling
to the scalar diquark current $\eta_s= \epsilon_{abc}(u^a C\gamma_5 
d^b)u^c$ is an order of magnitude bigger than the coupling to the
pseudo-scalar current $\eta_p=\epsilon_{abc}(u^a Cd^b)\gamma_5u^c$.
This is in agreement with the idea that the scalar diquark channel
is very attractive and that these configurations play an important
role in the nucleon wave function. 

\begin{table}[t]
\caption{\label{tab_bar_res}
Nucleon and delta parameters in the different instanton ensembles. All
quantities are given in units of GeV. The current quark mass is $m_u
=m_d=0.1\Lambda$. }
\begin{tabular}{crrrr}
                & unquenched         & quenched           & RILM       
                & ratio ansatz (unqu.)  \\  \tableline
$m_N$           &    1.019           &    1.013           & 1.040          
                &    0.983    \\
$\lambda_N^1$   &    0.026           &    0.029           & 0.037              
                &    0.021    \\
$\lambda_N^2$   &    0.061           &    0.074           & 0.093             
                &    0.048    \\
$m_\Delta$      &    1.428           &    1.628           & 1.584             
                &    1.372    \\
$\lambda_\Delta$&    0.027           &    0.040           & 0.036             
                &    0.026    \\
\end{tabular}
\end{table}

\subsubsection{Delta correlation functions}
\label{sec_del_cor}

   In the case of the delta resonance, there exists only one independent 
current, given by (for the $\Delta^{++}$) 
\be
\eta^\Delta_\mu =  \epsilon_{abc} (u^a C\gamma_\mu u^b)  u^c.  
\ee
However, the spin structure of the correlator $\Pi^\Delta_{\mu\nu;\alpha
\beta}(x)=\langle\eta^\Delta_{\mu\alpha}(0)\bar\eta^\Delta_{\nu\beta}(x)
\rangle$ is much richer. In general, there are ten independent tensor 
structures, but the Rarita-Schwinger constraint $\gamma^\mu\eta_\mu^\Delta
=0$ reduces this number to four, see Tab. \ref{tab_bar_def}. The OPE 
predicts 
\be
 \frac{\Pi_1^\Delta} {\Pi_2^{\Delta 0}} &=& \;\;\, 4\frac{\pi^2}{12}
       |\langle\bar qq\rangle| \tau^3, \\
 \frac{\Pi_2^\Delta} {\Pi_2^{\Delta 0}} &=& 1 -
 \frac{25}{18}\frac{1}{768} \left\langle G^2 \right\rangle\tau^4 +
 6\frac{\pi^4}{72} |\langle\bar qq\rangle|^2 \tau^6 ,
\ee
which implies that power corrections, in particular due to the quark
condensate, are much larger in the Delta as compared to the nucleon. 
On the other hand, non-perturbative effects due to instantons are 
much smaller! The reason is that while there are large direct 
instanton contributions in the nucleon, there are none in the 
Delta.

  The Delta correlation function in the instanton liquid is shown
in Fig. \ref{fig_bar_cor}. The result is qualitatively different 
from the nucleon channel, the correlator at intermediate distance
$x\simeq 1$ fm is significantly smaller and close to perturbation 
theory. This is in agreement with the results of the lattice 
calculation \cite{CGHN_93b}. Note that, again, this is a quenched
result which should be compared to the predictions of the random 
instanton model. 
  
  The mass of the delta resonance is too large in the random model,
but closer to experiment in the unquenched ensemble. Note that similar
to the nucleon, part of this discrepancy is due to the value of the 
current mass. Nevertheless, the Delta-nucleon mass splitting in the
unquenched ensemble is $m_\Delta-m_N=409$ MeV, still too large as 
compared to the experimental value 297 MeV. Similar to the $\rho$ 
meson, there is no interaction in the Delta channel to first order 
in the instanton density. However, if we compare the correlation 
function with the mean field approximation based on the full
propagator, see Fig. \ref{fig_bar_cor}, we find evidence for
substantial attraction between the quarks. Again, more detailed
checks, for example concerning the coupling to the $\pi N$ 
continuum, are necessary. 

\subsection{Correlation functions on the lattice}
\label{sec_cool_cor}

\begin{figure}[t]
\begin{center}
\leavevmode
\epsfxsize=11cm
\epsffile{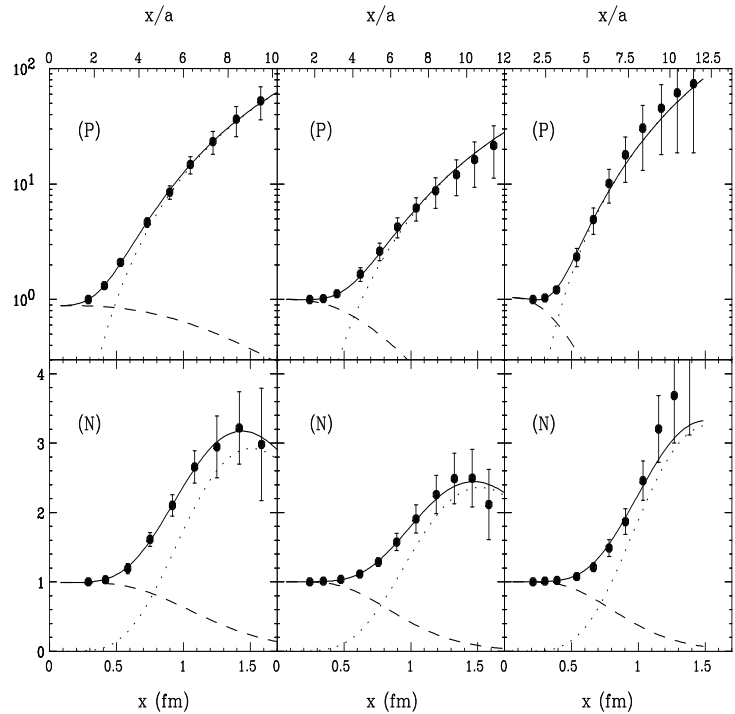}
\end{center}
\caption{\label{fig_cool_cor}
Behavior of pion (P) and nucleon (N) correlation functions under cooling, 
from \protect\cite{CGHN_94}. The left, center, and right panels show the 
results in the original ensemble, and after 25 and 50 cooling sweeps. 
The solid lines show fits to the data based on a pole plus continuum
model for the spectral function. The dotted and dashed lines show
the individual contributions from the pole and the continuum part.}
\end{figure}

 The study of hadronic (point-to-point) correlation functions on the
lattice was pioneered by the MIT group \cite{CGHN_93a,CGHN_93b} which 
measured correlation functions of the $\pi,\delta,\rho,a_1,N$ and 
$\Delta$ in quenched QCD. The correlation functions were calculated 
on a $16^3\times 24$ lattice at $6/g^2=5.7$, corresponding to a lattice 
spacing of $a\simeq 0.17$ fm. A more detailed investigation of baryonic 
correlation functions on the lattice can be found in \cite{Lei_95a,Lei_95b}. 
We have already shown some of the results of the MIT group in Figs. 
\ref{fig_pi_cor}-\ref{fig_bar_cor}. The correlators were measured 
for distances up to $\sim 1.5$ fm. Using the parametrization introduced 
above, they extracted ground state masses and coupling constants and
found good agreement with phenomenological results. What is even more
important, they found the {\em full correlation functions} to agree
with the predictions of the instanton liquid, even in channels (like
the nucleon and delta) where no phenomenological information is 
available. 

  In order to check this result in more detail, they also studied 
the behavior of the correlation functions under cooling \cite{CGHN_94}. 
The cooling procedure was monitored by studying a number of gluonic 
observables, like the total action, the topological charge and and 
the Wilson loop. From these observables, the authors conclude that 
the configurations are dominated by interacting instantons after 
$\sim 25$ cooling sweeps. Instanton-anti-instanton pairs are continually 
lost during cooling, and after $\sim 50$ sweeps, the topological charge 
fluctuations are consistent with a dilute gas. The characteristics of 
the instanton liquid were already discussed in Sec. \ref{sec_inst_lat}.
After 50 sweeps the action is reduced by a factor $\sim$300 while
the string tension (measured from $7\times 4$ Wilson loops) has 
dropped by a factor 6. 

  The behavior of the pion and nucleon correlation functions under 
cooling is shown in Fig. \ref{fig_cool_cor}. The behavior of the 
$\rho$ and $\Delta$ correlators is quite similar. During the cooling 
process the scale was readjusted by keeping the nucleon mass fixed. 
This introduces only a small uncertainty, the change in scale is 
$\sim$16\%. We observe that the correlation functions are {\em stable 
under cooling}, they agree almost within error bars. This is also 
seen from the extracted masses and coupling constants. While $m_N$ 
and $m_\pi$ are stable by definition, $m_\rho$ and $g_\rho$ change
by less than 2\%, $\lambda_\pi$ by 7\% and $\lambda_N$ by 1\%. 
Only the delta mass is too small after cooling, it changes 
by 27\%. 
 
\subsection{Gluonic correlation functions}
\label{sec_glue_cor}

   One of the most interesting problems in hadronic spectroscopy
is whether one can identify glueballs, bound states of pure glue, 
among the spectrum of observed hadrons. This question has two aspects. 
In pure glue theory, stable glueball states exist and they have 
been studied for a number of years in lattice simulations. In full 
QCD, glueballs mix with quark states, making it difficult to 
unambiguously identify glueball candidates. 
 
   Even in pure gauge theory, lattice simulations still require
large numerical efforts. Nevertheless, a few results appear to be
firmly established \cite{CSV_94}:
(i)   The lightest glueball is the scalar $0^{++}$, with a mass in the
      1.5-1.8 GeV range.
(ii)  The tensor glueball is significantly heavier $m_{2^{++}}
      /m_{0^{++}}\simeq 1.4$, and the pseudo-scalar is heavier still,
      $m_{0^{-+}}/m_{0^{++}}=1.5$-$1.8$ \cite{BSH_93}.
(iii) The scalar glueball is much smaller than other glueballs. The size 
      of the scalar is $r_{0^{++}}\simeq 0.2$ fm, while $r_{2^{++}}\simeq 
      0.8$ fm \cite{FL_92}. For comparison, a similar measurement for the 
      $\pi$ and $\rho$ mesons gives 0.32 fm and 0.45 fm, indicating that 
      spin-dependent forces between gluons are stronger than between quarks.

\begin{figure}[t]
\begin{center}
\leavevmode
\epsfxsize=12cm
\epsffile{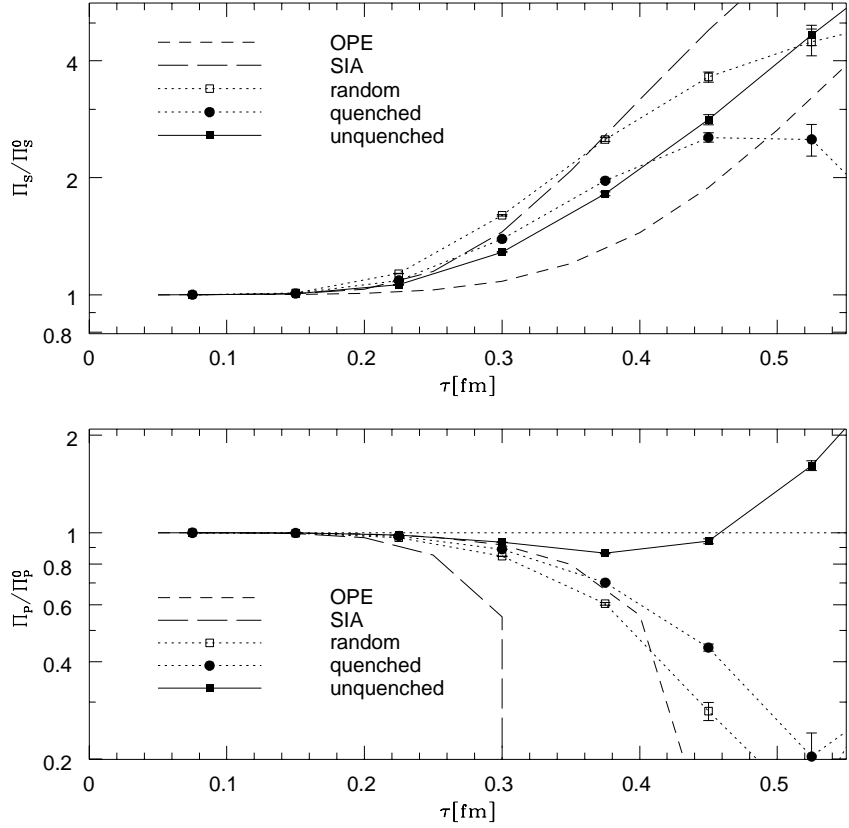}
\end{center}
\caption{\label{fig_glue}
Scalar and pseudo-scalar glueball correlation functions. Curves labeled
as in Fig. \ref{fig_pi_cor}.}
\end{figure}

   Gluonic currents with the quantum numbers of the lowest glueball 
states are the field strength squared ($S=0^{++}$), the topological 
charge density ($P=0^{-+}$), and the energy momentum tensors ($T=2^{++}$):
\be
\label{glue_cur}
j_S=  (G^a_{\mu\nu})^2, \hspace{0.5cm}
j_P= \frac 12 \epsilon_{\mu\nu\rho\sigma} G^a_{\mu\nu}G^a_{\rho\sigma},
     \hspace{0.5cm}
j_T= \frac 14 (G^a_{\mu\nu})^2- G^a_{0\alpha}G^a_{0\alpha}\, .
\ee
The short distance behavior of the corresponding correlation
functions is determined by the OPE \cite{NSVZ_80}
\be
\label{sglue_ope}
\Pi_{S,P}(x) &=& \Pi_{S,P}^0 \left( 1 \pm \frac{\pi^2}{192g^2}
  \langle f^{abc}G^a_{\mu\nu}G^b_{\nu\beta}G^c_{\beta\mu}\rangle x^6
 + \ldots \right) \\
\label{tglue_ope}
\Pi_{T}(x) &=& \Pi_{T}^0 \left( 1 + \frac{25\pi^2}{9216g^2}
 \langle 2{\cal O}_1 - {\cal O}_2\rangle \log (x^2) x^8 +\ldots \right)
\ee
where we have defined the operators ${\cal O}_1 = (f^{abc}
G_{\mu\alpha}^bG_{\nu\alpha}^c)^2,\;{\cal O}_2 = (f^{abc}
G_{\mu\nu}^bG_{\alpha\beta}^c)^2$ and the free correlation 
functions are given by
\be
\label{gb_cor_free}
  \Pi_{S,P}(x)\;=\; (\pm)\frac{384g^4}{\pi^4 x^8}, \hspace{1cm}
  \Pi_T(x)\; = \;\frac{24g^4}{\pi^4 x^8}.
\ee
Power corrections in the glueball channels are remarkably small. 
The leading-order power correction $O(\langle G_{\mu\nu}^2\rangle/x^4)$ 
vanishes\footnote{There is a $\langle G_{\mu\nu}^2 \rangle
\delta^4(x)$ contact term in the scalar glueball correlators
which, depending on the choice of sum rule, may enter momentum
space correlation functions.}, while radiative corrections of the 
form $\alpha_s \log(x^2)\langle G_{\mu\nu}^2\rangle/x^4$ (not 
included in (\ref{sglue_ope})), or higher order power corrections 
like  $\langle f^{abc}G_{\mu\nu}^a G_{\nu\rho}^b G_{\rho\mu}^c
\rangle/x^2$ are very small.

   On the other hand, there is an important low energy theorem 
that controls the large distance behavior of the scalar correlation
function \cite{NSVZ_79b}
\be
\label{sglue_let}
\int d^4x\,\Pi_S(x)&=&\frac{128\pi^2}{b}\langle G^2\rangle,
\ee
where $b$ denotes the first coefficient of the beta function. In 
order to make the integral well defined and we have to subtract 
the constant term $\sim \langle G^2\rangle^2$ as well as singular 
(perturbative) contributions to the correlation function. Analogously, 
the integral over the pseudo-scalar correlation functions is given by 
the topological susceptibility $\int d^4x\,\Pi_P(x)=\chi_{top}$. In  
pure gauge theory $\chi_{top} \simeq (32\pi^2)\langle G^2\rangle$, while 
in unquenched QCD $\chi_{top}=O(m)$, see Sec. \ref{sec_screen}. These 
low energy theorems indicate the presence of rather large non-perturbative 
corrections in the scalar glueball channels. This can be seen as follows: 
We can incorporate the low energy theorem into the sum rules by using a 
subtracted dispersion relation
\be
\label{sglue_sr}
  \frac{\Pi(Q^2)-\Pi(0)}{Q^2} &=& \frac{1}{\pi}
  \int ds\, \frac{{\rm Im}\Pi(s)}{s(s+Q^2)} .
\ee
In this case, the subtraction constant acts like a power correction. 
In practice, however, the subtraction constant totally dominates over
ordinary power corrections. For example, using pole dominance, the 
scalar glueball coupling $\lambda_S = \langle 0|j_S|0^{++}\rangle$ 
is completely determined by the subtraction, $\lambda_S^2/m_S^2 
\simeq (128\pi^2/b)\langle G^2\rangle$.
 
   For this reason, we expect instantons to give a large contribution
to scalar glueball correlation functions. Expanding the gluon operators
around the classical fields, we have
\be
\label{gluecor_exp} 
  \Pi_S(x,y) &=&  \langle 0| G^{2\, cl}(x) G^{2\, cl}(y) |0\rangle  +
   \langle 0| G^{a\, ,cl}_{\mu\nu}(x) \left[ D_\mu^x D_\alpha^y
   D_{\nu\beta}(x,y)\right]^{ab} G^{b\, ,cl}_{\alpha\beta}(y) |0\rangle 
   +\ldots ,
\ee
where $D^{ab}_{\mu\nu}(x,y)$ is the gluon propagator in the classical
background field. If we insert the classical field of an instanton, 
we find \cite{NSVZ_79b,Shu_82b,SS_95}
\be
\label{gb_SIA}
\Pi_{S,P}^{SIA}(x)&=&  \int d\rho\, n(\rho)
 \frac{8192\pi^2}{\rho^4}\frac{\partial^3}{\partial (x^2)^3}
 \left\{ \frac{\xi^6}{x^6} \left( \frac{10-6\xi^2}{(1-\xi^2)^2}
 +\frac{3}{\xi}\log\frac{1+\xi}{1-\xi} \right)\right\}
\ee
where $\xi$ is defined as in (\ref{ps_SIA}). There is no classical 
contribution in the tensor channel, since the stress tensor in the 
self-dual field of an instanton is zero. Note that the perturbative 
contribution in the scalar and pseudo-scalar channels have opposite sign, 
while the classical contribution has the same sign. To first order in the 
instanton density, we therefore find the three scenarios discussed 
in Sec. \ref{sec_mes_cor}: {\em attraction} in the scalar channel, 
{\em repulsion} in the pseudo-scalar and {\em no} effect in the tensor 
channel. The single-instanton prediction is compared with the OPE 
in Fig. \ref{fig_glue}. We clearly see that classical fields 
are much more important than power corrections.

  Quantum corrections to this result can be calculated from the  
second term in (\ref{gluecor_exp}) using the gluon propagator in
the instanton field \cite{BCC_78,LY_79}. The singular contributions
correspond to the OPE in the instanton field. There is an analog
of the Dubovikov-Smilga result for glueball correlators: In a
general self-dual background field, there are no power corrections
to the tensor correlator \cite{NSVZ_80}. This is consistent with 
the result (\ref{tglue_ope}), since the combination $\langle 
2{\cal O}_1-{\cal O}_2\rangle$ vanishes in a self-dual field.
Also, the sum of the scalar and pseudo-scalar glueball correlators
does not receive any power corrections (while the difference
does, starting at $O(G^3)$).

   Numerical calculations of glueball correlators in different 
instanton ensembles were performed in \cite{SS_95}. At short distances, 
the results are consistent with the single instanton approximation. At 
larger distances, the scalar correlator is modified due to the presence
of the gluon condensate. This means that (like the $\sigma$ meson), the
correlator has to be subtracted and the determination of the mass is 
difficult. In the pure gauge theory we find $m_{0^{++}}\simeq 1.5$ GeV and
$\lambda_{0^{++}} = 16\pm 2\,{\rm GeV}^3$. While the mass is consistent
with QCD sum rule predictions, the coupling is much larger than expected
from calculations that do not enforce the low energy theorem 
\cite{Nar_84,BS_90}.

   In the pseudo-scalar channel the correlator is very repulsive and there
is no clear indication of a glueball state. In the full theory (with 
quarks) the correlator is modified due to topological charge screening.
The non-perturbative correction changes sign and a light (on the glueball
mass scale) state, the $\eta'$ appears. Non-perturbative corrections
in the tensor channel are very small. Isolated instantons and anti-instantons 
have a vanishing energy momentum tensor, so the result is entirely due 
to interactions. 

   In \cite{SS_95} we also measured glueball wave functions. The most
important result is that the scalar glueball is indeed small, $r_{0^{++}}
= 0.2$ fm, while the tensor is much bigger, $r_{2^{++}}=0.6$ fm. The
size of the scalar is determined by the size of an instanton, whereas
in the case of the tensor the scale is set by the average distance 
between instantons. This number is comparable to the confinement
scale, so the tensor wave function is probably not very reliable. On 
the other hand, the scalar is much smaller than the confinement scale, 
so the wave function of the $0^{++}$ glueball may provide an important
indication for the importance of instantons in pure gauge theory.

\subsection{Hadronic structure and $n$-point correlators}
\label{sec_3pc}

\begin{figure}[t]
\begin{center}
\leavevmode
\epsfxsize=6cm
\epsffile{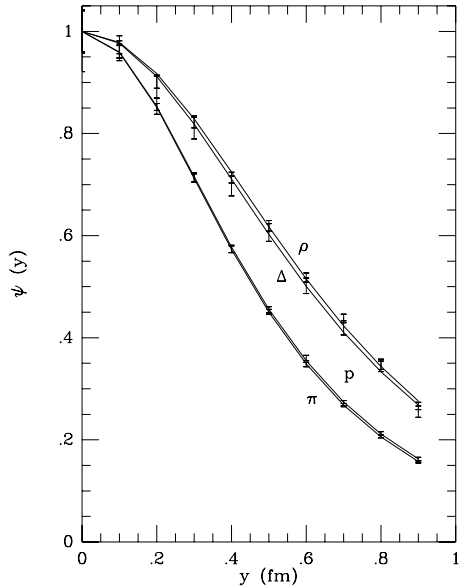}
\end{center}
\caption{\label{fig_wfct}
Hadronic wave functions of the pion, rho meson, proton and delta
resonance in the random instanton ensemble.}
\end{figure}

  So far, we have focussed on two-point correlation in the instanton 
liquid. However, in order to study hadronic properties like decay widths, 
form, structure functions etc., we have to calculate $n$-point 
correlators. There is no systematic study of these objects in the 
literature. In the following, we discuss two exploratory attempts 
and point out a number of phenomenologically interesting questions. 

  Both of our examples are related to the question of hadronic sizes.
Hadronic wave functions (or Bethe-Salpeter amplitudes) are defined by
three-point correlators of the type
\be
\label{pion_cor}
 \Pi_\pi(x,y)&=&\langle 0|T(\bar d(x)Pe^{i\int_x^{x+y}A(x')dx'}
 \gamma_5 u(x+y)\bar d(0)
 \gamma_5 u(0))|0\rangle  \; \stackrel{x\to\infty}{\longrightarrow}\;
 \psi(y) e^{-m_\pi x}.
\ee 
These wave functions are not directly accessible to experiment, but 
they have been studied in a number of lattice gauge simulations, both 
at zero \cite{VW_85,CLN_91,HG_92,GDG_93} and at finite temperature 
\cite{Ber_92,SC_93}. In the single instanton approximation, we find 
that light states (like the pion or the nucleon) receive direct 
instanton contributions, so their size is controlled by the typical 
instantons size. Particles like the $\rho$ or the $\Delta$, on the 
other hand, are not sensitive to direct instantons and are therefore 
less bound and larger in size.

   This can be seen from Fig. \ref{fig_wfct}, where we show
results obtained in the random instanton ensemble \cite{SS_95}.
In particular, we observe that the pion and the proton, as well
as the $\rho$ and the $\Delta$ have essentially the same size.
In the case of the pion and the proton, this in agreement with
lattice data reported in \cite{CLN_91}. These authors also find
that the $\rho$ meson is significantly larger (they did not study
the $\Delta$).

   A more detailed comparison with the lattice data reveals some
of the limitations of the instanton model. Lattice wave functions
are linear at the origin, not quadratic as in the instanton liquid.
Presumably, this is due to the lack of a perturbative Coulomb 
interaction between the quarks in the instanton model. Also,
lattice wave functions decay faster at large distance, which 
might be related to the absence of a string potential in the 
instanton liquid.

\begin{figure}[t]
\begin{center}
\leavevmode
\epsfxsize=10cm
\epsffile{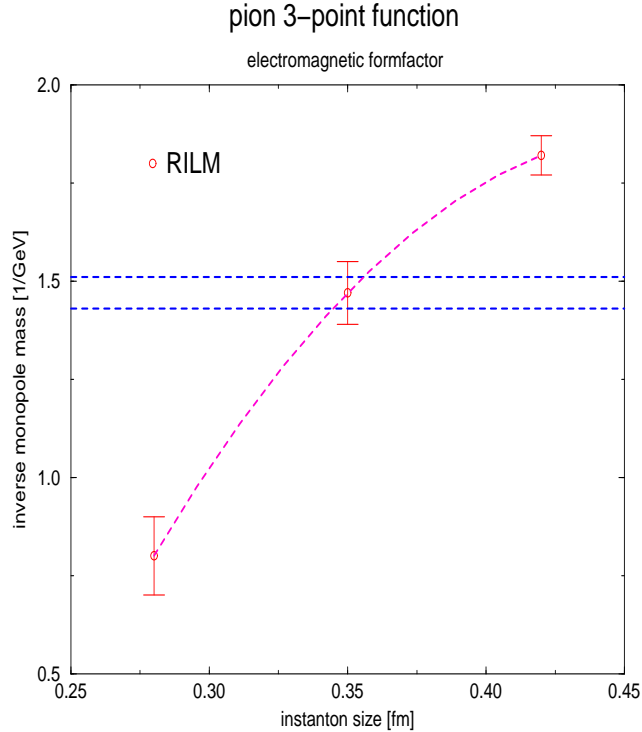}
\end{center}
\caption{\label{fig_pionff} 
The size of the pion, determined from the inverse monopole mass 
in the pion electromagnetic form, as a function of
the instanton size in the random instanton ensemble, from
\protect\cite{BS_96}. The horizontal lines show the experimental 
uncertainty. }
\end{figure}

   Let us now focus on another three-point function which is more
directly related to experiment. The pion electromagnetic form factor 
is determined by the following correlation function
\be
\label{PPV_3pt}
\Gamma_{\mu}(x,y)&=& \langle j^+_{5}(-x/2)j^3_{\mu}(y)j^-_{5}(x/2)\rangle ,
\ee
where $j_5^\pm$ are the pseudo-scalar currents of the charged pion
and $j^3_\mu$ is the third component of the isovector-vector
current. In this case, the correlator is completely determined 
by the triangle diagram. In the context of QCD sum rules, the 
pion form factor is usually analyzed using a three-point functions
build from axial-vector currents (because the pseudo-scalar sum
rule is known to be unreliable). In the single-instanton approximation 
the problem was recently studied in \cite{FN_95}, where it was
shown that including direct instantons (as in Sec. \ref{sec_cor_sia}),
the pion form factor can also be extracted from the pseud-scalar
correlator (\ref{PPV_3pt}). Using the standard instanton liquid 
parameters, these authors calculated the pion form factor for 
$Q^2\sim 1 GeV$, obtaining good agreement with experimental data.

   Recently, the three-point function (\ref{PPV_3pt}) in the instanton 
liquid was also calculated numerically \cite{BS_96}. In this case, one
can go to arbitrarily large distance and determine the pion form factor 
at small momentum. In this regime, the pion form factor has a simple 
monopole shape $F_\pi(Q^2)=M^2/(Q^2+M^2)$ with a characteristic mass 
close to the rho meson mass. However, at intermediate momenta $Q^2\sim 1$ 
the pion form factor is less sensitive to the meson cloud and largely
determined by the quark-anti-quark interaction inside the pion. In the 
instanton model, the interaction is dominated by single-instanton 
effects and the range is controlled by the instanton size. This can 
be seen from Fig. \ref{fig_pionff} where we show the fitted monopole 
mass as a function of the instanton size. Clearly, the two are strongly 
correlated and the instanton model reproduces the experimental value of 
the monopole mass if the mean instanton size is $\rho\simeq 0.35$ fm. 
 
  Finally, let us mention a few experimental observables that might be 
interesting in connection with instanton effects in hadronic structure.
These are observables that are sensitive to the essential features
of the 't Hooft interaction, the strong correlation between quark 
and gluon polarization and flavor mixing. The first is the flavor
singlet axial coupling constant of the nucleon $g_A^0$, which is 
a measure of the quark contribution to the nucleon spin. This 
quantity can be determined in polarized deep inelastic scattering
and was found to be unexpectedly small, $g_A^0\simeq 0.35$. The 
quark model suggests $g_A^0\simeq 1$, a discrepancy which is known 
as the ``proton spin crisis". 

  Since $g_A^0$ is defined by a matrix element of the flavor singlet
current, one might suspect that the problem is somehow related to the
anomaly. In fact, if it were not for the anomaly, one could not
transfer polarization from quarks to gluons and $g_A^0$ could not 
evolve. All this suggests that instantons might be crucial in 
understanding the proton spin structure. While some attempts in
this direction have been made \cite{FS_91,DKZ_93}, a realistic
calculation is still missing.

  Another question concerns the flavor structure of the unpolarized
quark sea in the nucleon. From measurements of the Gottfried sum
rule, the quark sea is known to be flavor asymmetric; there are
more $d$ than $u$ quarks in the proton sea. From a perturbative
point of view this is puzzling, because gluons are flavor-blind, 
so a radiatively generated sea is flavor symmetric. However, as 
pointed out by \cite{DK_93}, quark pairs produced by instantons 
are flavor asymmetric. For example, a valence $u$ quark generates 
$\bar dd,\bar ss$ sea, but no $\bar uu$ pairs. Since a proton 
has two valence $u$ quarks, this gives the right sign of the 
observed asymmetry.

\section{Instantons at finite temperature}
\label{sec_temp}


\subsection{Introduction}

\subsubsection{Finite temperature field theory and the caloron solution}
\label{sec_caloron}

  In the previous sections we have shown that the instanton liquid model
provides a mechanism for chiral symmetry breaking and describes a large 
number of hadronic correlation functions. Clearly, it is of interest to
generalize the model to finite temperature and/or density. This will 
allow us to study the behavior of hadrons in matter, the mechanism of 
the chiral phase transition and possible non-perturbative effects in 
the high temperature/density phase. Extending the methods described 
in the last three sections to non-zero temperature is fairly 
straightforward. In euclidean space, finite temperature corresponds
to periodic boundary conditions on the fields. Basically, all we
have to do is replace all the gauge potentials and fermionic wave 
functions by their $T\neq 0$ periodic counterparts. Nevertheless, 
doing so leads to a number of interesting and very non-trivial 
phenomena, which we will describe in detail below. Extending the
instanton model to finite density is more difficult. In Euclidean 
space a finite chemical potential corresponds to a complex weight 
in the functional integral, so applying standard methods from statistical 
mechanics is less straightforward. While some work on the subject
has been done \cite{Car_80,Bal_81,Che_81,Shu_82c,Abr_83}, many questions 
remain to be understood. 

   Before we study the instanton liquid at $T\neq 0$, we would like to 
give a very brief introduction to finite temperature field theory in 
Euclidean space \cite{Shu_80,GPY_81,McL_86,Kap_89}. The basic object 
is the partition function
\be
  Z &=& {\rm Tr}(e^{-\beta H}),
\ee
where $\beta=1/T$ is the inverse temperature, $H$ is the Hamiltonian
and the trace is performed over all physical states. In Euclidean space,
the partition function can be written as a functional integral
\be
\label{partfct}
  Z &=& \int_{per}DA_\mu \int_{aper}D\bar\psi D\psi\,
  \exp\left(-\int_0^\beta d\tau\int d^3 x\, {\cal L} \right),
\ee
where the gauge fields and fermions are subject to periodic/anti-periodic
boundary conditions
\be
\label{bc}
 A_\mu(\vec r,\beta) &=&\;\;A_\mu(\vec r,0), \\
 \psi(\vec r,\beta)  &=&-\psi(\vec r,0), \hspace{0.5cm}
 \bar\psi(\vec r,\beta)\,=\,-\bar\psi(\vec r,0).
\ee
The boundary conditions imply that the fields can be expanded in a 
Fourier series $\phi(\vec r,\tau)=e^{i\omega_n\tau}\phi_n(\vec r)$,
where $\omega_n=2\pi nT$ and $\omega_n=(2n +1)\pi T$ are the Matsubara
frequencies for boson and fermions. 

  As an example, it is instructive to consider the free propagator
of a massless fermion at finite temperature. Summing over all Matsubara 
frequencies we get
\be
\label{S_free_T}
 S_T(r,\tau) &=& \frac{i}{4\pi^2}\gamma\cdot\partial 
  \sum_n \frac{(-1)^n}{r^2+(\tau-n\beta)^2}.
\ee
The sum can easily be performed, and in the spatial direction one finds
\be
\label{s_T_spa}
 S_T(r,0) &=& \frac{i\vec\gamma\cdot\vec r}{2\pi^2r^4} z\exp(-z)
 \frac{(z+1)+(z-1)\exp(-2z)}{(1+\exp(-2z))^2},
\ee
where $z=\pi rT$. This result shows that at finite temperature, the 
propagation of massless fermion in the spatial direction is exponentially 
suppressed. The screening mass $m=\pi T$ is the lowest Matsubara frequency 
for fermions at finite temperature\footnote{
Let us mention another explanation for this fact which does not use the
Matsubara formalism. In the real-time formalism $S_T(r,0)$ is the 
spatial Fourier transform of $(1/2-n_f(E_k))$, where $n_f$ is the 
Fermi-Dirac distribution. For small momenta $k<T$, the two terms 
cancel and there is no powerlaw decay. In other words, the $1/r^3$
behavior disappears, because the contributions from real and virtual 
fermions cancel each other.}. The energy $\pi T$ acts like a (chiral) 
mass term for space-like propagation. For bosons, on the other hand, 
the wave functions are periodic, the lowest Matsubara frequency is zero
and the propagator is not screened. The propagator in the temporal
direction is given by
\be
\label{s_T_temp}
 S_T(0,\tau) &=& \frac{i\gamma_4}{2\pi^2\tau^3} 
  \frac{y^3}{2}\frac{1+\cos^2(y)}{\sin^3(y)},
\ee
with $y=\pi\tau T$. Clearly, there is no suppression factor for
propagation in the temporal direction. 

  Periodic instanton configurations can be constructed by lining up
zero temperature instantons along the imaginary time direction with 
the period $\beta$. Using 't Hoofts multi-instanton solution 
(\ref{A_n_inst}), the explicit expression for the gauge field 
is given by \cite{HS_78}
\be
\label{caloron}
  A_\mu^a &=& \overline \eta^a_{\mu\nu} \Pi(x)\partial_\nu\Pi^{-1}(x)
\ee
where
\be
 \Pi(x) &=& 1 + \frac{\pi\rho^2}{\beta r} \frac{\sinh (2\pi r/\beta)}
 {\cosh (2\pi r/\beta) - \cos (2\pi\tau/\beta) }
\ee
Here, $\rho$ denotes the size of the instanton. The solution 
(\ref{caloron}) is sometimes referred to as the caloron. It has 
topological charge $Q=1$ and action $S=8\pi^2/g^2$ independent 
of temperature\footnote{The topological classification of smooth
gauge fields at finite temperature is more complicated than at $T=0$
\cite{GPY_81}. The situation simplifies in the absence of magnetic
charges. In this case topological charge is quantized just like
at zero temperature.}. A caloron with $Q=-1$ can be constructed 
making the replacement $\overline\eta^a_{\mu\nu}\to\eta^a_{\mu\nu}$.
Of course, as $T\to 0$ the caloron field reduces to the field of an 
instanton. In the high temperature limit $T\rho\gg 1$, however, the field
looks very different \cite{GPY_81}. In this case, the caloron develops 
a core of size $O(\beta)$ where the fields are very strong $G_{\mu
\nu}\sim O(\beta^2)$. In the intermediate regime $O(\beta)<r<O(\rho^2
/\beta)$ the caloron looks like a ($T$ independent) dyon with unit
electric and magnetic charges,
\be
\label{dyon}
 E_i^a = B_i^a \simeq \frac{\hat r^a\hat r^i}{r^2} .
\ee 
In the far region, $r>O(\rho^2/\beta)$, the caloron resembles a three
dimensional dipole field, $E_i^a=B_i^a\sim O(1/r^3)$.

   At finite temperature, tunneling between degenerate classical 
vacua is related to the anomaly in exactly the same way as it is
at $T=0$. This means that during tunneling, the fermion vacuum
is rearranged and that the Dirac operator in the caloron field
has a normalizable left handed zero mode. This zero mode can 
be constructed from the zero modes of the exact $n$-instanton
solution \cite{Gro_77}. The result is
\be
\label{zm_T}
\psi_i^a &=& \frac{1}{2\sqrt{2}\pi\rho} \sqrt{\Pi(x)}\partial_\mu \left(
  \frac{\Phi(x)}{\Pi(x)}\right) \left(\frac{1-\gamma_5}{2}\gamma_\mu
  \right)_{ij} \epsilon_{aj}\, ,
\ee
where $\Phi(x)=(\Pi(x)-1)\cos(\pi\tau/\beta)/\cosh(\pi r/\beta)$.
Note that the zero-mode wave function also shows an exponential decay 
$\exp(-\pi rT)$ in the spatial direction, despite the fact the eigenvalue 
of the Dirac operator is exactly zero.  This will have important 
consequences for instanton interactions at non-zero temperature.

\subsubsection{Instantons at high temperature}
\label{sec_high_T}

  At finite temperature, just like at $T=0$, the instanton density
is controlled by fluctuations around the classical caloron solution.
In the high temperature plasma phase, the gluoelectric fields in 
the caloron are Debye screened, so we expect that instantons are 
strongly suppressed at high temperature \cite{Shu_78}. The perturbative 
Debye mass in the quark-gluon plasma is \cite{Shu_78b}
\be
m_D^2=  (N_c/3+N_f/6) g^2 T^2.
\ee
Normal $O(1)$ electric fields are screened at distances $1/(gT)$,
while the stronger $O(1/g)$ non-perturbative fields of the instantons 
should be screened for sizes $\rho>T^{-1}$. An explicit calculation of 
the quantum fluctuations around the caloron was performed by Pisarski 
and Yaffe \cite{PY_80}. Their result is 
\be
\label{pis}
  n(\rho,T) &=& n(\rho,T=0) \exp\left(-\frac{1}{3}(2N_c+N_f)
    (\pi\rho T)^2- B(\lambda) \right) \\
   B(\lambda) &=& \left(1+\frac{N_c}{6}-\frac{N_f}{6}\right)
    \left(-\log \left( 1+\frac{\lambda^2}{3} \right)
    + \frac{0.15}{(1+0.15 \lambda^{-3/2})^8} \right) \nonumber
\ee
where $\lambda=\pi\rho T$. As expected, large instantons $\rho\gg 1/T$ 
are exponentially suppressed. This means that the instanton contribution 
to physical quantities like the energy density (or pressure, etc.) is 
of the order
\be
\epsilon(T) \sim \int^{1/T}_0 {d\rho \over \rho^5} (\rho \Lambda)^b 
\sim T^4 (\Lambda/T)^b.
\ee
At high temperature, this is small compared to the energy density of 
an ideal gas $\epsilon(T)_{SB}\sim T^4$.

  It was emphasized in \cite{SV_94} that although the Pisarski-Yaffe result 
contains only one dimensionless parameter $\lambda$, its applicability is 
controlled by two separate conditions:
\be
\rho \ll 1/\Lambda, \hspace{1cm} T \gg \Lambda.
\ee
The first condition ensures the validity of the semi-classical approximation, 
while the second justifies the perturbative treatment of the heat bath.
In order to illustrate this point, we would like to discuss the derivation 
of the semi-classical result (\ref{pis}) in somewhat more detail. Our first 
point is that the finite temperature correction to the instanton density
can be split into two parts of different physical origin. For this 
purpose, let us consider the determinant of a scalar field in the 
fundamental representation\footnote{The quark and gluon (non-zero mode)
determinants can be reduced to the determinant of a scalar fields in the
fundamental and adjoint representation \cite{GPY_81}.}. The 
temperature dependent part of the one loop effective action 
\be
\label{S_eff_T}
 \left.\log\det \left( \frac{-D^2}{-\partial^2}\right)\right|_T 
  &=&\left.\log\det \left( \frac{-D^2}{-\partial^2}\right)\right|_{T=0}
  +\delta 
\ee
can be split into two pieces, $\delta =\delta_1+\delta_2$, where
\be
\label{del_1}
\delta_1 &=& {\rm Tr}_T\log\left(\frac{-D^2(A(\rho))}{-\partial^2}\right)-
{\rm Tr}\log\left(\frac{-D^2(A(\rho))}{ -\partial^2}\right), \\
\label{del_2}
\delta_2 &=& {\rm Tr}_T\log\left(\frac{-D^2(A(\rho,T))}{ -\partial^2}\right)-
{\rm Tr}_T\log\left(\frac{-D^2(A(\rho))}{ -\partial^2}\right).
\ee
Here $Tr_T$ includes an integration over $R^3\times [0,\beta]$, and 
$A(\rho,T)$, $A(\rho)$ are the gauge potentials of the caloron and 
the instanton, respectively. The two terms $\delta_1$ and $\delta_2$ 
correspond to the two terms in the exponent in the semi-classical 
result (\ref{pis}). 

   It was shown in \cite{Shu_82c} that the physical origin of the first 
term is scattering of particles in the heath bath on the instanton field. 
The forward scattering amplitude $T(p,p)$ of a scalar quark can be
calculated using the standard LSZ reduction formula
\be
\label{S_forw}
{\rm Tr}\left(T(p,p)\right) &=& \int d^4x d^4y \ e^{ip\cdot (x-y)} {\rm Tr}
 \left(\partial ^2_x \Delta(x,y) \partial^2_y\right)
\ee
where $\Delta(x,y)$ is the scalar quark propagator introduced in Sec.
\ref{sec_prop_sia}. By  subtracting the trace of the free propagator and going to the 
physical pole $p^2=0$ one gets a very simple answer
\be
\label{S_forw_r}
{\rm Tr}\left(T(p,p)\right) &=&  \,-4\pi^2\rho^2 .
\ee
Since the result is just a constant, there is no problem with analytic 
continuation to Minkowski space. Integrating the result over a thermal
distribution, we get
\be
\label{del1_res}
\delta_1&=&\int{ d^3p\over(2\pi)^3}{1\over 2p(\exp(p/T)-1)} 
 {\rm Tr}\left(T(p,p)\right)\, =\,-\frac{1}{6}(\pi\rho T)^2
 \, = \,  -\frac{1}{6}\lambda^2
\ee
The constant appearing in the result are easily interpreted: $\rho^2$ comes 
from the scattering amplitude, while the temperature dependence enters via 
the integral over the heat bath. Also note that the scattering amplitude
has the same origin (and the same dependence on $N_c,N_f$) as the Debye mass. 

   Formally, the validity of this perturbative calculation requires that
$g(T)\ll 1$, but in QCD this criterion is satisfied only at extremely 
high temperature. We would argue, however, that the accuracy of the 
calculation is controlled by the same effects that determine the 
validity of the perturbative result for the Debye mass. Available 
lattice data \cite{Irb_91} suggest that screening sets in right 
after the phase transition, and that the perturbative prediction 
for the Debye mass works to about 10\% accuracy for $T>3T_c \simeq 
500$ MeV. Near the critical temperature one expects that the Debye 
mass becomes small; screening disappears together with the plasma. 
Below $T_c$ the instanton density should be determined by the scattering 
of hadrons, not quarks and gluons, on the instanton. We will return to 
this point in the next section.

  The second term in the finite temperature effective action (\ref{del_2})
has a different physical origin. It is determined by quantum corrections 
to the colored current \cite{BC_78}, multiplied by the $T$-dependent 
variation of the instanton field, the difference between the caloron 
and the instanton. For $\lambda$ small, the correction is
given by

\be
\label{del2_res}
\delta_2=-\frac{1}{36}\lambda^2+O(\lambda^3).
\ee
The result has the same sign and the same dependence on the parameters 
as $\delta_1$, but a smaller coefficient. Thus, finite $T$ effects not 
only lead to the appearance of the usual (perturbative) heat bath, they 
also modify the strong $O(1/g)$ classical gauge field of the instanton. 
In Euclidean space, we can think of this effect as arising from the 
interaction of the instanton with its periodic ``mirror images" along 
the imaginary time direction. However, below $T_c$, gluon correlators 
are exponentially suppressed since glueballs states are very heavy.  
Therefore, we expect that for $T<T_c$, the instanton field is not 
modified by the boundary conditions. Up to effects of the order
$O(\exp(-M/T))$, where $M$ is the mass of the lowest glueball, there
is no instanton suppression due to the change in the classical 
field below $T_c$.

\subsubsection{Instantons at low temperature}
\label{sec_low_T}

    The behavior of the instanton density at low temperature was
studied in \cite{SV_94}. At temperatures well below the phase transition
the heat bath consists of weakly interacting pions. The interaction of
the pions in the heat bath with instantons can be determined from the
effective Lagrangian (\ref{Leff_nf2}). For two flavors, this Lagrangian 
is a product of certain four-fermion operators and the semi-classical 
single instanton density. In the last section, we argued that the 
semi-classical instanton density is not modified at small
temperature. The temperature dependence is then determined by the  
$T$ dependence of the vacuum expectation value of the four-fermion
operators. For small $T$, the situation can be simplified even 
further, because the wavelength of the pions in the heat bath is
large and the four-fermion operators can be considered local.
 
    The temperature dependence of the expectation value of a general
four-fermion operator $\langle (\bar q A q)(\bar q B q)\rangle$ can be 
determined using soft pion techniques \cite{GL_89}. To order $T^2/f_\pi^2$
the result is determined by the matrix element of the operator between
one-pion states, integrated over a thermal pion distribution. The one-pion 
matrix element can be calculated using the reduction formula and current 
algebra commutators. These methods allow us to prove the general formula
\cite{Ele_93}
\be
\label{fourq_T}
\langle (\bar q A q)(\bar q B q)\rangle_T &=& 
  \langle (\bar q A q)(\bar q B q)\rangle _0
 -{ T^2 \over 96 f^2_\pi}\langle (\bar q\{\Gamma^a_5,\{\Gamma_5^a, A\}\} q)
  (\bar q B q)\rangle _0 \\
 & & -{ T^2 \over 96 f^2_\pi}\langle (\bar q A q)(\bar q\{\Gamma^a_5, 
  \{\Gamma^a_5, B\}\} q)\rangle _0
  -{ T^2 \over 48 f^2_\pi}\langle (\bar q \{\Gamma^a_5, A\} q)(\bar q
  \{\Gamma^a_5, B\} q)\rangle _0 ,
\ee 
where $A,B$ are arbitrary flavor-spin-color matrices and $\Gamma_5^a=\tau^a 
\gamma_5$. Using this result, the instanton density at low temperature is 
given by
\be
\label{idens_lowT}
 n(\rho,T) &=&  n(\rho) \left(\frac{4}{3}\pi^2 \rho^3 \right)^2
\left[ \langle K_1\rangle_0 \frac{1}{4} \left(1-\frac{T^2}{6f^2_\pi}\right)
-\langle K_2\rangle_0\frac{1}{12}\left(1+\frac{T^2}{6f^2_\pi}\right)\right],
\ee
where we have defined the two operators ($\tau$ are isospin matrices):
\be
\label{K1_def}
K_1&=& \bar q_L q_L \bar q_L q_L
+{3\over 32} \bar q_L \lambda^a q_L \bar q_L \lambda^a q_L
-{9\over 128} \bar q_L \sigma_{\mu\nu} \lambda^a q_L 
 \bar q_L \sigma_{\mu\nu} \lambda^a q_L ,\\
\label{K2_def}
K_2&=& \bar q_L \tau^i q_L \bar q_L\tau^i q_L 
+{3\over 32}  \bar q_L \tau^i \lambda^a q_L \bar q_L \tau^i \lambda^a q_L
-{9\over 128} \bar q_L \sigma_{\mu\nu}\tau^i \lambda^a q_L 
 \bar q_L \sigma_{\mu\nu} \tau^i \lambda^a q_L .
\ee
Although the vacuum expectation values of these two operators are  
unknown, it is clear that (baring unexpected cancellations) the 
$T$ dependence should be rather weak, most likely inside the range
$n=n_0( 1 \pm T^2/(6f^2_\pi))$. Furthermore, if one estimates the 
expectation values using the factorization assumption (\ref{vac_dom}),
the $T$ dependence cancels to order $T^2/f_\pi^2$.

   Summarizing the last two sections we conclude that the 
instanton density is expected to be essentially constant below
the phase transition, but exponentially suppressed at large 
temperature. We will explore the physical consequences of this 
result in the remainder of this chapter. Numerical evidence 
for our conclusions that comes from lattice simulations will
be presented in Sec. \ref{sec_lgt_inst_T}.

\subsubsection{Instanton interaction at finite temperature}
\label{sec_int_T}

\begin{figure}[t]
\begin{center}
\leavevmode
\epsfxsize=12cm
\epsffile{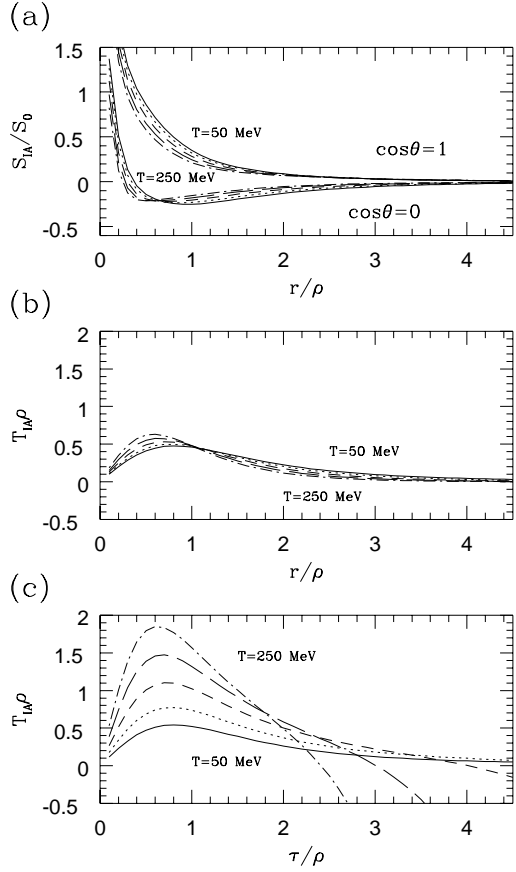}
\end{center}
\caption{\label{fig_SIA_T}
Instanton-anti-instanton interaction at finite temperature.
Fig.a) shows the pure gauge interaction in units of $S_0$
as a function of the spatial separation in units of $\rho$.
The temperature is given in MeV for $\rho=0.33$ fm. Fig. b)
shows the fermionic overlap matrix element $T_{IA}$ for
spatial separation $r$ and Fig. c) for temporal separation
$\tau$.} 
\end{figure}

    In order to study an interacting system of instantons at finite
temperature, we need to determine the bosonic and fermionic 
interaction between instantons at $T\neq 0$. This subject was
studied in \cite{DM_88} and later, in much greater detail, in
\cite{SV_91}, which we will follow here. As for $T=0$, the gluonic 
interactions between an instantons and an anti-instanton is defined 
as $S_{int}=S[A_{IA}]-2S_0$, where we have to specify an ansatz 
$A_{IA}$ for the gauge potential of the $IA$ configuration. The main 
difficulty at $T\neq 0$ is that Lorenz invariance is lost and
the interaction depends on more variables: the spatial and temporal 
separation $r,\tau$ of the instantons, the sizes $\rho_I,\rho_A$, 
the relative orientation matrix $U$ and the temperature $T$. Also,
there are more invariant structures that can be constructed from
the relative orientation vector $u_\mu$. In the case of color 
$SU(2)$, we have
\be
\label{S_int_T_dec}
S_{IA} &=& s_0 + s_1 (u\cdot \hat R)^2 + s_2 (u\cdot\hat R)^4
 + s_3 (u^2-(u\cdot\hat R)^2-u_4^2) \\ 
 & &+ s_4 (u^2-(u\cdot\hat R)^2-u_4^2)^2 
 + s_5 (u\cdot\hat R)^2 (u^2-(u\cdot\hat R)^2-u_4^2), \nonumber
\ee
where $s_i=s_i(r,\tau,\rho_I,\rho_A,\beta)$. 
Because of the reduced symmetry of the problem, it is difficult
to implement the streamline method. Instead, we will study the 
interaction in the ratio ansatz. The gauge field potential is 
given by a straightforward generalization of (\ref{ratio}). 
Except for certain limits, the interaction cannot be obtained 
analytically. We therefore give a parameterization of the numerical 
results obtained in \cite{SV_91}
\be
\label{S_IA_T}
 S_{IA} &=& \frac{8\pi^2}{g^2} \bigg\{
   \frac{4.0}{(r^2+2.0)^2} \frac{\beta^2}{\beta^2+5.21}|u|^2
   - \left[ \frac{1.66}{(1+1.68 r^2)^3}
   + \frac{0.72\log (r^2)}{(1+0.42 r^2)^4} \right]
     \frac{\beta^2}{\beta^2+0.75} |u|^2  \\
   & & \hspace{0.6cm} +  \left[ -\frac{16.0}{(r^2+2.0)^2}
         + \frac{2.73}{(1+0.33 r^2)^3}\right]
         \frac{\beta^2}{\beta^2+0.24+11.50 r^2/(1+1.14 r^2)}
         |u\cdot\hat R|^2
         \nonumber \\
   & & \hspace{0.6cm} +
         0.36\log \left( 1+\frac{\beta}{r}\right)
         \frac{1}{(1+0.013 r^2)^4}
         \frac{1}{\beta^2+1.73}
         (|u|^2-|u\cdot\hat R|^2-|u_4|^2) \bigg\}
         \nonumber
\ee
This parameterization is shown in Fig. \ref{fig_SIA_T}. We observe
that the qualitative form of the interaction does not change, but
it becomes more short range as the temperature increases. This is
a consequence of the core in the instanton gauge field discussed
above. At temperatures $T<1/(3\rho)$ the interaction is essentially
isotropic. As we will see below, this is not the case for the
fermionic matrix elements.

   The fermion determinant is calculated from the overlap matrix 
elements $T_{IA}$ with the finite temperature zero mode wave functions. 
Again, the orientational dependence is more complicated at $T\neq 0$. 
We have
\be
\label{T_IA_dec_T}
T_{IA} &=& u_4 f_1 + \vec u\cdot\hat r f_2
\ee
where $f_i=f_i(r,\tau,\rho_I,\rho_A,\beta)$. The asymptotic form
of $T_{IA}$ for large temperatures $\beta\to 0, R\to\infty$ can be
determined analytically. The result is \cite{SV_91,KY_91}
\be
\label{f1_as}
f_1^{as} &=& i\frac{\pi^2}{\beta}\sin\left(\frac{\pi\tau}{\beta}
 \right)\exp\left(-\frac{\pi r}{\beta}\right), \\
\label{f2_as}
f_2^{as} &=& i\frac{\pi^2}{\beta}\cos\left(\frac{\pi\tau}{\beta}
 \right)\exp\left(-\frac{\pi r}{\beta}\right). 
\ee
A parameterization of the full result is shown in Fig. \ref{fig_SIA_T}
\cite{SS_96b}. The essential features of the result can be seen 
from the asymptotic form (\ref{f1_as},\ref{f2_as}): At large 
temperature, $T_{IA}$ is very anisotropic. The matrix element  
is exponentially suppressed in the spatial direction and periodic 
along the temporal axis. The exponential decay in spatial direction 
is governed by the lowest Matsubara frequency $\pi T$. The qualitative
behavior of $T_{IA}$ has important consequences for the structure of 
the instanton liquid at $T\neq 0$. In particular, the overlap matrix
element favors the formation of instanton-anti-instanton molecules
oriented along the time direction. 

\begin{figure}[t]
\begin{center}
\leavevmode
\epsfxsize=12cm
\epsffile{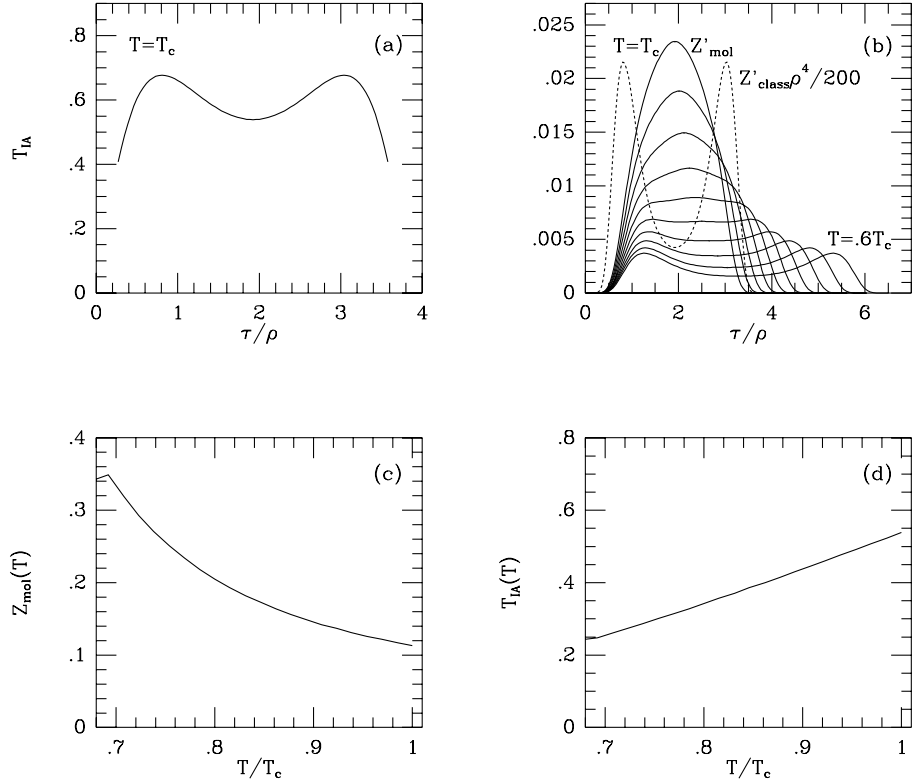}
\end{center}
\caption{\label{fig_momtau} 
(a) Fermionic overlap matrix element $T_{IA}(\tau,r=0,T=T_c)$ 
as a function of the distance $\tau$ between the centers of 
the instanton and anti-instanton in the time direction. (b) 
Same for the partially integrated partition function $Z_{mol}'
=\int d^{10}\Omega T^{2N_f}_{IA}e^{-S_{int}}$, ($Z$ is 
not yet integrated over Matsubara time) at different temperatures
$T=(0.6-1.0)T_c$ (in steps of $\Delta T=0.04 T_c$). The dashed line 
shows the behaviour of $T_{IA}$. (c,d) Fermionic overlap 
matrix element and partition function after integration over 
Matsubara time.}
\end{figure}

   These configurations were studied in more detail in \cite{SV_96}.
These authors calculate the IA partition function (\ref{Z_IA}) from 
the bosonic and fermionic interaction at finite temperature. The
integration over the collective coordinates was performed as follows:
The point $r=0$ (same position in space) and the most attractive
relative orientation ($U=1,\,\cos\theta=1$) are maxima of the
partition function, so one can directly use the saddle point method 
for 10 integrals out of 11 (3 over relative spatial distance between 
the centers and over 7 relative orientation angles). The remaining 
integral over the temporal distance $\tau$ is more complicated and
has to be done numerically.

   In Fig. \ref{fig_momtau}(a) we show the dependence of the overlap 
matrix element $T_{IA}$ on $\tau$ for $T=T_c$, $r=0$ and $U=1$. Even
at $T_c$, $T_{IA}$ does not have a maximum at the symmetric point
$\tau=1/(2T)$, but a minimum. However, when one includes the bosonic
interaction and the pre-exponential factor from taking into account
fluctuations around the saddle-point, the result looks different. 
The partition function after integrating over 10 of the 11 collective
coordinates is shown in Fig. \ref{fig_momtau}(b) for temperatures in
the range $(0.6-1.0)T_c$. We observe that there is a maximum in the 
partition function at the symmetric point $\tau=1/(2T)$ if the 
temperature is bigger than $T_{molec}\simeq 0.2\rho\simeq 120$ 
MeV. The temperature dependence of the partition function integrated
over all variables is shown in Fig. \ref{fig_momtau}(c,d).

   This means that there is a qualitative difference between 
the status of instanton-anti-instanton molecules at low and high 
($T>T_{molec}$) temperature. At low temperature, saddle points 
in the instanton-anti-instanton separation only exist if the 
collective coordinates are analytically continued into the complex 
plane\footnote{The two maxima at $\tau\simeq\rho$ and $\tau\simeq 3
\rho$ in Fig. \protect\ref{fig_momtau}a are not really physical, 
they are related to the presence of a core in the ratio ansatz 
interaction.} (as in Sec. \ref{sec_doublewell}). At high temperature  
there is a real saddle point in the middle of the Matsubara box 
($\tau=1/(2T)$), which gives a real contribution to the free energy.
It is important that this happens close to the chiral phase 
transition. In fact, we will argue that the phase transition 
is caused by the formation of these molecules. 
 
\subsection{Chiral symmetry restoration}
\label{sec_chi_res}

\subsubsection{Introduction to QCD phase transitions}
\label{sec_chi_res_intro}

   Before we come to a detailed discussion of the instanton liquid at
finite temperature, we would like to give a brief summary of what is
known about the phase structure of QCD at finite temperature. It 
is generally believed that at high temperature (or density) QCD 
undergoes a phase transition from hadronic matter to the quark-gluon 
plasma (QGP). In the plasma phase color charges are screened \cite{Shu_78b} 
rather then confined, and chiral symmetry is restored. At sufficiently
high temperature perturbation theory should be applicable and the physical
excitations are quarks and gluons. In this case, the thermodynamics
of the plasma is governed by the Stefan-Boltzmann law, just like 
ordinary black body radiation. 

\begin{figure}[t]
\begin{center}
\leavevmode
\epsfxsize=12cm
\epsffile{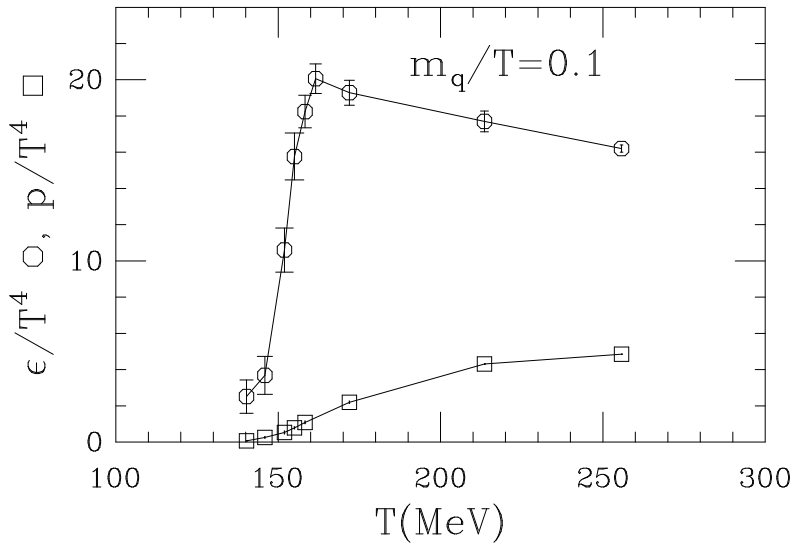}
\end{center}
\caption{\label{fig_eos}
Equation of state for $N_f=2$ QCD with Kogut-Susskind fermions, 
from \protect\cite{BKT_95}.}
\end{figure}

   This basic scenario has been confirmed by a large number of 
lattice simulations. As an example, we show the equation of state 
for $N_f=2$ (Kogut-Susskind) QCD in Fig. \ref{fig_eos} \cite{BKT_95}. 
The transition temperature is $T_c(N_f=2)\simeq 150$ MeV. The 
energy density and pressure are small below the phase transition, 
but the energy density rises quickly to its perturbative (Stefan 
Boltzmann) value. The pressure, on the other hand, lags behind 
and remains small up to $T\simeq 2T_c$.

   In Fig. \ref{fig_eos} the energy density and pressure are 
measured with respect to the perturbative vacuum. However, we have
repeatedly emphasized that in QCD, there is a non-perturbative 
vacuum energy density (the bag pressure) even at $T=0$. In order
to compare a non-interacting gas of quarks and gluons with the 
hadronic phase, we have to shift the pressure in the high $T$ phase 
by this amount, $p_{QGP}\to p_{QGP}-B$. Since the pressure from
thermal hadrons below $T_c$ is small, a rough estimate of the 
transition temperature can be obtained from the requirement 
that the (shifted) pressure in the plasma phase is positive, 
$p_{QGP}>0$. From this inequality we expect the plasma phase to 
be favored for $T>T_c=[(90B)/(N_d\pi^2)]^{1/4}$, where $N_d$ 
is the effective number of degrees of freedom in the QGP phase. 
For $N_f=2$ we have $N_d=37$ and $T_c\simeq 180$ MeV. 

   Lattice simulations indicate that in going from quenched QCD
to QCD with two light flavors the transition temperature drops
by almost a factor of two, from $T_c(N_f=0)\simeq 260$ MeV to 
$T_c(N_f=2)\simeq 150$ MeV. The number of degrees of freedom in 
the plasma phase, on the other hand, increases only from 16 to 37 
(and $T_c$ does not vary very much with $N_d$, $T_c\sim N_d^{-1/4}$). 
This implies that there are significant differences between the 
pure gauge and unquenched phase transitions.

 In particular, the low transition temperature observed for $N_f=2,3$ 
suggests that the energy scales are quite different. We have already 
emphasized that the bag pressure is directly related to the gluon 
condensate \cite{Shu_78}. This relation was studied in more detail 
in \cite{AHZ_91,Den_89,KB_93}. From the canonical energy momentum 
tensor and the trace anomaly, the gluonic contributions to the energy 
density and pressure are related to the electric and magnetic field 
strengths
\be
\label{eps}
\epsilon &=& {1\over 2}\langle B^2-E^2\rangle - {g^2b \over 128\pi^2}
\langle E^2+B^2 \rangle,\\
\label{pres}
 p &=&{1\over 6}\langle B^2 - E^2 \rangle + {g^2b \over 128\pi^2}
\langle E^2+B^2 \rangle .
\ee
Using the available lattice data one finds that in pure gauge theory
the gluon condensate essentially disappears in the high temperature 
phase, while in full QCD ($N_f=2,3$) about half of the condensate
remains. G. Brown has referred to this part of the gluon condensate 
as the ``hard glue" or ``epoxy". It plays an important role in 
keeping the pressure positive despite the low transition temperature.

\begin{figure}[t]
\begin{center}
\leavevmode
\epsfxsize=8cm
\epsffile{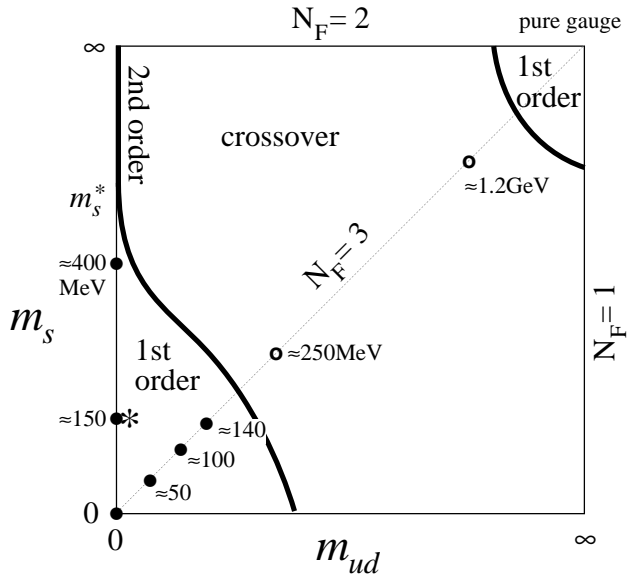}
\end{center}
\caption{\label{fig_nf3_phase}
Schematic phase diagram of $N_f=3$ QCD with dynamical quark masses $m_{ud}
\equiv m_u=m_d$ and $m_s$, from \protect\cite{IKK_96}. Solid and open points 
mark simulations (with Wilson fermions) that found a first order transition
or a smooth crossover, respectively. The (approximate) location of QCD 
with realistic masses is shown by a star.}
\end{figure}

   More general information on the nature of the phase transition
is provided by universality arguments. For this purpose, we have
to identify an order parameter that characterizes the transition. 
In QCD, there are two limits in which this can be achieved.
For infinitely heavy quarks, the Polyakov line \cite{Pol_78} 
provides an order parameter for the deconfinement transition, while 
for massless quarks the fermion condensate is an order parameter 
for spontaneous chiral symmetry breaking. The Polyakov line is 
defined by
\be
\label{Pol_line}
\langle L(\vec{x})\rangle  &=& \langle  
P\exp\left(i\int_0^\beta  d\tau\, A_0(\vec x,\tau) 
 \right)\rangle ,
\ee
which can be interpreted as the propagator of a heavy quark. The free
energy of a single static quark (minus the free energy of the vacuum)
is given by $F_q-F_0=-T\log(|\langle L(\vec x)\rangle|)$. For $\langle 
L\rangle =0$, the free energy of an isolated quark is infinite and the 
theory is in the confining phase. For $\langle L\rangle \neq 0$ the 
free energy is finite, and quarks are screened rather than confined. 

   The Polyakov line has a nontrivial symmetry. Under gauge transformations
in the center of the gauge group $Z_{N_c}\subset SU(N_c)$ local observables
are invariant but the Polyakov line picks up a phase $L\to zL$ with
$z\in Z_{N_c}$. The deconfinement transition was therefore related with
spontaneous breakdown of the $Z_{N_c}$ center symmetry. Following 
Landau, long wave excitations near the phase transition should be
governed by an effective $Z_{N_c}$ symmetric Lagrangian for the 
Polyakov line \cite{SY_82}. Since long range properties are determined
by the lowest Matsubara modes, the effective action is defined in three
dimensions. For $N_c=2$, this means that the transition is in the same 
universality class as the $d=3$ Ising model, which is known to have a 
second order phase transition. For $N_c\geq 3$, on the other hand, we 
expect the transition to be first order\footnote{Under certain 
assumptions, the $N_c>3$ phase transition can be second order, see
\cite{PT_97}.}. These expectations are supported by lattice results.

   For massless quarks chiral symmetry is exact, and the quark condensate 
$\langle \bar qq\rangle$ provides a natural order parameter. The symmetry 
of the order parameter is determined by the transformation properties of 
the matrix $U_{ij}=\langle\bar q_iq_j\rangle$. For $N_f=2$ flavors, this 
symmetry is given by $SU(2)\times SU(2)=O(4)$, so the transition is governed 
by the effective Lagrangian for a four dimensional vector field in three
dimension \cite{PW_84,Wil_92,RW_93}. The $O(4)$ Heisenberg magnet is known 
to have a second order phase transition and the critical indices have been 
determined from both numerical simulations and the $\epsilon$ expansion. 
For $N_f\geq 3$, however, the phase transition is predicted to be first 
order.

   From these arguments, one expects the schematic phase structure 
of QCD in the $m_{ud}-m_s$ (with $m_{ud}\equiv m_u=m_d$) plane to 
look as shown in Fig. \ref{fig_nf3_phase} \cite{BBC_90,IKK_96}. The 
upper right hand corner corresponds to the first order purge gauge 
phase transition. Presumably, this first order transition extends to 
lower quark masses, before it ends in a line of second order phase 
transitions. The first order $N_f=3$ chiral phase transition is located 
in the lower left corner, and continues in the mass plane before it 
ends in another line of second order phase transitions. At the left 
edge there is a tri-critical point where this line meets the line of 
$N_f=2$ second order phase transitions extending down from the upper 
left corner. 

   Simulations suggest that the gap between the first order chiral
and deconfinement transitions is very wide, extending from $m\equiv 
m_u=m_d=m_s\simeq 0.2$ GeV to $m\simeq 0.8$ GeV. This is in line with
the arguments given above, indicating that there are important 
differences between the phase transitions in pure gauge and full
QCD. Nevertheless, one should not take this distinction to literally.
In the presence of light quarks, there is no deconfinement phase
transition in a strict, mathematical, sense. From a practical 
point of view, however, deconfinement plays an important role 
in the chiral transition. In particular, the equation of state
shows that the jump in energy density is dominated by the 
release of 37 quark and gluon degrees of freedom.

  There are many important questions related to the phase diagram
that still have to be resolved. First of all, we have to determine
the location of real QCD (with two light, one intermediate mass and 
several heavy flavors) on this phase diagram. While results using 
staggered fermions \cite{BBC_90} seem to suggest that QCD 
lies outside the range of first order chiral phase transitions 
and only shows a smooth crossover, simulations using Wilson fermions 
\cite{Iwa_91} with realistic masses find a first order transition.

  A more general is problem is to verify the structure of the phase 
diagram and to check the values of the critical indices near second
order transitions. While earlier studies confirmed (within errors)
the $O(4)$ values of the critical indices for the $N_f=2$ chiral
transition with both staggered \cite{KL_94} and Wilson fermions
\cite{IKK_96}, more recent work concludes that $1/\delta$ (defined 
as $\langle \bar q q \rangle _{T_c} \sim m^{1/\delta}$) is consistent 
with zero \cite{Ukawa_96}. If this result is correct, it would
imply that there is no $N_f=2$ second order phase transition and the
only second order line in Fig. \ref{fig_nf3_phase} corresponds
to the boundary of the first order region, with standard Ising 
indices.

  Are there any possible effects that could upset the expected $O(4)$ 
scenario for the $N_f=2$ transition? One possibility discussed in the 
literature is related to the fate of the $U(1)_A$ symmetry at $T_c$.
At zero temperature the axial $U(1)_A$ symmetry is explicitly broken 
by the chiral anomaly and instantons. If instantons are strongly 
suppressed \cite{PW_84} or rearranged \cite{Shu_94} at $T_c$, then
the $U(1)_A$ symmetry might effectively be restored at the phase
transition. In this case, the masses of the $U(1)_A$ partners of 
the $\pi,\sigma$, the $\delta,\eta'$, could become sufficiently
small so that fluctuations of these modes affect the universality 
arguments given above. In particular, if there are eight rather 
than four light modes at the phase transition, the transition is 
expected to be first order. Whether this is the correct interpretation
of the lattice data remains unclear at the moment. We will return 
to the question of $U(1)_A$ breaking at finite temperature in 
Sec. \ref{sec_cor_ua1} below. Let us only note that the 
simulations are very difficult and that there are several possible
artefacts. For example, there could be problems with the extrapolation 
to small masses. Also, first order transitions need not have an
order parameter, and it is difficult to distinguish very weak
first order transitions from second order transitions. 

  Completing this brief review of general arguments and lattice
results on the chiral phase transition, let us also comment on some 
of the theoretical approaches. Just like perturbative QCD describes 
the thermodynamics of the plasma phase at very high temperature, 
effective chiral lagrangian provide a very powerful tool at low 
temperature. In particular, chiral perturbation theory predicts 
the temperature dependence of the quark and gluon condensates as
well as of the masses and coupling constants of light hadrons
\cite{GL_89}. These results are expressed as an expansion in 
$T^2/f_\pi^2$. It is difficult to determine what the range of 
validity of these predictions is, but the approach certainly
has to fail as one approaches the phase transition. 

  In addition to that, the phase transition has been studied  
in various effective models, for example in the linear 
sigma model, the chiral quark model or the Nambu-Jona-Lasinio 
model \cite{HK_85b,BM_88,KLW_90}. In the NJL model, the chiral
transition connects the low temperature phase of massive 
constituent quarks with the high temperature phase of massless 
quarks (but of course not gluons). The mechanism for the transition
is similar to the transition from a superconductor to a normal metal:
The energy gain due to pairing disappears once a sufficient number of 
positive energy states is excited. We have seen that at zero 
temperature, the instanton liquid leads to a picture of chiral 
symmetry breaking which closely resembles the NJL model. Nevertheless,
we will argue that the mechanism for the phase transition is quite
different. In the instanton liquid, it is the ensemble itself 
which is rearranged at $T_c$. This means that not only do we 
have non-zero occupation numbers, but the effective interaction 
itself changes. 

  Finally, a number of authors have extended random matrix models
to finite temperature and density \cite{JV_96,WSW_96,NPZ_96}. It 
is important to distinguish these models from random matrix methods 
at $T,\mu=0$. In this case, there is evidence that certain 
observables\footnote{On the other hand, macroscopic observables, 
like the average level density (and the quark condensate) 
are not expected to be universal.}, like scaled correlations between 
eigenvalues of the Dirac operator, are universal and can be described 
in terms of suitably chosen random matrix ensembles. The effects of 
non-zero temperature and density, on the other hand, are included 
in a rather schematic fashion, by putting terms like $(\pi T+i\mu)$ 
into the Dirac operator\footnote{See \protect\cite{WSW_96} for an
attempt to model the results of the instanton liquid in terms of a 
random matrix ensemble.}. This procedure is certainly not universal. 
From the point of view of the instanton model, the entries of the 
random matrix correspond to matrix elements of the Dirac operator 
in the zero mode basis. If we 
include the effects of a non-zero $\mu,T$ in the schematic form 
mentioned above, we assume that at $\mu,T\neq 0$ there are no zero
modes in the spectrum. But this is not true, we have explicitly
constructed the zero-mode for the caloron configuration in 
Sec. \ref{sec_caloron}, for $\mu\neq 0$ see \cite{Car_80,Abr_83}. 
As we will see below, in the instanton model the phase transition 
is not caused by a constant contribution to the overlap matrix 
elements, but by specific correlations in the ensemble.

\subsubsection{The instanton liquid at finite temperature and  
$IA$ molecules}
\label{sec_mol}
 
   In Sec. \ref{sec_low_T}, we argued that the instanton density 
remains roughly constant below the phase transition. This means 
that the chiral phase transition has to be caused by a rearrangement 
of the instanton ensemble. Furthermore, we have shown that the 
gluonic interaction between instantons remains qualitatively 
unchanged even at fairly high temperatures. This suggests that 
fermionic interactions between instantons drive the phase 
transition \cite{IS_94,SSV_95,SS_96}.

   The mechanism for this transition is most easily understood by 
considering the fermion determinant for one instanton-anti-instanton
pair \cite{SSV_95}. Using the asymptotic form of the overlap matrix
elements specified above, we have
\be
\label{det_IA}
\det(D\!\!\!\!/\,) &\sim& \left| \frac{\sin(\pi T\tau)}{\cosh(\pi Tr)}
\right|^{2N_f}.
\ee
This expression is maximal for $r=0$ and $\tau=1/(2T)$, which is 
the most symmetric orientation of the instanton-anti-instanton pair 
on the Matsubara torus. Since the fermion determinant controls the
probability of the configuration, we expect polarized molecules
to become important at finite temperature. The effect should be 
largest, when the IA pairs exactly fit onto the torus, i.e. $4\rho
\simeq\beta$. Using the zero temperature value $\rho\simeq 0.33$ fm,
we get $T\simeq 150$ MeV, close to the expected transition temperature
for two flavors.

   In general, the formation of molecules is controlled by the 
competition between minimum action, which favors correlations,
and maximum entropy, which favors randomness. Determining the
exact composition of the instanton liquid as well as the transition
temperature requires the calculation of the full partition function,
including the fermion induced correlations. We will do this using
a schematic model in Sec. \ref{sec_cocktail} and using numerical
simulations in Sec. \ref{sec_liquid_T}. 

   Before we come to this we would like to study what physical effects
are caused by the presence of molecules. Qualitatively, it is clear 
why the formation of molecules leads to chiral symmetry restoration. 
If instantons are bound into pairs, then quarks mostly propagate 
from one instanton to the corresponding anti-instanton and back. In 
addition to that, quarks propagate mostly along the imaginary time 
direction, so all eigenstates are well localized in space, and no quark 
condensate is formed. Another way to see this is by noting that the 
Dirac operator essentially decomposes into $2\times 2$ blocks corresponding 
to the instanton-anti-instanton pairs. This means that the eigenvalues
will be concentrated around some typical $\pm \lambda$ determined
by the average size of the pair, so the density of eigenvalues 
near $\lambda=0$ vanishes. We have studied the eigenvalue
distribution in a schematic model of random and correlated 
instantons in \cite{SSV_95}, and a random matrix model of the 
transition based on these ideas was discussed in \cite{WSW_96}.

   The effect of molecules on the effective interaction between 
quarks at high temperature was studied in \cite{SSV_95}, using   
methods similar to the ones introduced in Sec. \ref{sec_hfa}.
In order to determine the interaction in a quark-anti-quark
(meson) channel with given quantum numbers, it is convenient to 
calculate both the direct and the exchange terms, and Fierz-rearrange 
the exchange term into an effective direct interaction. The resulting 
Fierz symmetric Lagrangian reads \cite{SSV_95}
\be
\label{lmol}
 {\cal L}_{mol\,sym}&=& G
     \left\{ \frac{2}{N_c^2}\left[
     (\bar\psi\tau^a\psi)^2-(\bar\psi\tau^a\gamma_5\psi)^2
      \right]\right. \nonumber \\
     & & - \;\,\frac{1}{2N_c^2}\left. \left[
     (\bar\psi\tau^a\gamma_\mu\psi)^2+(\bar\psi\tau^a\gamma_\mu\gamma_5
     \psi)^2 \right] + \frac{2}{N_c^2}
     (\bar\psi\gamma_\mu\gamma_5\psi)^2 \right\} + {\cal L}_8,
\ee
with the coupling constant
\be
 G &=& \int n(\rho_1,\rho_2)\,d\rho_1 d\rho_2\,
        \frac{1}{8T_{IA}^2}(2\pi\rho_1)^2(2\pi\rho_2)^2\, .
\label{gmol}
\ee
Here, ${\cal L}_8$ is the color-octet part of the interaction and
$\tau^a$ is a four-vector with components $(\vec\tau,1)$. Also, 
$n(\rho_1,\rho_2)$ is the tunneling probability for the IA pair 
and $T_{IA}$ the corresponding overlap matrix element. The 
effective Lagrangian (\ref{lmol}) was determined by averaging 
over all possible molecule orientations, with the relative 
color orientation $\cos(\theta)=1$ fixed. Near the phase transition,
molecules are polarized in the temporal direction, Lorentz invariance 
is broken, and vector interactions are modified according to $(\bar
\psi\gamma_\mu\Gamma\psi)^2\to 4(\bar\psi\gamma_0\Gamma\psi)^2$.

   Like the zero temperature effective Lagrangian (\ref{Leff_nf2}),
the interaction (\ref{lmol}) is $SU(2)\times SU(2)$ symmetric.
Since molecules are topologically neutral, the interaction is
also $U(1)_A$ symmetric. This does not mean that $U(1)_A$ symmetry 
is restored in the molecular vacuum. Even a very small $O(m^{N_f})$ 
fraction of random instantons will still lead to $U(1)_A$ breaking
effects of order $O(1)$, see Sec. \ref{sec_cor_ua1}. If chiral
symmetry is restored, the effective interaction (\ref{lmol}) is 
attractive not only in the pion channel, but also in the other 
scalar-pseudo-scalar channels $\sigma,\delta$ and $\eta'$. Furthermore, 
unlike the 't Hooft interaction, the effective interaction in the 
molecular vacuum also includes an attractive interaction in the vector
and axial-vector channels. If molecules are unpolarized, the 
corresponding coupling constant is a factor 4 smaller than
the scalar coupling. If they are fully polarized, only the 
longitudinal vector components are affected. In fact, the
coupling constant is equal to the scalar coupling. A more 
detailed study of the quark interaction in the molecular
vacuum was performed in \cite{SSV_95,SS_95b}, where hadronic
correlation functions in the spatial and temporal direction 
were calculated in the schematic model mentioned above. We
will discuss the results in more detail below.

\subsubsection{Mean field description at finite temperature}
\label{sec_cocktail}

   In the following two sections we wish to study the statistical
mechanics of the instanton liquid at finite temperature. This is 
necessary not only in order to study thermodynamic properties of 
the system, but also to determine the correct ensemble for calculations
of hadronic correlation functions. We will first extend the mean field
calculation of Sec. \ref{sec_hfa} to finite temperature. In the next
section we will study the problem numerically, using the methods 
introduced in Sec. \ref{sec_free}.

  For pure gauge theory, the variational method was extended to finite 
temperature in \cite{DM_88} and \cite{Kan_88}. The gluonic interaction 
between instantons changes very little with temperature, so we will 
ignore this effect. In this case, the only difference as compared to 
zero temperature is the appearance of the perturbative suppression factor 
(\ref{pis}) (for $T>T_c$, although Diakonov and Mirlin used it for all 
$T$). Since the interaction is unchanged, so is the form of the single 
instanton distribution
\be
\label{var_T}
 \mu(\rho) &=& n(\rho,T) \exp\left[ \frac{-\beta\gamma^2 NT}{V_3}
 \overline{\rho^2}\rho^2\right],
\ee
where $n(\rho,T)$ is the semi-classical result (\ref{pis}) and the
four dimensional volume is given by $V=V_3/T$. The $T$ dependence 
of the distribution functions modifies the self-consistency equations 
for $\overline{\rho^2}$ and $N/V$. Following \cite{DM_88}, we can 
expand the coefficient $B(\lambda)$ in (\ref{pis}) to order $T^2$: 
$n(\rho,T)\simeq \exp(-\frac{1}{3}(\frac{11}{6}N_c-1)(\pi\rho T)^2) 
n(\rho,T=0)$. In this case, the self-consistency condition for the
average size is given by
\be
\label{rho2_T}  
 \overline{\rho^2} &=& \nu \left[ \frac{1}{3}(\frac{11}{6}N_c-1)
\pi^2T^2 +\frac{N\beta\gamma^2}{V_3}T\overline{\rho^2}\right]^{-1}.
\ee
For $T=0$, this gives $\overline{\rho^2}=[(\nu V)/(\beta\gamma^2N)]^{1/2}$ 
as before, while for large $T>(\pi\rho)^{-1}$ we have $\overline{\rho^2}
\sim 1/T^2$. The instanton density follows from the self-consistency
equation for $\mu_0$. For large $T$ we have
\be
\label{dens_T}
 N/V &=& C_{N_c} \beta^{2N_c} \Lambda^4 \Gamma(\nu)
 \left[ \frac{1}{3}(\frac{11}{6}N_c-1)\pi^2\frac{T^2}{\Lambda^2}
 \right]^{-\nu},
\ee
so that $N/V\sim 1/T^{b-4}$, which is what one would expect from
simply cutting the size integration at $\rho\simeq 1/T$. 

  The situation is somewhat more interesting if one extends the
variational method to full QCD \cite{IS_89,NVZ_89d}. In this case,
an additional temperature dependence enters through the $T$
dependence of the average fermionic overlap matrix elements. 
More importantly, the average determinant depends on the 
instanton size,
\be
\label{av_det}
 \det(D\!\!\!\!/\,) = \prod_I (\rho m_{det}), \hspace{1cm}
 m_{det} =\rho^{3/2} \left[\frac{1}{2} I(T) \int d\rho n(\rho)\rho
  \right]^{1/2},
\ee
where $I(T)$ is the angle and distance averaged overlap matrix
element $T_{IA}$. The additional $\rho$ dependence modifies the
instanton distribution (\ref{var_T}) and introduces an additional
nonlinearity into the self-consistency equation. As a result, the 
instanton density at large $T$ depends crucially on the number of
flavors. For $N_f=0,1$, the density drops smoothly with $N/V \sim
1/T^{2a}$ and $a=(b-4+2N_f)/(2-N_f)$ for large $T$. For $N_f=2$, 
the instanton density goes to zero continuously at the critical 
temperature $T_c$, whereas for $N_f>2$, the density goes to zero 
discontinuously at $T_c$. This behavior can be understood from
the form of the gap equation for the quark condensate. We have
$\langle\bar qq\rangle \sim const \langle\bar qq\rangle^{N_f-1}$,
which, for $N_f>2$, cannot have a solution for arbitrarily small
$\langle\bar qq\rangle$.

\begin{figure}[t]
\begin{center}
\leavevmode
\epsfxsize=8cm
\epsffile{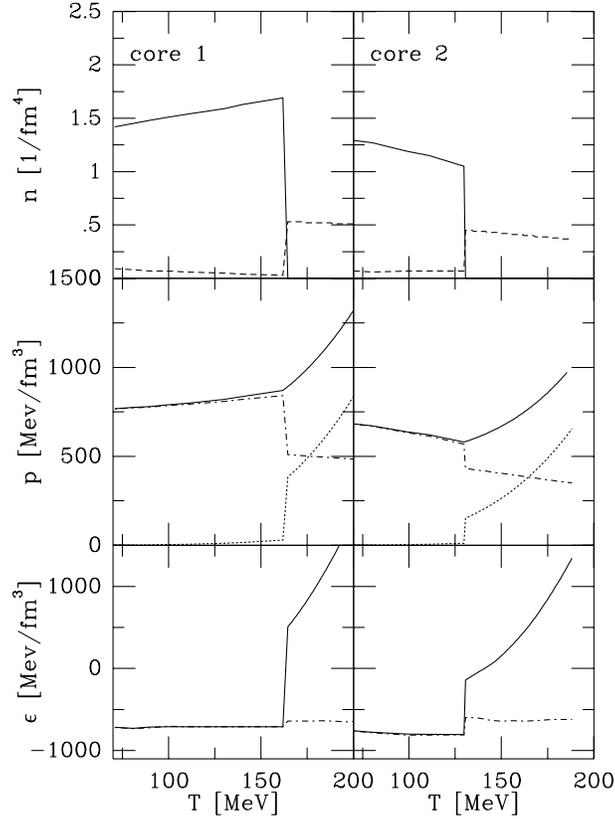}
\end{center}
\caption{\label{fig_cocktail}
Chiral restoration phase transitions for 2 massless flavors and two
different core parameters $R_c$. The upper panel shows the $T$ dependence 
of the densities $n_a$ (solid) and $n_m$ (dotted). In the middle panel 
the $T$ dependence of the total pressure $p$ (solid) is shown, including 
the contributions of the pion gas/quark-gluon plasma (dotted) and of 
instantons (dash-dotted). The energy density is presented in the lower 
panel (solid) which is modified by the instanton contribution (dash-dotted).}
\end{figure}

  In the mean-field approximation, the instanton ensemble remains 
random at all temperatures. This of course implies that instantons
cannot exist above $T_c$. We already argued that this is not correct, 
and that instantons can be present above $T_c$, provided they are 
bound in molecules (or larger clusters). In order to include this
effect into a variational calculation, Ilgenfritz and Shuryak 
introduced a ``cocktail" model \cite{IS_89,IS_94}, in which
the instanton ensemble consists of a random and a molecular 
component. The composition of the instanton liquid is determined
by minimizing the free energy with respect to the two 
concentrations.
 
  As above, the distribution functions for random instantons is
given by
\be
\label{mu_rand}
\mu(\rho) &=& n(\rho) \exp\left[ -\kappa \rho^2 (\overline{\rho^2_a}
n_a +2\overline{\rho^2_m} n_m)\right] (m_{det}\rho)^{N_f},
\ee
where $n_{a,m}$ are the densities of the random and molecular 
contributions and $\overline{\rho^2}_{a,m}$ are the corresponding
average radii. The parameter $\kappa=\beta\gamma^2$ characterizes
the average repulsion between instantons. The distribution of 
instantons bound in molecules is given by
\be
\label{mu_mol}
\mu(\rho_1,\rho_2) &=& n(\rho_1)n(\rho_2)
\exp\left[ -\kappa(\rho^2_1+\rho^2_2)(\overline{\rho^2_a} n_a
+2\overline{\rho^2_m} n_m)\right] 
\left \langle (T_{IA}T_{IA}^*)^{N_f} \right\rangle ,
\ee
where $I_m(N_f,T)\equiv\langle (T_{IA}T_{IA}^*)^{N_f} \rangle$ is the 
average determinant for an instanton-anti-instanton pair, with the 
relative orientation $\cos\theta=1$ fixed. Summing the contributions 
from both the random and molecular components, the self-consistency 
condition for the instanton size becomes
\be 
\frac{\overline{\rho^2_m}}{\overline{\rho^2_a}} = 
\frac{\alpha}{\beta}, \hspace{1cm}
\frac{1}{\kappa} = \frac{2(\overline{\rho^2_a})^2 n_a}{\beta} +
\frac{4(\overline{\rho^2_m})^2 n_m}{\alpha} . 
\ee
where $\alpha=b/2-1$ and $\beta=b/2+3N_f/4-2$. Using this result, one
can eliminate the radii and determine the normalizations  
\be
\mu_{0,m}&=&\frac{A}{[n_a+(2\alpha/\beta) n_m]^\alpha},\hspace{1.5cm}
A=\frac{I_m(N_f,T) C^2 \Gamma^2(\alpha)}{(4\kappa\beta)^\alpha},
\\
\mu_{0,a}&=&\frac{B n_a^{N_f/2}}{[n_a+(2\alpha/\beta) n_m]^{\beta/2+N_f/8}}
\hspace{0.5cm}  
B=\frac{ C\Gamma(\beta)}{(2\kappa)^\beta}\left(\frac{I(T)}{2}\right)^{N_f/2}
\left(\frac{\beta}{\kappa}\right)^{N_f/8-\beta/2} .
\ee
Finally, the free energy is given by
\be
F &=&-\frac{1}{V_4}\log Z = \frac{N_a}{V_4} \log\left(
 \frac{e \mu_{0,a} V_4}{N_a}\right) 
+ \frac{N_m}{V_4} \log\left(\frac{e \mu_{0,m} V_4}{N_m}\right),
\ee
and the instanton density is determined by minimizing $F$ with respect 
to the densities $n_{a,m}$ of random and correlated instantons. The 
resulting free energy determines the instanton contribution to the 
pressure $p=-F$ and the energy density $\epsilon=-p+T\frac{\partial 
p}{\partial T}$. In order to provide a more realistic description
of the thermodynamics of the chiral phase transition, Ilgenfritz 
and Shuryak added the free energy of a noninteracting pion gas in 
the broken phase and a quark gluon plasma in the symmetric phase.
 
   Minimizing the free energy gives a set of rather cumbersome 
equations. It is clear that in general there will be two phases, a low 
temperature phase containing a mixture of molecules and random instantons, 
and a chirally restored high temperature phase consisting of molecules only.
The density of random instantons is suppressed as the temperature increases
because the average overlap matrix element decreases. Molecules, on the 
other hand are favored at high $T$, because the overlap for the most
attractive orientation increases. Both of these results are simple
consequences of the anisotropy of the fermion wave functions.

   Numerical results for $N_f=2$ are shown in Fig. \ref{fig_cocktail}. In
practice, the average molecular determinant $I_m(N_f,T)$ was calculated
by introducing a core into the $IA$ interaction. In order to assess the 
uncertainty involved, we show the results for two different cores 
$R_c=\rho$ and $2\rho$. The overall normalization was fixed such
that $N/V=1.4 fm^{-4}$ at $T=0$. Figure \ref{fig_cocktail} shows
that the random component dominates the broken phase and that
the density of instantons is only weakly dependent on $T$ below
the phase transition. The number of molecules is small for $T<T_c$
but jumps up at the transition. The total instanton density above 
$T_c$, $(N/V)=2n_m$, turns out to be comparable to that at $T=0$. 

   The importance of the molecular component above $T_c$ can be seen
from the temperature dependence of the pressure. For $T=(1-2)T_c$, the 
contribution of molecules (dash-dotted line) is crucial in order to 
keep the pressure positive. This is the same phenomenon we already 
mentioned in our discussion of the lattice data. If the transition
temperature is as low as $T_c=150$ MeV, then the contribution of 
quarks and gluons is not sufficient to match the $T=0$ bag pressure.
The lower panel in Fig. \ref{fig_cocktail} shows the behavior of the
energy density. The jump in the energy density at $T_c$ is $\Delta
\epsilon\simeq (0.5-1.0){\rm GeV}/{\rm fm}^3$, depending on the size
of the core. Although most of the latent heat is due to the liberation 
of quarks and gluons, a significant part is generated by molecules. 

\subsubsection{Phase transitions in the interacting instanton model}
\label{sec_liquid_T}

\begin{figure}[t]
\begin{center}
\leavevmode
\epsfxsize=8cm
\epsffile{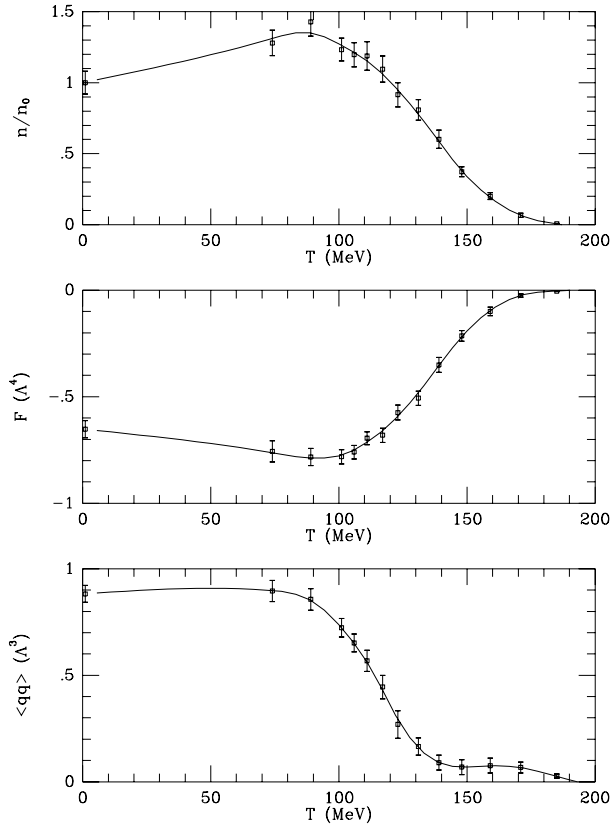}
\end{center}
\caption{\label{fig_eos_liq}
Instanton density, free energy and quark condensate as a function of 
the temperature in the instanton liquid with two light and one intermediate 
mass flavor, from \protect\cite{SS_96}. The instanton density is given in
units of the zero temperature value $n_0=1\,{\rm fm}^{-4}$, while the 
free energy and the quark condensate are given in units of the Pauli-Vilars
scale parameter, $\Lambda=222$ MeV.}
\end{figure}

   In this section, we will go beyond this schematic model and 
study the phase transition using numerical simulations 
of the interacting instanton liquid \cite{SS_96}. This means that 
we do not have to make any assumptions about the nature of the
important configurations (molecules, larger clusters, $\ldots$), 
nor do we have to rely on variational estimates to determine their 
relative importance. Also, we are not limited to a simple two phase 
picture with a first order transition. 

   In Fig. \ref{fig_eos_liq} we show the instanton density, free energy 
and quark condensate for the physically relevant case of two light and 
one intermediate mass flavor. In the ratio ansatz the instanton 
density at zero temperature is given by $N/V=0.69\Lambda^4$. Taking the 
density to be $1\,{\rm fm}^{-4}$ at $T=0$ fixes the scale parameter 
$\Lambda=222$ MeV and determines the absolute units. The temperature 
dependence of the instanton density\footnote{The instanton density
is of course sensitive to our assumptions concerning the role of 
the finite temperature suppression factor. In practice, we have 
chosen a functional form that interpolates between no suppression
below $T_c$ and the full suppression facto above $T_c$ with a 
width $\Delta T=0.3 T_c$.} is shown in Fig. \ref{fig_eos_liq}a. 
It shows a slight increase at small temperatures, starts to drop around 
115 MeV and becomes very small for $T>175$ MeV. The free energy closely 
follows the behavior of the instanton density. This means that the 
instanton-induced pressure first increases slightly, but then drops 
and eventually vanishes at high temperature. This behavior is expected 
for a system of instantons, but if all fluctuations are included, the 
pressure should always increase as a function of the temperature. 

\begin{figure}[t]
\begin{center}
\leavevmode
\epsfxsize=8cm
\epsffile{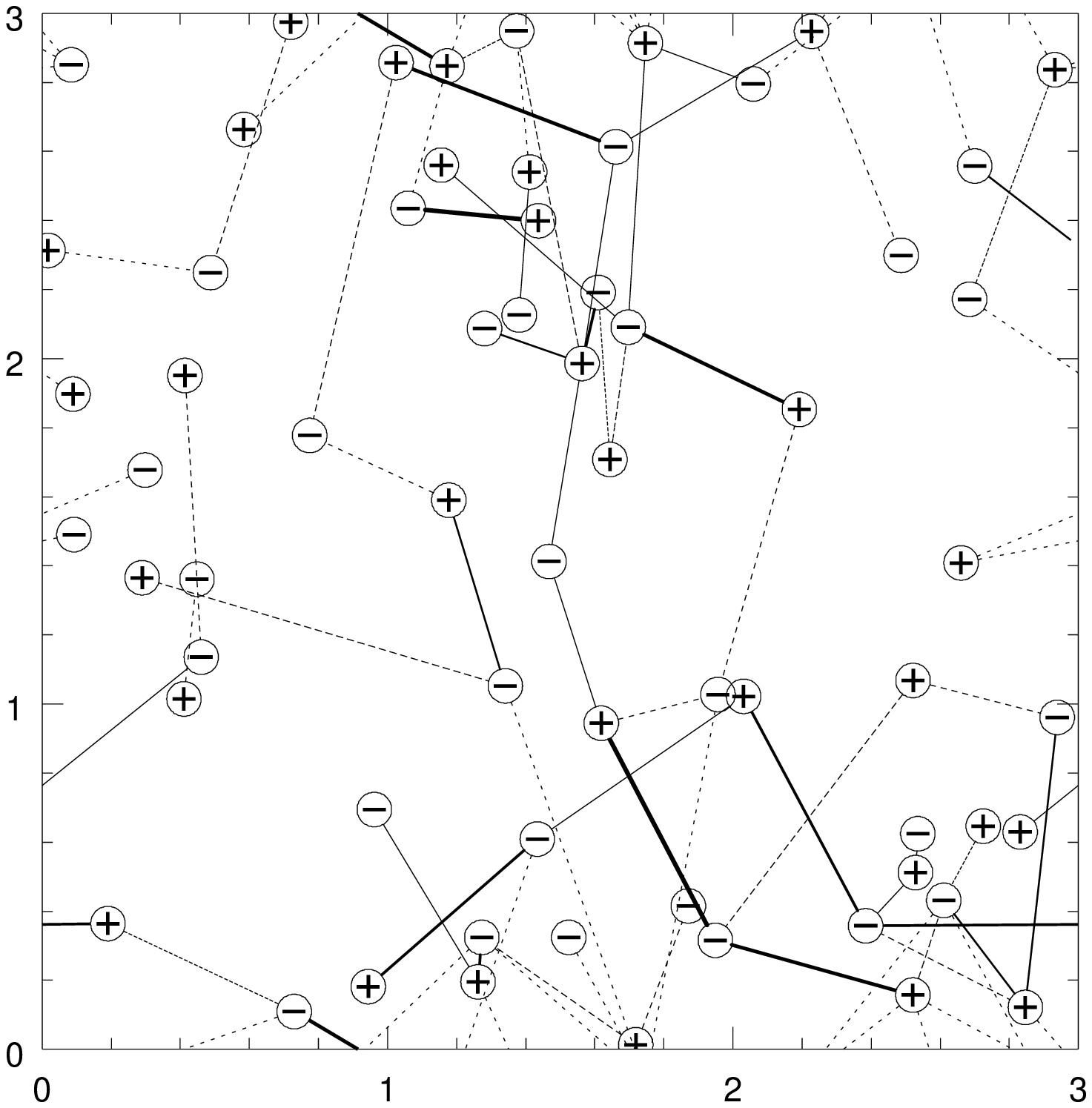}
\end{center}
\vspace*{-5cm}
\begin{center}
\leavevmode
\epsfxsize=8cm
\epsffile{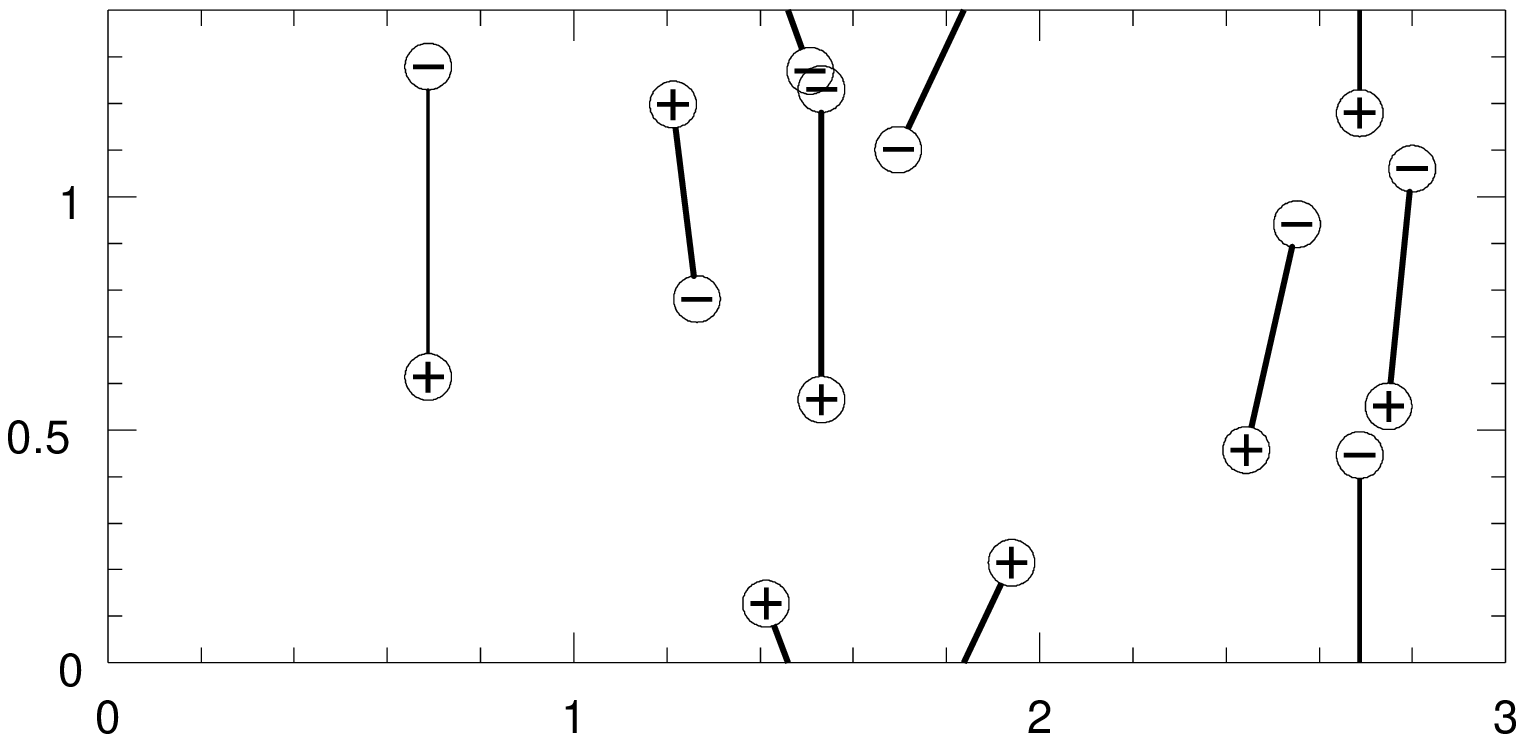}
\end{center}
\caption{\label{fig_hop}
Typical instanton ensembles for $T=75$ and 158 MeV, from
\protect\cite{SS_96}. The plots show projections of a four dimensional 
$(3.0\Lambda^{-1})^3\times T^{-1}$ box into the 3-4 plane. The positions
of instantons and anti-instanton are indicated by $+$ and $-$ symbols. 
Dashed, solid and thick solid lines correspond to fermionic overlap 
matrix elements $T_{IA}\Lambda>0.40,\,0.56,\,0.64$, respectively.}
\end{figure}

   The temperature dependence of the quark condensate is shown in Fig. 
\ref{fig_eos_liq}c. At temperatures below $T=100$ MeV it is practically 
temperature independent. Above that, $\langle\bar qq\rangle$ starts to drop 
and becomes very small above the critical temperature $T\simeq 140$ MeV. 
Note that at this point the instanton density is $N/V=0.6\,{\rm fm}^{-4}$, 
slightly more than half the zero temperature value. This means that the 
phase transition is indeed caused by a transition within the instanton 
liquid, not by the disappearance of instantons. This point is illustrated 
in Fig. \ref{fig_hop}, which shows projections of the instanton liquid at 
$T=74$ MeV and $T=158$ MeV, below and above the phase transition. The 
figures are projections of a cube $V=(3.0\Lambda^{-1})^3\times T^{-1}$ 
into the 3-4 plane. The positions of instantons and anti-instantons are 
denoted by $+/-$ symbols. The lines connecting them indicate the strength 
of the fermionic overlap (``hopping") matrix elements. Below the phase 
transition, there is no clear pattern. Instantons are unpaired, part of 
molecules or larger clusters. Above the phase transition, the ensemble 
is dominated by polarized instanton-anti-instanton molecules.

\begin{figure}[t]
\begin{center}
\leavevmode
\epsfxsize=8cm
\epsffile{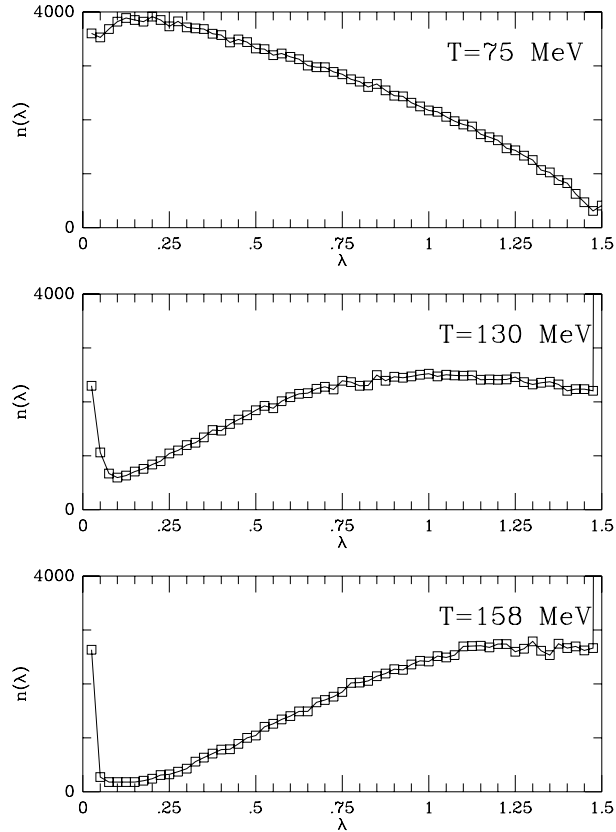}
\end{center}
\caption{\label{fig_dirac_T}
Spectrum of the Dirac operator for different temperatures $T=75,\,130,
\,158$ MeV, from \protect\cite{SS_96}. Eigenvalues are given in units
of the scale parameter. The normalization of the spectra is arbitrary 
(but identical).}
\end{figure}

  Fig. \ref{fig_dirac_T} shows the spectrum of the Dirac operator. Below 
the phase transition it has the familiar flat shape near the origin and 
extrapolates to a non-zero density of eigenvalues at $\lambda=0$. Near the 
phase transition the eigenvalue density appears to extrapolate to 0 as 
$\lambda\to 0$, but there is a spike in the eigenvalue density at $\lambda=0$.
This spike contains the contribution from unpaired instantons. In the
high temperature phase, we expect the density of random instantons to
be $O(m^{N_f})$, giving a contribution of the form $\rho(\lambda)\sim 
m^{N_f}\delta(\lambda)$ to the Dirac spectrum. These eigenvalues
do not contribute to the quark condensate in the chiral limit, but 
they are important for $U(1)_A$ violating observables.

   The nature of the phase transition for different numbers of 
flavors and values of the quark masses was studied in \cite{SS_96}. 
The order of the phase transition is most easily determined by
studying order parameter fluctuations near $T_c$. For a first 
order transition one expects two (meta) stable phases. The
system tunnels back and forth between the two phases, with 
tunneling events signaled by sudden jumps in the order parameter.
Near a second order phase transition, on the other hand, the  
order parameter shows large fluctuations. These fluctuations 
can be studied in more detail by measuring the scaling behavior
of the mean square fluctuations (the scalar susceptibility) with 
the current mass and temperature. Universality makes definite
predictions for the corresponding critical exponents. 
The main conclusion in \cite{SS_96} was that the transition
in QCD is consistent with a nearby ($N_f=2$) second order phase
transition with $O(4)$ critical indices. For three flavors with
physical masses, the transition is either very weakly first 
order or just a rapid crossover. As the number of flavors is
increased, the transition becomes more strongly first order.

\begin{figure}[t]
\begin{center}
\leavevmode
\epsfxsize=12cm
\epsffile{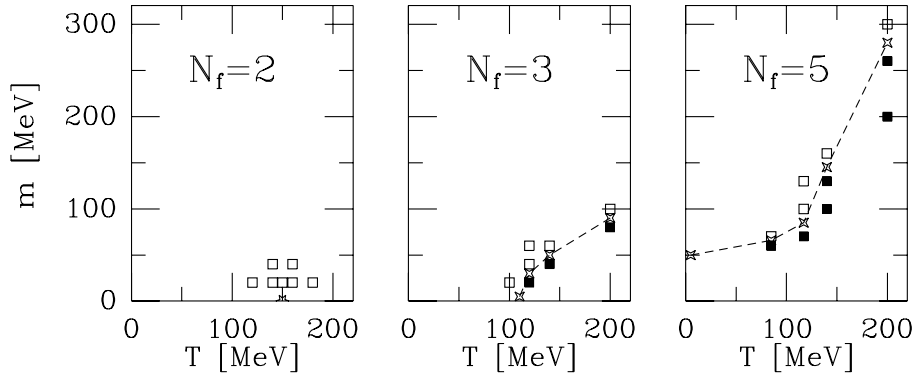}
\end{center}
\caption{\label{fig_phase_diag}
Phase diagram of the instanton liquid in the $T-m$ plane for 
different numbers of flavors. The open and closed squares show
points on the phase diagram where we have performed simulations.
For $N_f=2$, the open squares mark points where we found large 
fluctuations of the order parameter, indicative of a nearby second 
order phase transition (marked by a star). In the cases $N_f=3$
and 5, open and closed squares mark points in the chirally 
broken and restored phases, respectively. The (approximate)
location of the discontinuity is shown by the dashed line.}
\end{figure}

  The results of simulations with\footnote{The case $N_f=4$ is omitted 
because the contribution of small-size (semi-classical) instantons to
the quark condensate is very small and the precise location of the 
phase boundary hard to determine.} $N_f=2,3,5$ flavors with equal 
masses are summarized in the phase diagram \ref{fig_phase_diag} 
\cite{SS_96}. For $N_f=2$ there is second order phase transition 
which turns into a line of first order transitions in the $m-T$ plane
for $N_f>2$. If the system is in the chirally restored phase ($T>T_c$) 
at $m=0$, we find a discontinuity in the chiral order parameter if 
the mass is increased beyond some critical value. Qualitatively, the 
reason for this behavior is clear. While increasing the temperature 
increases the role of correlations caused by fermion determinant, 
increasing the quark mass has the opposite effect. We also observe 
that increasing the number of flavors lowers the transition temperature. 
Again, increasing the number of flavors means that the determinant
is raised to a higher power, so fermion induced correlations become
stronger. For $N_f=5$ we find that the transition temperature drops
to zero and the instanton liquid has a chirally symmetric ground state, 
provided the dynamical quark mass is less than some critical value. 

  Studying the instanton ensemble in more detail shows that in this 
case, all instantons are bound into molecules. The molecular vacuum 
at $T=0$ and large $N_f$ has a somewhat different theoretical status 
as compared to the molecular vacuum for small $N_f$ and large $T$. 
In the high temperature phase, large instantons are suppressed
and long range interactions are screened. This is not the case 
at $T=0$ and $N_f$ large, where these effects may contribute to 
chiral symmetry breaking, see the discussion in Sec. \ref{sec_big_pic}. 

   Unfortunately, little is known about QCD with different numbers 
of flavors from lattice simulations. There are some results on the 
phase structure of large $N_f$ QCD that we will discuss in Sec.
\ref{sec_big_pic}. In addition to that, there are some recent data 
by the Columbia group \cite{CM_97} on the hadron spectrum 
for $N_f=4$. The most important result is that chiral symmetry
breaking effects were found to be drastically smaller as compared 
to $N_f=0,2$. In particular, the mass splittings between chiral
partners such as $\pi-\sigma,\, \rho-a_1,\, N(\frac{1}{2}^+)-N
(\frac{1}{2}^-)$, were found to be 4-5 times smaller. This agrees 
well with what was found in the interacting instanton model, but 
more work in this direction is certainly needed.

\subsection{Hadronic correlation functions at finite temperature}
\label{sec_cor_T}

\subsubsection{Introduction}
\label{sec_cor_T_intro}

  Studying the behavior of hadronic correlation functions at 
finite temperature is of great interest in connection with possible
modifications of hadronic properties in hot hadronic matter. 
In addition to that, the structure of correlation functions
at intermediate distances directly reflects on changes in 
the interaction between quarks and gluons. There is very 
little phenomenological information on this subject, but the 
problem has been studied extensively in the context of QCD 
sum rules \cite{BS_86,EI_93,HKL_93}. At finite temperature,
however, the predictive power of QCD sum rules is very limited,
because additional assumptions about the temperature dependence 
of the condensates and the shape of the spectrum are needed. 
There is an extensive literature on spacelike screening 
masses on the lattice, but only very limited information on 
temporal point-to-point correlation functions \cite{BGK_94}.

  At finite temperature, the heat bath breaks Lorenz invariance, 
and correlation functions in the spatial and temporal direction
are independent. In addition to that, mesonic and baryonic 
correlation functions have to obey periodic or anti-periodic 
boundary conditions, respectively, in the temporal direction.
In the case of space-like correlators one can still go to large 
$x$ and filter out the lowest exponents known as screening masses. 
While these states are of theoretical interest and have been studied
in a number of lattice calculations, they do not correspond to poles
of the spectral function in energy. In order to look for real bound 
states, one has to study temporal correlation functions. However, at 
finite temperature the periodic boundary conditions restrict the useful
range of temporal correlators to the interval $\tau<1/(2T)$ (about 0.6
fm at $T=T_c$). This means that there is no direct procedure to extract 
information about the ground state. The underlying physical reason is
clear: at finite temperature excitations are always present. In the 
following we will study how much can be learned from temporal correlation
functions in the interacting instanton liquid. In the next section 
we will also present the corresponding screening masses.

\subsubsection{Temporal correlation functions}
\label{sec_cor_temp}

\begin{figure}[t]
\begin{center}
\leavevmode
\epsfxsize=12cm
\epsffile{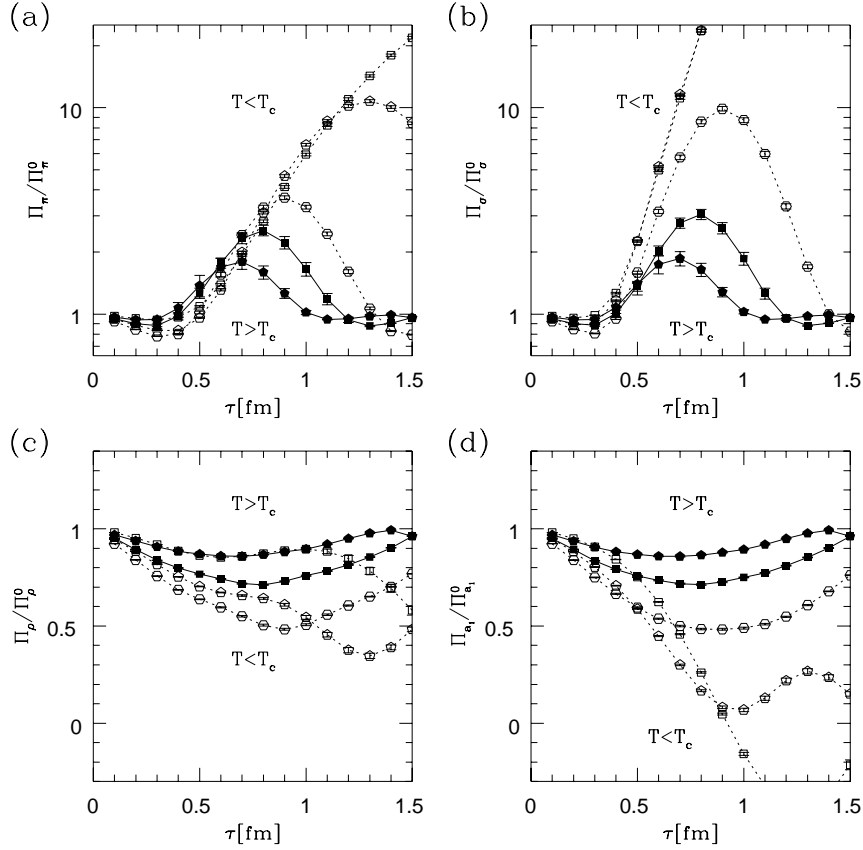}
\end{center}
\caption{\label{fig_tempcor}
Temporal correlation functions at $T\neq 0$, normalized to free
thermal correlators. Figure (a) shows the pseudo-scalar (pion) correlator, 
(b) the isoscalar scalar $\sigma$, (c) the isovector vector ($\rho$) 
and (d) the axial vector ($a_1$).  Correlators in the chirally symmetric  
phase ($T \geq T_c$) are shown as solid points, below $T_c$ as open points. 
The open triangles, squares and hexagons correspond to $T=0.43$, 0.60 and 
0.86 $T_c$, while the closed triangles and squares show the data at $T=1.00 
\,T_c$ and 1.13 $T_c$.}
\end{figure}

   Finite temperature correlation functions in the temporal direction 
are shown in Fig. \ref{fig_tempcor}. These correlators were obtained 
from simulations of the interacting instanton liquid \cite{SS_95b}. 
Correlators in the random phase approximation were studied in \cite{SV_96}. 
The results shown in Fig. \ref{fig_tempcor} are normalized to the 
corresponding non-interacting correlators, calculated from the free 
$T\neq 0$ propagator (\ref{S_free_T}). Figures \ref{fig_tempcor}a,b 
show the pion and sigma correlators for different temperatures below 
(open symbols) and above (closed symbols) the chiral phase transition. 
The normalized $\pi$ and $\sigma$ correlators are larger than one
at all temperatures, implying that there is an attractive interaction
even above $T_c$. In particular, the value of the pion correlator at 
$\tau\simeq 0.6$ fm, which is not directly affected by the periodic
boundary conditions, is essentially temperature independent. This
suggests that there is a strong $(\pi,\sigma)$-like mode present even 
above $T_c$. In \cite{SS_96b} we tried to determine the properties 
of this mode from a simple fit to the measured correlation function,
similar to the $T=0$ parametrization (\ref{pi_rep}). Above $T_c$, the 
mass of the $\pi$-like mode is expected to grow, but the precise 
value is hard to determine. The coupling constant is $\lambda_\pi
\simeq 1\,{\rm fm}^{-2}$ at $T=170$ MeV, as compared to $\lambda_\pi
\simeq 3\,{\rm fm}^{-2}$ at $T=0$.

  The $T$-dependence of the $\sigma$ correlator is more pronounced
because there is a disconnected contribution which tends to the square 
of the quark condensate at large distance. Above $T_c$, chiral symmetry 
is restored and the $\sigma$ and $\pi$ correlation functions are 
equal up to small corrections due to the current quark masses. 

   Vector and axial-vector correlation functions are shown in 
Fig. \ref{fig_tempcor}(c,d). At low $T$ the two are very different 
while above $T_c$ they become indistinguishable, again in accordance 
with chiral symmetry restoration. In the vector channel, the changes
in the correlation function indicate the ``melting" of the resonance 
contribution. At the lowest temperature, there is a small enhancement 
in the correlation function at $\tau\simeq 1$ fm, indicating the presence 
of a bound state separated from the two-quark (or two-pion) continuum. 
However, this signal disappears at $T\simeq 100 MeV$, implying that the 
$\rho$ meson coupling to the local current becomes small\footnote{Note 
however that in the instanton model, there is no confinement and the 
amount of binding in the $\rho$ meson channel is presumably small. In 
full QCD, the $\rho$ resonance might therefore be more stable as the 
temperature increases.}. This is consistent with the idea that hadrons 
``swell'' in hot and dense matter. At small temperature, the dominant
effect is mixing between the vector and axial-vector channels \cite{DEI_90}. 
In particular, there is a pion contribution to the vector correlator at 
finite $T$, which is most easily observed in the longitudinal vector 
channel $\Pi^V_{44}(\tau)$ (in Fig. \ref{fig_tempcor} we show the trace 
$\Pi^V_{\mu\mu}(\tau)$ of the vector correlator).  

    In \cite{SS_95b} we also studied baryon correlation functions
at finite temperature. Because the different nucleon and Delta 
correlation functions have different transformation properties
under the chiral $SU(2)\times SU(2)$ and $U(1)_A$ symmetries,
one can distinguish different modes of symmetry restoration.
The main possibilities are (i) that all resonances simply disappear, 
(ii) that all states become massless as $T\to T_c$, or (iii) that 
above $T_c$ all states occur in parity doublets. In the nucleon channel, 
we find clear evidence for the survival of a massive nucleon mode. There 
is also support for the presence of a degenerate parity partner above 
$T_c$, so the data seem to favor the third possibility. 

    In summary, we find that correlation functions in strongly 
attractive, chiral even channels are remarkably stable as a function 
of temperature, despite the fact that the quark condensate disappears. 
There is evidence for the survival of $(\pi,\sigma)$-like modes
even above $T_c$. These modes are bound by the effective quark
interaction generated by polarized instanton molecules, see Equ. 
(\ref{lmol}). In channels that do not receive large non-perturbative
contributions at $T=0$ (like the $\rho,a_1$ and $\Delta$), the 
resonances disappear and the correlators can be described in terms 
of the free quark continuum (possibly with a residual ``chiral" mass). 

\subsubsection{$U(1)_A$ breaking}
\label{sec_cor_ua1}

\begin{figure}[t]
\begin{center}
\leavevmode
\epsfxsize=12cm
\epsffile{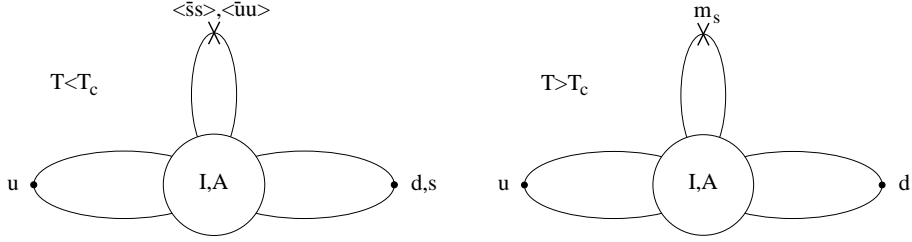}
\end{center}
\caption{\label{fig_anomaly}
Leading contributions to flavor mixing in the $\eta-\eta'$ system
below and above the chiral phase transition.}
\end{figure}

  So far, we have not discussed correlation functions related to
the $U(1)_A$ anomaly, in particular the $\eta$ and $\eta'$ channels.
A number of authors have considered the possibility that the $U(1)_A$ 
symmetry is at least partially restored near the chiral phase 
transition \cite{Shu_94,KKL_96,HW_96}. Given the fact that the
$\eta'-\pi$ mass difference is larger than all other meson mass 
splittings, any tendency towards $U(1)_A$ restoration is expected
to lead to rather dramatic observable effects. Possible signatures
of a hot hadronic phase with partially restored $U(1)_A$ symmetry
that might be produced in heavy ion collisions are anomalous
values of the $\eta/\pi$ and $\eta'/\pi$ ratios, as well as 
an enhancement in the number of low-mass dilepton pairs from 
the Dalitz decay of the $\eta'$. 

   In general, we know that isolated instantons disappear above
the chiral phase transition. In the presence of external sources,
however, isolated tunneling events can still occur, see Sec. 
\ref{sec_tun_qu}. The density of random instantons above $T_c$ 
is expected to be $O(m^{N_f})$, but the contribution of isolated 
instantons to the expectation value of the 't Hooft operator $\det_f
(\bar\psi_l\psi_R)$ (and other $U(1)_A$ violating operators) is 
of order $O(1)$ \cite{LH_96,EHS_96}. The presence of isolated 
zero modes in the spectrum of the Dirac operator above $T_c$ can 
be seen explicitly in Fig. \ref{fig_dirac_T}. The problem 
was studied in more detail in \cite{Sch_96}, where it was 
concluded that the number of (almost) zero modes above $T_c$ 
scales correctly with the dynamical quark mass and the volume. 

  A number of groups have measured $U(1)_A$ violating observables
at finite temperature on the lattice. Most works focus on the
the susceptibility $\chi_\pi-\chi_\delta$ \cite{Cha_95,Boy_96,Ber_97}. 
Above $T_c$, when chiral symmetry is restored, this quantity is
a measure of $U(1)_A$ violation. Most of the published results 
conclude that $U(1)_A$ remains broken, although recent results by
the Columbia group have questioned that conclusion\footnote{In
addition to that, it is not clear whether lattice simulations 
observe the peak in the spectrum at $\lambda=0$ which is due to
isolated instantons. The Columbia group has measured the valence 
mass dependence of the quark condensate, which is a folded version 
of the Dirac spectrum \cite{Cha_95}. The result looks very smooth,
with no hint of an enhancement at small virtuality.} \cite{Chr_96}.
In any case, one should keep in mind that all results involve an 
extrapolations to $m=0$, and that both instanton and lattice 
simulations suffer from certain artefacts in this limit.

  Phenomenological aspects of the $U(1)_A$ anomaly at finite 
temperature are usually discussed in terms of the effective 
lagrangian \cite{PW_84}
\be
\label{l_eff}
 {\cal L} &=& \frac{1}{2}{\rm Tr}\left((\partial_\mu\Phi)
  (\partial_\mu\Phi^\dagger)\right) - {\rm Tr}\left({\cal M}
  (\Phi+\Phi^\dagger)\right) + V(\Phi\Phi^\dagger)
  + c \left(\det\Phi+\det\Phi^\dagger\right),
\ee
where $\Phi$ is a meson field in the $(3,3)$ representation of 
$U(3)\times U(3)$, $V(\Phi\Phi^\dagger)$ is a $U(3)\times U(3)$
symmetric potential (usually taken be quartic), ${\cal M}$ is a 
mass matrix and $c$ controls the strength of the $U(1)_A$ breaking
interaction. If the coupling is chosen as $c=\chi_{top}/(12f_\pi^3)$,
the effective lagrangian reproduces the Witten-Veneziano relation 
$f_\pi^2m_{\eta'}^2=\chi_{top}$. In a quenched ensemble, we can 
further identify $\chi_{top}\simeq (N/V)$. The temperature dependence
of $c$ is usually estimated from the semi-classical tunneling
amplitude $n(\rho)\sim \exp(-(8/3)(\pi\rho T)^2)$. As a result, 
the strength of the anomaly is reduced by a factor $\sim 5$ 
near $T_c$. If the anomaly becomes weaker, the eigenstates are 
determined by the mass matrix. In that case, the mixing angle is 
close to the ideal value $\theta=-54.7^\circ$ and the non-strange 
$\eta_{NS}$ is almost degenerate with the pion. 

\begin{figure}[t]
\begin{center}
\leavevmode
\epsfxsize=11cm
\epsffile{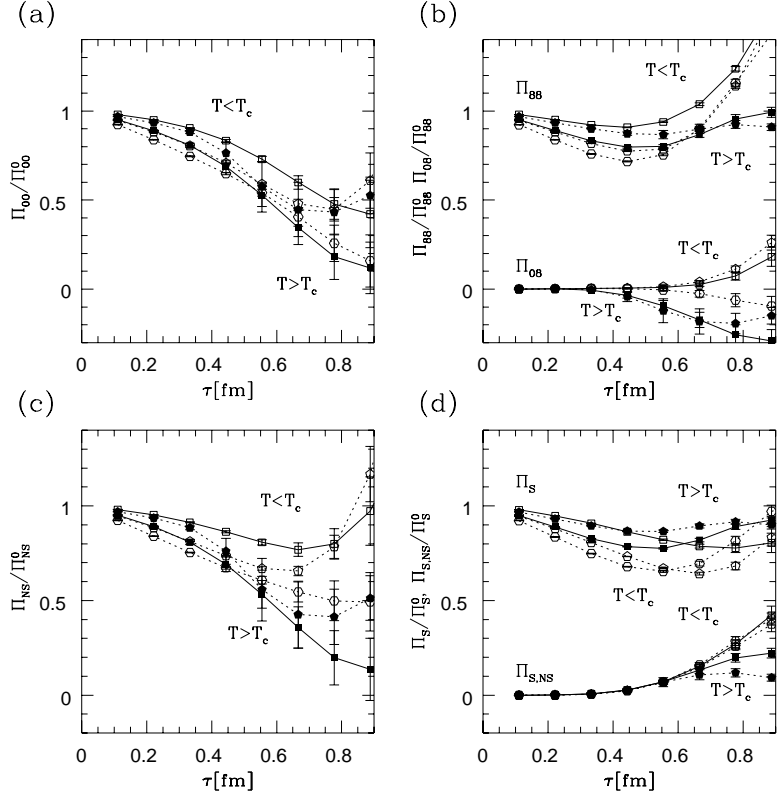}
\end{center}
\caption{\label{fig_teta}
Eta meson correlation functions at $T\neq 0$, normalized to free
thermal correlators. Figure (a) shows the flavor singlet, (b) the
octet and off-diagonal singlet-octet, (c) the non-strange and 
(d) the strange and off-diagonal strange-non-strange eta correlation
functions. The correlators are labeled as in figure
\protect\ref{fig_tempcor}.}
\end{figure}

   There are several points in this line of argument that are not 
entirely correct. The strength of the 't Hooft term is not 
controlled by the topological susceptibility ($\chi_{top}=0$
in massless QCD!), $\chi_{top}$ is not proportional to the instanton
density (for the same reason), and, at least below $T_c$, the
semi-classical estimate for the instanton density is not 
applicable. Only above $T_c$ do we expect instantons to be 
suppressed. However, chiral symmetry restoration affects the 
structure of flavor mixing in the $\eta-\eta'$ system (see 
Fig. \ref{fig_anomaly}). The mixing between the strange and 
non-strange eta is controlled by the light quark condensate, 
so $\eta_{NS}$ and $\eta_S$ do not mix above $T_c$. As a result, 
the mixing angle is not close to zero, as it is at $T=0$, but 
close to the ideal value. Furthermore, the anomaly can only 
affect the non-strange $\eta$, not the strange one. Therefore, 
if the anomaly is sufficiently strong, the $\eta_{NS}$ will be 
{\em heavier} than the $\eta_S$.

  This phenomenon can also be understood from the effective 
lagrangian (\ref{l_eff}). The determinant is third order in
the fields, so it only contributes to mass terms if some of
the scalar fields have a vacuum expectation value. Above $T_c$,
only the strange scalar has a VEV, so only light flavors mix
and only the $\eta_{NS}$ receives a contribution to its mass
from the anomaly. This effect was studied more quantitatively in 
\cite{Sch_96}. Singlet and octet, as well as strange and non-strange
eta correlation functions in the instanton liquid are shown in 
Fig. \ref{fig_teta}. Below $T_c$ the singlet correlation function 
is strongly repulsive, while the octet correlator shows some attraction 
at larger distance. The off-diagonal correlator is small and positive, 
corresponding to a negative mixing angle. The strange and non-strange 
eta correlation functions are very similar, which is a sign for
strong flavor mixing. This is also seen directly from the off-diagonal 
correlator between $\eta_S$ and $\eta_{NS}$.

   Above $T_c$, the picture changes. The off-diagonal singlet-octet 
correlator changes sign and its value at intermediate distances $\tau
\simeq 0.5$ fm is significantly larger. The strange and non-strange
eta correlators are very different from each other. The non-strange
correlation function is very repulsive, but no repulsion is seen
in the strange channel. This clearly supports the scenario
presented above. Near $T_c$ the eigenstates are essentially
the strange and non-strange components of the $\eta$, with the
$\eta_S$ being the lighter of the two states. This picture 
is not realized completely, $\Pi_{S,NS}$ does not vanish and
the singlet eta is still somewhat more repulsive than the 
octet eta correlation function. This is due to the fact that
the light quark mass does not vanish. In particular, in this 
simulation the ratio $(m_u+m_d)/(2m_s)=1/7$, which is about 
three times larger than the physical mass ratio.

   It is difficult to provide a quantitative analysis of temporal   
correlation functions in the vicinity of the phase transition. At 
high temperature the temporal direction in a euclidean box becomes 
short and there is no unique way to separate out the contribution
from excited states. Nevertheless, under some simplifying assumptions
one can try to translate the correlation functions shown in Fig.
\ref{fig_teta} into definite predictions concerning the masses 
of the $\eta$ and $\eta'$. For definiteness, we will use ideal
mixing above $T_c$ and fix the threshold for the perturbative 
continuum at 1 GeV. In this case, the masses of the strange
and non-strange components of the $\eta$ at $T=126$ MeV are 
given by $m_{\eta_S}=(0.420\pm 0.120)$ GeV and $m_{\eta_S}=
1.250\pm 0.400)$ GeV.

\subsubsection{Screening masses}
\label{sec_cor_scr}

\begin{figure}[t]
\begin{center}
\leavevmode
\epsfxsize=6cm
\epsffile{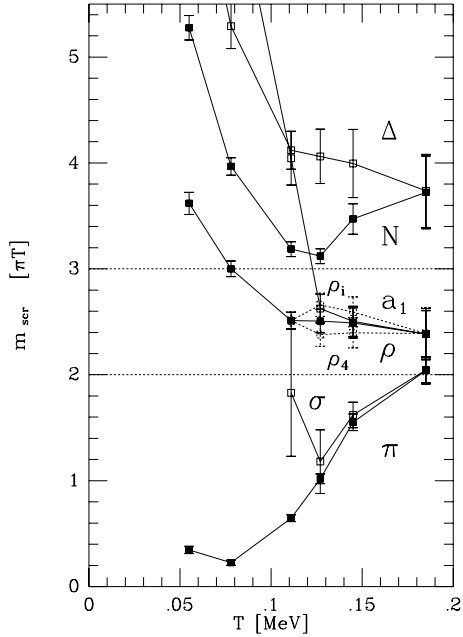}
\end{center}
\caption{\label{fig_scrm}
Spectrum of space-like screening masses in the instanton liquid as 
a function of the temperature, from \protect\cite{SS_96b}. The masses 
are given in units of the lowest fermionic Matsubara frequency $\pi T$.
The dotted lines correspond to the screening masses $m=2\pi T$ and
$3\pi T$ for mesons and baryons in the limit when quarks are 
non-interacting.}
\end{figure}

   Correlation functions in the spatial direction can be studied
at arbitrarily large distance, even at finite temperature. This 
means that (contrary to the temporal correlators) the spectrum of 
the lowest states can be determined with good accuracy. Although it
is not directly related to the spectrum of physical excitations, the 
structure of space-like screening masses is of theoretical interest 
and has been investigated in a number of lattice \cite{TK_87,Goc_91} 
and theoretical \cite{EI_88,HZ_92,KSB_92,HSZ_94} works.

   At finite temperature, the anti-periodic boundary conditions in the  
temporal direction require the lowest Matsubara frequency for fermions
to be $\pi T$. This energy acts like a mass term for propagation in the
spatial direction, so quarks effectively become massive. At asymptotically
large temperatures, quarks only propagate in the lowest Matsubara mode,
and the theory undergoes dimensional reduction \cite{AP_81}. The spectrum 
of space-like screening states is then determined by a 3-dimensional 
theory of quarks with chiral mass $\pi T$, interacting via the 
3-dimensional Coulomb law and the non-vanishing space-like string 
tension \cite{Bor_85,MP_87}. 

   Dimensional reduction at large $T$ predicts almost degenerate 
multiplets of mesons and baryons with screening masses close to 
$2\pi T$ and $3\pi T$. The splittings of mesons and baryons with 
different spins can be understood in terms of the non-relativistic
spin-spin interaction. The resulting pattern of screening 
states is in qualitative agreement with lattice results even at 
moderate temperatures $T\simeq 1.5 T_c$. The most notable exception
is the pion, whose screening mass is significantly below $2\pi T$.

   Screening masses in the instanton liquid are summarized in 
Fig. \ref{fig_scrm}. First of all, the screening masses clearly 
show the restoration of chiral symmetry as $T\to T_c$: chiral 
partners like the $\pi$ and $\sigma$ or the $\rho$ and $a_1$ 
become degenerate. Furthermore, the mesonic screening masses 
are close to $2\pi T$ above $T_c$, while the baryonic ones are 
close to $3\pi T$, as expected. Most of the screening masses are 
slightly higher than $n\pi T$, consistent with a residual chiral 
quark mass on the order of 120-140 MeV. The most striking observation 
is the strong deviation from this pattern seen in the scalar channels 
$\pi$ and $\sigma$, with screening masses significantly below $2\pi T$ 
near the chiral phase transition. The effect disappears around  
$T\simeq 1.5 T_c$. We also find that the nucleon-delta splitting 
does not vanish at the phase transition, but decreases smoothly. 

   In summary, the pattern of screening masses seen in the 
instanton liquid is in agreement with the results of lattice 
calculations\footnote{There is one exception which concerns
the screening masses in the longitudinal $\Pi_{44}$ and
transverse $\Pi_{ii}$ vector channels. In agreement with
dimensional reduction, we find $m_{\rho_i}>m_{\rho_4}$,
while lattice results reported in \protect\cite{TK_87}
find the opposite pattern.}. In particular, the attractive 
interaction provided by instanton molecules accounts for 
the fact that the $\pi,\sigma$ screening masses are much 
smaller than $2\pi T$ near $T_c$.

\subsection{Instantons at finite temperature: lattice studies}
\label{sec_lgt_inst_T}

\begin{figure}[tbp]
\begin{center}
\leavevmode
\epsfxsize=11cm
\epsffile{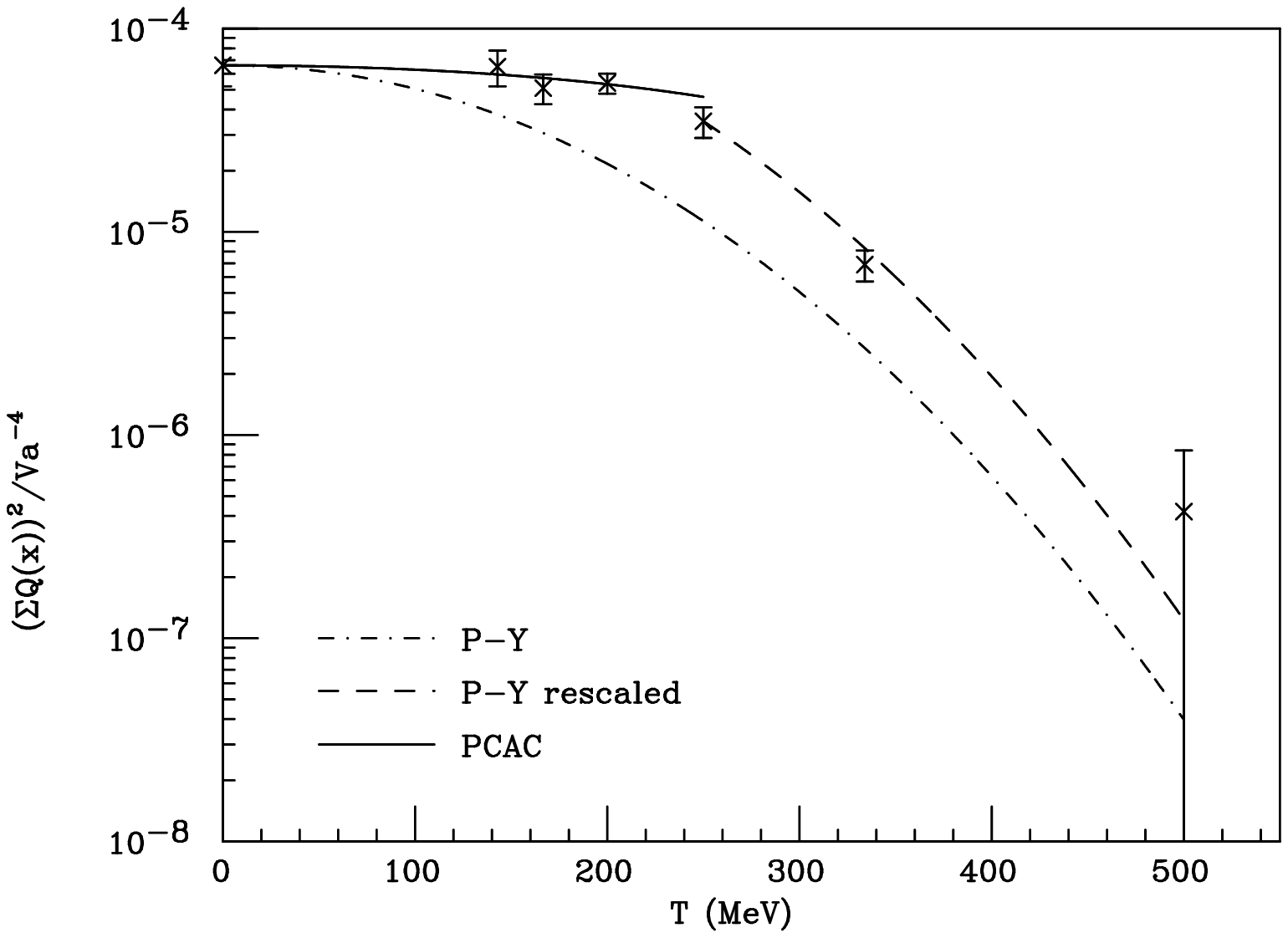}
\end{center}
\begin{center}
\leavevmode
\epsfxsize=11cm
\epsffile{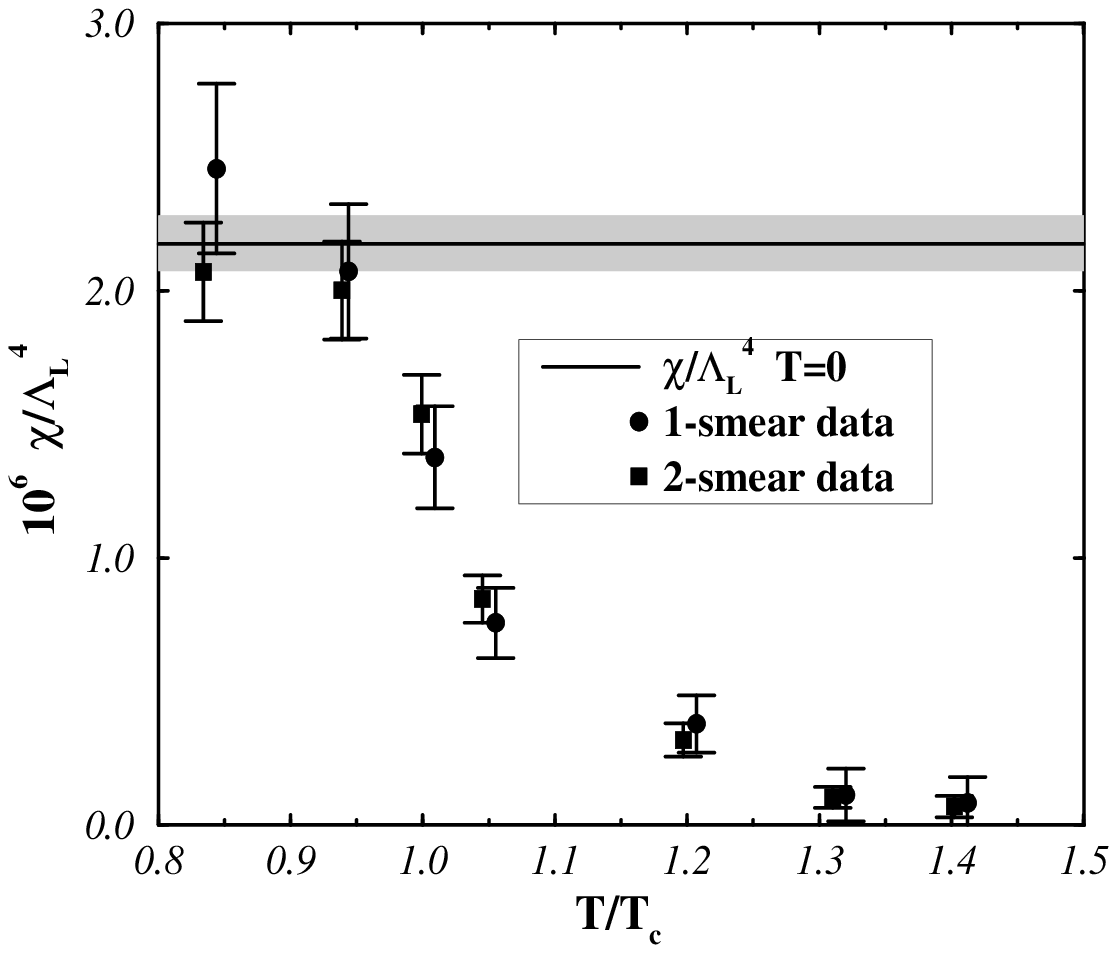}
\end{center}
\caption{\label{fig_lat_chiT}
Lattice measurements of the temperature dependence of the topological 
susceptibility in pure gauge $SU(3)$, from (a)\protect\cite{CS_95} 
and (b) \protect\cite{ADD_96}. The dash-dotted line labeled P-Y in 
Fig. (a) corresponds to the semi-classical (Pisarski-Yaffe) result 
(\protect\ref{pis}), while the dashed curve shows the rescaled formula
discussed in the text. The solid line labeled PCAC is not very relevant
for pure gauge theory. In Fig. (b), the circles and squares correspond
to two different topological charge operators. }
\end{figure}

   Very few lattice simulations have focused on the role of
instantons at finite temperature, and all the results that have
been reported were obtained in the quenched approximation
\cite{Tep_86,HTW_87,CS_95,IMM_95}. In this section we will 
concentrate on results obtained by the cooling method, in 
particular \cite{CS_95}. In this work, the temperature was varied 
by changing the number of time slices $N_\tau=2,4,\ldots,16$, while 
the spatial extent of the lattice and the coupling constant $\beta=6$ 
were kept fixed. 

  The topological susceptibility was calculated from
topological charge fluctuations $\langle Q^2\rangle/V$ 
in the cooled configurations. In the quenched theory, correlations 
between instantons are not very important, and the topological
susceptibility provides a good estimate of the instanton density. 
Figure \ref{fig_lat_chiT}a shows their results as a function of $T$.
At $T=0$, $\chi_{top}\simeq (180\,{\rm MeV})^4$, in agreement 
with the phenomenological value. The topological susceptibility 
is almost temperature independent below the critical temperature 
($T_c\simeq 260$ MeV in quenched QCD), but drops very fast above 
$T_c$. The temperature dependence of $\chi_{top}$ above $T_c$ is
consistent with the Debye-screening suppression factor (\ref{pis}),
but with a shifted temperature, $T^2\to(T^2-T_c^2)$. Clearly,
these results support the arguments presented in Sec. \ref{sec_low_T}. 
The conclusions of Chu and Schramm are consistent with results reported 
in \cite{IMM_95,ADD_96}. We show the  results of Alles et al.~in 
Fig. \ref{fig_lat_chiT}b.

  Chu and Schramm also extracted the average instanton size
from the correlation function of the topological charge density.
Below $T_c$, they find $\rho=0.33$ fm independent of temperature
while at $T=334$ MeV, they get a smaller value $\rho=0.26$ fm. This 
result is in good agreement with the 
Debye-screening  dependence discussed above. 

   Finally, they considered instanton contributions to the 
pressure and space-like hadronic wave functions. They find that 
instantons contribute roughly 15\% of the pressure at $T=334$ MeV
and 5\% at $T=500$ MeV. While space-like hadronic wave functions
are dominated by instantons at $T=0$ (see Sec. \ref{sec_3pc}),
this is not true at $T>T_c$. This is consistent with the idea
that space-like wave functions above $T_c$ are determined by the 
space-like string tension \cite{KSB_92}, which disappears during
cooling. 
 
   Clearly, studies with dynamical fermions are of great interest.
Some preliminary results have been obtained by Ilgenfritz et al. 
(private communication). Using a ``gentle cooling'' algorithm
with only a few cooling iterations in order to prevent instantons
and anti-instantons from annihilating each other, they found 
evidence for an anticorrelation of topological charges in 
the time direction above $T_c$. This would be the first direct
lattice evidence for the formation of polarized instanton molecules 
in the chirally symmetric phase.


\section{Instantons in related theories}
\label{sec_other}

\subsection{Two-dimensional theories}
\label{sec_twodim}

  Although two-dimensional theories should logically be placed between 
the simplest quantum mechanical systems and Yang-Mills theories, we have 
postponed their discussion up to now in order not to disrupt the main 
line of the review. Nevertheless, topological objects play an important 
role in many low dimensional models. We do not want to give an exhaustive
survey of these theories, but have selected two examples, the $O(2)$ 
and $O(3)$ models, which, in our opinion, provide a few interesting 
lessons for QCD. As far as other theories are concerned, we refer 
the reader to the extensive literature, in particular on the Schwinger
model \cite{Smi_94a,SSZ_95} and two dimensional QCD with fundamental 
or adjoint fermions \cite{Smi_94b}.

\subsubsection{The $O(2)$ sigma model}
\label{sec_O(2)}
 
   The $O(2)$ model is also known as a the $d=2$ Heisenberg magnet
or the $XY$ model. It describes a two dimensional spin vector $\vec S$ 
governed by the Hamiltonian 
\be 
\label{E_O(2)}
\frac{E}{T} &=& \frac{1}{2t} \int d^2x\, 
  \left(\partial_\mu \vec S\right)^2, 
\ee  
together with the constraint ${\vec S}^2=1$. Here, $t=T/J$ is a 
dimensionless parameter, and $J$ the coupling constant. In this 
section we will follow the more traditional language of statistical 
mechanics rather than euclidean quantum field theory. This means 
that the coordinates are $x,y$ and the weight factor in the functional 
integral is $\exp(-E/T)$. Of course, one can always switch to field 
theory language by replacing the energy by the action and the 
temperature by the coupling constant $g^2$.

  The statistical sum is Gaussian except for the constraint. We can
make this more explicit by parameterizing the two dimensional spin 
vector $\vec S$ in the form $S_1=\cos\theta,\,S_2=\sin\theta$. In
this case, the energy is given by $E/T = \frac{1}{2t} \int d^2x\, 
(\partial_\mu \theta)^2$, which would describe a non-interacting
scalar field if it were not for the fact that $\theta$ is a 
periodic variable. It is often useful to define the theory 
directly on the lattice. The partition function is given by
\be
\label{O(2)_lat}
 Z &=& \int\left(\prod_x \frac{d\theta_x}{2\pi}\right)\, \exp\left(
 \beta \sum_{x,\hat e_\mu} \left[ \cos(\theta_x-\theta_{x+\hat e_\mu})
 - 1 \right] \right) ,
\ee
which is automatically periodic in $\theta$. The lattice spacing 
$a$ provides an ultraviolet cutoff. We should also note that the
$O(2)$ model has a number of physical applications. First of all,
the model obviously describes a two dimensional magnet. In addition 
to that, fluctuations of the order parameter in superconducting 
films of liquid $^4$He and the dynamics of dislocations in the 
melting of two-dimensional crystals are governed by effective 
$O(2)$ models.

   A two dimensional theory with a continuous symmetry cannot have an 
ordered phase at non-zero temperature. This means that, under ordinary
circumstances, two dimensional models cannot have a phase transition at 
finite temperature \cite{MW_66}. The $O(2)$ model is special because it 
has a phase transition at $T_c\simeq (\pi J)/2>0$ (although, in agreement 
with the Mermin-Wagner theorem, the transition is not characterized by a 
local order parameter). Both the low and the high temperature phase are 
disordered, but the functional form of the spin correlation function 
$K(x)=\langle\vec S(x)\vec S(0)\rangle$ changes. In some sense, the 
whole region $T<T_c$ is critical because the correlation functions 
exhibits a power-law decay \cite{Ber_71}. For $T>T_c$ the spin correlator 
decays exponentially, and the theory has a mass gap.

   The mechanism of this phase transition was clarified in the 
seminal paper by \cite{KT_73}, see also the review \cite{Kog_79}. 
Let us start with the low temperature phase. In terms of the 
angle variable $\theta$, the spin correlation function is given
by 
\be 
K(x) &=& N^{-1}\int d\theta(x)\, e^{i\theta(x)} e^{-i\theta(0)}
 e^{-1/(2t)\int d^2x (\partial_\mu \theta)^2} .
\ee 
At small temperature, the system is dominated by spin waves
and we expect that fluctuations in $\theta$ are small. In this
case we can ignore the periodic character of $\theta$ for the 
moment. Using the propagator for the $\theta$-field, $G(r)=-1/
(2\pi)\log(r/a)$, we have
\be 
K(r) &=& \exp(tG(r))
 \, \sim \, r^{-\frac{t}{2\pi}} .
\ee
The correlator shows a power law, with a temperature-dependent
exponent $\eta=T/(2\pi J)$. 
 
  In order to understand the phase transition in the $O(2)$ model, 
we have to go beyond Gaussian fluctuations and include topological 
objects. These objects can be classified by a winding number
\be
 q&=& \frac{1}{2\pi} \oint d\vec x\cdot\vec\nabla\theta. 
\ee
Solutions with $q=\pm 1$ are called (anti) vortices. A solution with 
$q=n$ is given by $\theta=n\alpha$ with $\alpha=\arctan(y/x)$. The 
energy of a vortex is
\be 
\frac{E(q=1)}{T}&=& \frac{\pi}{t} \int\frac{dr}{r} 
 \left(\frac{\partial\theta}{\partial \alpha}\right)^2 
 \;=\;  \frac{\pi n^2}{t}\log(R/a),
\ee
where $R$ is the IR cutoff. Since the energy is logarithmically  
divergent, one might think that vortex configurations are irrelevant. 
In fact, they are crucial for the dynamics of the phase transition. 

    The reason is that it is not the energy, but the free energy 
$F=E-TS$ which is relevant for the statistical sum. The entropy of 
an isolated vortex is essentially the logarithm of all possible
vortex positions, given by $S=\log [(R/a)^2]$. For temperatures
$T>T_c\simeq (\pi J)/2$, entropy dominates over energy and 
vortices are important. The presence of vortices implies that
the system is even more disordered and the spin correlator
decays exponentially.

   What happens to the vortex gas below $T_c$? Although isolated
vortices have infinite energy, vortex-anti-vortex pairs (molecules)
have finite energy $E\simeq \pi/(2t)[\log(R/a)+ {\rm const}]$, 
where $R$ is the size of the molecule. At low temperature, molecules 
are strongly suppressed, but as the temperature increases, they 
become more copious. Above the critical temperature, molecules 
are ionized and a vortex plasma is formed. 

   There is yet another way to look at this transition. We can decompose
any field configuration into the contribution of vortices and a smooth 
field. Using this decomposition, one can see that the $O(2)$ sigma model 
is equivalent to a two dimensional Coulomb gas. The Koesterlitz-Thouless 
transition describes the transition from a system of dipoles to an ionized 
plasma. The correlation length of the spin system is nothing but the Debye 
screening length in the Coulomb plasma. At the transition point, the 
screening length has an essential singularity \cite{KT_73} 
\be
\label{eq_KTcorr}
\xi \sim \exp\left(\frac{\rm const}{|T-T_c|^{1/2}}\right)
\ee
rather than the power law divergence observed in ordinary phase
transitions\footnote{The interested reader can find further details in 
review talks on spin models given at the annual Lattice meetings. 
Using cluster algorithms to fight critical slowdown one can obtain 
very accurate data. In \cite{Wol_89} the correlation length reaches 
about 70 lattice units, confirming (\ref{eq_KTcorr}). Nevertheless,
not all Koesterlitz-Thouless results are reproduced: e.g. the value 
of index $\eta$ (defined by $K(r)\sim 1/r^{\eta}$ as $T\rightarrow 
T_c$, ) is not 1/4 but noticeably larger.}.

  There is an obvious lesson we would like to draw from this: the 
two phases of the $O(2)$ model resemble the two phases found in QCD
(Sec. \ref{sec_liquid_T}). In both cases the system is described by
an (approximately) random ensemble of topological objects in one phase, 
and by bound pairs of pseudo-particles with opposite charge in the other. 
The only difference is that the high and low temperature phase are 
interchanged. This is essentially due to the fact that temperature 
has a different meaning in both problems. In the $O(2)$ model, instantons
are excited at high temperature, while in QCD they are suppressed.
 
\subsubsection{The $O(3)$ sigma model}
\label{sec_O(3)}
  
  The $O(2)$ model can be generalized to a $d=2$ spin system with
a three (or, more generally, $N$) dimensional spin vector $\vec S$ 
of unit length. Unlike the $O(2)$ model, the $O(3)$ model is not
just a free field theory for a periodic variable. In fact, in many 
ways the model resembles QCD much more than the $O(2)$ model
does. First of all, for $N>2$ the rotation group is non-abelian 
and spin waves (with $(N-1)$ polarizations) interact. In order to treat
the model perturbatively, it is useful to decompose the vector field
in the form $\vec S=(\sigma,\vec\pi)$, where $\vec\pi$ is an $(N-1)$
dimensional vector field and $\sigma=\sqrt{1-{\vec\pi}^2}$. The 
perturbative analysis of the $O(N)$ model gives the beta function
\cite{Pol_75}
\be
\label{O(N)_beta}
 \beta(t) &=& -\frac{N-2}{2\pi}t^2 + O(t^3),
\ee
which shows that just like QCD, the $O(N)$ model is asymptotically free
for $N>2$. In the language of statistical mechanics, the beta function 
describes the evolution of the effective temperature as a function of 
the scale. A negative beta function then implies that if the temperature 
is low at the atomic scale $a$, the effective temperature grows as we go 
to larger scales. Eventually, the reduced temperature is $t=O(1)$ ($T$ 
is comparable to $J$), fluctuations are large and long range order is 
destroyed. Unlike the $N=2$ model, the $O(3)$ model has only one critical 
point, $t=0$, and correlation functions decay exponentially for all 
temperatures $T>0$. 
 
  Furthermore, like non-abelian gauge theory, the $O(3)$ model has 
classical instantons solutions. The topology is provided by the fact 
that field configurations are maps from two dimensional space-time 
(compactified to a sphere) into another sphere which describes the 
orientation of $\vec S$. The winding number is defined by
\be
\label{O(3)_q} 
 q &=& \frac{1}{8\pi} \int d^2x \, \epsilon_{\mu\nu}
  \vec S\cdot (\partial_\mu\vec S\times\partial_\nu\vec S) .
\ee
Similar to QCD, solutions with integer winding number and energy 
$E=4\pi J|q|$ can be found from the self-duality equation
\be
\label{O(3)_dual}
 \partial_\mu \vec S &=& \pm\epsilon_{\mu\nu} \vec S\times
 \partial_\nu \vec S .
\ee
A solution with $q=1$ can be constructed by means of the stereographic 
projection
\be 
\label{O(3)_inst}
\vec S &=& \left(\frac{2 \rho x_1}{x^2+\rho^2},
 \frac{2 \rho x_2}{x^2+\rho^2},
 \frac{x^2-\rho^2}{x^2+\rho^2}\right) .
\ee
At large distances all spins point up, in the center the spin points 
down, and at intermediate distances $x\sim\rho$ all spins are horizontal, 
pointing away from the center. Like QCD, the theory is scale invariant on 
the classical level and the energy of an instanton is independent of the 
size $\rho$. Instantons in the $O(2)$ model are sometimes called Skyrmions,
in analogy with the static (three-dimensional) solitons introduced by 
Skyrme in the $d=4$ $O(4)$ model \cite{Sky_61}.   

 The next logical step is the analog of the 't Hooft calculation of
fluctuations around the classical instanton solution. The result
of the semi-classical calculation is \cite{Pol_75}
\be 
dN_{inst}\sim {d^2x d\rho\over \rho^3}\exp(-E/T)
\sim {d^2x d\rho\over \rho} 
\ee
which is divergent both for large and small radii\footnote{This result
marks the first non-perturbative ultraviolet divergence ever discovered. 
It is similar to the divergence of the density of instanton molecules 
for large $N_f$ discussed in Sec. \protect\ref{sec_big_pic}.}. The result 
for large $\rho$ is of course not reliable, since it is based on the 
one-loop beta function.  

  Because the $O(3)$ model shows so many similarities with QCD,
it is natural to ask whether one can learn anything of relevance
for QCD. Indeed, the $O(3)$ model has been widely used as a
testing ground for new methods in lattice gauge theory. The 
numerical results provide strong support for the renormalization
group analysis. For example, \cite{CEP_95} studied the $O(3)$ model
at correlation lengths as large as $\xi/a\sim 10^5$! The results 
agree with state-of-the-art theoretical predictions (based on the 
3-loop beta function and an overall constant determined from the 
Bethe ansatz) with a very impressive accuracy, on the order of few 
percent.

  Unfortunately, studies of instantons in the $O(3)$ model have not
produced any significant insights. In particular, there are no
indications that small-size (semi-classical) instantons play any role in 
the dynamics of the theory. Because of the divergence in the instanton 
density, the topological susceptibility in the $O(3)$ has no continuum 
limit. This conclusion is supported by lattice simulations \cite{MS_94}. 
The simulations also indicate that for large size instantons, the size 
distribution is $dN \sim d\rho/\rho^{-3}$. This power differs from the 
semi-classical result, but it agrees with the $\rho$ dependence coming 
from the Jacobian alone, without the running coupling in the action. 
This result supports the idea of a frozen coupling constant discussed 
in Sec. \ref{sec_beta}.

\subsection{Instantons in electroweak theory}
\label{sec_ew}

  In the context of this review we cannot provide a detailed 
discussion of instantons and baryon number violation in electroweak
theory. Nevertheless, we briefly touch on this subject, because 
electroweak theory provides an interesting theoretical laboratory.
The coupling constant is small, the instanton action is large and 
the semi-classical approximation is under control. Unfortunately, 
this means that under ordinary conditions tunneling events are too 
rare to be of physical importance. Interesting questions arise 
when one tries to increase the tunneling rate, e.g. by studying
scattering processes with collision energies close to the barrier
height, or processes in the vicinity of the electroweak phase transition. 
Another interesting problem is what happens if we consider the 
Higgs expectation value to be a free parameter. When the Higgs 
VEV is lowered, electroweak theory becomes a strongly interacting
theory, and we encounter many of the problems we have to deal with 
in QCD.

  Electroweak theory is an $SU(2)_L\times U(1)$ chiral gauge theory
coupled to a Higgs doublet. For simplicity we will neglect the 
$U(1)$ interactions in the following, i.e. set the Weinberg angle
to zero. The most important difference as compared to QCD is the
fact that gauge invariance is spontaneously broken. In the ground
state, the Higgs field acquires an expectation value, which gives
masses to the $W$ bosons as well as to the quarks and charged
leptons. If the Higgs field has a non-zero VEV the instanton is,
strictly speaking, not a solution of the equations of motion. 
Nevertheless, it is clear that if $\rho\ll v^{-1}$ (where $v$ is 
the Higgs VEV), the gauge fields are much stronger than the Higgs 
field and there should be an approximate instanton solution\footnote{
This notion can be made more precise using the constrained instanton 
solution \protect\cite{Aff_81}. This technique is similar to the 
construction that defines the streamline solution for an 
instanton-anti-instanton pair.}. In the central region $x<m_W^{-1},
m_{H}^{-1}$, the solution can be found by keeping the instanton gauge 
field fixed and solving the equations of motion for the Higgs field. 
The result is 
\be 
\phi &=& \frac{x^2}{x^2+ \rho^2} U\left( 0\atop{v/\sqrt{2}}\right),
\ee
where $U$ is the color orientation matrix of the instanton. The 
result shows that the instanton makes a hole of size $\rho$ in the 
Higgs condensate. Scale invariance is lost and the instanton action 
depends on the Higgs VEV \cite{tHo_76b}
\be 
\label{S_I_EW}
S &=& {8\pi^2 \over g^2} + 2\pi^2\rho^2 v^2.
\ee  
The tunneling rate is $p\sim \exp(-S)$ and large instantons with
$\rho\gg v^{-1}$ are strongly suppressed. 

   The loss of scale invariance also implies that in electroweak
theory, the height of the barrier separating different topological
vacua can be determined. There is a static solution with winding 
number 1/2, corresponding to the top of the barrier, called the
sphaleron \cite{KM_84}. The sphaleron energy is $E_{sph}\simeq
4 m_W/\alpha_W\simeq 10$ TeV. 

   Elektroweak instantons also have fermionic zero modes, and as usual
the presence of these zero modes is connected with the axial anomaly. 
Since only left-handed fermions participate in weak interactions, both 
vector and axial-vector currents are not conserved. The 't Hooft 
vertex contains all 12 weak doublets
\be 
\label{ew_doub}
 (\nu_e,e),\; (\nu_\mu,\mu),\; (\nu_\tau,\tau),\; 3*(u,d),\; 
 3*(c,s),\; 3*(t,b),
\ee
where the factors of three come from color. Each doublet provides one 
fermionic zero mode, the flavor depending on the isospin orientation 
of the instanton. The 't Hooft vertex violates both baryon and lepton
number. These processes are quite spectacular, because all families
have to be involved, for example
\be
 u + d &\rightarrow & \bar d + \bar s + 2\bar c + 3\bar t + e^+ +
\mu^+ + \tau^+ ,  \\
 u + d &\rightarrow & \bar u + 2\bar s + \bar c + \bar t + 2\bar b
+ \nu_e + \nu_\mu + \tau^+ .
\ee
Note that $\Delta B=\Delta L=-3$, so $B+L$ is violated, but $B-L$ is
conserved. Unfortunately, the probability of such an event is tiny,
proportional to the square of the tunneling amplitude $P\sim \exp(
-16\pi^2/g^2_w)\sim 10^{-169}$ \cite{tHo_76}, many orders of magnitude 
smaller than any known radioactive decay.

   Many authors have discussed the possibility of increasing the
tunneling rate by studying processes near the electroweak phase
transition \cite{KRS_85} or scattering processes involving energies
close to the sphaleron barrier $E\simeq E_{sph}$. Since
$E_{sph}\simeq 10$ TeV, this energy would have been accessible at 
the SSC and will possibly be within reach at the LHC. The latter 
idea became attractive when it was realized that associated multi-Higgs 
and $W$ production increases the cross section \cite{Rin_90,Esp_90}. 
On general grounds one expects
\be
\label{holy_grail}
 \sigma_{\Delta (B+L)} &\sim & \exp\left[-\frac{4\pi}{\alpha_W}
 F\left(\frac{E}{E_{sph}}\right)\right],
\ee
where $F(\epsilon)$ is called the ``holy grail" function. At low
energy, $F(0)=1$ and baryon number violation is strongly suppressed.
The first correction is $F(\epsilon)=1-\frac{9}{8}\epsilon^{4/3}+
O(\epsilon^2)$, indicating that the cross section rises with energy.
Clearly, the problem is the behavior of $F(\epsilon)$ near $\epsilon=1$. 
Most authors now seem to agree that $F(\epsilon)$ will not drop below 
$F\simeq 1/2$, implying that baryon number violation in $pp$ collisions 
will remain unobservable \cite{Zak_92,MS_92,MS_92b,Ven_92,DP_94}. The 
question is of interest also for QCD, because the holy grail function 
at $\epsilon=1$ is related to the instanton-anti-instanton interaction 
at short distance. In particular, taking into account unitarity in the 
multi-$W$ (or multi-gluon) production process correspond to an effective 
instanton-anti-instanton repulsion, see Sec. \ref{sec_int_bos}.

   The other question of interest for QCD is what happens if the 
Higgs VEV $v$ is gradually reduced. As $v$ becomes smaller the theory
moves from the weak-coupling to the strong-coupling regime. The vacuum 
structure changes from a very dilute system of $IA$ molecules 
to a more dense (and more interesting) non-perturbative vacuum. 
Depending on the number of light fermions, one should eventually
reach a confining, QCD-like phase. In this phase, leptons are 
composite but the low-energy effective action is probably similar
to the one in the Higgs phase \cite{AF_81,CFJ_86}. Unfortunately, 
the importance of non-perturbative effects has never been studied.

  Let us comment only on one element which is absent in QCD, the 
scalar-induced interaction between instantons. Although the scalar
interaction is order $O(1)$ and therefore suppressed with respect 
to the gauge interaction $O(g^{-2})$, it is long range if the Higgs 
mass is small. For $m_H^{-1}\gg R \gg \rho$ the $IA$ interaction is 
\cite{Yun_88}
\be 
\label{eq_yung}
S_{Higgs}=4\pi^2 a^2\rho^2\left[1+\frac{\rho^2}{R^2}
 \left( 2(u\cdot\hat R)^2-1\right)\right]+
O\left(\frac{1}{R^4}\right)
\ee
Unlike the gluonic dipole interaction, it does not vanish if averaged
over all orientations, $\langle (u\cdot\hat R)^2\rangle =1/4$. This 
means that the scalar interaction can provide coherent attraction 
for distances $R m_H <1$, which is of the order $v^2 \rho^4 n /m_H^2$ 
where $n$ is the instanton density. This is large if the Higgs mass 
is small.

  Another unusual feature of Yung interaction (\ref{eq_yung}) is that
it is repulsive for $u\cdot\hat R=1$ (which is the most attractive 
orientation for the dipole interaction). This would suggest that for 
a light Higgs mass, there is no small $R$ problem. This question was 
studied in \cite{VS_93}. For the complete Yung ansatz (which is 
a good approximation to the full streamline solution) the 
approximate result (\ref{eq_yung}) is only valid for $R>10\rho$, 
while for smaller separation the dependence on $u\cdot\hat R$ is 
reversed.

\subsection{Supersymmetric QCD}
\label{sec_susy_qcd}

  Supersymmetry (SUSY) is a powerful theoretical concept which
is of great interest in constructing field theories beyond the
standard model. In addition to that, supersymmetric field 
theories provide a very useful theoretical laboratory, and
a number of important advances in understanding the ground
state of strongly interacting gauge theories. We have already
seen one example for the usefulness of supersymmetry in
isolating instanton effects in the context of SUSY quantum
mechanics, Sec. \ref{sec_susy_qm}. In that case, SUSY implies 
that perturbative contributions to the vacuum energy vanish and
allows a precise definition of the instanton-anti-instanton
contribution. In the following, we will see that supersymmetry 
can be used in very much the same way in the context of gauge
field theories.

  In general, one should keep in mind that SUSY theories are just 
ordinary field theories with a very specific matter content and 
certain relations between different coupling constants. Eventually,
we hope to understand non-abelian gauge theories for all possible 
matter sectors. In particular, we want to know how the structure 
of the theory changes as one goes from QCD to it supersymmetric 
generalizations, where many exact exact statements about instantons 
and the vacuum structure are known. Deriving these results often
requires special techniques that go beyond the scope of this review.
For details of the supersymmetric instanton calculus we refer the
reader to the extensive review \cite{AKM_88}. Nevertheless, we have 
tried to include a number of interesting results and explain them 
in standard language.

   As theoretical laboratories SUSY theories have several advantages 
over ordinary field theories. We have already mentioned one of them:
Non-renormalization theorems imply that many quantities do not receive
perturbative contributions, so instanton effects are more easily 
identified. In addition to that, SUSY gauge theories usually have 
many degenerate classical vacua. These degeneracies cannot be 
lifted to any order in perturbation theory, and instantons 
often play an important role in determining the ground state. In 
most cases, the classical vacua are characterized by scalar field
VEVs. If the scalar field VEV is large, one can perform reliable 
semi-classical calculations. Decreasing the scalar VEV, one moves 
towards strong coupling and the dynamics of the theory is non-trivial.
Nevertheless, supersymmetry restricts the functional dependence of 
the effective potential on the scalar VEV (and other parameters,
like masses or coupling constants), so that instanton calculations
can often be continued into the strong coupling domain. 

  Ultimately, we would like to understand the behavior of SUSY
QCD as we introduce soft supersymmetry breaking terms and send
the masses of the gluinos and squarks to infinity. Not much 
progress has been achieved in this direction, but at least 
for small breaking the calculations are feasible and some 
lessons have been learned.

\subsubsection{The instanton measure and the perturbative beta function}
\label{sec_NSVZ_beta}

  The simplest ($N=1$) supersymmetric non-abelian gauge theory is
$SU(2)$ SUSY gluodynamics, defined by
\be
\label{susy_gluo}
{\cal L} &=& -\frac{1}{4g^2}F_{\mu\nu}^a F^a_{\mu\nu} 
 + \frac{i}{2g^2}\lambda^a(D\!\!\!\!/\,)_{ab}\lambda^b
\ee
where the gluino field $\lambda^a$ is a Majorana fermion in the 
adjoint representation of $SU(2)$. More complicated theories can be
constructed by adding additional matter fields and scalars. $N$ 
extended supersymmetry has $N$ gluino fields as well as additional
scalars. Clearly, supersymmetric gluodynamics has $\theta$ vacua
and instanton solutions. The only difference as compared to QCD   
is that the fermions carry adjoint color, so there are twice as 
many fermion zero modes. If the model contains scalars fields 
that acquire a vacuum expectation value, instantons are approximate 
solutions and the size integration is automatically cut off by the 
scalar VEV. These theories usually resemble electroweak theory  
more than they do QCD. 

   At first glance, instanton amplitudes seem to violate supersymmetry:
the number of zero modes for gauge fields and fermions does not match, 
while scalars have no zero modes at all. However, one can rewrite the 
tunneling amplitude in manifestly supersymmetric form \cite{NSVZ_83b}.
We will not do this here, but stick to the standard notation. The
remarkable observation is that the determination of the tunneling 
amplitude in SUSY gauge theory is actually simpler than in QCD.
Furthermore, with some additional input, one can determine the 
complete perturbative beta function from the tunneling amplitude.

  The tunneling amplitude is given by
\be
\label{susy_amp_1l}
 n(\rho) \sim  \exp\left(-\frac{2\pi}{\alpha}\right)
 M^{n_g-n_f/2} \left( \frac{2\pi}{\alpha} \right)^{n_g/2}
 d^4x \frac{d\rho}{\rho^5} \rho^k \prod_f d^2\xi_f ,
\ee 
where all factors can be understood from the 't Hooft calculation 
discussed in Sec. \ref{sec_thooft}. There are $n_g=4N_c$ bosonic zero 
modes that have to be removed from the determinant and give one power of 
the regulator mass $M$ each. Similarly, each of the $n_f$ fermionic zero 
modes gives a factor $M^{1/2}$. Introducing collective coordinates 
for the bosonic zero modes gives a Jacobian $\sqrt{S_0}$ for every
zero mode. Finally, $d^2\xi$ is the integral over the fermionic 
collective coordinates and $\rho^k$ is the power of $\rho$ needed 
to give the correct dimension. Supersymmetry now ensures that all
non-zero mode contributions exactly cancel. More precisely, the 
subset of SUSY transformations which does not rotate the instanton 
field itself, mixes fermionic and bosonic modes non-zero modes 
but annihilates zero modes. This is why all non-zero modes cancel
but zero modes can be unmatched. Note that as a result of this
cancellation, the power of $M$ in the tunneling amplitude is 
an integer.

   Renormalizability demands that the tunneling amplitude is 
independent of the regulator mass. This means that the explicit 
$M$-dependence of the tunneling amplitude and the $M$ dependence
of the bare coupling have to cancel. As in QCD, this allows us to 
determine the one-loop coefficient of the beta function $b=(4-N)
N_c-N_f$. Again note that $b$ is an integer, a result that would 
appear very mysterious if we did not know about instanton zero modes.

   In supersymmetric theories one can even go one step further 
and determine the beta function to all loops \cite{NSVZ_83b,VZNS_86}. 
For that purpose let us write down the renormalized instanton 
measure
\be
\label{susy_amp}
 n(\rho) \sim  \exp\left(-\frac{2\pi}{\alpha}\right)
 M^{n_g-n_f/2} \left( \frac{2\pi}{\alpha_R} \right)^{n_g/2}
 Z_g^{n_g/2} \left( \prod_f Z_f^{-1/2} \right)
 d^4x \frac{d\rho}{\rho^5} \rho^k \prod_f d^2\xi_f ,
\ee
where we have introduced the field renormalization factors $Z_{g,f}$
for the bosonic/fermionic fields. Again, non-renormalization theorems
ensure that the tunneling amplitude is not renormalized at higher
orders (the cancellation between the non-zero mode determinants 
persists beyond one loop). For gluons the field renormalization
(by definition) is the same as the charge renormalization $Z_g=\alpha_R/
\alpha_0$. Furthermore, supersymmetry implies that the field 
renormalization is the same for gluinos and gluons. This means
that the only new quantity in (\ref{susy_amp}) is the anomalous
dimension of the quark fields, $\gamma_\psi=d \log Z_f/d \log M$. 

   Again, renormalizability demands that the amplitude is independent
of $M$. This condition gives the NSVZ beta function \cite{NSVZ_83b}
which, in the case $N=1$, reads
\be
\label{NSVZ_beta} 
\beta(g)&=&-\frac{g^3}{16\pi^2} 
  \frac{3N_c-N_f+N_f\gamma_\psi(g)}{1-N_c g^2/8\pi^2}.
\ee
The anomalous dimension of the quarks has to be calculated 
perturbatively. To leading order, it is given by
\be
\gamma_\psi(g) &=&-\frac{g^2}{8\pi^2} \frac{N_c^2-1}{N_c} +O(g^4).
\ee 
The result (\ref{NSVZ_beta}) agrees with explicit calculations
up to three loops \cite{JJN_97}. Note that the beta function is 
scheme dependent beyond two loops, so in order to make a comparison
with high order perturbative calculations, one has to translate 
from the Pauli-Vilars scheme to a more standard perturbative 
scheme, e.g. $\overline{MS}$.

   In theories without quarks, the NSVZ result determines the 
beta function completely. For $N$-extended supersymmetric 
gluodynamics, we have
\be
\label{NSVZ_beta_N}
\beta(g) &=& -\frac{g^3}{16\pi^2}\frac{N_c(4-N)}{1-(2-N)N_c g^2/(8\pi^2)}.
\ee
One immediately recognizes two interesting special cases. For $N=4$,
the beta function vanishes and the theory is conformal. In the case 
$N=2$, the denominator vanishes and the one loop result for the 
beta function is exact.

\subsubsection{$N=1$ supersymmetric QCD}
\label{sec_N=1_susy}

  In the previous section we used instantons only as a tool to
simplify a perturbative calculation. Naturally, the next question 
is whether instantons cause important dynamical effects in SUSY 
theories. Historically, there has been a lot of interest in SUSY 
breaking by instantons, first discovered in the quantum mechanical 
model discussed in Sec. \ref{sec_susy_qm}. The simplest field theory
in which SUSY is broken by instantons is the $SU(2)\times SU(3)$
model \cite{ADS_84c,ADS_85,VZS_85}. However, non-perturbative SUSY 
breaking does not take place in supersymmetric QCD, and we will not 
discuss this phenomenon any further.

  An effect which is more interesting in our context is gluino 
condensation in SUSY gluodynamics (\ref{susy_gluo}) \cite{NSV_83}. 
For $SU(2)$ color, the gluino condensate is most easily determined 
from the correlator $\langle\lambda^a_\alpha\lambda^{a\,\alpha}(x)
\lambda^b_\beta\lambda^{b\,\beta}(0)\rangle$. In $SU(N)$, one has 
to consider the $N$-point function of $\lambda^a_\alpha\lambda^{a\,
\alpha}$. In supersymmetric theories, gluinos have to be massless, 
so the tunneling amplitude is zero. However, in the $N$-point correlation 
function all zero modes can be absorbed by external sources, similar 
to the axial anomaly in QCD. Therefore, there is a non-vanishing 
one-instanton contribution. In agreement with SUSY Ward-identities, 
this contribution is $x$-independent. Therefore, one can use cluster 
decomposition to extract the gluino condensate $\langle\lambda\lambda
\rangle=\pm A\Lambda^3$ (in $SU(2)$), where $A$ is a constant that is
fixed from the single instanton calculation. 

  There are a number of alternative methods to calculate the gluino 
condensate in SUSY gluodynamics. For example, it has been suggested 
that $\langle\lambda\lambda\rangle$ can be calculated directly using 
configurations with fractional charge \cite{CG_84}. In addition to
that, one can include matter fields, make the theory Higgs-like,
and then integrate out the matter fields \cite{NSV_83b,SV_88b}. This 
method gives a different numerical coefficient for the 
gluino condensate, a problem that was recently discussed in \cite{KS_97}.

  The next interesting theory is $N=1$ SUSY QCD, where we add $N_f$ 
matter fields (quarks $\psi$ and squarks $\phi$) in the fundamental 
representation. Let us first look at the NSVZ beta function. For $N=1$, 
the beta function blows up at $g^2_*=8\pi^2/N_c$, so the renormalization 
group trajectory cannot be extended beyond this point. Recently, Kogan 
and Shifman suggested that at this point the standard phase meets the 
renormalization group trajectory of a different (non-asymptotically 
free) phase of the theory \cite{KS_95}. The beta function vanishes 
at $g^2_*/(8\pi^2)= [N_c(3N_c-N_f)]/[N_f(N_c^2-1)]$, where we have 
used the one-loop anomalous dimension. This is reliable if $g_*$ is 
small, which we can ensure by choosing $N_c\to\infty$ and $N_f$ in 
the conformal window $3N_c/2<N_f<3N_c$. Recently, Seiberg showed 
that the conformal point exists for all $N_f$ in the conformal 
window (even if $N_c$ is not large) and clarified the structure of 
the theory at the conformal point \cite{Sei_94}.
 
\begin{figure}[t]
\begin{center}
\leavevmode
\epsfxsize=12cm
\epsffile{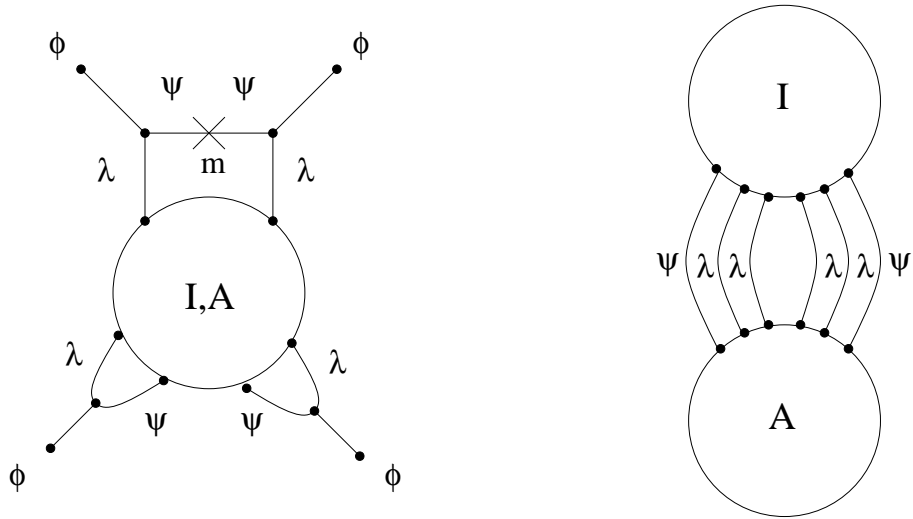}
\end{center}
\caption{\label{fig_susy_qcd} 
Instanton contributions to the effective potential in supersymmetric 
QCD.}
\end{figure}

   Let us now examine the vacuum structure of SUSY QCD for $N_c=2$ 
and $N_f=1$. In this case we have one Majorana fermion (the gluino), 
one Dirac fermion (the quark) and one scalar squark (or Higgs) field. 
The quark and squark fields do not have to be massless. We will denote 
their mass by $m$, while $v$ is the vacuum expectation value of the 
scalar field. In the semi-classical regime $v\gg\Lambda$, the tunneling
amplitude (\ref{susy_amp}) is $O(\Lambda^5)$. The 't Hooft effective
lagrangian is of the form $\lambda^4\psi^2$, containing four quark 
and two gluino zero modes. As usual, instantons give an expectation
value to the 't Hooft operator. However, due to the presence of
Yukawa couplings and a Higgs VEV, they can also provide expectation
values for operators with less than six fermion fields. In particular,
combining the quarks with a gluino and a squark tadpole we can construct
a two gluino operator. Instantons therefore lead to gluino condensation
\cite{ADS_84b}. Furthermore, using two more Yukawa couplings one can 
couple the gluinos to external quark fields. Therefore, if the quark 
mass is non-zero we get a finite density of individual instantons 
$O(m\Lambda^5/v^2)$, see Fig. \ref{fig_susy_qcd}.  

   As in ordinary QCD, there are also instanton-anti-instanton molecules.
The contribution of molecules to the vacuum energy can be calculated 
either directly or indirectly, using the single instanton result and 
arguments based on supersymmetry. This provides a nice check on the 
direct calculation, because in this case we have a rigorous definition 
of the contribution from molecules. Calculating the graph shown in 
Fig. \ref{fig_susy_qcd}b, one finds \cite{Yun_88}
\be
\label{eps_susy_IA}
\epsilon^{IA}_{vac} &=& \frac{32\Lambda^{10}}{g^8 v^6},
\ee   
which agrees with the result originally derived in \cite{ADS_84b}
using different methods. The result implies that the Higgs expectation 
value is driven to infinity and $N_f=1$ SUSY QCD does not have a stable 
ground state. The vacuum can be stabilized by adding a mass term. For 
non-zero quark mass $m$ the vacuum energy is given by
\be
\label{eps_susy}
\epsilon_{vac} &=& 2m^2v^2- \frac{16m\Lambda^{5}}{g^4v^2}
 + \frac{32 \Lambda^{10}}{g^8 v^6} 
 \; =\;  2\left( mv - \frac{4\Lambda^{5}}{g^4 v^3}\right)^2 
\ee   
where the first term is the classical Higgs mass term, the second
is the one-instanton contribution and the third is due to molecules. 
In this case, the theory has a stable ground state at $v^2=\pm 2
\Lambda^{5/2}/(g^2 m^{1/2})$. Since SUSY is unbroken, the vacuum 
energy is exactly zero. Let us note that in the semi-classical 
regime $m\ll\Lambda\ll v$ everything is under control: instantons 
are small $\rho\sim v^{-1}$, individual instantons are rare, molecules 
form a dilute gas, $nR^4\sim (v/\Lambda)^{10}$, and the instanton and 
anti-instanton inside a molecule are well separated, $R_{IA}/\rho\sim 
1/\sqrt{g}$. Supersymmetry implies that the result remains correct 
even if we leave the semi-classical regime. This means, in particular, 
that all higher order instanton corrections ($O(\Lambda^{15})$ etc.) 
have to cancel exactly. Checking this explicitly might provide a very 
non-trivial check of the instanton calculus.
  
   Recently, significant progress has been made in determining
the structure of the ground state of $N=1$ supersymmetric QCD 
for arbitrary $N_c$ and $N_f$ \cite{IS_95}. Seiberg showed that
the situation discussed above is generic for $N_f<N_c$:
A stable ground state can only exist for non-zero quark mass. 
For $N_f=N_c-1$ the $m\neq 0$ contribution to the potential 
is an instanton effect. In the case $N_f=N_c$, the theory has
chiral symmetry breaking (in the chiral limit $m\to 0$) and 
confinement\footnote{There is no instanton contribution to the 
superpotential, but instantons provide a constraint on the 
allowed vacua in the quantum theory. To our knowledge, the 
microscopic mechanism for chiral symmetry breaking in this 
theory has not been clarified.}. For $N_f=N_c+1$, there is
confinement but no chiral symmetry breaking. Unlike QCD, the 
't Hooft anomaly matching conditions can be satisfied with
massless fermions. These fermions can be viewed as elementary 
fields in the dual (or ``magnetic'') formulation of the theory.
For $N_c+1<N_f<3N_c/2$ the magnetic formulation of the theory is 
IR free, while for $3N_c/2<N_f<3N_c$ the theory has an infrared 
fixed point. Finally, for $N_f>3N_c$, asymptotic freedom is lost 
and the electric theory is IR free. 

\subsubsection{$N=2$ supersymmetric gauge theory}
\label{sec_N=2_susy}

   The work of Seiberg has shown that supersymmetry sufficiently
constrains the effective potential in $N=1$ supersymmetric QCD
so that the possible vacuum states for all $N_c$ and $N_f$ can
be determined. For $N=2$ extended supersymmetric QCD, these
constraints are even more powerful. Witten and Seiberg 
have been able to determine not just the effective potential,
but the complete low energy effective action \cite{SW_94}. Again, 
the techniques used in this work are outside the scope of this 
review. However, instantons play an important role in this theory, 
so we would like to discuss a few interesting results. 

   $N=2$ supersymmetric gauge theory contains two Majorana fermions 
$\lambda^a_\alpha,\,\psi^a_\alpha$ and a complex scalar $\phi^a$, 
all in the adjoint representation of $SU(N_c)$. In the case of $N=2$
SUSY QCD, we add $N_f$ multiplets $(q_\alpha,\tilde q_\alpha,Q,\tilde Q)$
of quarks $q_\alpha,\tilde q_\alpha$ and squarks $Q,\tilde Q$ in the 
fundamental representation. In general, gauge invariance is broken
and the Higgs field develops an expectation value $\langle\phi
\rangle = a\tau^3/2$. The vacua of the theory can be labeled 
by a (gauge invariant) complex number $u=(1/2)\langle {\rm tr}\phi^2
\rangle$. If the Higgs VEV $a$ is large ($a\gg\Lambda$), the 
semi-classical description is valid and $u=(1/2)a^2$. In this 
case, instantons are small $\rho\sim a^{-1}$ and the instanton 
ensemble is dilute $n \rho^4\sim\Lambda^4/a^4 \ll 1$.

  In the semi-classical regime, the effective lagrangian is given by
\be
\label{L_eff_N=2_SUSY}
 {\cal L}_{eff} &=& \frac{1}{4\pi}{\rm Im} \left[
 -{\cal F}''(\phi) \left( \frac{1}{2}(F_{\mu\nu}^{sd})^2
 +(\partial_\mu\phi)(\partial^\mu\phi^\dagger)
 +i\psi\partial\!\!\!/\,\overline\psi 
 +i\lambda\partial\!\!\!/\,\overline\lambda  \right)\right.
 \nonumber \\
 & & \hspace{2cm}\left.
 +\frac{1}{\sqrt{2}} {\cal F}'''(\phi)
     \lambda\sigma^{\mu\nu}\psi F_{\mu\nu}^{sd}
 +\frac{1}{4}{\cal F}''''(\phi)\psi^2\lambda^2 \right] 
 + {\cal L}_{aux} + \ldots,
\ee 
where $F^{sd}_{\mu\nu}=F_{\mu\nu}+i\tilde F_{\mu\nu}$ is the
self dual part of the field strength tensor, ${\cal L}_{aux}$
contains auxiliary fields, and $\ldots$ denotes higher derivative
terms. Note that the effective 
low energy lagrangian only contains the light fields. In the 
semi-classical regime, this is the $U(1)$ part of the gauge 
field (the ``photon") and its superpartners. Using arguments based 
on electric-magnetic duality, Seiberg and Witten determined 
the exact prepotential ${\cal F}(\phi)$. From the effective lagrangian, 
we can immediately read off the effective charge at the scale $a$
\be
 {\cal F}''(a) = \frac{\tau(a)}{2} 
        = \frac{4\pi i}{g^2(a)}+\frac{\theta}{2\pi}
\ee 
which combines the coupling constant $g$ and the $\theta$ angle.
Also, the Witten-Seiberg solution determines the anomalous 
magnetic moment ${\cal F}'''$ and the four-fermion vertex 
${\cal F}''''$. In general, the structure of the prepotential
is given by 
\be
 {\cal F}(\phi) &=& \frac{i(4-N_f)}{8\pi} \phi^2 
  \log\left(\frac{\phi^2}{\Lambda^2}\right)
  - \frac{i}{\pi} \sum_{k=1}^{\infty} {\cal F}_k \,\phi^2
 \left(\frac{\Lambda}{\phi}\right)^{(4-N_f)k}.
\ee
The first term is just the perturbative result with the one-loop
beta functions coefficient. As noted in Sec. \ref{sec_NSVZ_beta},
there are no corrections from higher loops. Instead, there is an
infinite series of power corrections. The coefficient ${\cal F}_k$
is proportional to $\Lambda^{(4-N_f)k}$, which is exactly what 
one would expect for a $k$-instanton contribution. 

  For $k=1$, this was first checked by \cite{FP_95} in the case
of $SU(2)$ and by \cite{IS_96} in the more general case of $SU(N_c)$.
The basic idea is to calculate the coefficient of the 't Hooft 
interaction $\sim\lambda^2\psi^2$. The gluino $\lambda$ and the 
Higgsino $\psi$ together have 8 fermion zero modes. Pairing zero 
modes using Yukawa couplings of the type $(\lambda \psi)\phi$ and 
the non-vanishing Higgs VEV we can see that instantons induce a 
4-fermion operator. In an impressive tour de force the calculation 
of the coefficient of this operator was recently extended to the 
two-instanton level\footnote{Supersymmetry implies that there is no 
instanton-anti-instanton contribution to the prepotential.}
\cite{DKM_96,AHS_96}. For $N_f=0$, the result is 
\be 
\label{eq_4fN=2}
S_{4f} &=& \int dx \left(\psi^2\lambda^2\right)
   \left[ \frac{15 \Lambda^4}{8\pi^2 a^6} 
 +  \frac{9!\Lambda^8}{3\times 2^{12}\pi^2a^{10}} + \ldots\right],
\ee     
which agrees with the Witten-Seiberg solution. This is also true
for $N_f\neq 0$, except in the case $N_f=4$, where a discrepancy
appears. This is the special case where the coefficient of the 
perturbative beta function vanishes. Seiberg and Witten assume that 
the non-perturbative $\tau(a)$ is the same as the corresponding 
bare coupling in the lagrangian. But the explicit two-instanton 
calculation shows that even in this theory the charge is renormalized 
by instantons \cite{DKM_96b}. In principle, these calculations 
can be extended order by order in the instanton density. The
result provides a very non-trivial check on the instanton calculus. 
For example, in order to obtain the correct two instanton contribution 
one has two use the most general (ADHM) two-instanton solution,  
not just a linear superposition of two instantons.

   Instantons also give a contribution to the expectation value of 
$\phi^2$. Pairing off the remaining zero modes, the semi-classical
relation $u=a^2/2$ receives a correction \cite{FP_95}
\be
u &=& \frac{a^2}{2} + \frac{\Lambda^4}{a^2} + 
  O\left(\frac{\Lambda^8}{a^6}\right). 
\ee
More interesting are instanton corrections to the effective charge 
$\tau$. The solution of Seiberg and Witten can be written in terms
of an elliptic integral of the first kind
\be 
\label{eq_SW}
\tau(u) &=& i \frac{K(\sqrt{(1-k^2))}}{K(k)}, \hspace{0.5cm} 
   k^2\;=\; \frac{ u-\sqrt{u^2-4\Lambda^4}}{u+\sqrt{u^2-4\Lambda^4}} 
\ee

In the semi-classical domain, this result can be written as the
one-loop perturbative contribution plus an infinite series of
$k$-instanton terms. Up to the two-instanton level, we have
\be
\label{eq_SW2}
\frac{8\pi}{g^2} &=& \frac{2}{\pi}\left(
  \log\left(\frac{2 a^2}{\Lambda^2}\right)
 - \frac{3\Lambda^4}{a^4} - \frac{3\times 5\times 7\Lambda^8}{8a^8}
 + \ldots \right).
\ee
It is interesting to note that instanton corrections tend to
accelerate the growth of the coupling constant $g$.
This is consistent with what was found 
in QCD by considering how small-size instantons renormalize the charge
of a larger instanton \cite{CDG_78}. However, the result is opposite 
to the trend discussed in Sec. \ref{sec_beta} (based on the instanton 
size distribution and lattice beta function) which suggests that in
QCD the coupling runs more slowly than suggested by perturbation theory.
 
 If the Higgs VEV is reduced the instanton corrections in (\ref{eq_SW}) 
start to grow and compensate the perturbative logarithm. At this point
the expansion (\ref{eq_SW2}) becomes unreliable, but the exact solution
of Seiberg and Witten is still applicable. In the semi-classical regime,
the spectrum of the theory contains monopoles and dyons with masses
proportional to $\tau$. As the Higgs VEV is reduced, these particles
can become massless. In this case, the expansion of the effective 
lagrangian in terms of the original (electrically charged) fields 
breaks down, but the theory can be described in terms of their
(magnetically charged) dual partners.


\section{Summary and discussion}
\label{sec_sum}
\subsection{General remarks}
\label{sec_sum_gen}

  Finally, we would like to summarize the main points of this review,
discuss some of the open problems and provide an outlook. In general,
semi-classical methods in quantum mechanics and field theory are well
developed. We can reliably calculate the contribution of small-size
(large action) instantons to arbitrary Green's functions. Problems
arise when we leave this regime and attempt to calculate the 
contribution from large instantons or close instanton-anti-instanton 
pairs. While these problem can be solved rigorously in some theories 
(as in quantum mechanics or in some SUSY field theories), in QCD-like
theories we still face a number of unresolved problems and therefore
have to follow a somewhat more phenomenological approach. Nevertheless, 
the main point of this review is that important progress has been made 
in this context. The phenomenological success of the instanton liquid 
model is impressive, and initial attempts to explicitly check the 
underlying assumptions on the lattice are very encouraging.

   With this review, we not only want to acquaint the reader with 
the theory of instantons in QCD, we also want to draw attention 
to the large number of observables, in particular hadronic 
correlation functions at zero and finite temperature, that 
have already been calculated in the instanton liquid model.
While some of these predictions have been compared with
phenomenological information or lattice results, many 
others still await confrontation with experiment or the 
lattice. The instanton liquid calculations were made possible
by a number of technical advances. We now have a variety of 
approaches at our disposal, including the single instanton,
the mean field and random phase approximation, as well as 
numerical calculations which take the 't Hooft interaction
into account to all orders.

  The progress in understanding the physics of instantons made in 
lattice calculations has been of equal importance. We now have data 
concerning the total density, the typical size, the size distribution 
and correlations between instantons. Furthermore, there are detailed 
checks on the mechanism of $U(1)_A$ violation, and on the behavior of 
many more correlation functions under cooling. Recent investigations have 
begun to focus on many interesting questions, like the effects of
quenching, correlations of instantons with monopoles, etc.  

    In the following we will first summarize the main results concerning 
the structure of the QCD vacuum and its hadronic excitations, then discuss 
the effects of finite temperature and finally try to place QCD in a broader 
context, comparing the vacuum structure of QCD with other non-abelian field 
theories.

\subsection{Vacuum and hadronic structure}
\label{sec_sum_had}

   The instanton liquid model is based on the assumption that  
non-perturbative aspects of the gluonic vacuum, like the gluon
condensate, the vacuum energy density or the topological 
susceptibility are dominated by small-size ($\rho\simeq 1/3$
fm) instantons. The density of tunneling events is $n\simeq
1\,{\rm fm}^{-4}$. These numbers imply that the gauge 
fields are very inhomogeneous, with strong fields ($G_{\mu\nu} 
\sim g^{-1}\rho^{-2}$) concentrated in small regions of  
space-time. In addition to that, the gluon fields are strongly
polarized, the field strength locally being either self-dual or
anti-self-dual. 

   Quark fields, on the other hand, cannot be localized inside 
instantons. Isolated instantons have unpaired chiral zero modes, 
so the instanton amplitude vanishes if quarks are massless. In
order to get a non-zero probability, quarks have to be exchanged
between instantons and anti-instantons. In the ground state, zero 
modes become completely delocalized and chiral symmetry is broken.
As a consequence, quark-anti-quark pairs with the quantum numbers
of the pion can travel infinitely far, and we have a Goldstone 
pion. 
 
   This difference in the distribution of vacuum fields leads to
significant differences in gluonic and fermionic correlation
functions. Gluonic correlators are much more short range, and
as a result the mass scale for glueballs $m_{0^{++}}\simeq
1.5$ GeV is significantly larger than the typical mass of 
non-Goldstone mesons, $m_{\rho}=0.77$ GeV. The polarized gluon
fields lead to large spin splittings for both glueballs and
ordinary mesons. In general, we can group all hadronic
correlation functions in three classes: Those that receive
direct instanton contributions that are attractive $\pi, K, 0^{++}$ 
glueball, $N,\ldots$, those with direct instantons effects that are
repulsive $\eta',\delta,0^{+-}$ glueball, $\ldots$, and correlation
functions with no direct instanton contributions $\rho, a_1, 2^{++}$
glueball, $\Delta,\ldots$. As we repeatedly emphasized throughout
this review, already this simple classification based on first order 
instanton effects gives a non-trivial understanding of the bulk 
features of hadronic correlation functions. 

   In addition to that, the instanton liquid allows us to go into 
much more detail. Explicit calculations of the full correlation 
functions in a large number of hadronic channels have been 
performed. These calculations only require two parameters
to be fixed. One is the scale parameter $\Lambda$ and the
other one characterizes the scale at which the effective 
repulsion between close pairs sets in. With these parameters
fixed from global properties of the vacuum, we not only 
find a very satisfactory description of the masses and
couplings of ground state hadrons, but we also reproduce 
the full correlation function whenever they are available. 
Of course, the instanton model reaches its limits as
soon as perturbative or confinement effects become 
dominant. This is the case, for example, when one attempts
to study bound states of heavy quarks or tries to resolve 
high lying radial excitations of light hadrons. 

   How does this picture compare with other approaches to hadronic
structure? As far as the methodology is concerned, the instanton 
approach is close to (and to some extent a natural outgrowth of) 
the QCD sum rule method. Moreover, the instanton effects explain why 
the OPE works in some channels and fails in others (those with direct 
instanton contributions). The instanton liquid provides a complete
picture of the ground state, so that no assumptions about higher
order condensates are required. Also, it allows the calculations 
to be extended to large distances, so that no matching is needed. 

   In the quark sector, the instanton model provides a picture
which is similar to the Nambu and Jona-Lasinio model. There is an 
attractive quark-quark interaction that causes quarks to condense 
and binds them into light mesons and baryons. However, the 
instanton liquid provides a more microscopic mechanism, with a 
more direct connection to QCD, and relates the different coupling
constants and cutoffs in the NJL model. 

   Instead of going into comparisons with the plethora of hadronic 
models that have been proposed over the years, let us emphasize
two points that we feel are important. Hadrons are not cavities which
are empty inside (devoid of non-perturbative fields) as the bag 
model suggests. Indeed, matrix elements of electric and magnetic fields 
inside the nucleon can be determined from the trace anomaly 
\cite{Shu_78,Ji_95}, showing that the density of instantons 
and the magnitude of the gluon condensate inside the nucleon 
is only reduced by a few percent. Hadrons are excitations of a 
very dense medium, and this medium should be understood first. 
Furthermore, spin splittings in glueballs ($2^{++}-0^{++},\ldots$)
and light hadrons ($\rho-\pi,\ldots$) are not small, so it makes no 
sense to treat them perturbatively. This was directly checked on 
the lattice: spin splittings are not removed by cooling, which 
quickly eliminates all perturbative contributions.

    What is the perspective for future work on hadronic structure? 
Of course, our understanding of hadronic structure is still very
far from being complete. Clearly, the most important question 
concerns the mechanism of confinement and its role in hadronic
structure. In the mean time, experiments continue to provide  
interesting new puzzles: the spin of the nucleon, the magnitude  
and polarization of the strange sea, the isospin asymmetry in  
the light $u,d$ quark sea, etc.

\subsection{Finite temperature and chiral restoration}
\label{sec_sum_temp}

  Understanding the behavior of hadrons and hadronic matter
at high temperature is the ultimate goal of the experimental
heavy ion program. These studies complement our knowledge of 
hadronic structure at zero temperature and density and provide 
an opportunity to directly observe rearrangements in the structure 
of the QCD vacuum. 

  Generalizing the instanton liquid model to finite temperature 
is straightforward in principle. Nevertheless, the role of instantons
at finite temperature has been reevaluated during the past few years. 
There is evidence that instantons are not suppressed near $T_c$, but 
disappear only at significantly higher temperatures. Only after
instantons disappear does the system become a perturbative 
plasma\footnote{We should mention that even at asymptotically high 
temperature there are non-perturbative effects in QCD, related to the 
physics of magnetic (three dimensional) QCD. However, these effects 
are associated with the scale $g^2 T$, which is small compared to
the typical momenta of the order $T$. This means that the corrections
to quantities like the equation of state are small.}. 

  In addition to that, we have argued that the chiral transition 
is due the dynamics of the instanton liquid itself. The phase 
transition is driven by a rearrangement of the instanton liquid, 
going from a (predominantly) random phase at small temperature 
to a correlated phase of instanton-anti-instanton molecules at 
high temperature. Without having to introduce any additional 
parameters, this picture provides the correct temperature scale 
for the transition and agrees with standard predictions concerning 
the structure of the phase diagram. 

  If instantons are bound into topologically neutral pairs at $T>T_c$, 
they no longer generate a quark condensate. However, they still contribute
to the gluon condensate, the effective interaction between quarks 
and the equation of state. Therefore, instanton effects are potentially
very important in understanding the plasma at moderate temperatures
$T=(1-3) T_c$. We have begun to explore some of these consequences
in more detail, in particular the behavior of spatial and temporal
correlation functions across the transition region. While spacelike
screening masses essentially agree with the results of lattice 
calculations, interesting phenomena are seen in temporal correlation
functions. We find evidence that certain hadronic modes survive in 
the high temperature phase. Clearly, much work remains do be done 
in order to improve our understanding of the high temperature phase.

\subsection{The big picture}
\label{sec_big_pic}

\begin{figure}[t]
\begin{center}
\leavevmode
\epsfxsize=12cm
\epsffile{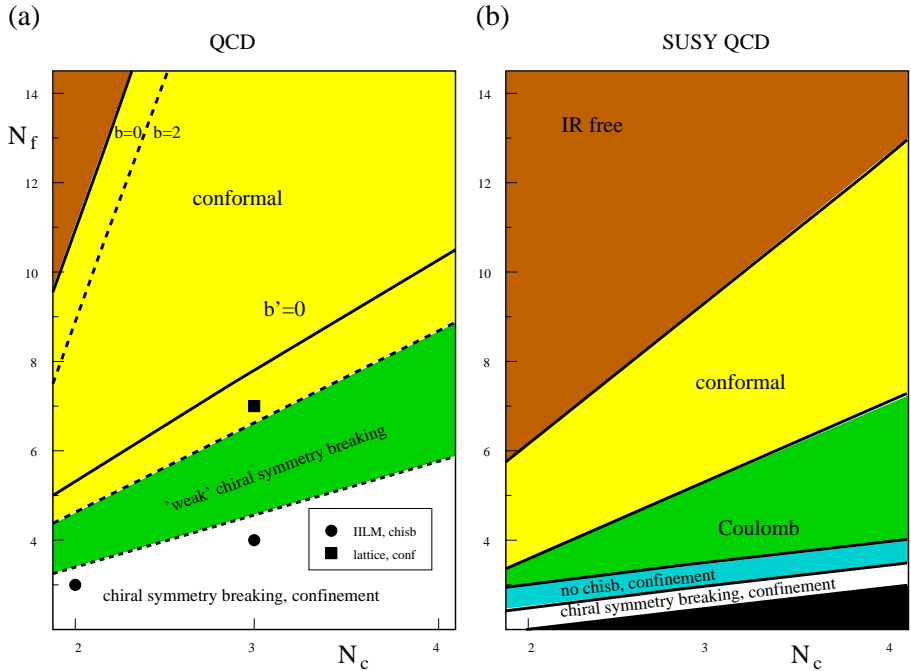}
\end{center}
\caption{\label{fig_map} 
Schematic phase diagram of QCD (a) and supersymmetric QCD (b)
as a function of the number of colors $N_c$ and the number 
of flavors $N_f$.}
\end{figure}

   Finally, we would like to place QCD in a broader perspective 
and discuss what is known about the phase structure of non-abelian 
gauge theories (both ordinary and supersymmetric) based on the gauge 
group $SU(N_c)$ with $N_f$ quark flavors. For simplicity, we will 
restrict ourselves to zero temperature and massless fermions. This 
means that the theory has no dimensional parameters other than $\Lambda$. 
The phase diagram of ordinary and SUSY QCD in the $N_c-N_f$ plane 
is shown in Fig. \ref{fig_map}. For simplicity, we have plotted $N_c$
and $N_f$ as if they were continuous variables\footnote{In a sense,
at least the number of flavors is a continuous variable: One can 
gradually remove a massless fermion by increasing its mass.}. We 
should emphasize that while the location of the phase boundaries can 
be rigorously established in the case of SUSY QCD, the phase diagram 
of ordinary QCD is just a guess, guided by some of the results mentioned 
below. 

  Naturally, we are mostly interested in the role of instantons in 
these theories. As $N_f$ is increased above the value 2 or 3 (relevant
to real QCD) the two basic components of the instanton ensemble, random 
(individual) instantons and strongly correlated instanton-anti-instanton 
pairs (molecules) are affected in very different ways. Isolated instantons
can only exist if the quark condensate is non-zero, and the instanton
density contains the factor $(|\langle\bar qq\rangle|\rho^3)^{N_f}$
which comes from the fermion determinant. As a result, small-size 
instantons are strongly suppressed as $N_f$ is increased. This 
suppression factor does not affect instantons with a size larger 
than $\rho \sim |\langle\bar qq\rangle|^{-1/3}$. This means that 
as $N_f$ is increased,  random instantons are pushed to larger 
sizes. Since in this regime the semi-classical approximation 
becomes unreliable, we do not know how to calculate the rate
of random instantons at large $N_f$.
 
  For strongly correlated pairs (molecules) the trend is exactly 
opposite. The density of pairs is essentially independent of the 
quark condensate and only determined by the interaction of the
two instantons. From purely dimensional considerations one expects
the density of molecules to be $dn_m\sim d\rho \Lambda^{2b}\rho^{2b-5}$,
which means that the typical size becomes smaller as $N_f$ is 
increased \cite{Shu_87b}. If $N_f>11N_c/2$, we have $b<2$ and 
the density of pairs is ultraviolet divergent (see the dashed 
line in Fig. \ref{fig_map}(a)). This phenomenon is similar to 
the UV divergence in the $O(3)$ non-linear $\sigma$ model. Both 
are examples of UV divergencies of non-perturbative nature. Most 
likely, they do not have significant effects on the physics of 
the theory. Since the typical instanton size is small, one would 
expect that the contribution of molecules at large $N_f$ can be 
reliably calculated. However, since the binding inside the pair 
increases with $N_f$, the separation of perturbative and non-perturbative 
fluctuations becomes more and more difficult. 

   Rather than speculate about these effects, let us go back 
and consider very large $N_f$. The solid line labeled $b=0$ in Fig. 
\ref{fig_map} corresponds to a vanishing first coefficient of the 
beta function, $b=(11/3)N_c-(2/3)N_f=0$ in QCD and $b=3N_c-N_f=0$ 
in SUSY QCD. Above this line, the coupling constant decreases at
large distances, and the theory is IR-free. Below this line, the 
theory is expected to have an infrared fixed point \cite{BM_74,BZ_82}. 
As discussed in Sec. \ref{sec_beta}, this is due to the fact 
that the sign of the the second coefficient of the beta function
$b'=34N_c^2/3-13N_cN_f/3+N_f/N_c$ is negative while the first 
one is positive. As a result, the beta function has a zero at 
$g^2_*/(16\pi^2) = -b/b'$. This number determines the limiting 
value of the charge at large distance. Note that the fixed point
is not destroyed by higher order perturbative effects, since we
can always choose a scheme where higher order coefficients vanish. 
The presence of an IR fixed point implies that the theory is conformal, 
that means correlation functions show a power law decay at large distance. 
There is no mass gap and the long distance behavior is characterized by 
the set of critical exponents.

 Where is the lower boundary of the conformal domain? A recent perturbative 
study based on the $1/N_f$ expansion \cite{Gra_96} suggests that the IR 
fixed point may persist all the way down to $N_f=6$ (for $N_c=3$). However, 
the critical coupling becomes larger and non-perturbative phenomena may
become important. For example, \cite{ATW_96} argue that if the coupling 
constant reaches a critical value the quark-anti-quark interaction is
sufficiently strong to break chiral symmetry. In their calculation,
this happens for $N_f^{c}\simeq 4 N_c$ (see the dashed line in Fig.
\ref{fig_map}(a)).
  
  It was then realized that instanton effects can also be important 
\cite{AS_97}. If the critical coupling is small, even large instantons 
have a large action $S=8\pi^2/g^2_*\gg 1$ and the semi-classical 
approximation is valid. As usual, we expect random 
instantons to contribute to chiral symmetry breaking. According to
estimates made in \cite{AS_97}, the role of instantons is comparable
to perturbative effects in the vicinity of $N_f=4 N_c$. Chiral symmetry 
breaking is dominated by large instantons with size $\rho \sim 
|\langle \bar qq\rangle|^{-1/3}>\Lambda^{-1}$, while the perturbative 
regime $\rho<\Lambda^{-1}$ contributes very little. For even larger 
instantons $\rho\gg |\langle\bar qq\rangle|^{-1/3}$ fermions acquire 
a mass due to chiral symmetry breaking and effectively decouple from 
gluons. This means that for large distances the charge evolves
as in pure gauge theory, and the IR fixed point is only an 
approximate feature, useful for analyzing the theory above
the decoupling scale. 

  Little is known about the phase structure of multi-flavor QCD  
from lattice simulations. Lattice QCD with up to 240 flavors was
studied in \cite{IKK_96} and it was shown that, as expected, the 
theory is trivial for $b>0$. The paper also confirms the existence 
of an infrared fixed point for $N_f\geq 7$ ($N_c=3$). In Fig. 
\ref{fig_map}(a) we have marked these results by open squares. 
Other groups have studied QCD with $N_f$= 8 \cite{Brown:1992}, 
12 \cite{Kogut:1988} and 16 \cite{Damgaard:1997} flavors. All 
of these simulations find a chirally asymmetric and confining 
theory at strong coupling, and a bulk transition to a chirally 
symmetric phase (at $\beta=2N_c/g^2>4.73$, 4.47, 4.12 respectively). 

  It may appear that these results are in contradiction with the
results of Appelquist et al. mentioned above, according to 
which chiral symmetry should be broken for $N_f<12$ ($N_c=3$),
but this is not the case since the condensate is expected 
to be exponentially small. In other words, in order to reproduce 
the subtle mechanism of chiral breaking by large-distance Coulomb 
exchanges or large-size instantons, the lattice has to include the 
relevant scales\footnote{This cannot be done by simply tuning
the bare coupling to the critical value for chiral symmetry
breaking $\beta \simeq 1$, because lattice artifacts create 
a chirally asymmetric and confining phase already at $\beta
\simeq 4-5$. Therefore, one has to start at weak coupling, 
and then go to sufficiently large physical volumes in order
to reach the chiral symmetry breaking scale.}, which is not  
the case in present lattice simulations. 
 
  Present lattice results resemble more closely the results
in the interacting instanton liquid discussed in section 
\ref{sec_liquid_T}. There we found a line of (rather robust  
first order) transitions which touches $T=0$ near $N_f\simeq 5$. 
A large drop in the quark condensate in going from $N_f=2,3$ to 
$N_f=4$ observed by the Columbia group \cite{CM_97} 
may very well be the first indication of this phenomenon.
 
  Again, there is no inconsistency between the interacting instanton
calculation and the results of Appelquist et al. The IILM 
calculation takes into account the effects of small instantons
only. If small instantons do not break chiral symmetry, then 
long-range Coulomb forces or large instantons can still be 
responsible for chiral symmetry breaking\footnote{The situation
is different at large temperature, because in that case both 
Coulomb forces and large instantons are Debye screened.}.
This mechanism was studied by a number of authors, e.g.  
\cite{aoki,barducci}, and the corresponding quark condensate is 
about an order of magnitude smaller\footnote{This can be seen 
from the fact that these authors need an unrealistically large 
value of $\Lambda_{QCD}\simeq 500 MeV$ to reproduce the 
experimental value of $f_\pi$.} than the one observed for
$N_f=2$. We would therefore argue that, in a practical sense,  
QCD has two different phases with chiral symmetry breaking.
One, where the quark condensate is large and generated by 
small size instantons and one where the condensate is 
significantly smaller and due to Coulomb forces or
large instantons. The transition regime is indicated by 
a wavy line in Fig. \ref{fig_map}a. Further studies of the 
mechanisms of chiral symmetry breaking for different $N_f$ 
are needed before final conclusions can be drawn. 

   Finally, although experiment tells us that confinement and 
chiral symmetry breaking go together for $N_f=2,3$, the two
are independent phenomena and it is conceivable that there
are regions in the phase diagram where only one of them takes 
place. It is commonly believed that confinement implies 
chiral symmetry breaking, but not even that is entirely
clear. In fact, SUSY QCD with $N_f=N_c+1$ provides a 
counterexample\footnote{It is usually argued that anomaly
matching shows that confinement implies chiral symmetry 
breaking for $N_c>2$, but again SUSY QCD provides examples
where anomaly matching conditions work out in subtle and
unexpected ways.}.

   For comparison we also show the phase diagram of ($N=1$) 
supersymmetric QCD (Fig. \ref{fig_map}b). As discussed in 
section \ref{sec_susy_qcd}, the phase structure of these theories 
was recently clarified by Seiberg and collaborators. In this case, 
the lower boundary of the conformal domain is at $N_f=(3/2)N_c$.
Below this line the dual theory based on the gauge group $SU(N_f-N_c)$
loses asymptotic freedom. In this case, the excitations are IR free 
``dual quarks'', or composite light baryons in terms of the original
theory. Remarkably, the t' Hooft matching conditions between the 
original (short distance) theory and the dual theory based on a 
completely different gauge group work out exactly. The theory 
becomes confining for $N_f=N_c+1$ and $N_f=N_c$. In the first 
case chiral symmetry is preserved and the low energy excitations
are massless baryons. In the second case, instantons modify the 
geometry of the space of possible vacua, the point where the 
excitations are massless baryons is not allowed, and chiral 
symmetry has to be broken. Note that in ordinary QCD, the 
't Hooft matching conditions cannot be satisfied for a confining
phase without chiral symmetry breaking (for $N_c\neq 2$). For
an even smaller number of flavors, $0< N_f\leq N_c-1$, massless
SUSY QCD does not have a stable ground state. The reason is
that instanton-anti-instanton molecules generate a positive
vacuum energy density which decreases with the Higgs expectation 
value, so that the ground state is pushed to infinitely large
Higgs VEVs.

  We do not shown the phase structure of SUSY QCD with 
$N>1$ gluinos. In some cases (e.g. $N=4,\,N_f=0$ or $N=2,\,N_f=4$) 
the beta function vanishes and the theory is conformal, although 
instantons may still cause a finite charge renormalization. As 
already mentioned, the low energy spectrum of the $N=2$ theory 
was recently determined by Seiberg and Witten. The theory does 
not have chiral symmetry breaking or confinement, but it contains  
monopoles/dyons which become massless as the Higgs VEV is decreased. 
This can be used to trigger confinement when the theory is perturbed 
to $N=1$.

  To summarize, there are many open questions concerning the phase
structure of QCD-like theories, and many issues to be explored in 
future studies, especially on the lattice. The location of the lower 
boundary of the conformal domain, and the structure of the chirally 
asymmetric phase in the domain $N_f=4-12$ should certainly be 
studied in more detail. Fascinating results have clarified the 
rich (and sometimes rather exotic) phase structure of SUSY QCD.
To what extent these results will help our understanding of 
non-supersymmetric theories remains to be seen. In any event,
it is certainly clear  that instantons and anomalies play a very 
important, if not dominant, role in both cases.

\section{Acknowledgments}

 It is a pleasure to thank our friends and collaborators who have 
contributed to this review or the work presented here. In particular, 
we would like to thank our collaborators J. Verbaarschot, M. Velkovsky
and A. Blotz. We would like to tank M. Shifman for many valuable 
comments on the original manuscript. We have also benefitted from 
discussions with G. E. Brown, D. I. Diakonov, J. Negele, M. Novak, 
A. V. Smilga, A. Vainshtein and I. Zahed.

\section{Notes added in proof}
               
  Since this manuscript was prepared, the subject of instantons
in QCD has continued to see many interesting developments. We would
like to briefly mention some of these, related to instanton searches 
on the lattice, the relation of instantons with confinement, instantons 
and charm quarks, instantons at finite chemical potential, and instantons 
in supersymmetric theories. 

  Significant progress was made studying topology on the lattice
using improved operators, renormalization group techniques, and
fermionic methods. Also, first results of studying instantons in 
the vicinity of the phase transition in full QCD were reported. 
As an example for the use of improved actions we mention results 
of Colorado group \cite{DHK_97b} in pure gauge $SU(2)$. They find
that chiral symmetry breaking and confinement are preserved by 
inverse blocking transformation used to smooth the fields. 
Instantons are easily identified, and inverse blocking (unlike 
``cooling") preserves even close instanton-antiinstanton pairs. 
The instanton size distribution is peaked at $\rho\simeq 0.2$ fm, 
and large instantons are suppressed.

  Low-lying eigenstates of the Dirac operator were studied by the 
MIT group \cite{IN_97}. They find that the corresponding wave functions 
are spatially correlated with the locations of instantons, providing 
support for the picture of the quark condensate as a collective state 
built from instanton zero modes. In addition to that, they studied the 
importance of low-lying states in hadronic correlation functions. 
They demonstrate that the lowest $\sim 100$ modes (out of $\sim 10^6$) 
are sufficient to quantitatively reproduce the hadronic ground state 
contribution to the $\rho$ and $\pi$ meson correlation functions.

  In addition to that, first attempts were made to study instantons at 
in the vicinity of the finite temperature phase transition in full QCD
\cite{FGH_97}. Using the cooling technique to identify instantons, they
verified the $T$-dependence of the instanton density discussed in Sect. 
VII.A. They observe polarized instanton-antiinstanton pairs above $T_c$, but 
these objects do not seem to dominate the ensemble. This point definitely
deserves further study, using improved methods and smaller quark masses. 

  In the main text we stated that there is no confinement in the 
instanton model. Recently, it was claimed that instantons generate
a linear potential with a slope close to the experimental value
1 GeV/fm \cite{Fuk_97}. This prompted \cite{CNS_98} to reinvestigate
the issue, and perform high statistics numerical calculations of the 
heavy-quark potential in the instanton liquid at distances up to 
3 fm. The main conclusions are: (i) the potential is larger and
significantly longer range than the dilute gas result Equ. (208);
(ii) a random ensemble with a realistic size distribution leads
to a potential which is linear even for large $R\simeq 3$ fm;
(iii) the slope of the potential is still too small, $K\simeq 200$
MeV/fm. This means that the bulk of the confining forces still
has to come from some (as of yet?) unidentified objects with small
action. These objects may turn out to be large instantons, but that 
would still imply that the main contribution is not semi-classical. 
Nevertheless, the result that the heavy quark potential is larger
than expected is good news for the instanton model. It implies that
even weakly bound states and resonances can be addressed within in 
the model. 

  We also did not discuss the role of charm quarks in the
QCD vacuum. However, the color field inside a small-size instantons
$G_{\mu\nu}\simeq 1\,{\rm GeV}^2$ is comparable to the charm quark 
mass squared $m_c^2\simeq 2\,{\rm GeV}^2$, so one might expect 
observable effects due to the polarization of charm quark inside
ordinary hadrons. Recently, it was suggested that CLEO observations
of an unexpectedly large branching ratio $B\rightarrow \eta' K$ (as 
well as inclusive $B\rightarrow \eta'+\ldots$) provide a smoking
gun for such effects \cite{HZ_98}. The basic idea is that 
these decays proceed via Cabbibo-unsuppressed $b\to \bar ccs$ 
transition, followed by $\bar cc\to\eta'$. The charm content of
the $\eta'$ in the instanton liquid was estimated in \cite{SZ_97},
and the result is consistent with what is needed to understand the 
CLEO data. Another interesting possibility raised in \cite{HZ_97} is 
that polarized charm quarks give a substantial contribution to the spin 
$\Sigma$ of the nucleon. A recent instanton calculation only 
finds a contribution in the range $\Delta c/\Sigma=-(0.08-0.20)$ 
\cite{BS_97}, but the value of $\Delta c$ remains an interesting 
question for future deep inelastic scattering experiments (e.g. 
Compass at CERN). In the context of (polarized) nucleon structure
functions we should also mention interesting work on leading and 
non-leading twist operators, see \cite{BPW_97} and references 
therein. 

  Finally, Bjorken discussed the possibility that instantons 
contribute to the decay of mesons containing real $\bar cc$ pairs
\cite{bj_97}. A particularly interesting case is the $\eta_c$, which
has three unusual 3-meson decay channels ($\eta'\pi\pi$, $\eta\pi\pi$, 
and $KK\pi$), which contribute roughly 5\% each to the total width.
This fits well with the typical instanton vertex $\bar uu \bar dd 
\bar ss$. In general, all of these observables offer the chance to
detect non-perturbative effects deep inside the semi-classical
domain. 

  Initial efforts were made to understand the instanton liquid
at finite chemical potential \cite{Sch_97}. The suggestion made
in this work is that the role that molecules play in the high
temperature phase is now played by more complicated ``polymers"
that are aligned in the time direction. More importantly, it
was suggested that instanton lead to the formation of diquark
condensates in high density matter \cite{RSS_97,ARW_97}. In 
the high density phase chiral symmetry is restored, but $SU(3)$
color is broken by a Higgs mechanism. 

  Instanton effects in SUSY gauge theories continue to be a very active 
field. For $N=2$ (Seiberg-Witten) theories the $n$-instanton contribution
was calculated explicitly \cite{DKM_97,DHK_97}, and as a by-product these
authors also determined the (classical) $n$-instanton measure in the 
cases $N=1$ and $N=0$ (non-SUSY). Yung also determined the one-instanton
contribution to higher derivative operators beyond the SW effective 
lagrangian \cite{Yun_97}. Another very interesting result is the 
generalization of the Seiberg-Witten solution to an arbitrary number
of colors $N_c$ \cite{DS_95}. This result sheds some light on the 
puzzling problem of instantons in the large $N_c$ limit. The large 
$N_c$ limit is usually performed with $g^2N_c$ held fixed, and in 
that case instantons amplitudes are suppressed by $\exp(-N_c)$.
However, \cite{DS_95} found that (at least in the case $N=2$) this 
is not the correct way to take the large $N_c$ limit (if we want 
to keep the physics unchanged).

 A comparison of the running of the effective charge in $N=2,1$ SUSY
QCD and QCD was performed in \cite{RRS_98}. In $N=2$ SUSY QCD, the
Seiberg-Witten solution shows that instantons accelerate the growth
of the coupling. As a result, the coupling blows up at a scale 
$\Lambda_{NP}>\Lambda_{QCD}$ where the perturbative coupling is 
still small. A similar phenomenon takes place in QCD if the 
instanton correction to the running coupling is estimated from
the formula of Callan, Dashen, and Gross. This might help to explain 
why in QCD the non-perturbative scale $\Lambda_{NP}\simeq\Lambda_{\chi SB}
\simeq 1\, GeV>\Lambda_{QCD}\simeq 200$ MeV.

\newpage
\appendix

\section{Basic instanton formulas}
\label{app_basics}
\subsection{Instanton gauge potential}
\label{app_inst_pot}  

  We use the following conventions for Euclidean gauge fields:
The gauge potential is $A_\mu = A_\mu^a\frac{\lambda^a}{2}$, 
where the $SU(N)$ generators satisfy $[\lambda^a,\lambda^b]=
2if^{abc}\lambda^c$ and are normalized according to ${\rm tr}
[\lambda^a\lambda^b] = 2\delta^{ab}$. The covariant derivative is 
given by $D_\mu = \partial_\mu -i A_\mu$ and the field strength tensor 
is
\be
 F_{\mu\nu} = \partial_\mu A_\nu - \partial_\nu A_\mu
         -i [A_\mu,A_\nu] .
\ee
In our conventions, the coupling constant is absorbed into the 
gauge fields. Standard perturbative notation corresponds to
the replacement $A_\mu\to gA_\mu$. The single instanton solution 
in regular gauge is given by
\be 
 A_\mu^a = \frac{2 \eta_{a\mu\nu} x_\nu}{x^2+\rho^2},
\ee
and the corresponding field strength is
\be
 G_{\mu\nu}^a  &=& -\frac{4 \eta_{a\mu\nu}\rho^2}
                          {(x^2+\rho^2)^2}, \\
 (G_{\mu\nu}^a)^2 &=& \frac{192\rho^4}{(x^2+\rho^2)^4}.
\ee
The gauge potential and field strength in singular gauge are
\be 
\label{A_sing}
 A_\mu^a &=&  \frac{2 \bar\eta_{a\mu\nu}x_\nu\rho^2}
      {x^2(x^2+\rho^2)},\\
\label{G_sing}
 G_{\mu\nu}^a &=& -\frac{4\rho^2}{(x^2+\rho^2)^2} 
  \left( \bar\eta_{a\mu\nu} -2\bar\eta_{a\mu\alpha}
    \frac{x_\alpha x_\nu}{x^2} -2\bar\eta_{a\alpha\nu}
    \frac{x_\mu x_\alpha}{x^2} \right)\; .
\ee
Finally, an $n$-instanton solution in singular gauge is given by
\be
\label{A_n_inst}
 A_\mu^a &=&  \bar\eta_{a\mu\nu} \partial_\nu\ln\Pi(x) ,\\
 \Pi(x)  &=& 1+\sum_{i=1}^n \frac{\rho_i^2}{(x-z_i)^2} .
\ee
Note that all instantons have the same color orientation. For
a construction that gives the most general $n$-instanton
solution, see \cite{AHDM_77}.

\subsection{Fermion zero modes and overlap integrals}
\label{app_zm}

  In singular gauge, the zero mode wave function $iD\!\!\!\!/\,\phi_0 =0$
is given by 
\be
 \phi_{a\nu} = \frac{1}{2\sqrt{2}\pi\rho} \sqrt{\Pi}
\left[ \partial\!\!\!/ \left(\frac\Phi\Pi \right)
\right]_{\nu\mu} U_{ab}\epsilon_{\mu b} ,
\ee
where $\Phi=\Pi-1$. For the single instanton solution, we get
\be
\phi_{a\nu} (x) = \frac{\rho}{\pi}\frac{1}{(x^2+\rho^2)^{3/2}}
 \left[\left( \frac{1-\gamma_5}{2}\right) \frac{x\!\!\! /}{\sqrt{x^2}}
 \right]_{\nu\mu} U_{ab}\epsilon_{\mu b}  .
\ee
The instanton-instanton zero mode density matrices are
\be
\phi_I(x)_{i\alpha}\phi_J^\dagger(y)_{j\beta} &=& 
  \frac{1}{8} \varphi_I(x)\varphi_J(y) 
  \left( x\!\!\! / \gamma_\mu\gamma_\nu y\!\!\! / 
  \frac{1-\gamma_5}{2} \right)_{ij}
  \otimes \left( U_I \tau_\mu^-\tau_\nu^+ U_J^\dagger\right)_{\alpha\beta}, \\
\phi_I(x)_{i\alpha}\phi^\dagger_A(y)_{j\beta} &=& 
 -\frac{i}{2} \varphi_I(x)\varphi_A(y) 
  \left( x\!\!\! / \gamma_\mu y\!\!\! / 
  \frac{1+\gamma_5}{2} \right)_{ij}
  \otimes \left( U_I \tau_\mu^- U_A^\dagger \right)_{\alpha\beta}, \\
\phi_A(x)_{i\alpha}\phi^\dagger_I(y)_{j\beta} &=& 
  \frac{i}{2} \varphi_A(x)\varphi_I(y) 
  \left( x\!\!\! / \gamma_\mu y\!\!\! / 
  \frac{1-\gamma_5}{2} \right)_{ij}
  \otimes \left( U_A \tau_\mu^+ U_I^\dagger \right)_{\alpha\beta} ,
\ee
with
\be
\varphi (x)= \frac{\rho}{\pi}\frac{1}{\sqrt{x^2}(x^2+\rho^2)^{3/2}}.
\ee
The overlap matrix element is given by
\be
 T_{AI} &=& \int d^4x\, \phi_A^\dagger (x-z_A)
 iD\!\!\!\! /\, \phi_I(x-z_I) \nonumber \\
        &=& r_\mu\,{\rm Tr}(U_I\tau_\mu^- U^\dagger_A)\,
            \frac{1}{2\pi^2 r}\frac{d}{dr} M(r) ,
\ee
with
\be
 M(r) = \frac{1}{r}\, \int\limits_0^\infty dp\, p^2|\varphi(p)|^2 J_1(pr) .
\ee 
The fourier transform of zero mode profile is given by
\be
 \varphi(p) = \pi\rho^2 \left.\frac{d}{dx}\left(I_0(x)K_0(x)-
 I_1(x)K_1(x)\right)\right|_{x=\frac{p\rho}{2}}.
\ee 

\subsection{ Properties of $\eta$ symbols}
\label{app_eta}

 We define 4-vector matrices
\be
    \tau_\mu^{\pm} = (\vec\tau,\mp i),
\ee
where $\tau^a\tau^b=\delta^{ab}+i\epsilon^{abc}\tau^c$ and
\be
 \tau_\mu^+\tau_\nu^- &=& \delta_{\mu\nu}+i\eta_{a\mu\nu}\tau^a ,\\
 \tau_\mu^-\tau_\nu^+ &=& \delta_{\mu\nu}+i\bar\eta_{a\mu\nu}\tau^a ,
\ee
with the $\eta$-symbols given by
\be
 \eta_{a\mu\nu} &=& \epsilon_{a\mu\nu} + \delta_{a\mu}\delta_{\nu 4}
                       - \delta_{a\nu}\delta_{\mu 4}, \\
 \bar\eta_{a\mu\nu} &=& \epsilon_{a\mu\nu} - \delta_{a\mu}\delta_{\nu 4}
                       + \delta_{a\nu}\delta_{\mu 4} .
\ee
The $\eta$-symbols are (anti) self-dual in the vector indices 
\be
 \eta_{a\mu\nu} = \frac 12 \epsilon_{\mu\nu\alpha\beta}
    \eta_{a\alpha\beta}, \hspace{1cm}
 \bar\eta_{a\mu\nu} = -\frac 12 \epsilon_{\mu\nu\alpha\beta}
    \bar\eta_{a\alpha\beta} \hspace{1cm} 
 \eta_{a\mu\nu} = -\eta_{a\nu\mu} . 
\ee
We have the following useful relations for contractions involving
$\eta$ symbols
\be
 \eta_{a\mu\nu}\eta_{b\mu\nu} &=& 4\delta_{ab},\\ 
 \eta_{a\mu\nu}\eta_{a\mu\rho} &=& 3\delta_{\nu\rho}, \\
 \eta_{a\mu\nu}\eta_{a\mu\nu} &=& 12, \\
 \eta_{a\mu\nu}\eta_{a\rho\lambda} &=&
       \delta_{\mu\rho}\delta_{\nu\lambda}
      -\delta_{\mu\lambda}\delta_{\nu\rho}
      +\epsilon_{\mu\nu\rho\lambda},\\
 \eta_{a\mu\nu}\eta_{b\mu\rho} &=& \delta_{ab}\delta_{\nu\rho} 
      + \epsilon_{abc}\eta_{c\nu\rho}, \\
 \eta_{a\mu\nu}\bar\eta_{b\mu\nu} &=& 0.
\ee
The same relations hold for $\bar\eta_{a\mu\nu}$, except for
\be
\bar \eta_{a\mu\nu}\bar\eta_{a\rho\lambda} =
       \delta_{\mu\rho}\delta_{\nu\lambda}
      -\delta_{\mu\lambda}\delta_{\nu\rho}
      -\epsilon_{\mu\nu\rho\lambda} .
\ee
Some additional relations are
\be
 \epsilon_{abc}\eta_{b\mu\nu}\eta_{c\rho\lambda} &=&
 \delta_{\mu\rho}\eta_{a\nu\lambda} -
 \delta_{\mu\lambda}\eta_{a\nu\rho} +
 \delta_{\nu\lambda}\eta_{a\mu\rho} -
 \delta_{\nu\rho}\eta_{a\mu\lambda} ,\\
 \epsilon_{\lambda\mu\nu\sigma}\eta_{a\rho\sigma} &=&
  \delta_{\rho\lambda}\eta_{a\mu\nu} +
  \delta_{\rho\nu}\eta_{a\lambda\mu} +
  \delta_{\rho\mu}\eta_{a\nu\lambda}.
\ee

\subsection{Group integration}
\label{app_group}

    In order to perform averages over the color group, we need
the following integrands over the invariant $SU(3)$ measure
\be
 \int dU\, U_{ij}U^\dagger_{kl} &=& \frac{1}{N_c}
 \delta_{jk}\delta_{li} ,\\
 \int dU\, U_{ij}U^\dagger_{kl}U_{mn}U^\dagger_{op} &=& \frac{1}{N_c^2}
 \delta_{jk}\delta_{li}\delta_{no}\delta_{mp}
 + \frac{1}{4(N_c^2-1)}(\lambda^a)_{jk}(\lambda^a)_{li}
 (\lambda^b)_{no}(\lambda^b)_{mp}. 
\ee
Additional results can be found in \cite{Cre_83}. These results 
can be rearranged using $SU(N)$ Fierz transformation
\be
 (\lambda^a)_{ij}(\lambda^a)_{kl} &=& -\frac{2}{N_c}
  \delta_{ij}\delta_{kl} + 2 \delta_{jk}\delta_{il} .
\ee

\newpage
\bibliographystyle{rmp} 
\bibliography{rev,new} 

\end{document}